\documentclass[aps,rmp,reprint,amsmath,amssymb,graphicx,longbibliography]{revtex4-1}

\usepackage{bm}

\input{Qcircuit}

\newcommand{\calG}{{\cal G}}
\newcommand{\calC}{{\cal C}}
\newcommand{\calP}{{\cal P}}
\newcommand{\calS}{{\cal S}}
\newcommand{\la}{\langle}
\newcommand{\ra}{\rangle}

\newcommand{\CC}{\mathbb{C}}
\newcommand{\calL}{{\cal L }}
\usepackage{url}
\usepackage{amssymb}
\usepackage{graphicx}

\newtheorem{theorem}{Theorem}

\begin{document}

\title{Quantum Error Correction for Quantum Memories}

\author{Barbara M. Terhal}
\affiliation{JARA Institute for Quantum Information, RWTH Aachen University, 52056 Aachen, Germany}

\date{\today{}}

\begin{abstract}
Active quantum error correction using qubit stabilizer codes has emerged as a promising, but experimentally challenging, engineering program for building a universal quantum computer. In this review we consider the formalism of qubit stabilizer and subsystem stabilizer codes and their possible use in protecting quantum information in a quantum memory. We review the theory of fault-tolerance and quantum error-correction, discuss examples of various codes and code constructions, the general quantum error correction conditions, the noise threshold, the special role played by Clifford gates and the route towards fault-tolerant universal quantum computation. The second part of the review is focused on providing an overview of quantum error correction using two-dimensional (topological) codes, in particular the surface code architecture. We discuss the complexity of decoding and the notion of passive or self-correcting quantum memories. The review does not focus on a particular technology but discusses topics that will be relevant for various quantum technologies. 
\end{abstract}

\pacs{03.67.Pp, 03.67.Lx, 03.67.Lx, 42.50.-p}

\maketitle

\tableofcontents

\section{Introduction}

Physics in the past century has demonstrated the experimental viability of macroscopic quantum {\em states} such as the superconducting state or a Bose-Einstein condensate. Quantum error correction which strives to preserve not a single macroscopic quantum state but the macroscopic states in a small subspace can be viewed as a natural but challenging extension to this. At the same time storing macroscopic quantum information is a first step towards the more ambitious goal of manipulating quantum information for computational purposes. 

When the idea of a quantum computer took hold in the '90s it was immediately realized that its implementation would require some form of robustness and error correction. Alexei Kitaev proposed a scheme in which the physical representation of quantum information and realization of logical gates would be naturally robust due to the topological nature of the 2D physical system \cite{kitaev:top}. Around the same time Peter Shor formulated a first quantum error-correcting code and proved that a quantum computer could be made fault-tolerant \cite{shor:faulttol}. Several authors then established the fault-tolerance threshold theorem (see Theorem~\ref{theo:ft}, Sec. \ref{sec:ec_ft}) which shows that in principle one can realize almost noise-free quantum computation using noisy components at the cost of a moderate overhead. \\

The goal of this review is to discuss the basic ideas behind active quantum error correction with stabilizer codes for the purpose of making a quantum memory. In this review we also discuss how Clifford group gates (Section \ref{sec:tt}) are realized on the stored quantum data. In this sense the review goes beyond a pure quantum memory perspective, but for stabilizer codes these Clifford group gates play an essential role. Clifford gates are by themselves not sufficient for realizing universal fault-tolerant quantum computation.

We distinguish schemes of active quantum error correction from forms of passive quantum error correction or self-correction. In the latter quantum information is encoded in physical degrees of freedom which are naturally protected or have little decoherence, either through topology (topological order) or physical symmetries (symmetry-protected order) at sufficiently low temperature. Even though our review focuses on active quantum error correction, we will discuss some aspects of passive protection of quantum information using quantum error correcting codes in Section \ref{sec:self}.\\

In an actively corrected quantum memory, quantum information is distributed among many elementary degrees of freedom, e.g. qubits, such that the dominant noise and decoherence processes affect this information in a reversible manner. This means that there exists an error reversal procedure that allows one to undo the decoherence. The choice of how to represent the quantum information in the state space of many elementary qubits is made through the choice of quantum error correcting code. In order to execute the error reversal, active quantum error correction proceeds by continuously gathering information about which errors took place (for example by quantum measurement), classical processing of this data and applying a corrective quantum operation on the quantum data. The active gathering of information takes place, at least for stabilizer codes, via quantum measurements which measure the parity of subsets of qubits in Pauli matrix bases. These measurements are called parity check measurements. By the active gathering of error information, entropy is effectively removed from the computation and dumped into ancilla degrees of freedom which are supplied in known states to collect the error information. This active cycling of entropy from the computation into ancillary degrees of freedom which are further processed in a classical world makes active quantum error correction very different from the notion of passively storing quantum information in a low-temperature thermal environment.\\

In Section \ref{sec:stabform} we start by discussing Shor's code as the most basic example of a quantum error correction code. Using Shor's code we illustrate the ideas behind the general framework of stabilizer codes \cite{thesis:gottesman}, including subsystem stabilizer codes. We then treat stabilizer and subsystem stabilizer codes on qubits more formally in Sections \ref{sec:ssc1} and \ref{sec:ssc2}. In section \ref{sec:CSS} we will also discuss various small examples of quantum error correcting codes and the construction due to Calderbank, Steane and Shor by which two classical codes can be used to construct one quantum code. In Section \ref{sec:phys_codes} we widen our perspective beyond stabilizer codes and discuss the general quantum error correction conditions as well as some codes which encode qubit(s) into bosonic mode(s) (oscillators). In Section \ref{sec:Ddim} we define $D$-dimensional stabilizer codes and give various examples of such codes. Roughly speaking, $D$-dimensional stabilizer codes are quantum error correcting codes where the elementary qubits are laid out on a $D$-dimensional lattice and all the quantum operations for quantum error correction can be executed by coupling qubits only locally on this lattice. \\

As the procedure of detecting and correcting errors itself is subject to noise, the existence of quantum error correcting codes does by itself not yet show that one can store or compute with quantum information for an arbitrary long time. In Section \ref{sec:ec_ft} we review how one can, through a procedure called code concatenation, arrive at the fault-tolerance threshold Theorem. In essence, the threshold theorem says that in order to combat noise and decoherence we can add redundancy, a poly-logarithmic overhead in the total number of qubits and overall computation time, provided that the fundamental noise rate on the elementary qubits is below some critical value which is called the noise threshold. Topological quantum error correction discussed in Section \ref{sec:topo} provides a different route for establishing such threshold theorem.\\

In Section \ref{sec:ec_ft} we also discuss various proposals for realizing quantum error correction, including the idea of dissipative engineering. The topic of the realization of quantum error correction is again picked up in Section \ref{sec:disc_pract}, but in the later section the emphasis is on $D$-dimensional (topological) codes. In the last section of our introductory Chapter, Section \ref{sec:tt}, we review constructions for obtaining a universal set of logical gates for qubits encoded with stabilizer codes and motivate our focus on 2D topological stabilizer codes for the use as a quantum memory. \\

For stationary, non-flying, qubits, an important family of codes are quantum codes in which the elementary qubits can be laid out on a two-dimensional plane such that only local interactions between small numbers of nearest-neighbor qubits in the plane are required for quantum error correction. The practical advantage of such 2D geometry over an arbitrary qubit interaction-structure is that no additional noisy operations need to be performed to move qubits around. Elementary solid-state qubits require various electric or magnetic control fields per qubit, both for defining the qubit subspace and/or for single- and two qubit control and measurement. The simultaneous requirement that qubits can interact sufficiently strongly and that space is available for these control lines imposes technological design constraints, see e.g. \cite{levy+:elec}. A two-dimensional layout can be viewed as a compromise between the constraints coming from coding theory and control-line and material fabrication constraints: since quantum error correcting codes defined on one-dimensional lines have poor error-correcting properties (Section \ref{sec:Ddim}), it is advantageous to use a two-dimensional or a more general non-local lay-out of qubits. The qubits in such a lay-out should be individually addressable and/or defined by local electrostatic or magnetic fields and thus 2D structures would be favored over 3D or general non-local interaction structures. \\

These considerations are the reason that we focus in Section \ref{sec:topo} on 2D (topological) codes, in particular the family of 2D topological surface codes which has many favorable properties. For the surface code we show explicitly in Section \ref{sec:surfcode} how many noisy elementary qubits can be used to represent one `encoded' qubit which has a much lower noise rate, assuming that the noise rate of the elementary qubits is below a critical value, the noise threshold. For the surface code this threshold turns out to be very high. We will review two possible ways of encoding qubits in the surface code. We will also discuss how logical gates such as the CNOT and the Hadamard gate (see Section \ref{sec:tt} for definitions of these gates) can be realized in a resource-efficient way.

In Section \ref{sec:alt} we review a few interesting alternatives to the surface code which are the non-topological Bacon-Shor code, a surface code with harmonic oscillators and a subsystem version of the surface code. Section \ref{sec:disc_pract} discusses the physical locality of the process of decoding as well as recent ideas on the realization of so-called direct parity measurements.  In Section \ref{sec:self} we discuss the ideas behind passive or self-correction and its relation with topological order. 

We conclude our review with a discussion on some future challenges for quantum error correction. We recommend the book \cite{book:lidar_brun} as a broad, comprehensive, reference on quantum error correction.

\subsection{Error Mitigation}
\label{sec:mit}
Active quantum error correction is not the only way to improve the coherence properties of elementary physical quantum systems and various well-known methods of error mitigation exist. In a wide variety of systems there is 1/f noise affecting the parameters of the qubit with a noise power spectral density $S(\omega) \sim 1/\omega^{\alpha}$, $\alpha \approx 1$, favoring slow fluctuations of those parameters \cite{weissman:rmp} which lead to qubit dephasing. Standard NMR techniques \cite{VC:rmp} have been adapted in such systems to average out these fluctuations using rapid pulse sequences (e.g. spin-echo). More generally, dynamical decoupling is a technique by which the undesired coupling of qubits to other quantum systems can be averaged out through rapid pulse sequences \cite{lidar:review}. Aside from actively canceling the effects of noise, one can also try to encode quantum information in so-called decoherence-free subspaces which are effectively decoupled from noise; a simple example is the singlet state $\frac{1}{\sqrt{2}} (\ket{\uparrow, \downarrow}-\ket{\downarrow,\uparrow})$ which is invariant under a joint (unknown) evolution $U\otimes U$. This example is not yet a code as it encodes only one state, not a qubit. One can more generally formulate decoherence-free subspaces in which the encoded qubits are protected against collectively-acting noise given by a set  of error operators, see Chapter 3 in \cite{book:lidar_brun}. 

In Section \ref{sec:self} we discuss another form of error mitigation which is to encode quantum information in a many-body quantum system with a Hamiltonian corresponding to that of a $D$-dimensional quantum (stabilizer) code.

\subsection{Some Experimental Advances}

Experimental efforts have not yet advanced into the domain of {\em scalable} quantum error correction. Scalable quantum error correction would mean (1) making encoded qubits with decoherence rates which are genuinely below that of the elementary qubits and (2) demonstrate how, by increasing coding overhead, one can reach even lower decoherence rates, scaling in accordance with the theory of quantum error correction.

Several experiments exist of the 3-qubit (or 5-qubit) repetition code in liquid NMR, ion-trap, optical and superconducting qubits. Four qubit stabilizer pumping has been realized in ion-trap qubits \cite{barreiro+:stab}. Some topological quantum error correction has been implemented with eight-photon cluster states in \cite{yao+:topo} and a continuous-variable version of Shor's 9-qubit code was implemented with optical beams \cite{aoki+:CVshor}.  In \cite{bell+:graph} the authors did an implementation of the [[4,1,2]] code in an all-optical set-up using a five-qubit polarization-based optical cluster state.

In \cite{nigg:steanecode} seven trapped-ion qubits were used to represent, using Steane's 7-qubit code (see Section \ref{sec:CSS}), one effective, encoded, qubit and several logical gates were performed on this encoded qubit via the transversal execution of gates on the seven elementary qubits. 

The book \cite{book:lidar_brun} has a chapter with an overview of experimental quantum error correction. Given the advances in coherence times and ideas of multi-qubit scalable design, in particular in ion-trap and superconducting qubits e.g. \cite{barends+:jos_surf}, one may hope to see scalable error correction, fine-tuned to experimental capabilities and constraints, in the years to come.

\section{Concepts of Quantum Error Correction}
\label{sec:conc}

\subsection{Shor's Code and Stabilizer Codes}
\label{sec:stabform}

The goal of this section is to introduce the concepts and terminology of stabilizer codes in an informal way illustrated by Shor's 9 qubit code. In the later Sections \ref{sec:ssc1}-\ref{sec:CSS} we discuss the formalism of stabilizer codes and give further examples.\\

The smallest classical code which can correct a single bit-flip error (represented by Pauli $X$\footnote{The Pauli matrices are $\sigma_x \equiv X=\left(\begin{array}{cc} 0 & 1 \\ 1 & 0 \end{array}\right)$,  $\sigma_z \equiv Z=\left(\begin{array}{cc} 1 & 0 \\ 0 & -1 \end{array}\right)$ and $\sigma_y \equiv Y=\left(\begin{array}{cc} 0 & -i \\ i & 0 \end{array}\right)=i XZ$.})  is the 3-(qu)bit repetition code where we encode $\ket{\overline{0}}=\ket{000}$ and $\ket{\overline{1}}=\ket{111}$. A single error can be corrected by taking the majority of the three bit values and flipping the bit which is different from the majority. In quantum error correction we don't want to measure the 3 qubits to take a majority vote, as we would immediately lose the quantum information. This quantum information is represented in the amplitude $\cos(\theta)$ and the phase $e^{i \phi}$ of an encoded qubit state $\ket{\overline{\psi}}=\cos(\theta)\ket{\overline{0}}+\sin(\theta)e^{i \phi} \ket{\overline{1}}$. If we measure the three qubits in the $\{\ket{0},\ket{1}\}$ basis, we may get answers which depend on $\cos(\theta)$ and $e^{i\phi}$, but we also decohere the quantum state, leaving just three classical bits and losing all information about $\cos(\theta)$ and $e^{i\phi}$. \\

But let us imagine that we can measure the parity checks $Z_1 Z_2$ and $Z_2 Z_3$ {\em without learning the state of each individual qubit}, that is, without the measurement revealing any information about the eigenvalues of $Z_1$ or $Z_2$ or $Z_3$ individually.  If the parity checks $Z_1 Z_2$ and $Z_2 Z_3$ have eigenvalues +1, one concludes no error as the encoded states $\ket{000}$ and $\ket{111}$ have eigenvalue $+1$ with respect to these checks. An outcome of, say, $Z_1 Z_2=-1$ and $Z_2 Z_3=1$ is consistent with the erred state $X_1 \ket{\overline{\psi}}$ where $\ket{\overline{\psi}}$ is any encoded state. And $Z_1 Z_2=1$ and $Z_2 Z_3=-1$ points to the error $X_3$. But how can we measure $Z_1 Z_2$ and $Z_2 Z_3$ without measuring the individual $Z_i$ operators and destroy the encoded qubit? Essentially, through making sure that the `signals' from different qubits are indistinguishable and only a global property like parity is communicated to the outside world. One can realize this with a quantum circuit as follows. \\

One uses an extra `ancilla' qubit which will interact with the 3 qubits such that the value of the parity check is copied onto the ancilla qubit. A general circuit which measures a 'parity check', represented by a multi-qubit Pauli operator $P$, using an ancilla is given in Fig.~\ref{fig:paritycheck}(a). One can verify the action of the circuit by writing an arbitrary input state as a superposition of a $+1$ eigenstate $\psi_{+1}$ and a  $-1$ eigenstate $\psi_{-1}$ of the Pauli operator $P$ to be measured ($P \ket{\psi_{\pm}}=\pm \ket{\psi_{\pm}}$). A concrete example for $P=X_1 X_2 X_3 X_4$ is given in Fig.~\ref{fig:paritycheck}(c), where we have decomposed the 5-qubit controlled-P gate into 4 2-qubit controlled-X or CNOT gates. In such a circuit, the parity information is collected in steps, via several CNOT gates, so that the state of the ancilla qubit during the execution of the gates {\em does} contain information about the individual qubits. It is thus important that this partial information on the ancilla qubit does not leak to the environment as it leads to decoherence on the encoded qubits during the parity check measurement. One can see that the parity check measurement using an ancilla qubit initially set to a fixed, known, state is actively letting us remove entropy from the computation by providing us information about what errors have taken place. \\

It may be clear that the 3-qubit repetition code does not protect or detect $Z$ (dephasing) errors as these parity checks only measure information in the $Z$-basis ($M_Z$). More precisely, any single qubit $Z$ error will harm the quantum information. We encode the qubit state $\ket{+} \equiv \frac{1}{\sqrt{2}}(\ket{0}+\ket{1})$ as $\ket{\overline{+}}=\frac{1}{\sqrt{2}}(\ket{000}+\ket{111})$ and $\ket{-} \equiv \frac{1}{\sqrt{2}}(\ket{0}-\ket{1})$ as $\ket{\overline{-}}=\frac{1}{\sqrt{2}}(\ket{000}-\ket{111})$. We can verify that any single qubit $Z$ error, say $Z_1$, maps $\ket{\overline{+}} \leftrightarrow \ket{\overline{-}}$, corrupting the quantum information. \\

Having seen this simple example, let us informally introduce some of the notions used in describing a quantum (stabilizer) code. In general we will denote logical or encoded states as $\ket{\overline{\psi}}$ and logical operators as $\overline{X}, \overline{Z}$ etc. where by definition $\overline{X} \ket{\overline{0}} \leftrightarrow \ket{\overline{1}}$ and $\overline{Z} \ket{\overline{+}} \leftrightarrow \ket{\overline{-}}$. The logical operators $\overline{X},\overline{Z}$ can always be expressed in terms of their action as Pauli operators on the elementary qubits. For a code $C$ encoding $k$ qubits, one defines $k$ pairs of logical Pauli operators $(\overline{X}_i,\overline{Z}_i)$, $i=1,\ldots k$, such that $\overline{X}_i \overline{Z}_i=-\overline{Z}_i \overline{X}_i$ while logical Pauli operators with labels $i$ and $i'$ mutually commute. The logical Pauli operators simply realize the algebra of the Pauli operators acting on $k$ qubits. For the 3-qubit code we have $\overline{X}=X_1 X_2 X_3$ (flipping all the bits) and $\overline{Z}=Z_1$. \\

The {\em code space} of a code $C$ encoding $k$ qubits is spanned by codewords $\ket{\overline{x}}$ where $x$ is a $k$-bitstring. In general these codewords $\ket{\overline{x}}$ will be highly entangled states. All states in the code space obey the parity checks, meaning that the parity check operators have eigenvalue $+1$ for all states in the code space (we say that the parity checks act trivially on the code space). The parity checks are all represented by mutually commuting multi-qubit Pauli operators. The logical operators of a quantum error-correcting code are non-unique as we can multiply them by the trivially-acting parity check operators to obtain equivalent operators. For example, $\overline{Z}$ for the 3-qubit code is either $Z_1$ or $Z_2$, or $Z_3$ or $Z_1 Z_2 Z_3$ as all these operators have the same action $\ket{\overline{+}} \leftrightarrow \ket{\overline{-}}$.

(a)

\begin{figure}[h!]
\centering
\mbox{
\Qcircuit @C=1em @R=1em {
&\qw & \multigate{1}{P} & \qw \\
&\ustick{\vdots} \qw & \ghost{P} & \ustick{\vdots} \qw & \\
&\lstick{\ket{+}} &  \ctrl{-1} & \gate{H} & \meter  
}
}
\end{figure}

(b)
\begin{figure}[h!]
\centering
\mbox{
\Qcircuit @C=1em @R=1em {
& \qw & \multigate{1}{P} & \qw & \multigate{1}{P} & \qw \\
& \ustick{\vdots} \qw & \ghost{P} & \ustick{\vdots} \qw & \ghost{P} & \qw \\
\lstick{\ket{+}} & \qw & \ctrl{-1} & \gate{R_X(\theta)} & \ctrl{-1} & \qw
}
}
\end{figure}

(c)
\begin{figure}[h!]
\centering
\mbox{
\Qcircuit @C=1em @R=1em {
& \targ & \qw & \qw & \qw & \qw \\
& \qw & \targ & \qw & \qw & \qw \\
& \qw & \qw & \targ & \qw & \qw \\
& \qw & \qw & \qw & \targ & \qw \\
\lstick{\ket{+}} & \ctrl{-4} & \ctrl{-3} & \ctrl{-2} & \ctrl{-1} & \gate{H} & \meter
}
}
\end{figure}

(d)\begin{figure}[h!]
\centering
\mbox{
\Qcircuit @C=1em @R=1em {
& \ctrl{4} & \qw & \qw & \qw & \qw \\
& \qw & \ctrl{3} & \qw & \qw & \qw \\
& \qw & \qw & \ctrl{2} & \qw & \qw \\
& \qw & \qw & \qw & \ctrl{1} & \qw \\
\lstick{\ket{0}} & \targ & \targ & \targ & \targ & \meter
}
}
\caption{Measuring parity checks the quantum-circuit way. The meter denotes measurement in the $\{\ket{0},\ket{1}\}$ basis or $M_Z$. (a) Circuit to measure the $\pm1$ eigenvalues of a unitary multi-qubit Pauli operator $P$. The gate is the controlled-$P$ gate which applies $P$ when the control qubit is 1 and $I$ if the control qubit is 0. (b) Realizing the evolution $\exp(-i \theta P/2)$ itself (with $R_x(\theta)=\exp(-i \theta X/2)$). (c) Realization of circuit (a) using CNOTS when $P=X_1 X_2 X_3 X_4$. (d) Realization of circuit (a)  using CNOTs when $P=Z_1Z_2Z_2Z_4$.}
\label{fig:paritycheck}
\end{figure}

Shor's 9-qubit code was the first quantum error-correcting code which encodes a single qubit and corrects any single qubit Pauli error, i.e. single qubit bit-flip errors $X$, phase flip errors $Z$ and bit+phase-flip errors $Y$.  As it turns out, if one wants to correct against any single qubit error, it is sufficient to be able to correct against any single qubit Pauli error. Let us thus assume for now that the only possible errors are multi- or single-qubit Pauli errors and afterwards we show that correcting against such Pauli errors is indeed sufficient. 

Shor's code is obtained from the 3-qubit repetition code by {\em concatenation}. Code concatenation is a procedure in which we take the elementary qubits of the codewords of a code $C$ and replace them by encoded qubits of a new code $C'$. In Shor's construction we choose the first code $C$ as a `rotated' 3-bit repetition code, that is, we take $\ket{\overline{+}}=\ket{+} \ket{+} \ket{+}$ and $\ket{\overline{-}}=\ket{-}\ket{-}\ket{-}$ with $\ket{\pm}=\frac{1}{\sqrt{2}}(\ket{0}\pm \ket{1})$.  One can verify that the parity checks of $C$ are $X_1 X_2$ and $X_2 X_3$ and the logical operators are $\overline{Z}_{C}=Z_1 Z_2 Z_3$ and $\overline{X}_{C}=X_1$. As the second code $C'$ we choose the normal 3-qubit repetition code, i.e. we replace $\ket{+}$ by $\ket{\overline{+}}=\frac{1}{\sqrt{2}}(\ket{000}+\ket{111})$ etc. 

We get all the parity checks for the concatenated 9-qubit code by taking all the parity checks of the codes $C'$ and the $C'$-encoded parity checks of $C$. For Shor's code this will give: the $Z$-checks $Z_1 Z_2$, $Z_2 Z_3$, $Z_4 Z_5$, $Z_5 Z_6$, $Z_7 Z_8$ and $Z_8 Z_9$ (from three uses of the code $C'$) and the $X$-checks $X_1 X_2 X_3 X_4 X_5 X_6$, $X_4 X_5 X_6 X_7 X_8 X_9$ (from the parity checks $\overline{X}_1 \overline{X}_2$ and $\overline{X}_2 \overline{X}_3$ where $\overline{X}$ is the logical operator of the code $C'$). The non-unique logical operators of the encoded qubit are $\overline{Z}=Z_1 Z_4 Z_7$ and $\overline{X}=X_1 X_2 X_3$.

This code can clearly correct any $X$ error as it consists of three qubits each of which is encoded in the repetition code which can correct an $X$ error. What happens if a single $Z$ error occurs on any of the qubits? A single $Z$ error will anti-commute with one of the parity $X$-checks or with both. For example, the error $Z_1$ anti-commutes with $X_1 X_2 X_3 X_4 X_5 X_6$ so that the state $Z_1 \ket{\overline{\psi}}$ has eigenvalue $-1$ with respect to this parity check. The error $Z_2$ or error $Z_3$ would have the same syndrome: these errors have the same effect on the code space as $Z_1 Z_2$ and $Z_2 Z_3$ act trivially on the code space. The same holds for the 3 qubits in the second block and the 3 qubits in the third block. Thus the $X$-parity check will only tell you whether there is a $Z$ error in the first, the second or the third block but this is ok and any single error in the block applies the proper correction on the code space. Thus the code can correct against a single qubit $X$ error, or $Z$ error and therefore also $Y$ error.

The eigenvalues of the parity check operators are called the {\em error syndrome}. Aside from detecting errors (finding $-1$ syndrome values) the error syndrome should allow one to infer which error occurred. How do we make this inference in general? We could assign a probability to each possible error: this assignment is captured by the {\em error model}.  Then our decoding procedure can simply choose an error, consistent with the syndrome, which has highest probability given our error model. Typically, the error model would assign a lower probability to errors which act on many qubits, and so the decoding could consist of simply picking a Pauli error, which could be responsible for the given syndrome, which acts on the fewest number of qubits. This kind of decoding is called minimum-weight decoding. It is important to note that the decoding procedure does not necessarily have to point to a unique error. For example: for the 9-qubit code, the error $Z_1$ and the error $Z_2$ have an equivalent effect on the code space as $Z_1 Z_2$ is a parity check which acts trivially on the code space. The syndromes for errors which are related by parity checks are always identical: the syndrome of a Pauli error $E$ is determined by the parity checks with which it anti-commutes. Multiplying $E$ by parity checks, which are by definition all mutually commuting operators, does not change the syndrome therefore. This means that the classical algorithm which processes the syndrome to infer an error --this procedure is called {\em decoding}-- does not need to choose between such {\em equivalent} errors. 

But there is further ambiguity in the error syndrome. For Shor's code the error $Z_1$ and the error $Z_4 Z_7$ have an identical syndrome as $Z_1 Z_4 Z_7$ is the $\overline{Z}$ operator which commutes with all parity checks.  If we get a single non-trivial ($-1$) syndrome for the parity check $X_1 X_2 X_3 X_4 X_5 X_6$ we could decide that the error is $Z_1$ or $Z_4 Z_7$.  But if we make a mistake in this decision and correct with $Z_4 Z_7$ while $Z_1$ happened then we have effectively performed a $\overline{Z}$ without knowing it! This means that the decoding procedure should decide between errors, all of which are consistent with the error syndrome, which are mutually related by logical operators. We will discuss the procedure of decoding more formally in Section \ref{sec:ssc1}. For Shor's code we can decode the syndrome by picking a single qubit error which is consistent with the syndrome. If a two-qubit error occurred we may thus have made a mistake. However, for Shor's code there are no two single-qubit errors $E_1$ and $E_2$ with the same syndrome whose product $E_1 E_2$ is a logical operator as each logical operator acts on at least 3 qubits. This is another way of seeing that Shor's code can correct any single qubit Pauli error. It is a $[[n,k,d]]=[[9,1,3]]$ code, encoding $k=1$ qubit into $n=9$ ($n$ is called the {\em block size} of the code) and having {\em distance} $d=3$. Having seen how this works for Shor's code, we can understand the role of the distance of a code more generally as follows.\\

The distance $d$ of the code is defined as the minimum weight of any logical operator (see the formal definition in Eq.~(\ref{def:dist})). The weight of a Pauli operator is the number of qubits on which it acts non-trivially, i.e. $Z_4 Z_7$ has weight 2. 
The definition of distance refers to a minimum weight of any logical operator as there are several logical operators, i.e. $\overline{X}, \overline{Z}$ etc. and we want any of them to have high weight, and secondly, the weight of each one of them can be varied by multiplication with parity checks. 

It is simple to understand why a code with distance $d=2t+1$ can correct $t$ errors. Namely, errors of weight at most $t$ have the property that their products have weight at most $2t < d$. Therefore the product of these errors can never be a logical operator as those have weight $d$ or more. Thus if one of these errors $E_1$ occurs and our decoding procedure picks another error $E_2$ of weight at most $t$  (both giving rise to the same syndrome) and applies $E_2$ to the encoded qubits, then effectively we have the state $E_2 E_1 \ket{\overline{\psi}}$. This state has trivial syndrome as all parity checks commute with $E_1 E_2$ (they either anti-commute with both $E_1$ and $E_2$ or commute with both), but $E_1 E_2$ has weight $2t < d$. Thus $E_1 E_2$ cannot be a logical operator but has to be some product of trivially-acting parity checks as $E_1 E_2$ commutes with all parity checks.

Another direct consequence of the distance of the code is how the code can handle so-called {\em erasure} errors. If errors only take place on some {\em known} subset of qubits, then a code with distance $d$ can correct (errors on) subsets of size $d-1$ as the product of any two Pauli errors on this subset has weight at most $d-1$. In other words, if $d-1$ or fewer qubits of the codeword are lost or their state completely {\em erased} by other means, one can still recover the entire codeword from the remaining set of qubits. One could do this as follows. First one replaces the lost $d-1$ qubits by the completely-mixed state $I/2^{d-1}$ \footnote{The erasure of a qubit, i.e. the qubit state $\rho$ is replaced by $I/2$, can be written as the process of applying a $I, X, Y$ resp. $Z$ error with probability $1/4$: $I/2=(\rho+X \rho X+Z \rho Z+Y \rho Y)/4$.}. Then one measures the parity checks on all qubits which gives us a syndrome which is only non-trivial for the parity checks which act on the $d-1$ qubits which had been erased. The syndrome points to a (non-unique) Pauli operator acting on these $d-1$ qubits or less and applying this Pauli corrects the error.

\subsubsection{Error Modeling}

Clearly, the usefulness of error correction is directly related to the error model; it hinges on the assumption that low-weight errors are more likely than high-weight errors. Error-correcting a code which can perfectly correct errors with weight at most $t$, will lead to failure with probability roughly equal to the total probability of errors of weight larger than $t$. This probability for failure of error correction is called the {\em logical error probability}. The goal of quantum error correction is to use redundancy and correction to realize logical qubits with logical error rates below the error rate of the elementary constituent qubits.\\

It may seem rather simplistic and limiting to use error models which assign $X$, $Z$ and $Y$ errors probabilistically to qubits as in real quantum information, through the interaction with classical or quantum systems, the amplitude and phase of a qubit will fluctuate over time: bare quantum information encoded in atomic, photonic, spin or other single quantum systems is barely information as it is undergoing continuous changes. It is important to note that the ideal parity check measurement
provides a discretization of the set of errors which is not naturally present in such elementary quantum systems. 

Consider for example noise on a single qubit due the fact that its time evolution (in a rotating frame) is not completely canceled and equals $\exp(-i \delta \omega Z t/2)$ for some probability distribution over frequencies ${\rm Prob}(\delta \omega)$ centered around $\delta \omega=0$. If this qubit is, say, the first qubit which is part of a multi-qubit encoded state $\ket{\overline{\psi}}$ we can write $\exp(-i \delta \omega Z_1 t/2)\ket{\overline{\psi}}=(\cos(\delta \omega \,t)I+i Z_1 \sin(\delta \omega \,t)) \ket{\overline{\psi}}$, i.e. we expand the small error of strength $\delta \omega$ in a basis of Pauli errors which occur with some amplitude related to $\delta \omega$. Consider then measuring a parity $X$-check which involves qubit 1. One obtains eigenvalue +1 with probability $\cos^2(\delta \omega \,t)$, close to 1 for small $\delta \omega \,t$, and we project onto the error-free state $\ket{\overline{\psi}}$. One obtains eigenvalue -1 with small probability $\sin^2(\delta \omega\, t)$ while we project onto the state with Pauli-error $Z_1 \ket{\overline{\psi}}$. Since any operator $E$ on $n$-qubits can be expanded in a basis of Hermitian Pauli matrices, this simple example illustrates the general principle that the correction of Pauli errors of weight less than $t$ suffices for the correction of any error of weight less than $t$. This property holds in fact for arbitrary quantum codes (including non-stabilizer codes for which we may gather error information through different means than parity check measurements), as it follows from the quantum error correction conditions, see Section \ref{sec:phys_codes}.\\

Ideal parity measurement can induce such discrete error model stated in terms of probabilities, but as parity measurements themselves will be inaccurate in a continuous fashion, such a fully digitized picture is an oversimplification. The theory of quantum fault-tolerance, see Section \ref{sec:ec_ft}, has developed a framework which allows one to establish the results of quantum error correction and fault-tolerance for very general quantum dynamics obeying physical locality assumptions (see the comprehensive results in \cite{AGP:ft}). However, for numerical studies of code performance it is impossible to simulate such more general open system dynamics and several simple error models are used to capture the expected performance of the codes. 

Two further important remarks can be made with this general framework in mind. Firstly, errors can be correlated in space and time arising from non-Markovian dynamics, but as long as (a) we use the proper estimate of the strength of the noise (which may involve using amplitudes and norms rather than probabilities) and (b) the noise is sufficiently short-ranged (meaning that noisy interactions between distant uncoupled qubits are sufficiently weak \cite{AKP:shortrange}), fault-tolerance threshold results can be established. The second remark is that qubit coding does not directly deal with leakage errors. As many elementary qubits are realized as two-level subspaces of higher-dimensional systems to which they can leak, other protective mechanisms such as cooling (or teleporting to a fresh qubit) will need to be employed in order to convert a leakage error into a regular error which can be corrected. In \cite{AT:leakage} it was shown one can derive general fault-tolerance threshold results for leakage errors by invoking the use of leakage reduction units (LRU) such as quantum teleportation. 

\subsubsection{Shor's Code as a Subsystem Code}

\begin{figure}[htb]
    \centering
    \includegraphics[width=0.5\hsize]{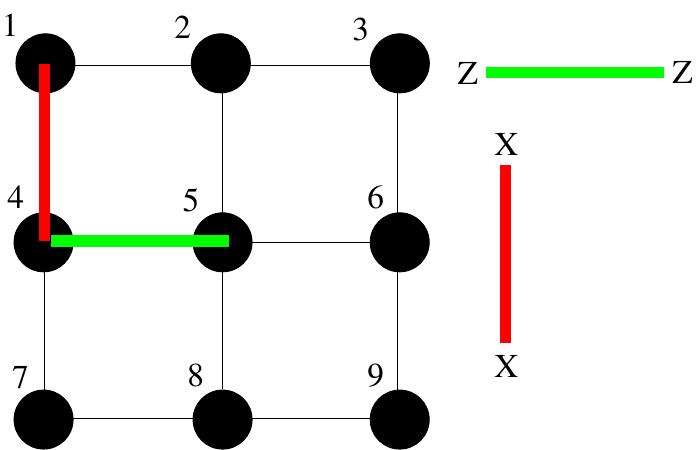} \caption{(Color Online) The 9-qubit $[[9,1,3]]$ Shor code with black qubits on the vertices. The stabilizer of Shor's code is generated by the weight-2 $Z$-checks as well as two weight-6, double row, $X$-checks ${\bf X}_{=,1}=X_1 X_2 X_3 X_4 X_5 X_6$ and ${\bf X}_{=,2}$. An alternative way of measuring ${\bf X}_{=,1}$ and ${\bf X}_{=,2}$ is by measuring the weight-2 $X$-checks in the Figure. One can similarly define two weight-6, double column, $Z$-checks, ${\bf Z}_{||,1}$ and ${\bf Z}_{||,2}$ as products of elementary weight-2 $Z$ checks. See also Fig.~\ref{fig:BS}.}
\label{fig:BS1}
\end{figure}

Let us come back to Shor's code and imagine that the nine qubits are laid out in a $3 \times 3$ square array as in Fig.~\ref{fig:BS1}. It looks relatively simple to measure the parity $Z$-checks locally, while the weight-6 $X$-checks would require a larger circuit using 6 CNOT gates between ancilla and data qubits. But why should there be such asymmetry between the $X$- and $Z$-checks? Imagine that instead of measuring the `double row' stabilizer operator ${\bf X}_{=,1} \equiv X_1 X_2 X_3 X_4 X_5 X_6$, we measure (in parallel or sequentially) the eigenvalues of $X_1 X_4$, $X_2 X_5$ and $X_3 X_6$ and take the product of these eigenvalues to obtain the eigenvalue of ${\bf X}_{=,1}$. The important property of these weight-2 operators is that they all individually commute with the logical operators $\overline{X}$ and $\overline{Z}$ of the Shor code, hence measuring them does not change the expectation values of $\overline{X}$ and $\overline{Z}$. These weight-2 $X$-checks do not commute with the weight-2 $Z$-checks however. If we first measure all the weight-2 $X$-checks and then measure the $Z$-checks, then with the second step the eigenvalues of individual $X$-checks are randomized but correlated. Namely, their product $X_1 X_2 X_3 X_4 X_5 X_6$ remains fixed as $X_1 X_2 X_3 X_4 X_5 X_6$ commutes with the weight-2 $Z$-checks. By symmetry, the weight-2 $X$-checks commute with the double column operators ${\bf Z}_{||,1}=Z_1 Z_2 Z_4 Z_5 Z_7 Z_8$ and ${\bf Z}_{||,2}=Z_2 Z_3 Z_5 Z_6 Z_8 Z_9$. Viewing the Shor code this way we can imagine doing 
error correction and decoding using the stable commuting parity checks ${\bf X}_{=,1}, {\bf X}_{=,2},{\bf Z}_{||,1},{\bf Z}_{||,2}$ while we deduce their eigenvalues from measuring 12 weight-2 parity checks. 

Shor's code in this form is the smallest member in the family of Bacon-Shor codes $[[n^2,1,n]]$ \cite{bacon:mem, AC:bs} whose qubits can be laid out in a $n \times n$ array as in Fig. \ref{fig:BS}, see Section \ref{sec:BS}. The Bacon-Shor code family in which non-commuting (low-weight) parity checks are measured in order to deduce the eigenvalues of commuting parity checks is an example of a (stabilizer) subsystem code.

\subsection[Formalism of Stabilizer Codes]{Formalism of Stabilizer Codes\footnote{Readers less interested in this general framework can skip the next two sections without major inconvenience.}}
\label{sec:ssc1}

Shor's code and many existing codes defined on qubits are examples of {\em stabilizer} codes \cite{thesis:gottesman}. Stabilizer codes are attractive as (i) they are the straightforward quantum generalization of classical binary linear codes, (ii) their logical operators and distance are easily determined, and it is relatively simple to (iii) understand how to construct universal sets of logical gates and (iv) execute a numerical analysis of the code performance. \\

The main idea of stabilizer codes is to encode $k$ logical qubits into $n$ physical qubits using a subspace, the code space, $\calL\subseteq (\CC^2)^{\otimes n}$ spanned by
states $|\psi\ra$ that are invariant under the action of a {\it stabilizer group} $\calS$,
\[
\calL=\{|\psi\ra\in (\CC^2)^{\otimes n}\, : \, P\, |\psi\ra=
|\psi\ra\quad \forall P\in \calS\}.
\]
Here $\calS$ is an Abelian subgroup of the Pauli group $\calP_n=\la iI, X_1,Z_1,\ldots,X_n,Z_n\ra$  such that $-I\notin \calS$ \footnote{$G=\langle g_1, \ldots, g_m \rangle$ denotes a group $G$ generated by elements $g_1, \ldots, g_m \in G$.}. For any stabilizer group $\calS$ one can always choose a set of generators $S_1,\ldots, S_m$, i.e. 
$\calS=\la S_1,\ldots,S_m\ra$, such that $S_a\in \calP_n$ are Hermitian Pauli operators. The nice thing about this stabilizer formalism is that instead of specifying the code-space by a basis of $2^n$-dimensional vectors, we specify the code-space by the generators of the stabilizer group which fix (or stabilize) these vectors. The mutually commuting parity checks that we considered before are the generators of the stabilizer group. If there are $n-k$ linearly independent generators (parity checks) then the code space $\calL$ is $2^k$-dimensional, or encodes $k$ qubits. This description of the code space in a $2^n$-dimensional vector space is thus highly efficient as it requires specifying at most $n$ linearly independent parity checks.\\

The {\em weight} $|P|$ of a Pauli operator $P=P_1 \ldots P_n \in \calP_n$ is the number of single-qubit Pauli operators $P_i$ which are unequal to $I$, in other words, the number of qubits on which $P$ acts nontrivially. If the code encodes $k$ logical qubits, it is always possible to find $k$ pairs of logical operators $(\overline{X}_j,\overline{Z}_j)_{j=1,...,k}$. These logical operators commute with all the parity checks, i.e. they commute with all elements in $\calS$ as they preserve the code space. However they should not be generated by the parity checks themselves otherwise their action on the code space is trivial. Thus these logical operators are elements of the Pauli group $\calP_n$ which are not elements in $\calS$ (otherwise their action is trivial), but which do commute with all elements in $\calS$. The set of operators in $\calP_n$ which commutes with $\calS$ is called the centralizer of $\calS$ in $\calP_n$, defined as $\calC(\calS)=\{P \in \calP_n |\forall s \in \calS, \;P s=sP\}$.
We thus have ${\cal C}(\mathcal{S})=\langle {\cal S}, \overline{X}_1,\overline{Z}_1,\ldots, \overline{X}_{k},\overline{Z}_{k}\rangle$, i.e. the logical operators of the code are elements of $\mathcal{C}(\mathcal{S}) \setminus \mathcal{S}$ as they are in $\calC(\calS)$ but not in $\calS$ \footnote{In some quantum error correction literature $\calC(\calS)\setminus \calS$ is denoted as $\calC(\calS)-\calS$. Also, the centralizer ${\cal C}(\calS)$ of $\calS$ in $\calP$ is also sometimes referred to as the normalizer ${\cal N}(\calS)$: for Pauli operators which either commute or anti-commute these groups coincide.}. The distance $d$ of a stabilizer code can then be defined as
\begin{equation}
d=\min_{P\in \calC(\calS)\backslash \calS} |P|,
\label{def:dist}
\end{equation}
i.e. the minimum weight that any logical operator can have. As the logical operators $\overline{P} \in {\cal C}(\calS)\setminus \calS$ commute with all parity check operators both the code state $\ket{\overline{\psi}}$ and $\overline{P} \ket{\overline{\psi}}$
have $+1$ eigenvalues with respect to the parity checks. Measuring the parity checks does thus not reveal whether a $\overline{P}$ has taken place or not while quantum information is drastically changed. Clearly, these logical operators $\overline{P}$ should be prevented from happening! A good stabilizer code will have high distance $d$ so that it is unlikely that local low-rate decoherence processes acting on a few qubits at the time will lead to a logical operator $\overline{P}$ which one cannot undo.\\

\subsubsection{Decoding}
\label{sec:decode}

Error correction proceeds by measuring the error syndrome ${\bf s}$ which is a vector of $\pm 1$ eigenvalues of the generators of $\calS$. As we mentioned in Section \ref{sec:stabform} this syndrome will not point to a unique Pauli error but all $E'=E P$ where $P \in {\cal C}({\cal S})$ give rise to the same syndrome. Let us now describe the formal procedure of decoding.

We can define an {\em equivalence class of errors} $[E]$ consisting of errors $E'=E P$ where $P \in \calS$, that is, elements in $[E]$ are related to $E$ by a (trivially-acting) element in the stabilizer group \footnote{[E] is a coset of the group ${\cal S}$ in $\calP_n$. Note that left and right cosets are the same modulo trivial errors proportional to $I$.}. If error $E$ occurs and we decide to correct this error by applying $E'$, the sequence $E E' \in \calS$ has been applied to the codeword leaving it unchanged. We can associate a total error probability with such class, ${\rm Prob}([E])=\sum_{s \in S} {\rm Prob}(Es)$, depending on some error model which assigns a probability ${\rm Prob}(P)$ to every Pauli operator $P \in \calP_n$. Given an error $E$ and a syndrome, one can similarly define a discrete number of classes $[E \overline{P}]$ where the logicals $\overline{P}\in {\cal C}(\calS)\backslash \calS$. 

The procedure which maximizes the success probability of reversing the error while making no logical error is called {\em maximum likelihood decoding}. Given a syndrome ${\bf s}$ and some error $E(\bf s)$ which is consistent with the syndrome, a maximum likelihood decoder compares the values of ${\rm Prob}([E \overline{P}])$ for the various $\overline{P}$ and chooses the one with maximal value pointing to some $\overline{P}$. Then it applies the corrective operator $E \overline{P}$ which is by definition the most likely correction.

If $E(\bf s)$ happens to be the error $E$ which actually took place, then such decoding procedure is thus successful when ${\rm Prob}([E]) > {\rm Prob}([E\overline{P}])$ for any non-trivial $\overline{P}$.\\

It is important to consider how efficiently (in the number $n$ of elementary qubits) maximum likelihood decoding can be done since ${\rm Prob}([E\overline{P}])$ is a sum over the number of elements in $\calS$ which is exponential in $n$. For a simple depolarizing error model where each qubit undergoes a $X$, $Y$ or $Z$ error with probability $p/3$ and no error with probability $1-p$, one has ${\rm Prob}([E\overline{P}])=(1-p)^n \sum_{s \in \calS} \exp(-\beta |E\overline{P}s|)$ with inverse `temperature' $\beta=\ln(3(1-p)/p)$. 

We can define a classical Hamiltonian $H_{E\overline{P}}(s) \equiv |E\overline{P}s|$ which acts on spin variables $s_i \in \{-1,1\}$ each of which corresponds to a generator $S_i$ of the stabilizer group $\calS$. The Hamiltonian will be a sum of terms, each corresponding to a single qubit in the code and contributing either 0 or 1. Each term can be written as a function of the stabilizer generators $s_i=\pm 1$, $E$ and $\overline{P}$ which act on the particular qubit making it trivial (weight 0) or non-trivial (weight 1). We can view $Z_{E\overline{P}}\equiv \sum_{s \in \calS} \exp(-\beta H_{E\overline{P}}(s))$ as a partition function of the Hamiltonian $H_{E \overline{P}}(s)$ at a temperature related to the error probability. 

For small error rates $p \ll 1$ corresponding to low temperatures $\beta \rightarrow \infty$, the value of this partition function is dominated by the spin configuration $s$ which minimizes $H_{E\overline{P}}(s)=|E \overline{P}s|$. Thus for sufficiently low error rates, instead of maximum likelihood decoding which compares the relative values of ${\rm Prob}([E\overline{P}])$, one can also opt for {\em minimum-weight} decoding. In minimum-weight decoding one simply picks an error $E(\bf s)$, consistent with the syndrome ${\bf s}$, which has minimum weight $|E|$. We will discuss this decoding method for the surface code in Section \ref{sec:topo}.  \\

For topological codes, the criterion for successful maximum likelihood decoding and the noise threshold of the code can be related to a phase-transition in a classical statistical model with quenched disorder \cite{dennis+:top}, \cite{KA:glass}. This can be readily understood as follows. A probabilistic noise model such as the depolarizing noise model induces a probability distribution ${\rm Prob}(E)$ over the errors $E$. 
For a given error $E$ we can decode successfully when $Z_E > Z_{\overline{P} E}$ where $Z_E$ is the partition function of the quenched-disorder Hamiltonian $H_E(s)=|Es|$ defined in the previous paragraph. We want to be able to decode successfully for typical errors $E$, hence we are interested in looking at averages over the disorder $E$. Assume that one has a family of codes for which one can define a thermodynamic limit in which the number of qubits $n \rightarrow \infty$. One can define a critical, say depolarizing error rate $p_c$ by the following condition
\begin{eqnarray}
p < p_c \rightarrow \lim_{n \rightarrow \infty} \sum_E {\rm Prob}(E) \log \left(\frac{Z_E}{Z_{\overline{P}E}}\right)=\infty, \nonumber \\
p > p_c \rightarrow \lim_{n \rightarrow \infty} \sum_E {\rm Prob}(E)  \log \left(\frac{Z_E}{Z_{\overline{P}E}}\right)=0.
\end{eqnarray}
One thus studies the behavior of the free energy of the statistical model with quenched disorder (which is determined by the error probability $p$), i.e. $\langle \log Z_E \rangle_p=\sum_E {\rm Prob}(E) \log Z_E$ to determine the value of $p_c$. The temperature $\beta$ and the quenched disorder are not independent but directly depend on the same error probability $p$. For this reason one identifies $p_c$ with a phase-transition of the quenched-disorder model along the so-called Nishimori line on which $\beta$ is a function of the strength of the error probability which is also the disorder parameter. One of the first studies of this sort was done in \cite{WHP:threshold}.

\subsubsection{Stabilizer Code Examples and The CSS Construction}
\label{sec:CSS} 
We discuss a few small examples of stabilizer codes to illustrate the formalism. For classical error correction the smallest code which can detect an $X$ error is a two-bit code and the smallest code which can correct any $X$ error is the 3-qubit code. As a quantum error correcting code has to correct both $X$ and $Z$ errors, the smallest quantum error correcting code will have more qubits. 

Let us consider first the two-qubit code. For the two-qubit code with $\ket{\overline{0}}=\frac{1}{\sqrt{2}}(\ket{00}+\ket{11})$ and $\ket{\overline{1}}=\frac{1}{\sqrt{2}}(\ket{01}+\ket{10})$ we have $\overline{X}=X_1$ or $\overline{X}=X_2$ and $\overline{Z}=Z_1 Z_2$. The code can detect any single $Z$ error as such error maps the two codewords onto the orthogonal states $\frac{1}{\sqrt{2}}(\ket{00}-\ket{11})$ and $\frac{1}{\sqrt{2}}(\ket{01}-\ket{10})$ (as $\overline{Z}$ is of weight-2). The code can't detect single $X$ errors as these are logical operators.\\

The smallest non-trivial quantum code is the $[[4,2,2]]$ error-detecting code. Its linearly independent parity checks are $X_1 X_2 X_3 X_4$ and $Z_1 Z_2 Z_3 Z_4$: the code encodes $4-2=2$ qubits. One can verify that one can choose $\overline{X}_1=X_1 X_2, \overline{Z}_1=Z_1 Z_3$ and $\overline{X}_2=X_2 X_4$, $\overline{Z}_2=Z_3 Z_4$ as the logical operators which commute with the parity checks. The code distance is 2 which means that the code cannot correct a single qubit error. The code can however still {\em detect} any single qubit error as any single qubit error anti-commutes with at least one of the parity checks which leads to a nontrivial $-1$ syndrome. Alternatively, we can view this code as a subsystem code (see Section \ref{sec:ssc2}) which has one logical qubit, say, qubit 1, and one gauge qubit, qubit 2. In that case $\calG=\langle X_1 X_2 X_3 X_4, Z_1 Z_2 Z_3 Z_4, Z_3 Z_4, X_2 X_4 \rangle=\langle Z_1 Z_2, Z_3 Z_4, X_1 X_3, X_2 X_4 \rangle$, showing that measuring weight-2 checks would suffice to detect single qubit errors on the encoded qubit 1. The smallest stabilizer code which encodes 1 qubit and corrects 1 error is  the $[[5,1,3]]$ code; one can find its parity checks in \cite{book:nielsen&chuang}.\\

In order to make larger codes out of small codes one can use the idea of code concatenation which we will first illustrate with an explicit example. 

We take a small stabilizer code $C_6$ (defined in \cite{knill:nature}) with parity checks $X_1 X_4X_5 X_6$, $X_1X_2 X_3 X_6$, $Z_1Z_4Z_5Z_6$ and $Z_1Z_2Z_3Z_6$ acting on 6 qubits. This code has 4 independent parity checks, hence it encodes $6-4=2$ qubits with the logical operators $\overline{X}_1=X_2 X_3$, $\overline{Z}_1=Z_3 Z_4 Z_6$ and $\overline{X}_2=X_1 X_3 X_4$, $\overline{Z}_2=Z_4 Z_5$.  As its distance is 2, it can only detect single $X$ or $Z$ errors.

One can concatenate this code $C_6$ with the code $[[4,2,2]]$ (called $C_4$ in \cite{knill:nature}) by replacing the three pairs of qubits, i.e. the pairs (12), (34) and (56), in $C_6$ by three sets of $C_4$-encoded qubits, to obtain a new code. This code has thus $n=12$ qubits and encodes $k=2$ qubits. We can represent these 12 qubits as 3 sets of 4 qubits such that the $X$-checks read
\begin{equation}
{\cal S}(X)=\left(\begin{array}{cccc|cccc|cccc}
 X & X & X & X &  I & I & I & I & I & I &  I & I \\
I & I & I & I & X & X & X & X & I & I & I & I\\
I & I & I & I & I & I & I  & I & X & X & X & X \\
X & X & I & I & I & X & I & X & X & I & I & X\\
X &  I & I & X & X & X & I & I & I & X & I & X \\
\end{array}\right)
\nonumber
\end{equation}
The $Z$-checks are 
\begin{equation}
{\cal S}(Z)=\left(\begin{array}{cccc|cccc|cccc}
Z & Z & Z & Z &  I & I & I & I & I & I &  I & I \\
I & I & I & I & Z & Z & Z & Z & I & I & I & I\\
I & I & I & I & I & I & I  & I & Z & Z & Z & Z \\
Z & I & Z & I & I & I & Z & Z & Z & I & I & Z\\
Z &  I & I & Z & Z & I & Z & I & I & I & Z & Z \\
\end{array} \right).
\nonumber
\end{equation}
and the logical operators are
\begin{equation}
\begin{array}{lcccc|cccc|cccc}
\overline{X}_1= & I & X & I & X  &  X & X & I  & I   & I & I &  I & I \\
\overline{Z}_1= & I & I & I &  I  &  Z & I & I & Z    & I & I & Z & Z\\
\overline{X}_2= & X & X & I & I  &  X & I & I & X    & I & I & I & I \\
\overline{Z}_2= & I & I & I & I   & I & I & Z & Z     & Z & I & Z & I\\
\end{array} 
\nonumber
\end{equation}
One can verify that the minimum weight of the logical operators of this concatenated code is $4$. Thus the code is a $[[12,2,4]]$ code, able to correct any single error and to detect any three errors. \\

One could repeat the concatenation step and recursively concatenate $C_6$ with itself (replacing a pair of qubits by three pairs of qubits etc.) as in Knill's $C_4/C_6$ architecture \cite{knill:nature} or, alternatively, recursively concatenate $C_4$ with itself as was considered in \cite{AP:fib}. 

In general when we concatenate a $[[n_1,1,d_1]]$ code with a $[[n_2,1,d_2]]$ code, we obtain a code which encodes one qubit into $n=n_1 n_2$ qubits and has distance $d=d_1 d_2$.  Code concatenation is a useful way to obtain a large code from smaller codes as the number of syndrome collections scales linearly with the number of concatenation steps while the number of qubits and the distance grows exponentially with the number of concatenation steps. In addition, decoding of a concatenated code is efficient in the block size $n$ of the code and the performance of decoding can be strongly enhanced by using message passing between concatenation layers \cite{poulin:message}. \\

Another way of constructing quantum error-correcting codes is by using two classical binary codes in the Calderbank-Shor-Steane (CSS) construction \cite{book:nielsen&chuang}. 

Classical binary linear codes are fully characterized by their parity check matrix $H$. The parity check matrix $H_1$ of a code $C_1$ encoding $k_1$ bits is a $(n-k_1) \times n$ matrix with 0,1 entries where linearly independent rows represent the parity checks. The binary vectors $c \in \{0,1\}^n$ which obey the parity checks, i.e. $H c=0$ (where addition is modulo 2), are the codewords. The distance $d=2t+1$ of such classical code is the minimum (Hamming) weight of any codeword and the code can correct $t$ errors. 

We can represent a row $r$ of $H_1$ of a code $C_1$ by a parity check operator $s(Z)$ such that for the bit $r_i=1$ we take $s(Z)_i=Z$ and for bit $r_i=0$ we set $s(Z)_i=I$. These parity checks generate some stabilizer group $\calS_1(Z)$. In order to make this into a quantum code with distance larger than one, one needs to add $X$-type parity checks. These could simply be obtained from the $(n-k_2) \times n$ parity check matrix $H_2$ of another classical code $C_2$. We obtain the stabilizer parity checks $\calS_2(X)$ by replacing the 1s in each row of this matrix by Pauli $X$ and $I$ otherwise.  But in order for $\calS=\langle \calS_1(Z), \calS_2(X) \rangle$ to be an Abelian group the checks all have to commute.
This implies that every parity $X$-check should overlap on an even number of qubits with every parity $Z$-check. In coding words it means that the rows of $H_2$ have to be orthogonal to the rows of $H_1$. This in turn can be expressed as $C_2^{\perp} \subseteq C_1$ where $C_2^{\perp}$ is the code dual to $C_2$ (codewords of $C_2^{\perp}$ are all the binary vectors orthogonal to all codewords $c \in C_2$). 

In total ${\cal S}=\langle \calS_1(Z), \calS_2(X) \rangle$ will be generated by $2n-k_1-k_2$ independent parity checks so that the quantum code encodes $k_1+k_2-n$ qubits. The distance of the quantum code is the minimum of the distance $d(C_1)$ and $d(C_2)$ as one code is used to correct $Z$ errors and the other code is used to correct $X$ errors. \\

A good example of this construction is Steane's 7-qubit code $[[7,1,3]]$ which is constructed using a classical binary code $C$ which encodes 4  bits into 7 bits and has distance 3. Its parity check matrix is
\begin{equation}
H=\left(\begin{array}{ccccccc} 0 &  0 & 0 & 1  & 1 & 1 & 1 \\
0 & 1 & 1 & 0 & 0 & 1 & 1 \\
1 & 0 & 1 & 0 & 1 & 0 & 1 \end{array} \right).
\end{equation}
The codewords $c$ which obey $Hc =0$ are linear combinations of the $7-3=4$ binary vectors $(1,1,1,0,0,0,0)$, $(0,0,0,1,1,1,1)$, $(0,1,1,0,0,1,1)$, $(1,0,1,0,1,0,1)$ where the last three are the rows of the parity check matrix: these are also codewords of $C^{\perp}$. Hence $C^{\perp} \subseteq C$ and we can use the CSS construction with $C_1=C$ and $C_2=C$ to get a quantum code. As $C^{\perp}$ (as well as $C$) has distance $3$, the quantum code will have distance $3$ and encodes one qubit.
The parity checks are $Z_4 Z_5 Z_6 Z_7$, $Z_2 Z_3 Z_6 Z_7$, $Z_1 Z_3 Z_5 Z_7$ and $X_4 X_5 X_6 X_7$, $X_2 X_3 X_6 X_7$, $X_1 X_3 X_5 X_7$. 

Steane's code is the smallest example in a family of two-dimensional color codes \cite{BM:colorcodes}. Codes obtained using the CSS construction have some useful properties in terms of what logical gates can be realized easily on the encoded qubits, see Section \ref{sec:tt}. Homological codes discussed in Section \ref{sec:homo} are another interesting class of CSS codes.

\subsection{Formalism of Subsystem Stabilizer Codes}
\label{sec:ssc2}

Subsystem stabilizer codes can be viewed as stabilizer codes in which some logical qubits, called gauge qubits, are not used to encode information \cite{poulin:stabsub}. The state of these extra qubits is irrelevant and is in principle left to vary. The presence of the gauge qubits sometimes lets one simplify the measurement of the stabilizer parity checks as the state of the gauge qubits is allowed to freely change under these measurements. \\

To define a subsystem code, one can thus takes a stabilizer code $\calS$ and split its logical operators $(\overline{X}_i,\overline{Z}_i)$ into two groups: the gauge qubit logical operators $(\overline{X}_i,\overline{Z}_i)$, $i=1\ldots m$ and the remaining logical operators $(\overline{X}_i,\overline{Z}_i)$ with $i=m+1,\ldots k$.
Then we define a new subgroup $\calG=\langle {\cal S},\overline{X}_1,\overline{Z}_1,\ldots \overline{X}_m,\overline{Z}_m \rangle$ which contains $\calS$ but also the logical operators of the irrelevant gauge qubits. We note that $\calG$ is non-Abelian as the logical $\overline{X}$ and $\overline{Z}$-operators of a gauge qubit do not mutually commute. However, all elements in the center of this group, defined as $\calG \cap \calC(\calG)=\{P \in {\cal G}|\;\forall g \in \calG, P g=g P\}=\calS$ (modulo trivial elements) will commute with all elements in $\calG$. 

One could do error correction by measuring the parity check operators in $\calS$ but imagine that instead we measure the (non-commuting) generators of the group $\calG$. As some of these operators are non-commuting, their $\pm 1$ eigenvalues cannot simultaneously be fixed. However by choosing the proper order to measure these non-commuting checks, we can determine the eigenvalues of the generators for $\calS$ since $\calS \subseteq \calG$. For example, for a code $\calG=\langle \calG_1(X), \calG_2(Z) \rangle$ where the group $\calG_1(X)$ ($\calG_2(Z)$) only consists of $X$-checks resp. $Z$-checks, one can first measure all the generators of $\calG_1(X)$ and then all the generators of $\calG_2(Z)$. For more general gauge groups $\calG$ which do not split up in an $X$ and a $Z$-part, there is a simple condition which constrains the order in which the gauge checks have to be measured (see e.g. \cite{SBT:top}) in order to derive stable values for the stabilizer checks in $\calS$.
Note that the $k-m$ logical operators $(\overline{X}_i,\overline{Z}_i)$, $i=m+1,\ldots ,k$ of the logical qubits in-use do commute with $\calG$ and so these logical operators are unaffected by the measurement of elements in $\calG$. \\

A priori, there is no reason why measuring the generators of $\calG$ would be simpler than measuring the generators of the stabilizer $\calS$. In the interesting constructions such as the Bacon-Shor code and the subsystem surface code discussed in Section \ref{sec:alt}, we gain because we measure very low-weight parity checks in $\calG$, while (often) we lose by allowing more qubit-overhead or declining noise threshold.  \\

The perspective of viewing a subsystem code merely as a partially-used stabilizer code is useful for understanding the role of $\calG$ versus $\calS$. It is not in general the way one wants to construct such a code as creating $\calG$ from an arbitrary stabilizer code $\calS$ (generated by low-weight checks) by adding some logical operators of gauge qubits gives no guarantee that $\calG$ is itself generated by low weight-parity checks.\\

When we measure the eigenvalues of the non-commuting generators of $\calG$, the gauge check operators, we are affecting the state of the gauge qubits. Consider a subsystem code $\calG=\langle \calG_1(X), \calG_2(Z) \rangle$. If we measure the gauge $Z$-checks we fix the state of the gauge qubits to be an eigenstate of these $Z$-checks, and hence the gauge qubits are eigenstates of their logical $\overline{Z}$ operators. If we then measure all the gauge $X$-checks, we project the gauge qubit states onto eigenstates of their logical $\overline{X}$ operators, thus actively changing their logical  state. We can make a new stabilizer code out of a subsystem code by `fixing the gauge' as follows. In order to fix, say, an $X$-gauge, we add all the logical $\overline{X}$ operators of the gauge qubits $\overline{X}_1, \ldots, \overline{X}_m$ to $\calS$, let's call this the stabilizer code $\calS_{\rm X-fix}= \langle S, \overline{X}_1, \ldots, \overline{X}_m \rangle$. 
For this stabilizer code, all the gauge qubits are prepared in their logical $\ket{\overline{+}}$ state.  Gauge fixing is a concept that can be useful in the efficient realization of a universal set of logical gates, see Section \ref{sec:tt}.\\

The distance of a subsystem code is not the same as that of a stabilizer code, Eq.~(\ref{def:dist}), as we should only consider the minimum weight of the genuine $k-m$ logical operators. These logical operators are not unique as they can be multiplied by elements in $\calS$ {\em but also} by the logical operators of the irrelevant gauge qubits (which change the state of the gauge qubits). This motivates the definition of the distance as $d=\min_{P \in {\cal C}({\cal S})\backslash{\cal G}} |P|$. 
We can further distinguish logical operators in being so-called bare or dressed logical operators. Bare logical operators do not change the state of the gauge qubits: they commute with all elements in $\calG$, in other words they are contained in $\calC(\calG)$. Dressed logical operators can be obtained from bare logical operators by multiplication with elements in $\calG$, in particular by multiplication with the gauge qubit logical operators which are in $\calG$ but not in $\calS$.

It is the distance and properties of the dressed logical operators which define the qualities of the code. For example, one can easily construct a `Heisenberg' subsystem code of $n$ qubits on a line with $n$ odd. Let $\calG=\langle X_1 X_{2}, Z_1 Z_{2}, \ldots , X_{n-1} X_n, Z_{n-1} Z_n \rangle$. The operators $\overline{X}=X^{\otimes n}$ and $\overline{Z}^{\otimes n}$ mutually anti-commute and commute with all elements in $\calG$ but are not elements of $\calG$ as they are of odd weight. Hence they are bare logical operators of weight $n$, but multiplying these operators by elements in $\calG$ will result in dressed logical operators which have weight 1, hence a low-distance code. \\

As errors on the gauge qubits are harmless, it means that equivalent classes of errors are those related to each other by elements in $\calG$ (instead of $\calS$ for stabilizer codes). Given the eigenvalues of the stabilizer generators, the syndrome ${\bf s}$, the decoding algorithm considers equivalence classes defined as $[E]=\{E'| \;\exists g \in \calG,\; gE'=E\}$. Maximum likelihood decoding (or minimum weight decoding) can proceed similar as for stabilizer codes: one determines which class $[E \overline{P}]$ has a maximum value for ${\rm Prob}([E \overline{P}])=\sum_{g \in \calG} {\rm Prob}(E \overline{P}g)$ where $\overline{P}$ varies over the possible logical operators. 

\subsection{QEC Conditions and Other Small Codes of Physical Interest}
\label{sec:phys_codes}

One may ask what properties a general (not necessarily stabilizer) quantum code, defined as some subspace $C$ of a physical state space, should have in order for a certain set of errors to be correctable. These properties are expressed as the quantum error correction conditions which can hold exactly or only approximately. \\

We encode some $k$ qubits into a code space $C$ which is a subspace of a $n$-qubit space such that $\ket{\overline{x}}$ are the codewords encoding the $k$-bit strings $x$. Assume there is a set of errors ${\cal E}=\{E_i\}_{i=1}^I$ against which we wish to correct. We can capture the action of these errors by a superoperator ${\cal S}(\rho)=\sum_{i=1}^I E_i \rho E_i^{\dagger}$ which is not necessarily trace-preserving. We are seeking a trace-preserving reversal superoperator ${\cal R}$ such that ${\cal R} \cdot {\cal S}(\overline{\rho}) \propto \overline{\rho}$ for any encoded density matrix $\overline{\rho}$ (which is only supported on the code space). The quantum error correction conditions \cite{bdsw, kl:qec} say that there exists such an error-correcting reversal operation ${\cal R}$ if and only if the following conditions are obeyed for all errors $E_i, E_j\in {\cal E}$ 
\begin{equation}
\;\forall x,x',\bra{\overline{x}}E_i^{\dagger} E_j \ket{\overline{x'}}=c_{ij}\delta_{xx'}.
\label{eq:QEC}
\end{equation}
Here $c_{ij}$ is a constant {\em independent} of the codeword $\ket{\overline{x}}$ with $c_{ij}=c_{ji}^*$. The condition for $x=x'$ informally says that the codewords are not distinguished by the error observables. The condition for $x \neq x'$ indicates that the orthogonal codewords need to remain orthogonal after the action of the errors (otherwise we could not undo the effect of the errors). One can understand these conditions as arriving from the requirement that in order for a reversal operation ${\cal R}$ to exist, no quantum information should leak to the environment. One can find a derivation of these conditions in \cite{book:nielsen&chuang} directly from demanding that ${\cal R} \cdot {\cal S}(\overline{\rho}) \propto \overline{\rho}$.\\

If a code can correct the error set $\{E_i\}$, it can also correct an error set $\{F_j\}$ where each $F_j$ is any linear combination of the elements $E_i$ as one can verify that the set $\{F_j\}$ will also obey the quantum error correction conditions in Eq.~(\ref{eq:QEC}). This means that if a code can correct against Pauli errors on any subset of $t$ qubits, it can correct against any error on $t$ qubits as the Pauli matrices form an operator basis in which one can expand the errors. Stabilizer codes are generally designed such that the code has distance $d=2t+1$ which implies that it can correct any Pauli error on any subset of $t$ qubits (and thus any other error on these subsets as well).\\

These QEC conditions can be generalized to the unified framework of operator quantum error correction \cite{KLP:qec_uni, NP:operator_qec} which covers both subsystem codes as well as error-avoidance techniques via the use of decoherence-free subspaces and noise-free subsystems. Another generalization of the stabilizer framework is the formalism of codeword-stabilized quantum codes \cite{cross+:cws} which also includes non-stabilizer codes.\\

\subsubsection{Physical Error Models}

How do we determine the set of error operators $\{E_i\}$ for a given set of qubits? In principle, one could start with a Hamiltonian description of the dynamics of the qubits, the system $S$, coupled to a physically-relevant part of the rest of the world, which we call the environment $E$. 

One has a Hamiltonian $H(t)=H_S(t)+H_{SE}(t)+H_E(t)$ where $H_S(t)$ ($H_E(t)$) acts on $S$ ($E$) and $H_{SE}(t)$ is the coupling term. We assume that the qubits of the system and environment are initially ($t=0$) in some product state $\rho_S \otimes \rho_E$ and then evolve together for time $\tau$. The dynamics due to the $U(0,\tau)={\cal T}\exp(-i \int_{0}^{\tau} dt' H(t'))$, for the system alone can then be described by the superoperator ${\cal S}_{\tau}$:
\begin{equation}
{\cal S}_{\tau}(\rho_S)={\rm Tr}_E \;U(0,\tau)\rho_S \otimes \rho_E U^{\dagger}(0,\tau)=\sum_i E_i \rho_S E_i^{\dagger},\nonumber
\end{equation}
where $\{E_i\}$ with $\sum_i E_i^{\dagger} E_i=I$ are the so-called Kraus operators determining the action of the superoperator. These Kraus operators $\{E_i\}$ will thus be the error operators. The Kraus operators of a given superoperator are not unique.  One can define a different set of error operators $F_j=\sum_{i} U_{ji} E_i$ with a unitary matrix $U$ which realizes the same superoperator (see e.g. \cite{book:nielsen&chuang}), but as we noted before if the set $\{E_i\}$ is correctable then the set $\{F_j\}$ is correctable as well.\\

This derivation of the error operators is appropriate when the system-environment interaction is memory-less or Markovian beyond a time-scale $\tau$ so that it is warranted that we start the evolution with an initial product state between system and environment. For general non-Markovian noise such description is not directly appropriate. Instead of finding a map describing the dynamics of the system by itself, one can always consider the unitary dynamics of the system and environment together and expand this in terms of errors. One writes the joint unitary transformation between time $t_1$ and $t_2$ as $U(t_1,t_2)=U_{ideal}(t_1,t_2)+E_{SE}$ with $U_{ideal}(t_1,t_2)$ is the ideal faultless evolution. The operator $E_{SE}$ can always be expanded as $E_{SE}=\sum_i E_i \otimes B_i$ where $\{E_i\}$ can be identified as a set of error operators on the system. We refer to Chapter 5 in the book \cite{book:lidar_brun} for a more extensive treatment of non-Markovian noise models.\\

Quite commonly one can describe the open system dynamics by a Markovian master equation of Lindblad form 
\begin{equation}
\frac{d \rho}{dt}=-i [H(t),\rho]+{\cal L}(\rho)\equiv {\cal L}_{\rm tot}(\rho),
\label{eq:lindblad}
\end{equation}
where ${\cal L}(\rho)=\sum_j L_j \rho L_j^{\dagger}-\frac{1}{2}\{L_j^{\dagger} L_j,\rho \}$ with quantum-jump or Lindblad operators $L_j$ \footnote{Using the definition of the anti-commutator $\{A,B\}=AB+BA$.}. Here $H(t)$ is the Hamiltonian of the quantum system which could include some time-dependent driving terms. For short times $\tau$ we have $\rho(\tau)={\cal  S}_{\tau}(\rho(0))=E_0 \rho E_0^{\dagger}+\sum_i E_i \rho E_i^{\dagger}$ with $E_0 \approx I-i \tau H-\frac{1}{2}\tau \sum_i L_i^{\dagger} L_i=I-O(\tau)$ and $E_i \approx \sqrt{\tau} L_i$. Thus the error set is given by the quantum jump operators $L_i$ and the no-error operator $E_0$ which is nontrivial in order $O(\tau)$.\\ 

A special simple case of such Lindblad equation leads to the Bloch equation which is used to describe qubit decoherence at a phenomenological level. We consider a qubit, described by a Hamiltonian $H=-\frac{\omega}{2} Z$, which exchanges energy with a large Markovian environment in thermal equilibrium at temperature $\beta=\frac{1}{kT}$. One can model such open-system dynamics using a master equation of Lindblad form with quantum jump operators with $L_- = \sqrt{\kappa_-} \sigma_-$ and $L_+ =\sqrt{\kappa_+} \sigma_+$ with $\sigma_-=\ket{0}\bra{1}$ and $\sigma_+=\ket{1}\bra{0}$. Here the rates $\kappa_+, \kappa_-$ obey a detailed balance condition $\frac{\kappa_+}{\kappa_-}=\exp(-\beta \omega)$. The resulting Lindblad equation has the thermal state $\rho_{\beta}=\frac{\exp(-\beta H)}{{\rm Tr}(\exp(-\beta H))}$ as its unique stationary point for which ${\cal L}_{tot}(\rho_{\beta})=0$. We can include additional physical sources of qubit dephasing modeled by quantum jump operator $L_Z =\sqrt{\gamma_Z} Z$ in the Lindblad equation; this, of course, does not alter its stationary point.

We can parametrize a qubit as $\rho=\frac{1}{2}(I+{\bf r} \cdot {\bf \sigma})$ with Bloch vector ${\bf r}$ and Pauli matrices ${\bf \sigma}=(X,Y,Z)$ and re-express such Lindblad equation as a differential equation for ${\bf r}$, the Bloch equation.  Aside from the process of thermal equilibration and dephasing, one may add time-dependent driving fields in the Bloch equation (which are assumed not to alter the equilibration process) so that the general Hamiltonian is $H(t)=\frac{1}{2}{\bf M}(t) \cdot {\bf \sigma}$. 

The Bloch equation then reads
\begin{equation}
\frac{d {\bf r}}{dt}={\bf r}(t) \times {\bf M}(t)+{\bf R} ({\bf r}(t)-{\bf r}_\beta),
\end{equation}
where the first (second) part describes the coherent (dissipative) dynamics. Here the equilibrium Bloch vector ${\bf r}_{\beta}=(0,0,\tanh(\beta \omega/2))$ and the diagonal relaxation matrix equals ${\bf R}={\rm diag}(-1/T_2, -1/T_2,-1/T_1)$ where the decoherence time $T_2$ and relaxation time $T_1$ characterize the basic quality of the qubit. \\

We will now consider two simple codes which approximately obey the conditions in Eq.~(\ref{eq:QEC}). The first code protects against amplitude damping which models $T_1$ relaxation, for qubits. In the second code, the cat code,  one encodes a qubit into a bosonic mode so as to be partially protected against photon loss. We continue in Section \ref{sec:qiq} with another stabilizer code which encodes a qubit in a bosonic mode which demonstrates that one can also apply the stabilizer formalism to phase space.

\subsubsection{Amplitude Damping Code}

Even though the $[[5,1,3]]$ code is the smallest code which can correct against any single qubit error, one can use 4 qubits to approximately correct any {\em amplitude-damping} error which can model energy loss \cite{leung+:approx}. The noise process for amplitude damping on a single qubit is given by the superoperator ${\cal S}(\rho)=\sum_i A_i \rho A_i^{\dagger}$ with $A_0=\left(\begin{array}{cc} 1 & 0 \\ 0 & \sqrt{1-\kappa} \end{array}\right)\approx I-O(\kappa)$ and $A_1=\sqrt{\kappa} \sigma_-$. The codewords for the 4-qubit amplitude damping code are $\ket{\overline{0}}=\frac{1}{\sqrt{2}}(\ket{0000}+\ket{1111})$ and $\ket{\overline{1}}=\frac{1}{\sqrt{2}}(\ket{0011}+\ket{1100})$. 

We assume that each qubit in this code is subjected to amplitude-damping noise. We wish to approximately correct against the error set $E_0=A_0^{\otimes 4}$, $E_1=A_1 \otimes A_0^{\otimes 3}$, $E_2=A_0 \otimes A_1 \otimes A_0^{\otimes 2}, E_3=A_0^{\otimes 2} \otimes A_1 \otimes A_0, E_4=A_0^{\otimes 3} \otimes A_1$, corresponding to no damping and single qubit damping on any of the four qubits respectively. The authors in \cite{leung+:approx} show that this code obeys the QEC conditions approximately with $O(\kappa^2)$ corrections which is a quadratic improvement over the basic error rate $\kappa$. 

Clearly, when one uses an approximate error correction code, one can only approximately undo the errors. Determining an optimal recovery (defined as optimizing a worst-case or average-case fidelity) is more involved, see the most recent results on this code and the general approach in \cite{MN:approximate, BO:approximate}.


\subsubsection{Qubit-into-Oscillator Codes}
\label{sec:qiq}

Another interesting example is that of a single bosonic mode (with creation and annihilation operators $a^{\dagger}$ resp. $a$) which is used to encode a qubit in two orthogonal states which are approximately protected against photon loss. The damping process can be modeled with the Lindblad equation, Eq.~(\ref{eq:lindblad}), with $L=\sqrt{\kappa} a$ while $H=\omega (a^{\dagger}a+\frac{1}{2})$ (which we can transform away by going to the rotating frame at frequency $\omega$). One can choose two Schr\"odinger cat states as encoded states $\ket{\overline{0}_{+}}$, $\ket{\overline{1}_+}$ with
\begin{eqnarray}
\ket{\overline{0}_{\pm}}=\frac{1}{\sqrt{N_{\pm}}}\left(\ket{\alpha}\pm \ket{-\alpha}\right), \nonumber \\ 
\ket{\overline{1}_{\pm}}=\frac{1}{\sqrt{N_{\pm}}}
\left(\ket{i\alpha}\pm \ket{-i\alpha}\right).
\label{eq:cat}
\end{eqnarray}
Here $\ket{\alpha}$ is a coherent state $\ket{\alpha}=\exp(-|\alpha|^2/2) \sum_{n=0}^{\infty} \frac{\alpha^n}{\sqrt{n!}} \ket{n}$ and $N_{\pm}=2(1\pm\exp(-2|\alpha|^2)) \approx 2$. For sufficiently large photon number $\langle n \rangle=|\alpha|^2$, the states $\ket{\pm \alpha}, \ket{\pm i \alpha}$, and therefore $\ket{\overline{0}_+}$ and $\ket{\overline{1}_+}$, are approximately orthogonal (as $|\bra{\alpha} \beta \rangle|^2=\exp(-|\alpha-\beta|^2)$). 

The creation and manipulation of cat states has been actively explored, see an extensive discussion on cavity-mode cats in microwave cavities \cite{book:haroche}. The code states are chosen such that loss of a photon from the cavity maps the states onto (approximately) orthogonal states. As $a \ket{\alpha}=\alpha \ket{\alpha}$, we have
\begin{equation}
a \ket{\overline{0}_+}=\alpha \sqrt{N_-/N_+} \ket{\overline{0}_-}, \; a\ket{\overline{1}_+}=i \alpha\sqrt{N_-/N_+} \ket{\overline{1}_-}\;,
\end{equation}
with $\ket{\overline{0}_-},\ket{\overline{1}_-}$ defined in Eq.~(\ref{eq:cat}). The preservation of orthogonality is a prerequisite for these code states to be correctable. More precisely, one can verify that in the unphysical limit $|\alpha| \rightarrow \infty$ one obeys the QEC conditions \footnote{If we were to use two coherent states as code states, say, $\ket{\overline{0}}=\ket{\alpha}$ and $\ket{\overline{1}}=\ket{-\alpha}$, the QEC conditions would not be obeyed, as $\bra{\alpha} E_1^{\dagger} E_0 \ket{\alpha} \neq \bra{-\alpha} E_1^{\dagger} E_0 \ket{-\alpha}$ for any $\alpha$.}, Eq.~(\ref{eq:QEC}), for $E_0=\sqrt{\kappa} a$ and $E_1=I-\frac{\kappa}{2} a^{\dagger}a$. \\

The code space (spanned by $\ket{\overline{0}_+}, \ket{\overline{1}_+}$) is distinguished from the orthogonal erred space (spanned by $\ket{\overline{0}_-}$ and $\ket{\overline{1}_-}$) by the photon parity operator $\exp(i \pi a^{\dagger} a)=\sum_{n} (-1)^n \ket{n}\bra{n}=P_{even}-P_{odd}$. This parity operator has $+1$ eigenvalue for the even photon number states $\ket{\overline{0}_+}, \ket{\overline{1}_+}$ and $-1$ eigenvalue for the odd photon number states $\ket{\overline{0}_-}, \ket{\overline{1}_-}$. By continuously monitoring the value of the parity operator one could track the occurrence of errors \cite{HBM:parity, sun+:cat}. Better even would be the realization of a restoring operation which puts back an erred state with decayed amplitude $\alpha e^{-\kappa t/2}$ into the code space while restoring the amplitude back to $\alpha$. However such restorative process will always add noise to the codewords as it is physically impossible to increase the distinguishability between (decayed) non-orthogonal code words. Thus starting with cat states with finite $\alpha$, after repeated cycles of errors followed by, let's assume, perfect error detection and correction, the cat states will gradually lose their intensity and thus their approximate protection. In \cite{leghtas+:QEC, mirra+:cats} the interaction of superconducting qubits coupled to 2D or 3D microwave cavities (circuit QED) is proposed to be used for encoding, correction and decoding of such cat states while \cite{sun+:cat} shows how this is done in an experiment. \\

One can generalize the stabilizer formalism to continuous-variable systems characterized by an infinite-dimensional Hilbert space \cite{braunstein:CV, LS:CV}. Of particular interest are codes which encode a discrete amount of information, a qubit say, in a harmonic oscillator \cite{GKP}. In \cite{thesis:harrington} more general `symplectic' codes are constructed which encode multiple qubits in multiple oscillators.\\

Given are two conjugate (dimensionless) variables $\hat{p}$  and $\hat{q}$ which represent a generalized momentum and position, obeying $[\hat{q},\hat{p}]=i$. The idea is to encode the information such that {\em small} shifts in position or momentum correspond to correctable errors while logical operators are represented as large shifts. For a harmonic oscillator space, the Pauli group $P_n$ can be generalized to the Weyl-Heisenberg group generated by the unitary shift operators $\exp(i t \hat{p})$ and $\exp(i s \hat{q})$ for real $s$ and $t$. These operators form a basis for the space of operators and thus any error $E$ (any operator) can be written as $E=\int ds \int dt \,c(s,t) e^{i t \hat{p}} e^{i s \hat{q}}$ with complex coefficients $c(s,t)$. Small shifts in $p$ and $q$ may not a priori seem like a very natural noise  model, but one can show that generic errors of low rate and low-degree in $\hat{p}$ and $\hat{q}$ can be expanded into linear combinations of products of these shifts. One should compare this to the similar expansion of any low-weight error in terms of low-weight Pauli errors.\\

In order to define a qubit in this infinite-dimensional space we select a set of commuting check generators whose $+1$ eigenvalue space is two-dimensional. We can observe that the operators $\exp(i t \hat{p})$ and $\exp(i s \hat{q})$ commute if and only if $s t =0 \mod 2 \pi$: this follows from the fact that $e^A e^B=e^{[A,B]} e^B e^A$ when $A,B$ are linear combinations of $\hat{q}$ and $\hat{p}$. We will consider two examples. \\

In our first (trivial) example the code space is a single state. We choose $S_q=e^{2 i \hat{q}}$ and $S_p=e^{-i \pi \hat{p}}$ as commuting check operators and we seek the states that have eigenvalue $+1$ with respect to the $S_p$ check operator. When $S_p=1$ the eigenvalues of $\hat{p}$ are even integers.  We can define $\hat{n}=\hat{p}/2$ and $\hat{\phi}=2\hat{q}$ so that for $S_p=1$ we have $\hat{n}=0,\pm 1,\ldots$.

The eigenvalue of the commuting operator $S_q$ can be simultaneously fixed to be $e^{i \phi}$ so that we should identify  $\phi= \phi \mod 2 \pi$. Thus we have the state space of a quantum rotor which is described by conjugate variables $\hat{n}$ taking integer values and $2\pi$-periodic phase $\hat{\phi}$ with $[\hat{\phi}, \hat{n}]=i$. 

A physical realization of these degrees of freedom is the quantization of a superconducting circuit where $\phi$ is the superconducting phase (difference phase across a Josephson junction) and $\hat{n}$ represents the number of Cooper pairs (difference number of Cooper pairs across a Josephson junction). Fixing both the eigenvalues for $S_p$ and $S_q$ leads to a single state characterized by its superconducting phase $\phi \mod 2 \pi$. Small shifts $\exp(i \epsilon \hat{\phi})$ for small $\epsilon$ do not commute with $S_p$ and gets one out of the `Cooper pair' code space fixed by $S_p=1$. This in some sense represents the phase-stability of the superconducting state at a purely mathematical level.

If we want to use this state space to represent a qubit, one has to use (linear combinations of) such states characterized by their phase. For example, the Hamiltonian of the multi-level transmon qubit \cite{koch+:transmon} equals $H_{\rm transmon}=4E_C \hat{n}^2-E_J \cos(\phi)$ where $E_C$ ($E_J$) are the capacitive energy (resp. inductive energy). This Hamiltonian has been interpreted as that of a charged quantum rotor in a magnetic field in \cite{koch+:transmon}. 
The lowest two energy levels of the system can define a qubit, the transmon qubit. If we expand $\cos(\phi) \approx 1-\phi^2/2+\phi^4/4!$, one obtains the Hamiltonian of an anharmonic (Duffing) oscillator with eigenstates which are superposition of 
$\phi$ eigenstates. This type of qubit has thus no intrinsic protection against dephasing, i.e. the value of the energy-level splitting is affected by charge and flux noise (representable as linear combinations of small shifts in $\hat{p}$ and $\hat{q}$).\\

A different choice of $S_q$ and $S_p$ leads to a real code which encodes a single qubit and has built-in protection. We choose as checks the operators $S_q=e^{2 i \hat{q}}$ and $S_p=e^{- 2 i  \pi \hat{p}}$. Fixing the eigenvalues of these operators to be $+1$ leads to the discretization $\hat{p}=0,\pm 1,\pm 2 \ldots$ and again $\hat{q}$ should have eigenvalues which are multiples of $\pi$. Now there are two operators which commute with $S_q$ and $S_p$ but which mutually anti-commute: these are $\overline{Z}=e^{i \hat{q}}$ and $\overline{X}=e^{-i \pi \hat{p}}$. 

\begin{figure}[htb]
    \centering
 \includegraphics[width=1\hsize]{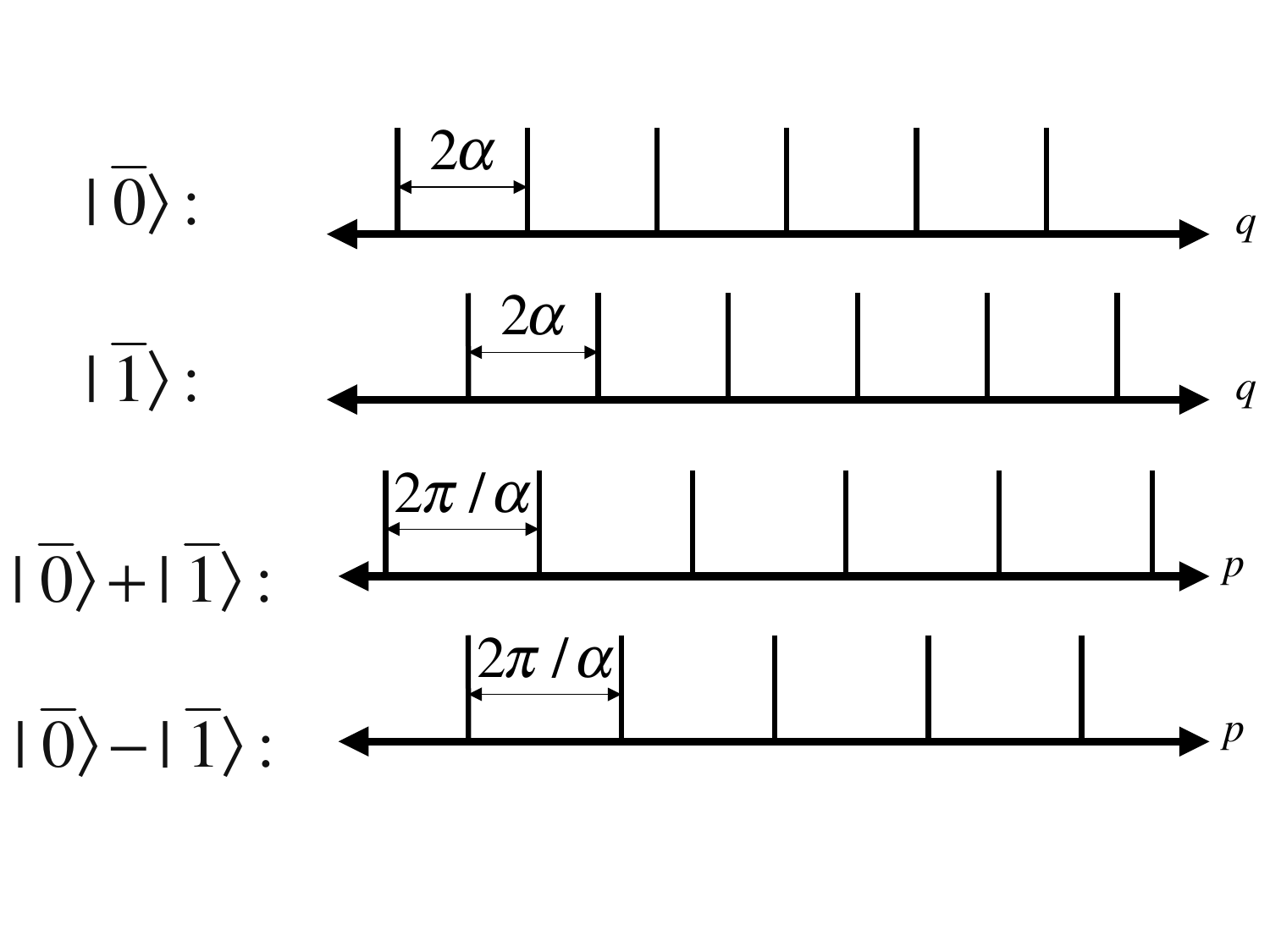} 
   \caption{Picture from \cite{GKP}: Amplitude of codewords for the stabilizer code with commuting checks $S_q(\alpha)=e^{2i \pi \hat{q}/\alpha}$ and $S_p(\alpha)=e^{-2i \hat{p} \alpha}$ which encodes a qubit in an oscillator.}
\label{fig:osc}
\end{figure}

The state $\ket{\overline{0}}$ (defined by $\overline{Z} \ket{\overline{0}}=\ket{\overline{0}}$ and $S_p \ket{\overline{0}}=\ket{\overline{0}}$) is  a uniform superposition of states with $\hat{q}=0,\pm 2\pi, \ldots$. Similarly, $\ket{\overline{1}}$ corresponds to a uniform superposition of $\hat{q}=\pm \pi,\pm 3 \pi, \ldots$, see Fig.~\ref{fig:osc} with $\alpha=\pi$. Consider the effect of shifts of the form $e^{i \delta \hat{p}}$ where $|\delta| < \pi/2$, which are correctable. Such shifts map the codewords outside of the code space as they do not commute with the stabilizer operator $S_q$. Error correction thus takes place by measuring $q \mod \pi$ and applying the smallest shift which resets $q=0 \mod \pi$. Similarly, the $\ket{\overline{+}}$ is a uniform superposition of states with $\hat{p}=0,\pm 2, \pm 4,\ldots$ while $\ket{\overline{-}}$ is a uniform superposition of states with $\hat{p}=\pm 1, \pm 3, \ldots$, see Fig.~\ref{fig:osc}.
The qubit is protected against shifts $e^{i \epsilon \hat{q}}$ with $|\epsilon| < 1/2$.\\

This code space can be viewed as the state space of a Majorana fermion qubit \cite{alicea:review} where $\hat{p}=\hat{n}$ counts the total number of electrons while $\hat{q}=\hat{\phi}$ is the $\pi$-periodic conjugate phase variable. The $\ket{\overline{+}}$ eigenstate of $\overline{X}$ with an even number of electrons correspond to the Majorana mode unoccupied while $\ket{\overline{-}}$ is the state with an odd number of electrons as the Majorana mode is occupied. The protection of the Majorana fermion qubit can thus also be understood from this coding perspective although the perspective sheds no light on how to physically realize this qubit nor does it shed light on the effect of noise which cannot be represented by shifts.
 
Another representation of this code space, which does not use Majorana fermion qubits, but superconducting circuits is the $0\mbox{-}\pi$ qubit (see e.g. \cite{kitaev:mirror}) which is designed such that the superconducting phase difference between terminals has degenerate energy minima at $0$ and $\pi$ corresponding to the approximate codewords $\ket{\overline{0}}$ and $\ket{\overline{1}}$.\\

More generally, we can parametrize this code by a real number $\alpha$ by taking the stabilizer checks as $S_q=e^{2 i \pi  \hat{q}/\alpha}$ and $S_p=e^{-2 i \hat{p} \alpha}$ (above we took $\alpha=\pi$). The logical operators are $\overline{Z}=e^{\pi i \hat{q}/\alpha}$ and $\overline{X}=e^{-i \hat{p} \alpha}$ \cite{GKP}, see the codewords in Fig.~\ref{fig:osc}. The code can correct against shifts $e^{i \epsilon \hat{q}}$ with $|\epsilon| < \frac{\pi}{2\alpha}$ and $e^{-i \delta \hat{p}}$ where $|\delta| < \frac{\alpha}{2}$.\\

One can use this code for encoding a qubit in a bosonic mode where $\hat{q}$ and $\hat{p}$ arise as quadrature variables, i.e. $\hat{q}=\frac{1}{\sqrt{2}}(a^{\dagger}+a)$ and $\hat{p}=\frac{i}{\sqrt{2}}(a^{\dagger}-a)$. The free Hamiltonian $H_0=\omega (a^{\dagger} a+\frac{1}{2})$ will periodically transform $\hat{q}$ into $\hat{p}$ and vice versa so it is natural to let $S_q$ be of the same form as $S_p$ and choose $\alpha=\sqrt{\pi}$.

It can be explicitly shown \cite{GKP} how errors such as photon loss  $L_- =\sqrt{\kappa_-} a$, photon gain $L_+=\sqrt{\kappa_+} a^{\dagger}$, dephasing (or decay) of the oscillator $e^{i \theta a^{\dagger} a}$ (or $e^{-\kappa a^{\dagger} a}$), or a non-linearity $e^{i K (a^{\dagger} a)^2}$ for sufficiently small parameters $\kappa_{\pm},\theta, K$ can be expanded into the small shift operators and can thus be corrected. The level of protection thus goes well beyond that of the cat state code.\\

However, the codewords of this code in Fig.~\ref{fig:osc} are not physically achievable as it requires an infinite amount of squeezing (and thus an infinite amount of energy) to prepare (superpositions of) of a quadrature eigenstates such as $\ket{q}$ or $\ket{p}$. \cite{GKP} has proposed to use approximate codewords, e.~g.~ the approximate codeword $\ket{\tilde{0}}$ is a superposition of Gaussian peaks in $q$-space, each one centered at integer multiples of $2\sqrt{\pi}$ with width $\Delta$, in a total Gaussian envelope of width $1/\kappa$. Viewed as a superposition of $p$-eigenstates, such state is a superposition of peaks with width $\kappa$ and total envelope of width $\Delta^{-1}$. An error analysis of this approximate encoding was done in \cite{GK:osc}, while \cite{VSG:all_opt} considered the preparation of the encoded states using cat states as in Eq.~(\ref{eq:cat}), squeezing and homodyne detection. \\

In \cite{menicucci:ft} the author shows how one can use a continuous-variable cluster state and homodyne measurements to perform quantum error correction on these approximate GKP (Gottesman-Kitaev-Preskill) codewords and realize a universal set of gates assuming that the noise is only due to the finite amount of squeezing in the preparation of the GKP codewords and the cluster state. For squeezing levels of 21dB, the author estimates that the (worst-case) effective gate error-rate is $10^{-6}$, sufficiently below the noise threshold of the surface code discussed in Section \ref{sec:surfcode}. In Section \ref{sec:surface_osc} we will consider a version of the surface or toric code which encodes an oscillator in a 2D coupled array of harmonic oscillators which can also be viewed as a way to concatenate the GKP code with the surface code.

\subsection{$D$-dimensional (Stabilizer) Codes}
\label{sec:Ddim}

Of particular practical interest are $D$-dimensional stabilizer codes. These are stabilizer code families on qubits located at vertices of some $D$-dimensional cubic lattice (with or without periodic boundary conditions). The parity checks involve $O(1)$ qubits which are within $O(1)$ distance of each other on this lattice where $O(1)$ means that this quantity is a constant independent of block size $n$. One can easily prove that one-dimensional stabilizer codes have distance $O(1)$, independent of block size \cite{BT:mem}, showing that without concatenation, such codes offer little fault-tolerant protection. Various two-dimensional topological stabilizer codes will be discussed in Section \ref{sec:topo}, while some 3D and 4D examples of topological codes are the Haah code \cite{haah:nostring}, the Chamon code \cite{BLT:chamon}, the 3D toric code \cite{CC:3Dtoric} and the 4D toric code \cite{dennis+:top}, to be discussed in Section \ref{sec:homo}.\\

There are of course many codes which are not captured by the stabilizer formalism. Here I would like to briefly mention the class of 2D topological qubit codes where the stabilizer checks are still commuting, but they are no longer simple Pauli operators. As Hamiltonians these correspond to the so-called 2D Levin-Wen models \cite{LW:models}, as codes they are called Turaev-Viro codes \cite{KKR}. The advantage of these codes which generalize the 2D surface code in Section \ref{sec:topo}, is that universal quantum computation can achieved by purely topological means. The disadvantage from the coding perspective is that (1) the stabilizer checks are more complicated as operators, e.g. for the so-called Fibonacci code on a hexagonal lattice, the stabilizer checks act on 3 and 12 qubits and (2) decoding and determining a noise-threshold for these codes has only recently begun \cite{wootton+:nonab, brell+:nonab}.

\subsection{Error Correction and Fault-Tolerance}
\label{sec:ec_ft}

We have understood from the previous sections that the crucial element of quantum error correction for stabilizer codes is the realization of the (parity) check measurement as in Fig. \ref{fig:paritycheck}. The immediate problem is that the parity check measurement suffers from the same imperfections and noise as any other gate or measurement that one may wish to do.  

In practice a parity check measurement may arise as a continuous weak measurement leaving a classical stochastic data record which hovers around the value $+1$ (pointing to the state being in the code space) while occasionally {\em jumping} to a value centered around $-1$, modeled using a stochastic master equation.
One can imagine that such continuously acquired record is immediately fed back to unitarily steer the qubits to the code space \cite{ADL:feedback}. The feedback cannot just rely on the instantaneously measured noisy signal but should integrate over a longer measurement record to estimate the current conditional quantum state of the system \cite{book:WM}. However, tracking the entire quantum state in real-time is computationally expensive and defeats the purpose of quantum computation. For the realization of quantum error correction, \cite{HM:filter} describes a filter in which one only tracks the probability that the quantum system at time $t$ is in a state with particular error syndrome ${\bf s}$ given the continuous measurement record in time. \cite{CLG:feedback_filter} improves on this construction by explicitly including the effect of the feedback Hamiltonian in the stochastic analysis.\\

Another model of feedback is one in which no (weak) measurements are performed and processed, but rather the whole control loop is a (dissipative) quantum computation. One could set up a simple local error correction mechanism by explicitly engineering a dissipative dynamics which drives/corrects the qubits towards the code space as proposed in \cite{barreiro+:stab, muller+:pumping}. We assume that the open-system dynamics of code qubits and environment is described by a Lindblad equation as in Eq.~(\ref{eq:lindblad}). For simplicity, let us consider the case in which we would like to {\em pump} or drive four qubits into a state with even parity so that the 4-qubit parity $Z$-check, $Z_1 Z_2 Z_3 Z_4$ has eigenvalue $+1$. Imagine that we can engineer the dissipation (in the interaction picture) such that there is a single quantum jump operator $L=\sqrt{\kappa} X_1 P_{odd}$ with $P_{odd}=\frac{1}{2}(I- Z_1 Z_2 Z_3 Z_4)$, the projector onto the odd parity space, and $H \propto Z_1 Z_2 Z_3 Z_4$. Integration of the Lindblad equation gives rise to the time-dynamics $\rho(t)=\exp(t {\cal L}_{tot})(\rho(t=0))$ with stationary states $\rho$ determined by ${\cal L}_{tot}(\rho)=0$. States supported on the even-parity subspace are `dark' states with ${\cal L}(\rho)=0$ and $[H, \rho]=0$. The odd-parity subspace is not stationary as the quantum jump operator $L$ flips the first qubit so that an odd parity state becomes an even parity state pumping the system towards the stationary dark subspace.

In \cite{muller+:pumping} one considers the following stroboscopic evolution using an ancillary dissipative qubit or mode which approximately gives rise to such Lindblad equation. The idea is to alternate (or {\em trotterize}) the coherent evolution with $H$ and the dissipative evolution with ${\cal L}$ for short periods of time $\tau$ so that $\exp(\tau {\cal L}_{tot}) \approx \exp(-i \tau [H,.]) \exp(\tau {\cal L})$. The dynamics of $H$ can be obtained by a small modification of the parity check measurement circuits in Fig.~\ref{fig:paritycheck}: for the evolution $\exp(-i \theta P)$ where $P$ is a multi-qubit Pauli operator we can use the circuit in Fig.~\ref{fig:paritycheck}(b). 

The dissipative evolution ${\cal L}$ could be implemented for short times $\tau \ll 1$ using a circuit consisting of a dissipative ancilla coupled to the four qubits as in Fig.~\ref{fig:paritycheck}(d). At the end of this circuit, instead of immediately measuring the ancilla qubit, we apply a CNOT with the ancilla qubit as control and qubit 1 as target (to change the parity of the odd states to even). This is then followed by natural dissipation of the ancilla qubit ($T_1$ process) so any amplitude in the $\ket{1}$ state is transferred to $\ket{0}$. This means that the ancilla qubit is effectively reset and can be used for the next round of application of $\exp(\tau {\cal L})$. 

These ideas of stabilizer pumping were experimentally tested on two and four ion-trap qubits in \cite{barreiro+:stab}. The use of this kind of extremely local feedback is limited as the dissipative evolution applies a correction only depending on the outcome of a single parity check whereas classical decoding in general makes decisions on the outcomes of many parity checks. We will continue the discussion on local decoders for the surface code in Section \ref{sec:disc_pract}. \\

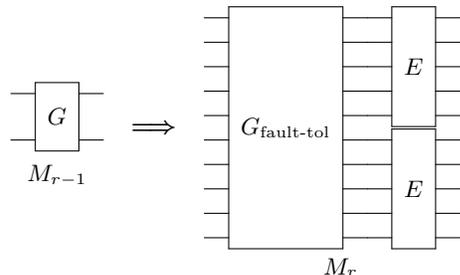
\begin{figure}[htb]
\centering
\parbox{150pt}{
\Qcircuit @C=1em @R=1em {
& \multigate{1}{G} & \qw  \\
& \ghost{G} & \qw\\
& \mbox{$M_{r-1}$} &
}
}
\parbox{32pt}{\centering $\boldsymbol{\Longrightarrow}$}
\parbox{150pt}{
\Qcircuit @C=1em @R=0.1em {
& \multigate{9}{G_{\textrm{fault-tol}}} & \qw & \multigate{4}{E} & \qw \\
& \ghost{G_{\textrm{fault-tol}}}        & \qw & \ghost{E}        & \qw \\
& \ghost{G_{\textrm{fault-tol}}}        & \qw & \ghost{E}        & \qw \\
& \ghost{G_{\textrm{fault-tol}}}        & \qw & \ghost{E}        & \qw \\
& \ghost{G_{\textrm{fault-tol}}}        & \qw & \ghost{E}        & \qw \\
& \ghost{G_{\textrm{fault-tol}}}        & \qw & \multigate{4}{E} & \qw \\
& \ghost{G_{\textrm{fault-tol}}}        & \qw & \ghost{E}        & \qw \\
& \ghost{G_{\textrm{fault-tol}}}        & \qw & \ghost{E}        & \qw \\
& \ghost{G_{\textrm{fault-tol}}}        & \qw & \ghost{E}        & \qw \\
& \ghost{G_{\textrm{fault-tol}}}        & \qw & \ghost{E}        & \qw \\
&                   \qquad \qquad \,\,\, \parbox[r][2em][s]{18pt}{\vfill $M_r$} & &
}
}
\caption{Code Concatenation: each qubit in the circuit on the left is replaced by an encoded block of qubits in the circuit on the right. The gate $G$ in the circuit $M_{r-1}$ is replaced by a rectangle consisting of the fault-tolerant encoded realization of the gate ($G_{\rm fault-tol}$) followed by error-correcting steps ($E$). The process can be repeated for every elementary qubit and gate in the new circuit $M_r$.}
\label{fig:recur}
\end{figure}

As any realization, closed or open-loop, of quantum error correction will suffer from inaccuracies there is no guarantee that one will improve coherence times by encoding a qubit in a code as it may introduce more errors that it takes away.  And if coding leads to a lower logical error rate, then how does one proceed to get an even lower logical error rate? In topological code families such as the surface code in Section \ref{sec:topo}, the logical error rate decreases exponentially as some function of the block size $n$, once one is below a critical error rate. This implies that the more qubit overhead one is willing to tolerate the smaller the logical error rate will be. Another way of obtaining a decreasing logical error rate is through recursively applied code concatenation of codes of a fixed block size $n$. The main ideas of this mathematical theory of quantum fault-tolerant computation by means of code concatenation are the following. \\

For simplicity, we assume that every elementary gate, idling step or measurement --these are called {\em locations} in the circuit-- can fail independently with some error probability $p$ (independent stochastic noise). In a concatenation scheme every qubit and operation in a quantum circuit is replaced by an encoded qubit and an encoded operation resp. and the process is recursively repeated. The encoded operation consists of an error correction step and a fault-tolerant realization of the operation, see Fig.~\ref{fig:recur}, which together constitute a {\em rectangle}. 

For a code such as Steane's $[[7,1,3]]$ code which can correct a single error, the fault-tolerance of the rectangle should be such that a single error in any of the locations of the rectangle cannot lead to two, {\em incorrectable}, errors in one code block. 
Then, if the elementary error rate scales as $p$, it follows that the encoded error rate scales as $ C p^2$ as two elementary errors are required for a logical error. Here $C$ is a constant which roughly counts the number of pairs of locations in the rectangle where failure can lead to a logical error. If $Cp^2 < p$ the concatenation step helps and $r$ steps of concatenation will drive down the error rate to $\sim p^{2^r}$ while the overhead in terms of qubits and gates increase only exponentially in $r$. The equality $Cp^2=p$ sets the {\em noise threshold} $p_c$. 

If we have a code with higher distance which can, say, correct $t$ errors, then fault-tolerance of a rectangle means that any error of weight $k \leq t$ in this rectangle spreads to at most $k$ qubits in a block. This will ensure that the logical error rate is $O(p^{t+1})$. \\

It is not trivial to make sure that a single error in a rectangle can lead to at most one error in the block for a code such as Steane's code. Consider the parity check circuit in Fig.~(\ref{fig:paritycheck})(c) which is used to measure the weight-4 $X$-check operators of this code where the data qubits are some subset of the 7 qubits. One needs to ensure that a single error on the ancilla qubit cannot spread to two errors in the block. However, a single $X$ error on the ancilla between the first and the last 2 CNOT gates will directly spread to two $X$ errors on the data. We can see this by commuting the Pauli $X$ on the ancilla through the two CNOT gates (and note that the $X$ error on the ancilla does not affect the outcome of the measurement). Thus the bare parity check measurement circuit is not fault-tolerant for the Steane code and one needs to modify this. Three methods have been devised to deal with making parity check circuits fault-tolerant. This first method is called Shor error correction which replaces the ancilla qubits by a $k$-qubit verified cat state $\frac{1}{\sqrt{2}}(\ket{00 \ldots 0}+\ket{11 \ldots 1})$ where $k$ is the weight of the check to be measured (see e.g. \cite{preskill:faulttol} for details). The second method is Steane error correction for CSS codes. In this method the ancilla is replaced by an encoded verified ancilla $\ket{\overline{0}}$ (or $\ket{\overline{+}}$) and a logical CNOT gate is executed between the encoded data qubit and the encoded ancilla \cite{steane:active, CDT:study}.  A third method is Knill error correction which uses quantum teleportation into a new encoded qubit such that the logical Bell measurement outcomes reveal the error syndrome \cite{knill:nature}. \\

The idea of repeated code concatenation was used in the early days of quantum error correction to prove the {\em Threshold Theorem} \cite{AGP:ft, AB:faulttol, KLZ:res, kitaev:survey}. This theorem says that fault-tolerant computation is possible with arbitrary small error rate if one is willing to tolerate an overhead which scales poly-logarithmically with the {\em size} $N$ of the computation to be performed (the size of a quantum circuit is the number of locations in it), that is

\begin{theorem}
An ideal circuit of size $N$ can be simulated with arbitrary small error $\delta$ by a noisy quantum circuit subjected to independent stochastic noise of strength less than $p < p_c$ where the noisy quantum circuit is of size $O (N (\log N)^{c})$ with some constant $c$.
\label{theo:ft}
\end{theorem}

It should be noted that this theorem assumes that `fresh' ancillas can be added during the quantum computation or quantum storage for doing parity check measurements. This means that these ancillas or qubit preparations have an error rate similar as other elementary components in the computation. The same assumption underlies the results on the asymptotic noise threshold for topological quantum error correction. Another assumption underlying the threshold results for concatenated and topological codes is that qubits can be acted upon in parallel. In practice simultaneous read-out or control of, say, multiple superconducting qubits using only a few microwave lines can be achieved by using qubits operating at sufficiently different microwave frequencies and frequency division multiplexing. \\


Another typical assumption is that classical processing of error information is fast and accurate imposing no delay in the execution of the quantum computation. We will return to the demands on classical processing in Section \ref{sec:backlog} and Sec \ref{sec:disc_pract}.\\

Practically relevant questions with respect to the threshold theorem are: how high is the value of the noise threshold $p_c$, how large is the constant $c$ and what is the value of the constant in $O(.)$. These numbers determine when quantum error correction will be useful and how large an overhead one should expect concretely. The constant $c$ in the Theorem roughly equals $c \approx \log_2 S$ where $S$ is the number of locations in a rectangle. 

The best performing concatenated coding scheme to date is the $C_4/C_6$ scheme of Knill \cite{knill:nature}. For this scheme which does assume non-local interactions between qubits, Knill has numerically estimated a noise threshold as high as $p_c \approx 3\%$ albeit at the cost of huge overheads. \cite{AGK:post} derives a rigorous lower bound of the noise threshold of this scheme of $0.1\%$. 

It was shown in \cite{gottesman:local} that the threshold theorem still holds if all interactions between elementary qubits are local on a one-, two- or higher-dimensional lattice. In such a scheme non-local interactions between elementary qubits are assumed to be realized via chains of noisy swap gates. This result means that, even though a one-dimensional quantum error correcting code (see Section \ref{sec:Ddim}) has a distance $O(1)$, one can use a small 1D code and concatenate it with itself to obtain a fully fault-tolerant one-dimensional scheme. The additional noisy movement via swap gates will negatively impact the noise threshold. For example, in an entirely 2D realization of the concatenated Steane $[[7,1,3]]$ code in which movement of data qubits via noisy swap gates is explicitly included \cite{SDT:local}, the fault-tolerant CNOT has $S=O(10^3)$ so that $c \approx 10$ demonstrating the potential inefficiency of code concatenation. For this scheme, the threshold was estimated as $p_c \approx 1.85 \times 10^{-5}$ while for the same non-local scheme the analysis resulted in a threshold of $3.61 \times 10^{-5}$.  These fairly low numbers should be contrasted with the noise threshold of about $1\%$ for the 2D surface code in Section \ref{sec:surfcode}. 

One may at first sight expect that the overhead incurred by code concatenation is worse than the overhead that is incurred with topological error correction (Section \ref{sec:topo}). One possible reason is that in topological quantum error correction parity check measurements are simply made robust by repeating the measurement needing no additional qubits. In contrast, in code concatenation the parity check measurements are realized using more complicated ancillas as in Steane error correction. However this picture is too simplistic: the comparative study in \cite{suchara:study} shows that for a computational task such as factoring the number 1024, concatenated Bacon-Shor codes perform better than the surface code at low error rates below $1 \times 10^{-7}$ while at high error rates the surface code performs better. Other studies of coding overhead for several families of codes were undertaken in \cite{steane:overhead} and \cite{CDT:study}. \cite{fowler:practical} estimates that in order to factor a 2000 bit number one needs about $10^{4}$ physical qubits per logical qubit using the double-defect encoding of the surface code described in Section \ref{sec:braiding}.\\

One can ask whether it is, in principle, possible to realize fault-tolerant computation with {\em constant overhead}, meaning that the number of qubits of the noisy fault-tolerant circuit scales with the number of qubits of the original circuit. This question was analyzed and answered in the affirmative in \cite{gottesman:overhead}. The fault-tolerant construction in \cite{gottesman:overhead} can be based upon any family of quantum LDPC (Low Density Parity Check) codes with constant {\em rate} $R=\frac{k}{n} \geq c$ and, loosely speaking, finite noise-threshold (when the block size $n \rightarrow \infty$) even if parity check measurements are faulty.  

LDPC stabilizer qubit codes are codes such that all parity checks (stabilizer generators) act on O(1) qubits, independent of block size. Several codes with such properties have recently been developed \cite{TZ:codes, FH:hypergraphs, GL:hyperbole} which have distances $d=O(n^{\alpha})$ with $0 < \alpha \leq 0.5$. For such LDPC codes it has been shown \cite{KP:badcodes} that having a distance scale as some function of $n$ guarantees the existence of a finite noise-threshold, assuming that we can do minimum weight decoding. In order to be of practical interest, decoding of such LDPC codes with 
constant rate should be computationally efficient. However, efficient minimum-weight decoders are not known to exist for quantum LDPC codes in general. In \cite{hastings:eff_decoder} it was shown how one can decode a 4D hyperbolic code with an efficient local decoder running in time $O(n \log n)$ to get a logical error rate $\overline{p} \sim p^{\log n}$ with $p$ representing a basic error rate, thus falling off only polynomially (instead of exponentially) with $n$. \\

It was proven in \cite{BPT:tradeoff} that 2D stabilizer codes (which are LDPC codes with qubits on a 2D regular lattice) obey the trade-off $kd^2 =O(n)$. This result demonstrates that 2D codes such as the surface codes discussed in Section \ref{sec:topo} do not allow for fault-tolerant computation with constant overhead. More generally, the results in \cite{BPT:tradeoff} show that any $D$-dimensional stabilizer code family which has distance scaling with lattice size will have a vanishing rate (when $n \rightarrow \infty$), showing that non-local parity checks (between $O(1)$ but distant qubits) on such lattices are necessary in order to achieve a constant overhead. We note that it is an open question whether there exist quantum LDPC stabilizer codes with constant rate and distance scaling as $n^{1/2+\beta}$ for some $\beta > 0$.

\subsection{Universal Quantum Computation}
\label{sec:tt}

In quantum error correction with stabilizer (subsystem) codes a special role is played by logical gates which are elements of the Clifford group. The Clifford group $\calC_n$ is a finite subgroup of the group of unitary transformations ${\cal U}(2^n)$ on $n$ qubits. It is defined as the normalizer of the Pauli group: ${\calC}_n=\{U \in {\cal U}(2^n)| \forall P \in \calP_n, \exists P',\; U P U^{\dagger}=P'\}$, meaning that its maps Pauli operators onto Pauli operators. An overcomplete set of generators of the Clifford group are the 2-qubit CNOT gate, the Hadamard $H$ gate, the phase gate $S$ \footnote{$H=\frac{1}{\sqrt{2}}\left(\begin{array}{cc} 1 & 1 \\ 1 & -1 \end{array}\right)$, $S=\left(\begin{array}{cc} 1 & 0 \\ 0 & i \end{array}\right)$, $T=\left(\begin{array}{cc} 1 & 0 \\ 0 & e^{i \pi/4} \end{array}\right)$.} and Pauli $X,Z$. Note that $S^2=Z$, so that $S$, $H$ and CNOT suffice to generate the whole group.

The Knill-Gottesman theorem \cite{gottesman:heisenberg} proves that one can efficiently classically simulate any quantum circuit which only employs gates from the Clifford group. One does this by tracking the stabilizer group, more precisely its generators, which has the input state of the quantum circuit as its unique $+1$ state. Every Clifford gate and measurement maps the stabilizer generators, which are Pauli operators, onto new stabilizer generators providing an efficient representation of the action of the quantum circuit. 
Thus if a quantum circuit with Clifford gates contains additional known Pauli errors, one can easily represent these Pauli errors by additional updates of the stabilizer generators in the classical simulation. \\

For universal quantum computation one needs additional gates such as the $T$ gate ($\pi/8$ rotation) $^{10}$.  Examples of universal gate sets are $\{H, T, \mbox{CNOT}\}$, $\{H, \mbox{Toffoli}\}$ and $\{H, \Lambda(S)\}$ where $\Lambda(S)$ is the two-qubit controlled-$S$ gate \footnote{ $\Lambda(S) \ket{b_1,b_2}=\ket{b_1} S^{b_1} \ket{b_2}$ for $b_1,b_2=0,1$.}. Even though Clifford group gates have no quantum computational power they can be used to develop a {\em quantum substrate} on which to build universal computation using stabilizer codes. This comes about by combining the following sets of ideas.

First of all, note that stabilizer error correction by itself only uses CNOT gates, preparations of $\ket{+},\ket{-},\ket{0},\ket{1}$ and measurements in the $Z$- and $X$-basis as  is clear from Fig.~\ref{fig:paritycheck}. The $T$, $\Lambda(S)$ and the Toffoli gate, each of which can be used with Clifford gates to get universality,  are special unitary gates as they map Pauli errors onto elements of the Clifford group. One can define a Clifford hierarchy \cite{GC:gatetele} $\calC(j)=\{U \in {\cal U}(2^n)| U \calP_n U^{\dagger} \subseteq C(j-1)\}$ such that $\calC(0)=\calC(1)=\calP_n$, $\calC(2)=\calC_n$. The $T, \Lambda(S)$ and Toffoli are thus members of $\calC(3)$. Such gates in $\calC(3)$ (and similarly gates in $C(j)$ for $j > 3$) can be realized with ancillas and Clifford group gates using quantum teleportation ideas \cite{GC:gatetele,zhou+:gatetele}. The idea is illustrated in Fig.~\ref{fig:gatereal} for the $T$ gate.

One teleports the qubit on which the $T$ gate has to act, prior to applying the gate, using the bottom one-bit teleportation circuit in Fig.~\ref{fig:onebit_tele}. We first put a $T$ gate at the end of that teleportation circuit so that the output is $T \ket{\psi}$. Now we want to modify this circuit and commute the $T$ gate backwards. In the quantum circuit we insert $I=T T^{\dagger}$ prior to the corrective Pauli $X$ so that we can use $T X T^{\dagger}=e^{-i \pi/4} SX$ \footnote{Note that in the quantum circuit gates are applied from the left to right while in equations gates are applied from the right to the left.}. Hence the correction (in case we measure $M_Z=-1$) is now the Clifford gate $SX$. As a last step, we note that the $T$ gate can be commuted through the control-line of the CNOT as both gates are diagonal in the $Z$-basis on the control qubit. In this way we obtain the circuit in Fig.~\ref{fig:gatereal}. Note that if we do not apply the correction, we obtain the state $XT^{\dagger} \ket{\psi}$.


\begin{figure}[htb]
\centering
\mbox{
\Qcircuit @C=0.5em @R=0.5em {
\lstick{\ket{0}} & \targ & \qw & \gate{Z} \cwx[1]&  \rstick{\ket{\psi}}\qw \\
\lstick{\ket{\psi}} & \ctrl{-1} & \gate{H} & \meter 
}
}
\end{figure}
\begin{figure}[htb]
\centering
\mbox{
\Qcircuit @C=1em @R=1em{
\lstick{\ket{0}} & \gate{H} & \ctrl{1} & \gate{X} \cwx[1] & \rstick{\ket{\psi}} \qw \\
\lstick{\ket{\psi}} & \qw & \targ & \meter \cwx[-1]
}
}
\caption{The so-called one-bit teleportation circuits \cite{zhou+:gatetele}. The measurement denoted by the meter is a measurement in the $Z$-basis and determines whether to do a Pauli on the output qubit: for outcome $M_Z=+1$ no correction is performed.}
\label{fig:onebit_tele}
\end{figure}

\begin{figure}[htb]
\centering
\mbox{
\Qcircuit @C=1em @R=1em {
&\lstick{\ket{0}} & \gate{H} & \gate{T} & \ctrl{1} & \gate{SX} & \rstick{T\ket{\psi}} \qw \\
&\lstick{\ket{\psi}} & \qw & \qw & \targ & \meter \cwx[-1] \gategroup{1}{1}{1}{4}{2em}{.}
}
}
\caption{Using the ancilla $T \ket{+}$ in the dashed box, one can realize the $T$ gate by doing a corrective operation $SX$.}
\label{fig:gatereal}
\end{figure}
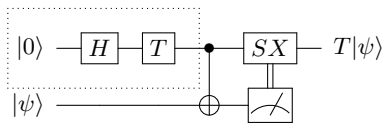

We can do the same trick for the $S = T^2$ gate, that is, we can reduce the $S$ gate to the preparation of a $\ket{\mbox{+}i}=\frac{1}{\sqrt{2}}(\ket{0}+i \ket{1})$ ancilla, a CNOT gate and a corrective Pauli $Y$. We get this from starting with the bottom circuit in Fig.~\ref{fig:onebit_tele} to which we apply the $S$ gate at the output. We insert $SS^{\dagger}$ in the quantum circuit before the corrective Pauli $X$ and use that $S X S^{\dagger} \propto Y$. We thus need the ancilla $S H \ket{0}=\frac{1}{\sqrt{2}}(\ket{0}+i \ket{1})$.

\subsubsection{Fault-Tolerant Logical Gates}

How do we realize a universal set of logical fault-tolerant gates for a code? Fault-tolerance means that such logical gates do not spread errors, ideally errors of weight $t$ remain errors of weight $t$. In principle, fault-tolerant gate constructions can be made for any stabilizer code \cite{thesis:gottesman}. The question is how to do computation with minimal resource requirements and overheads, that is, as close as possible to the resources needed for a quantum memory alone. Ideally, the computation threshold,  i.e. the performance of the code when used for computation, is close to the memory noise threshold, the performance of the code as a pure quantum memory. \\

An example of a gate which does not require additional qubits and does not spread errors is a {\em transversal} CNOT between two code blocks (each block encoding a single qubit into $n$ qubits). In such transversal CNOT  every qubit in the block is paired with a qubit in the other block in a CNOT gate such that the encoded CNOT is realized by doing $n$ two-qubit CNOTS in parallel. One can do a logical CNOT gate transversally for any CSS stabilizer code with ${\cal S}=\langle {\cal S}_1(X),{\cal S}_2(Z) \rangle$ \cite{thesis:gottesman}. One can understand this by observing that the product of the two stabilizer groups ${\cal S} \times {\cal S}$ of the encoded logical qubits is preserved under doing CNOTS between all elementary qubits in the blocks. Thus the code does not change by doing these gates. Secondly, one can always assume that the logical $\overline{X}$ of a CSS code is only a product of Pauli $X$s and the logical $\overline{Z}$ is only a product of Pauli $Z$s. Doing CNOTs transversally then has the same action on these logical operators as doing the CNOT on a pair of qubits.\\

In the CSS code construction when the classical codes $C_2=C_1=C$ (and thus the CSS constraint $C_2^{\perp} \subseteq C_1$ implies that $C^{\perp} \subseteq C$), the Hadamard gate $H$ on the code block encoding a single qubit is also transversal. For such code the stabilizer has an identical $X$ and $Z$-part, ${\cal S}=\langle {\cal S}(X),{\cal S}(Z) \rangle$. The gate $H^{\otimes n}$ maps these stabilizers onto each other and similarly $H^{\otimes n} \colon \overline{X} \leftrightarrow \overline{Z}$ as these operators have the same support. An example of a code with a transversal Hadamard and CNOT gate is Steane's $[[7,1,3]]$ code. \\

It has been shown \cite{EK:nogo} that if a quantum code can detect at least any error on a single qubit (meaning that it is a nontrivial code), then it does not have a transversally-realizable universal set of gates. A somewhat weaker version of this theorem, namely that qubit stabilizer codes do not allow for a universal set of gates to be realized via transversal unitary gates was proved in \cite{ZCC:trans}. 

In \cite{BK:uni} it was shown for any 2D stabilizer code that the logical gates which can be performed by constant-depth circuits employing only local gates (between neighboring qubits), are members of the Clifford group. The reason to focus on constant-depth local circuits is that such circuits are small, naturally fault-tolerant and provide a simple extension of the idea of a transversal gate. For a constant-depth local circuit any number of errors which occurs in the circuit will only affect a patch of $O(1)$ qubits on the 2D lattice, and such $O(1)$ error patches are correctable when the code distance scales with the lattice size. Hence we expect that such constant-depth implementation of gates does not negatively impact the noise threshold or qubit overhead. The result of \cite{BK:uni} is subtle as we can realize a fault-tolerant set of universal gates for any stabilizer code, but apparently we cannot do this by composing a sequence of constant-depth encoded gates. \\

The results of \cite{BK:uni} also hold if we try to do a gate by a constant-depth circuit while at the same time altering the stabilizer code to a new stabilizer code. Transforming one stabilizer code into a new one in a sequence of steps is sometimes called `code deformation'. The idea is that after the entire sequence of deformations one comes back to the original code but with a logical operation applied to the encoded qubits. Code deformation is a very useful concept for topological codes. As we will discuss in Section \ref{sec:topo}, one can use the code deformation technique to implement the logical $H$ and $\mbox{CNOT}$ in the 2D surface code.  For the surface code is not clear how one can do a logical $S$ gate in this manner, as $S^{\otimes n}$ maps the $X$-checks of the surface code onto $Y$-checks. The stabilizer code with $Z$-checks and $Y$-checks is not simply related to the original stabilizer code by some code deformation, translation or rotation.  For the surface code one can do the logical $S$ in the same fashion as the logical $T$, see below (for a different logical $S$ trick, see \cite{thesis:aliferis}). Other stabilizer codes, 2D color codes, have been found which do allow for an efficient fault-tolerant realization of the full Clifford group \cite{BM:colorcodes}.

\subsubsection{T gate}

The results in \cite{BK:uni} thus suggest that with 2D stabilizer codes it is not possible to realize a $T$ gate without large overhead. However, for gates such as the $T$ gate (or Toffoli gate, also in ${\cal C}(3)$) the method of magic-state-distillation has been developed \cite{BK:magicdistill}. This method shows how to realize these gates fault-tolerantly assuming that noiseless Clifford group operations and a supply of noisy unencoded $T \ket{+}$ ancillas are available. Thus, once we have built a low-noise Clifford computation substrate, universal quantum computation can be bootstrapped from it. In a nutshell, the ideas are as follows. We implement the $T$ gate at the logical level using Fig.~\ref{fig:gatereal} which requires the preparation of low-noise logical ancillas $\ket{A} \equiv \overline{T} \ket{\overline{+}}$. We can obtain such an ancilla in a non-fault-tolerant noisy manner by, for example, {\em injecting} several noisy unencoded ancillas into the code \cite{knill:nature}. From many of these noisy encoded ancillas we distill using logical $H$, CNOTs and measurements, a single low-error encoded ancilla. The strength of the distillation scheme is that the noise rate which one can tolerate on the unencoded $T \ket{+}$ ancillas for the scheme to work and produce very-low noise encoded $\ket{A}$ ancillas, is extremely high. Distillation can succeed if and only if the unencoded ancilla $\rho$ has a fidelity $F=\bra{A}\rho \ket{A}$ above approximately $0.854$ \cite{reichardt:distill}.\\

The downside of this scheme is that the qubit/gate overhead per logical $T$ gate is orders of magnitude larger than that of a `topological' CNOT (see e.g. Fig.~11 in \cite{RHG:threshold}). Current work is ongoing to design alternative schemes to reach universal computation with reduced overhead, see e.g. \cite{jones:dist}, \cite{thesis:jones} and references therein. 

Here it is worthwhile to mention a family of 3D color codes introduced in \cite{BM:transT} which allow for the transversal realization of the $T$ gate, thus requiring no additional qubit overhead. As these codes have stabilizers $\calS=\langle \calS_1(X), \calS_2(Z)\rangle$ one has a transversal CNOT. For the Hadamard gate one can use the gate teleportation ideas above if one can prepare an ancilla in a $\ket{\overline{+}}$ state. In Section \ref{sec:had} we will discuss how to prepare the state $\ket{\overline{+}}$ for the surface code; such a technique also works for these color codes. 

The smallest member of this class of color codes, $[[15,1,3]]$, is a quantum Reed-Muller code which has been known to have a transversal $T$ gate due to the special symmetry which is inherent in their construction via classical Reed-Muller codes \cite{steane:RM}. A possibly even more attractive family of 3D codes are the gauge color codes \cite{bombin:gauge} for which the Hadamard gate is transversal. By fixing the logical state of the gauge qubits, one obtains a 3D color code for which the $T$ gate is transversal. The idea of gauge-fixing as a means of getting around the Eastin-Knill theorem was first explored in \cite{PR:gauge}.

\subsubsection{The (Logical) Pauli frame}
\label{sec:backlog}

In this section we discuss how, during a fault-tolerant computation, one can handle the logical and elementary Pauli operators which are inferred from the syndrome measurement data. The idea is that the decoding procedure gives both a logical as well as a physical Pauli error which can be interpreted as a frame, the so-called Pauli frame \cite{knill:nature} which we can classically track during the quantum computation.

First of all, we note that in principle it is never necessary to physically correct Pauli errors to map back to the $+1$ eigenspace of the stabilizer ${\cal S}$. This is because any syndrome eigenspace of $\calS$ is a good code and we simply need to {\em know} which code space we are using. Secondly, note that if the quantum circuit entirely consists of Clifford gates, both at the logical as well as at the physical level, the classical information of the logical and physical Pauli frame does not necessarily need to be available during the execution of the circuit as the entire Pauli frame can be commuted through the circuit (as Clifford gates map Paulis onto Paulis). The Pauli frame simply alters the interpretation of the final measurement outcome of the computation. This is different if the circuit consists of non-Clifford gates as follows. \\

Imagine that syndrome data are collected and processed and every so often it is deduced that a logical (or physical) Pauli has happened on the coded data. It may also be that the quantum circuit we want to realize includes some (logical) Pauli operations. 
Consider what happens when a logical Pauli $X$ occurs on the data prior to doing a $T$ gate, as in Fig.~\ref{fig:gatereal}. The $X$ commutes through the CNOT gate and then effectively changes the way we should interpret the measurement outcome $M_Z$. Now, in case $M_Z=-1$ we have done $T \ket{\psi}$ and we don't need to do a correction. If $M_Z=+1$ we need to correct with $SX$. 

This means that the original Pauli $X$ is mapped onto a logical Clifford error on the data which will subsequently need to be corrected. If we do not correct the Clifford error it may spread and become a more complicated multi-qubit error which is even harder to correct. This implies that it is best to know the logical Pauli frame of the data qubit and the ancilla qubit before we move on to the next gate after the $T$ gate. 

Once we are done determining the logical Pauli frame, we will know which Clifford error took place and we should then try to correct right away. Concerning knowing the logical Pauli frame, what is important is that one only needs to only know the logical Pauli frame which influences the outcome of the measurement $M_Z$. For example, the results of parity check measurements after the CNOT on the ancilla qubit in Fig.~\ref{fig:gatereal} will not influence this logical Pauli frame and hence can be processed later.  The outcome of these measurements may of course cause a change in logical Pauli frame, but as $SX$ itself (but {\em not} controlled $SX$) is a Clifford gate, this change can be commuted through and remain a logical Pauli frame. 

If we know the logical Pauli frame on time, that is, before we move to a next gate, we can handle this logical Pauli frame in the classical control software as discussed for example in \cite{fowler:practical}. If we want to handle any logical Pauli in the classical control software, we just use this logical Pauli frame information to correct the interpretation of $M_Z$ (or $M_X$ measurement) thus changing which correction we do. In addition since we don't want to do any logical Paulis, we can replace the correction gate $SX$ by the correction $S$ (using that $S X T^{\dagger} \propto Y T$ so that we have realized $T$ modulo an additional $Y$).\\

In conclusion, one can argue that one does not need to physically implement any logical or elementary Pauli operation, but one does need to know the logical Pauli before one can proceed further with the computation. If determining the logical Pauli frame takes some time, as for example the quantum measurement is slow or the processing of the parity checks using classical computation is slow, one is thus required to wait before doing the classically-controlled $SX$ gate. This additional delay is not problematic as was observed in \cite{DA:slow} as parity checks are collected during this delay time and thus the qubits are protected.  \cite{DA:slow} has generally shown that a {\em slow} measurement will not lead to a lower noise threshold but can be accommodated by small modifications in the fault-tolerant (concatenated code) architecture. Slow measurement here means a quantum measurement with a long latency: the rate at which parity checks are collected is {\em not changed}, but it takes a while to measure an ancilla qubit which has been coupled to the data (as in Fig.~\ref{fig:paritycheck}). It may be clear that acquiring syndrome data at a slower rate will lead to a lower noise threshold as it effectively corresponds to a higher error rate. Thus in order to keep the rate of syndrome data acquisition high if the measurement of the ancillas is slow (say 10 times slower than the gate time) one has to couple 10 different ancillas to the data in sequence so that we get the measurement outcome of one of those 10 ancillas at the rate of (roughly) the inverse gate time.\\

Given these considerations concerning the logical Pauli frame, we see an important distinction between the complexity of building a quantum memory (including only Clifford gates) versus building a quantum fault-tolerant computer using stabilizer codes. In a fault-tolerant computer, one may allow for slow measurements with some latency, but the classical processing of the syndrome data record, the decoding, should never lead to a {\em increasing backlog} of syndrome data.  Let $r_{\rm proc}$ be the rate (in bauds) at which syndrome bits are processed and $r_{\rm gen}$ be the rate at which these syndrome bits are generated. We can argue that if $r_{\rm gen}/r_{\rm proc}=f > 1$, a small initial backlog in processing syndrome data will to {\em an exponential slow down} during the computation, by the following arguments.\\

Given that $f > 1$, there will be some time $t_0$ at which there is enough backlog in our syndrome record for us to have to delay executing the corrective gate after the $T$ gate as we don't know whether a logical Pauli happened which influences which correction we should do. Let $t_0^{\rm proc}$ be the time up to which we have processed the syndrome data at time $t_0$, so $\Delta_{\rm gen}=|t_0-t_0^{\rm proc}|$ is large enough so that it is likely that a logical Pauli error has happened in the time-interval $\Delta_{\rm gen}$. In this time interval we have generated an additional $D_1=r_{\rm gen} \Delta_{\rm gen}$ bits. We now process this record at a rate $r_{\rm proc}$, hence this takes time $\Delta_{\rm proc}= f\Delta_{\rm gen}$. The problem is that during this delay time $\Delta_{\rm proc}$ a new data record is generated of $D_2=\Delta_{\rm proc} r_{\rm gen}=D_1 f > D_1$ bits. If there was a sufficient possibility for a logical Pauli error in the original data record of size $D_1$, then this also holds for data record $D_2$. Hence at some next $T$ gate which is impacted by this Pauli frame information, we need to have at least processed the $D_2$ record. This implies again a delay in executing the gate during which one acquires a new data record $D_3=f D_2$ etc. Let us assume that the number of $T$ gates on a logical qubit is some polynomial in $n$, ${\rm poly}(n)$, e.g. for Shor's factoring algorithm $O(n^3)$ Toffoli gates are needed on $n$ qubits. Then the backlog data record that we have acquired at the $k={\rm poly}(n)^{\rm th}$ gate is $D_k= f^k D_1$ which is exponential in $n$. Hence in order to execute the $k^{\rm th}$ $T$ gate, one has to wait $r_{\rm proc} D_k$ time, an exponential amount of time in $n$.\\

The conclusion is that the syndrome data acquisition through quantum measurement and the classical processing should be fast enough to let the logical Pauli frame be `retarded' by only a constant amount of time, i.e. not increasing during the time of the computation. In order to achieve this one needs to decode using maximum classical parallelism, possibly using an on-chip decoder.  Whether the backlog question is a practical problem thus depends on how fast one can decode as compared to the physical error rate of the elementary qubits, see the further discussion in Section \ref{sec:disc_pract}. \\

We should contrast this backlog issue with the case of a quantum memory (including Clifford gates) in which the computation never has to wait for the classical processing of the logical Pauli frame. Such a stored qubit could be measured at the end of its storage time $T_{\rm store}$: in case of slow classical processing the outcome of the measurement may not be immediately available (as it depends on the syndrome record), but it would just mean that the computation, including the processing of syndrome data is finished in time $f T_{\rm store}$ which is just a constant slow-down.\\

The upshot of these considerations is that 2D and 3D stabilizer codes will be most suitable for building a quantum memory and performing Clifford group operations. The goal of universal quantum computation within the same platform can be reached using methods such as injection-and-distillation or using a code with a transversal $T$ gate but the additional overhead and complexity of distillation and demands for fast decoding are considerable.

\section{2D (Topological) Error Correction}
\label{sec:topo}

In this section we discuss several stabilizer/subsystem codes in which the parity checks act locally on qubits laid out on a 2D lattice. Before we discuss these codes, we make a few comments on noise models and noise thresholds.\\

In numerical or analytical studies of code performance, one uses simple error models such the independent {\em depolarizing noise model} to assess the performance of the code. Independent depolarizing noise assumes that every qubit {\em independently} undergoes a $X,Y$ or $Z$ error with equal probabilities $p/3$ and no error with probability $1-p$. Similarly, if qubits undergo single-, two-qubit gates or measurement and preparation steps, one assumes that the procedure succeeds with probability $1-p$ while with total probability $p$ (tensor products of) Pauli $X,Y,Z$ are applied. 

A related noise model is that of independent $X$ and $Z$ errors in which a qubit can undergo independently an $X$ error with probability $p$ and a $Z$ error with probability $p$ in each time-step. In all codes that we discuss in this section the parity checks are either $X$ or $Z$-like, detecting either $Z$ or $X$ errors. In addition, the parity $Z$- and $X$-checks have the same form, hence the simplest form of error correction is to correct $X$ and $Z$ errors in the same fashion but independently. For depolarizing noise, this means that we effectively neglect correlations between $X$ and $Z$ errors. It is also possible to decode the surface code taking these correlations into account, see \cite{fowler:correlated} and references therein.\\

We will consider codes which encode a single qubit in a block of $n$ qubits with $n=O(L^2)$ with $L$ the linear size of the 2D array. Several parameters can characterize the code performance.  One is the so-called pseudo-threshold $p_c(L)$ for which $p_c(L)=\overline{p}(p,L)$, i.e. the logical error rate equals the elementary error rate given a fixed block size. This logical error rate $\overline{p}(p,L)$ could be separately split into a logical, $X$, $Z$ or total, error rate, all being functions of the block size and the elementary error rate $p$.  In the definition of the pseudo-threshold we can assume that the elementary error rate is less than $50\%$ (otherwise the qubits would be completely randomized and no coding will help) and note that the logical error rate is also maximally equal to $50\%$. When the elementary error rate is less than the logical error rate, coding is not helpful. When the logical error rate is less than the elementary error rate, coding is helpful. The pseudo-threshold thus captures the crossover point. This crossover point will depend on $L$ and gives more information than the typically stated {\em asymptotic} threshold $p_c=\lim_{L \rightarrow \infty} p_c(L)$. In \cite{SCCA:flowmap} the behavior of pseudo-thresholds was considered for concatenated code schemes. \\

For the Bacon-Shor code in Section \ref{sec:BS}, the asymptotic threshold $p_c=0$, hence it is of interest to consider what is the optimal block size for this code. Another interesting class of 2D topological stabilizer codes are the color codes \cite{BM:colorcodes}. The color codes offer little practical advantage over the surface code if the goal is to build a quantum memory as some of the parity checks involve more than $4$ qubits. Having higher-weight parity checks negatively impacts the noise threshold as we assume each gate in the parity check measurement circuit can fail. This is likely to be the reason that the phenomenological threshold of the color code obtained as approximately $0.082\%$ in the detailed study in \cite{LAR:colorcodes} is lower than the surface code threshold (about $1\%$). The higher-dimensional color codes may be of interest in schemes for universal encoded computation, see the earlier discussion in Section \ref{sec:tt}.

\subsection{Surface Code}
\label{sec:surfcode}

The surface code is a version of Kitaev's toric code \cite{kitaev:top} in which the periodic boundaries of the torus have been replaced by open boundaries \cite{BK:surface, FM:surface}. Many of its properties and ideas for its use as a quantum memory were first analyzed in the seminal paper \cite{dennis+:top}. The topological 2D realization of the CNOT gate (Section \ref{sec:braiding}) was first proposed in \cite{BM:deform, RH:cluster2D}. 

There are several different ways of encoding and representing qubits in the surface code. We will start by discussing how to encode a single logical qubit in a sheet or patch and then show in Section \ref{sec:surg} how a CNOT can be performed between such qubits in patches using logical Pauli measurements. In Section \ref{sec:braiding} we discuss encoding single qubits into the surface code using so-called smooth and rough defects, here called smooth and rough qubits. The performance of this encoding in terms of logical error rate and overhead has been studied extensively in papers by Fowler {\em et al.}. The review \cite{fowler:practical} gives an excellent overview of how to do fault-tolerant quantum computation using this encoding. Another way of encoding qubits in the surface code is by means of pairs of distant dislocations \cite{bombin:twist}; the use of this scheme for fault-tolerant quantum computation has been analyzed in \cite{HG:dislocation}.\\

\begin{figure}[htb]
    \centering
    \includegraphics[width=1\hsize]{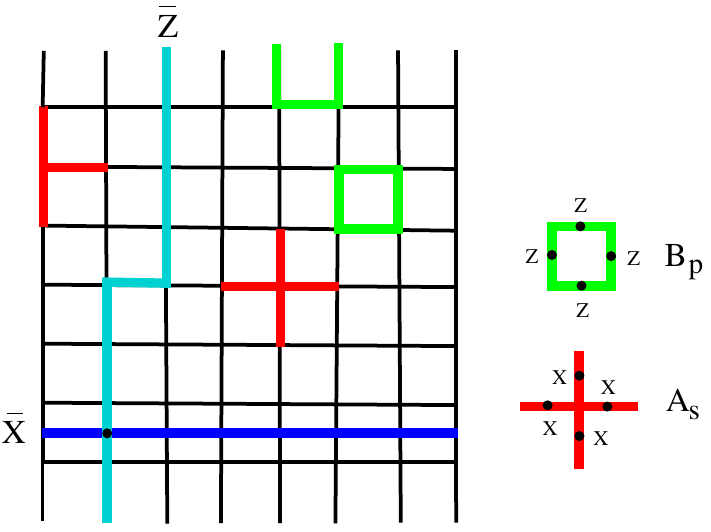} \caption{(Color Online) Surface code on a $L \times L$ lattice. On every edge of the (black) lattice there is a qubit, in total $L^2+(L-1)^2$ qubits (depicted is $L=8$). Two types of local parity checks, $A_s$ and $B_p$, each act on four qubits, except at the boundary where they act on three qubits. The subspace of states which satisfy the parity checks is two-dimensional and hence represents a qubit. $\overline{Z}$ is any $Z$-string connecting the north to the south boundary, which is referred to as `rough', while $\overline{X}$ is any $X$-string connecting the east to west `smooth' boundary running through vertices of the dual lattice.}
\label{fig:sur}
\end{figure}

A simple continuous sheet, depicted in Fig.~\ref{fig:sur}, can encode one logical qubit. The linearly-independent parity checks are weight-4 plaquette $Z$-checks $B_p$ and star $X$-checks $A_s$ which mutually commute and are modified at the boundary to act on 3 qubits, see Fig.~\ref{fig:sur}. Note that the star operators are just plaquette operators on the dual lattice when also interchanging $X \leftrightarrow Z$. The smallest surface code encoding 1 logical qubit which can correct 1 error is the code $[[13,1,3]]$ \footnote{One can minimize the qubit overhead while keeping the distance equal to 3 by rotating the lattice and chopping off some qubits at the corners to get a total of 9 qubits. This rotation+chopping, while leaving the distance unchanged, can be done for arbitrary-sized lattices, see e.g. \cite{horsman+:suture}.}. $\overline{Z}$ is any $Z$-string which connects the north and south {\em rough} boundaries; we can deform this string by multiplication by the trivially-acting plaquette operators. $\overline{X}$ is any $X$-string (on the dual lattice) connecting the {\em smooth} east and west boundary. As these strings have to connect boundaries in order to commute with the check operators, their minimum weight is $L$. Thus for general $L$, the code parameters are $[[L^2+(L-1)^2,1,L]]$. 

Using 13 qubits to correct 1 error does not seem very efficient, but the strength of the surface code is not fully expressed in its distance which only scales as the square root of the number of qubits in the block \footnote{One can prove that the distance of any 2D stabilizer code is at most $O(L)$ \cite{BT:mem}. However, one can also show \cite{BPT:tradeoff} that any block of size $R \times R$ where $R$ is less than some constant times the distance, is correctable, i.e. {\em all} errors in such $R \times R$ patch can be corrected. These arguments show that there are no other 2D stabilizer codes with better distance scaling and that this scaling allows one to correct failed blocks of size beyond the distance.}. \\

Kitaev's original toric code is defined on a 2D lattice with periodic boundary conditions (a torus). For the toric code, there is a linear dependency between all the $Z$-checks (the product of the $Z$-checks is $I$) and a similar linear dependency between all the $X$-checks. With this linear dependency it follows that the number of logical qubits is 2. The torus has two non-trivial loops: the logical $\overline{Z}_1$ is one non-trivial loop of $Z$s and the logical $\overline{Z}_2$ correspond to the other non-trivial loop of $Z$s. The matching logical $\overline{X}_1$ and $\overline{X}_2$ are similar loops running over the dual lattice. 

For the toric code it may be clear that the logical operators are directly connected to the homology of the torus. One can deform a logical $\overline{Z}$ by multiplying it with $Z$-checks $B_p$ but it will remain a non-contractible loop on the torus, as products of plaquette $B_p$ checks correspond to trivial, contractible, loops. This holds analogously for the $\overline{X}$ loops and products of star $A_s$ checks on the dual lattice.  

\subsubsection{Viewing The Toric Code as a Homological Quantum Code\footnote{Readers less interested in this mathematical framework can skip this section without major inconvenience.}}
\label{sec:homo}

The toric code is a simple example of a homological (CSS) quantum code \cite{kitaev:top, FM:surface, GL:hyperbole} in which the logical $\overline{Z}$ resp. $\overline{X}$ operators correspond to the homology and co-homology groups of the underlying manifold. In the surface code one can view the homology as being {\em relative} to a boundary \cite{BK:surface}. In this section we will discuss the framework of homological stabilizer codes and illustrate the concepts with the toric code in 2-, 3 and 4 dimensions. 

For the toric code one takes a flat two-dimensional manifold with periodic boundaries, a torus. One has to fix a triangulation of the manifold resulting in a so-called simplicial complex which consists of 0-simplices (vertices), 1-simplices (edges) and 2-simplices (faces) etc. The toric code just corresponds to taking a square lattice with faces which consist of four edges.  

In the general construction, with each type of object, e.g. vertices, edges or faces, or generally $i$-simplices, one associates a $\mathbb{Z}_2$-vector space $C_i$. Elements of $C_0$ are thus a collection of vertices, elements of $C_1$ are collections of edges etc.   In a $\mathbb{Z}_2$-vector space $C_i$, addition is mod 2. Two binary vectors $a$ and $b$ are orthogonal if and only if $\sum_i a_i b_i=0 \mod 2$ or the number of bits $i$ for which $a_i=b_i=1$ is even. If one represents such binary vector $a$ by a Pauli $X$ operator $P_X(a)=\Pi_i X_i^{a_i}$ and $b$ by a Pauli $Z$ operator $P_Z(b)=\Pi_i Z_i^{b_i}$, then the inner product between $a$ and $b$ is 0 iff $P_X(a)$ and $P_Z(b)$ commute.\\

For the toric code we associate the qubits with the 1-simplices (the edges), but for more general homological codes in higher dimensions one can associate qubits with $i$-simplices. The stabilizer generators and logical operators of the CSS code are then constructed using subspaces of the vector-space $C_i$ such that the required commutation relations between these operators hold and the logical operators directly relate to topological properties of the manifold. This comes about as follows. \\

One starts by defining boundary operators $\partial_i \colon C_i  \rightarrow C_{i-1}$ which act as the name suggests.  The boundary operator $\partial_2$ maps a face onto the collection of edges which are incident to the face, the boundary operator $\partial_1$ maps a collection of edges onto a collection of vertices, namely the end-points of these edges. For qubits associated with $1$-simplices, the $Z$-checks are obtained as the boundary space $B_1={\rm Im}(\partial_2)$, i.e. generating vectors in $B_1$ correspond to the boundary of a face. For the square lattice, these generators are thus the $Z$-plaquettes acting on the four edges of every face of the lattice. 

An important property of the boundary operator is that the boundary of an $i$-simplex 
does not have a boundary, mathematically this is expressed as $\partial_{i-1} \circ \partial_i=0$ applied to any vector in $C_i$.

A 1-cycle is defined to be a collection of edges without a boundary. This means that the vector space of 1-cycles equals $Z_1={\rm ker}(\partial_1)$. Any element of $B_1$ is a 1-cycle, or $B_1 \subseteq Z_1$, these are the trivial cycles which corresponds to products of the $Z$-check operators. The first homology group $H_1(T, \mathbb{Z}_2)=Z_1/B_1$ of the torus $T$ is generated by 1-cycles which are not the boundaries of plaquettes, i.e. the two non-trivial cycles around the torus. These cycles correspond to the logical $\overline{Z}$ operators. 

In the general construction when qubits are associated with $i$-simplices, the $Z$-checks correspond to the generators of $B_i={\rm Im}(\partial_{i+1})$,  the $i$-cycle space is $Z_i={\rm Ker}(\partial_i)$ and the $i$th homology group $H_i(M,\mathbb{Z}_2)=Z_i/B_i$ captures the logical $\overline{Z}$ operators. \\

In order to define the $X$-checks for the quantum code one does the same construction on the dual lattice or, equivalently, one uses cohomology. One can define the coboundary operator $\delta_i \colon C_i \rightarrow C_{i+1}$ which maps an $i$-simplex onto the set of $i+1$-simplices incident to it. Thus the coboundary operator $\delta_1$ maps an edge onto the faces which are incident to this edge, $\delta_0$ maps a vertex onto the edges emanating from it etc.  
With the coboundary operator one can define a cocycle space $Z^i={\rm Ker}(\delta_i)$ and a coboundary space $B^i={\rm Im}(\delta_{i-1})$. For qubits associated with $i$-simplices, the generators of $B^i$ will correspond to the $X$-checks and the generators of the $i$th cohomology group $H^i(M, \mathbb{Z}_2)=Z^i/B^i$ are the logical $\overline{X}$ operators. For the toric code one has $i=1$ and so the $X$-checks correspond to the generators of $B^1$, which are obtained from taking the edges incident to a vertex, hence the star operators. 

For the toric code it is easy to verify that the logical operators and the checks are all mutually commuting. For a general homological CSS quantum code, this essential property comes about from the fact that $\delta_i=\partial_{i+1}^T$ (where $T$ is matrix transposition if we view these linear maps as matrices acting on a finite-dimensional space). Then $B^i={\rm Im}(\delta_{i-1})={\rm Im}(\partial_i^T)=({\rm Ker}(\partial_i))^{\perp}=Z_i^{\perp}$ (and similarly $B_i=(Z^i)^{\perp}$). As $B_i \subseteq Z_i$, the spaces $B^i$ and $B_i$ will be orthogonal, so the check operators all commute and $B^i=Z_i^{\perp}$ implies that the $X$-checks commute with the logical $\overline{Z}$ etc.

Thus in general the $i$th homology groups $H_i(M, \mathbb{Z}_2)$ and cohomology groups $H^i(M,\mathbb{Z}_2)$ and their dimensions $\dim(H_i)=\dim(H^i)$ determine the number of logical qubits and also the character of the logical operators, meaning the dimensionality of their support (one-dimensional string-like or two-dimensional surface-like etc). \\

Instead of using the coboundary operator, one can also consider the dual of a simplicial complex of a $n$-dimensional manifold. Going to the dual means that an $i$-simplex is mapped onto a $n-i$-simplex, i.e. in two-dimensions a vertex becomes a face, while in three dimensions a vertex becomes a 3-simplex. For the toric code, a face (2-simplex) thus gets mapped onto a vertex (0-simplex) and vice-versa and edges (1-simplices) remain the same. In order to obtain the $X$-check operators and the logical $\overline{X}$, we can define boundary operators on the dual lattice.  If we associate qubits with $i$-simplices on the primal lattice, the boundary space $B_{n-i}^{\rm dual}$ generates the $X$-checks and $H_{n-i}^{\rm dual}$ (which is isomorphic to $H^i$) is generated by the logical $\overline{X}$ operators. \\

We can illustrate the construction with the 3D and the 4D toric codes defined on cubic lattices with periodic boundaries in all directions so that one has a 3-torus $T^3$ and a 4-torus $T^4$ respectively. For the 3D toric code \cite{CC:3Dtoric} the 3D cubes are the 3-simplices, their faces are the 2-simplices and we associate the qubits with the $1$-simplices or edges. For the 3D toric code, ${\rm Im}(\partial_2)$ is generated by the four-qubit $Z$-plaquette operators in $xy$, $xz$ and $yz$-planes. The logical $\overline{Z}$ operators are elements in $H_1(T^3,\mathbb{Z}_2)$, the three non-contractible $Z$-loops around the 3-torus.

If we take the boundary of the boundary of a 3D cube, i.e. apply the map $\partial_2 \circ \partial_3$ on the cube we get 0. This implies that product of $Z$ plaquettes which make up the boundary of the cube has no support on the edges, in other words the product of these stabilizer checks is $I$. This is a local linear dependency or a redundancy among the stabilizer checks which ensures that for any $X$ error, the $Z$-checks which are non-trivial form a connected string. To see this note that if one plaquette of the cube has nontrivial eigenvalue $-1$, some other plaquette of this cube must also have $-1$ eigenvalue as the product of all plaquettes which make up the cube is always $I$.  We can also understand this property as a Gauss' law for $\mathbb{Z}_2$-charges: $\mathbb{Z}_2$ flux-lines (lines of nontrivial syndromes) form closed loops which have no sources/do not terminate. This kind of redundancy is not present for the two-dimensional toric or surface code as no set of edges is the boundary of a boundary. We will discuss in Sections \ref{sec:disc_pract} and \ref{sec:self} how this redundancy and the lack thereof plays a role in the complexity of (locally) decoding and the question of self-correction and finite-temperature topological order.\\

Let us consider the $X$-checks of the 3D toric code.  The $X$-checks can be obtained from the coboundary operator $\delta_0$ (taking ${\rm Im}(\delta_0)$) which maps a vertex on the set of 6 edges emanating from this vertex. Hence the $X$-check is a star operator centered on a vertex acting on $6$ qubits. The logical $\overline{X}$ operators are elements in $H^1(T^3, \mathbb{Z}_2)$. i.e. they are a $xy$-oriented, $yz$-oriented or $xz$-oriented planes of $X$s on the dual lattice. 

We can observe that as the edges ($1$-simplices) become faces ($3-1$-simplices) on the dual lattice, one does not have a local linear dependency for the $X$-checks, as the faces are only the boundary of some three-dimensional objects and never the boundary of a boundary..... \\

One thus needs to go to four dimensions in order for there to be a local linear dependency for both $X$ and $Z$-checks. In the 4D toric code we associate qubits with the faces of a four-dimensional cubic lattice. Each $X$-check is associated with an edge such that the $X$-check acts on the qubits on the 6 faces which touch the edge (elements of ${\rm Im}(\delta_1)$). Similarly, the $Z$-checks are obtained as ${\rm Im}(\partial_3)$, i.e. as the collections of faces which forms the boundary of a three-dimensional cube. Hence the $Z$-check also acts on 6 qubits. The logical operators are associated with (co)homology groups $H_2(T^4,\mathbb{Z}_2)$ which has rank 6, so the code encodes 6 logical qubits, and 
$H^2(T^4,\mathbb{Z}_2) \simeq H_2(T^4, \mathbb{Z}_2)$. Now both $X$ and $Z$ checks have a local linear dependency, as $\partial_3 \circ \partial_4=0$ (the boundary of a four-dimensional cube is a collection of three-dimensional cubes which has no boundary) and $\delta_1 \circ \delta_0=0$. Both logical operators are surface-like (have a two-dimensional support) as they are elements in $H_2(T^4,\mathbb{Z}_2)$.

\subsection{Quantum Error Correction with the Surface Code}

We first consider how quantum error correction can take place for the surface code assuming that the parity check measurements are noise-free.
If a single $X$ error occurs on an edge in the bulk of the system, then the two plaquette operators next to it will have eigenvalue $-1$. The places where these plaquette eigenvalues are $-1$ are sometimes called defects. A connected string of $X$ errors will produce only two defects at its boundary. If the $X$ error rate $p$ per qubit is sufficiently small, one obtains a low {\em density} of close-by defects. Such errors are correctable as defects can be locally paired without much ambiguity.  As we know, inferring an error $E'$ which differs from the real error $E$ by only stabilizer operators (plaquette operators in this case) is harmless. Here it means that we decode correctly as long as $E' E$ does not form an $X$-string ($\overline{X}$) which goes from one smooth boundary to the other smooth boundary. For sufficiently low rate, the operators $E'E$ will instead form small closed loops which are products of check operators. From this picture it may be intuitively clear that there should be a {\em finite} asymptotic threshold $p_c$ for noise-free error correction.\\

For the bulk system the error syndrome thus gives us the location of the defects. A minimum-weight decoding algorithm then corresponds to finding a minimum-weight error string $E(X)$ which has these defects as end-points. This decoding algorithm can be implemented using Edmond's minimum-weight matching (Blossom) algorithm \cite{edmonds}. Open-source software Autotune has been developed specifically for surface code decoding \cite{fowler+:software} and is available on GitHub. We should note that at the boundary of the lattice an error can produce a single defect: in a minimum-weight matching decoder one can match these defects with a possible ghost defect beyond the boundary. In \cite{FWH:timing} it was demonstrated {\em empirically} that the number of steps in the minimum-weight matching algorithm scales as $O(L^2)$ per round of error correction.

Ideal decoding is not minimum weight decoding, but maximum-likelihood decoding as described in Section \ref{sec:decode}. As we argued in Section \ref{sec:decode}, one can estimate the maximally-achievable threshold $p_c$ with any decoder by relating the maximum likelihood decoding problem to a phase transition of a classical Hamiltonian with quenched-disorder. For the surface code this Hamiltonian is the 2D random-bond Ising model \cite{dennis+:top, WHP:threshold}. Assuming noise-free parity checks and independent $X$ errors with probability $p$, the critical value has been numerically estimated as $p_c \approx 11\%$ \cite{dennis+:top}. For a depolarizing noise model with error probability $p$, this threshold has been shown to increase to $p_c \approx 18.9\%$ in the numerical study in \cite{bombin+:depol_transition}. 

These thresholds for independent $X$ errors or depolarizing noise are extremely high as they come close to what one can achieve by using random codes as is expressed by the Hashing bound. The Hashing bound for a depolarizing channel equals $1-H_{\rm depol}(p)$ where $H_{\rm depol}(p)$ is the Shannon entropy of the depolarizing channel, $H_{\rm depol}(p)=-(1-p)\log_2 (1-p)-p \log_2 (p/3)$. This bound equals 0 at $p_c \approx 18.9\%$. \\

This picture gets modified when the parity checks are inaccurate. A simple way to model noisy parity checks is to assign a probability $q$ for the parity check outcome to be inaccurate while in between the parity checks qubits undergo $X$ and $Z$ errors with probability $p$ as before. In practice, one would expect the parity check measurements to induce some correlated errors between the qubits of which we take the parity. For example, for the parity $Z$-check one may expect additional qubit dephasing if more information than merely the parity is read out.\\

As the parity check measurements are no longer reliable, one needs to change their use as an error record. For example, a single isolated defect which appears for a few time-steps and then disappears for a long time is likely to be caused by a faulty parity measurement outcome instead of an error on the data qubits. The strength of topological codes for sufficiently large $L$ (as compared to using small codes and code concatenation) is that noisy parity checks can be dealt with by repeating their measurement as the additional noise which the parity checks produce on the code qubits is local and, at sufficiently low rate, correctable.\\

\begin{figure}[htb]
    \centering
    \includegraphics[width=1\hsize]{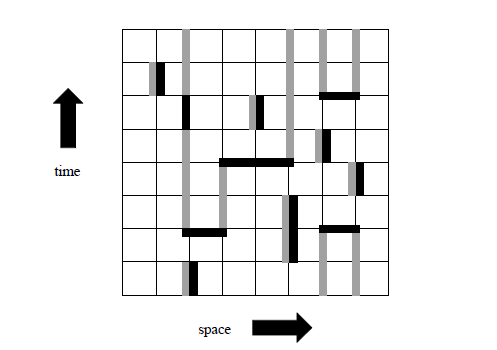} 
   \caption{Picture from \cite{dennis+:top}: 1D cross-section of the lattice in space, and time. Grey links correspond to non-trivial $-1$ syndromes. Errors which could have caused such a syndrome are represented by black links. Horizontal black links are qubit errors while vertical black links are parity check measurement errors. Note that a possible error $E$ has the same boundary as the gray defect links: a likely error $E$ (in the bulk) can be found by looking for a minimum-weighted matching of the end-points of the gray links.}
\label{fig:3ddec}
\end{figure}

Both minimum weight decoding and maximum likelihood decoding can be generalized to the noisy parity check measurement setting. We extend the lattice into the third (time) dimension \cite{dennis+:top}, see Fig.~\ref{fig:3ddec}. Vertical links, corresponding to parity check measurements, fail with probability $q$ while horizontal links fail with probability $p$. In minimum weight decoding the goal is now to find a minimum weight error $E$ which has vertical defect links, where the parity check is $-1$, as its boundary, see Fig.~\ref{fig:3ddec}. 
When we match the defects in 3D we obtain an inferred error $E'$ which can have a vertical time component (corresponding to a measurement error) as well as horizontal space components (corresponding to qubit errors). We can visualize the difference between errors $E$ which get properly corrected and errors for which decoding fails, by considering $E' E$. When error correction succeeds, $E'E$ is a trivial loop in the 3D lattice, but decoding fails when $E' E$ is some non-trivial space-time loop which winds around the torus (or for the surface code which connects the proper two boundaries).

If the parity check measurements are ongoing, one needs to decide how long a time-record to keep in which one matches defects  in the time-direction; this length depends on the failure probability $q$. In the simple case when $q=p$ the record length is taken as $L$ \cite{WHP:threshold}.

An analytical lower bound on the noise threshold for $q < p$ is derived in \cite{dennis+:top} with the value $p_c \geq 1.1\%$. Numerical studies in \cite{WHP:threshold} (using minimum weight-decoding) show a threshold of $p_c \approx 2.9\%$ for $p=q$.

If we assume that the parity check measurement errors are due to depolarizing noise on all elementary gates, measurement and preparations with depolarizing probability $p$, \cite{RHG:threshold} finds a threshold of $0.75\%$. Below the noise threshold the logical error rate $\overline{p}(p,L) \sim  \exp(-\kappa(p) L)$ where $\kappa(p) \approx 0.8-0.9$ at $p=p_c/3$ \cite{WHP:threshold, RHG:threshold}. \cite{wangetal:threshold} even estimates the depolarizing noise threshold to be in the range of $1.1-1.4\%$. All these results have been obtained for toric codes, assuming periodic boundary conditions: one may expect results to be somewhat worse for surface codes \cite{fowler:surface_bad}.

These results indicate that the surface code, even with noisy parity check measurements, has a very high threshold as compared to other coding schemes, see e.g. those studied in \cite{CDT:study} \footnote{One has to be careful in comparing noise threshold values across publications as slightly different methods, noise model, decoding strategy, code usage can impact the results.}. A practically relevant question is how much overhead $L$ is needed before one is in the scaling regime where the pseudo-threshold is close to the asymptotic threshold $p_c(L) \approx p_c$? 

The pseudo-threshold for a small code such as $[[13,1,3]]$ is very tiny, {\em certainly} no higher than $0.1\%$. Using the results in \cite{fowler:surface_bad}, one can estimate that the $[[25,1,4]]$ ($L=4$) surface code has a pseudo-threshold (defined by $\overline{p}=A p^{L/2}$ with $A=A_X,A_Z$ given in Table I in \cite{fowler:surface_bad}) of approximately $0.2\%$ and $[[61,1,6]]$ has a pseudo-threshold of approximately $0.7\%$. Thus with a depolarizing error rate $p=5 \times 10^{-4}$ [[25,1,4]] gives a logical $X$ or $Z$ error rate $\overline{p}_X \approx \overline{p}_Z \approx A p^2 \approx 1 \times 10^{-4}$ which is barely lower than the bare depolarizing rate. Even though small surface codes have worse performance than large codes they could still be used as testbeds for individual components and error scaling behavior.

Minimum-weight decoding with Edmonds' matching algorithm is a good decoding method if our goal is to realize a quantum memory (with or without encoded Clifford group operations). As one never needs to physically do any correction (see the notion of Pauli frame discussed in Section \ref{sec:tt}), the measurement record can be stored and the data record can be processed at leisure and used to interpret a final $M_{\overline{X}}$ or $M_{\overline{Z}}$ measurement on the qubits. The realization of such quantum memory will require that the record of parity check measurements is obtained at sufficiently high rate compared to the error rate, since a low rate stroboscopic picture of the defects (even if they are obtained perfectly) could potentially miss the occurrence of a logical error. In \cite{fowler:practical} the authors propose a 200 nanosec surface code cycle time (based on a 10-100 nanosec elementary gate time) meaning that every 200 nanosec both X and Z parity check measurements over the whole lattice are executed.\\

Researchers have developed potentially more efficient renormalization-group decoders \cite{DP:3Ddecoding, BH:3Dcube} which process the defects using parallel processing over the 2D or 3D lattice in time $O(\log L)$ (not taking into account a finite speed of communication). The idea of the simple decoder in \cite{BH:3Dcube} which works for any $D$-dimensional stabilizer code is to recursively match defects locally. For a 2D surface code with perfect parity check measurements, one starts by dividing up the defect record into local clusters of O(1) size. In each cluster the algorithm tries to find a local error which removes the defects. If a cluster contains a single defect for example, then no such local error can be found. Thus the next step is to enlarge the linear size of the cluster by a factor of $2$ and reapply the same procedure on the leftover defect record. The decoder stops when no more defects are present or when one has a reached a certain maximum number of iterations $r=O(\log L)$. For the toric code with perfect parity checks, \cite{BH:3Dcube} has obtained a noise threshold of $p_c=6.7\%$ using this RG decoder while the RG decoder in \cite{DP:renorm} achieves $9\%$ (minimum-weight decoding via matching gives $10.3\%$). \\

As we mentioned before, there are various ways in which we can encode multiple qubits in the surface code and do a logical Hadamard and CNOT gate. The simplest method is to encode multiple qubits in multiple separate sheets  (as in Fig.~\ref{fig:sur}) laid out next to each other in a 2D array as in Fig.~\ref{fig:horsman}. Using operations on a single sheet one can do a logical Hadamard gate, Section \ref{sec:had}. A CNOT between qubits in separate sheets can be realized using the idea of lattice surgery in which sheets are merged and split as proposed in \cite{horsman+:suture}. 

The important point of doing a CNOT and Hadamard gate using these code deformation methods is that their implementation does not affect the surface noise threshold as error correction is continuously taking place during the implementation of the gates and the single qubit noise-rate is not substantially changed. In addition, the realization of these gates does not require a large overhead in terms of space, meaning additional qubits, but the gates do require some overhead in time, as compared to transversal or constant-depth gates. 

Another method of encoding qubits is to have one sheet for all qubits in the computation such that logical qubits are represented by holes in the lattice, see Section \ref{sec:braiding}. Given this encoding, it is possible to disconnect and then deform the encoding of a single qubit so that it becomes a single disconnected sheet (see details in \cite{fowler:practical}) on which we can do the Hadamard gate or do a preparation or measurement step. 

\subsubsection{Preparation and Measurement of Logical Qubits}
\label{sec:prep}

How do we prepare the surface code memory in the states $\ket{\overline{0}}, \ket{\overline{1}}$ or $\ket{\overline{\pm}}$? And how do we read out information, that is, realize $M_{\overline{X}}$ and $M_{\overline{Z}}$ ? In order to prepare $\ket{\overline{0}}$, we initialize all elementary qubits in Fig.~\ref{fig:sur} to $\ket{0}$ and start measuring the parity checks. The state $\ket{00 \ldots 0}$ has $B_p=+1$ and $\overline{Z}=+1$ while the star operators $A_s$ have random eigenvalues $\pm 1$ corresponding to the presence of many $Z$ errors. Thus we choose some correction $E$ for these $Z$ errors (we pick a Pauli frame): the choice will not matter as $E$ commutes with $\overline{Z}$. If the preparation of $\ket{0}$ and the parity check measurements are noisy, one needs to measure the parity checks for a while before deciding on a Pauli frame for both $X$ and $Z$ errors. The preparation of $\ket{\overline{1}}$ and $\ket{\overline{\pm}}$ can be performed analogously using the ability to prepare the elementary qubits in $\ket{1}$ and $\ket{\pm}$ respectively. Instead of preparing the quantum memory in one of these four fixed states, there are also methods for encoding a single unencoded qubit $\ket{\psi}$ into a coded state $\ket{\overline{\psi}}$, see \cite{dennis+:top, horsman+:suture}. Of course, during this encoding procedure, the qubit to be stored is not fully protected, as the qubit starts off in a bare, unencoded state. \\

A projective destructive measurement in, say, the $\ket{\overline{0}},\ket{\overline{1}}$-basis ($M_{\overline{Z}}$) proceeds essentially in reverse order. One measures all qubits in the $Z$-basis. Using the past record of parity $Z$-check measurements and this last measurement, one infers what $X$ errors have taken place and corrects the outcome of $\overline{Z}=\pm 1$ accordingly.

\subsubsection{Hadamard Gate}
\label{sec:had}

\begin{figure}[htb]
\includegraphics[height=3cm]{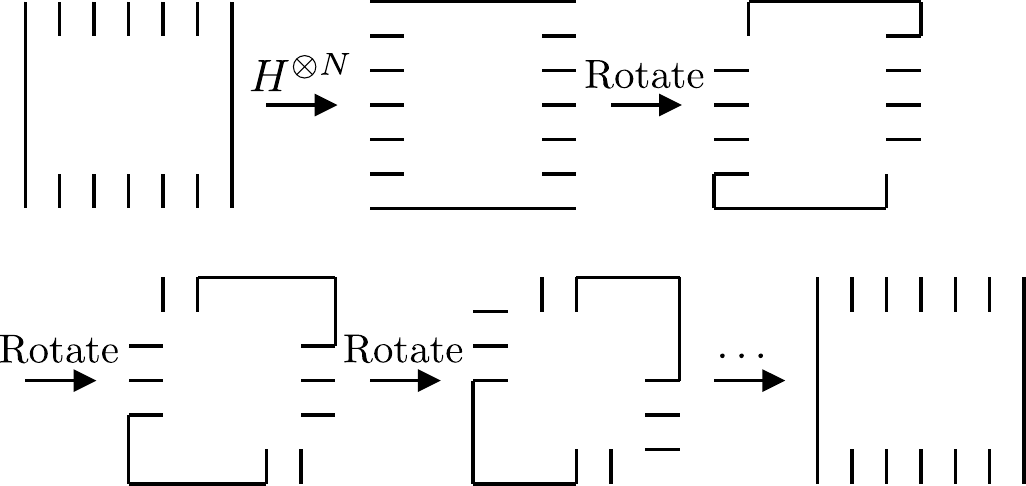}
\caption{Sketch of the Hadamard gate on a sheet which encodes a single logical qubit as in Fig.~\ref{fig:sur}, see also \cite{horsman+:suture}. After doing a Hadamard gate on each qubit, the rotated code lattice is gradually rotated back to its original orientation by adding and taking away qubits and stabilizer checks at the boundaries.}
\label{fig:hadamard}
\end{figure}

Consider doing a Hadamard rotation on every elementary qubit on a sheet encoding one logical qubit. The resulting state is a $+1$ eigenstate of the Hadamard-transformed parity checks $ H A_s H^{\dagger}$ (and $H B_p H^{\dagger}$) which are the plaquette $Z$-check (resp. the star $X$-check) of the code ${\cal S}_{dual}$ defined on the dual lattice. The dual lattice is defined by placing a vertex on each plaquette in Fig.~\ref{fig:sur} and connecting these vertices by edges on which the qubits are defined. On the dual lattice the rough and smooth boundaries are thus interchanged so that the lattice (code) is effectively rotated by $90^\circ$. The Hadamard gates map $\overline{Z}$ onto $\overline{X}_{dual}$ and $\overline{X}$ onto $\overline{Z}_{dual}$.  We have thus performed a Hadamard transformation but we have also rotated our code. In principle one can work with this rotated code as long as we can connect the qubits of another, say, non-rotated sheet, with this rotated sheet via some long-range interactions. However, it is more practical if we rotate the code back to its initial orientation. The original procedure described in \cite{dennis+:top} shows how by a sequence of ancilla preparations at the boundaries and local CNOT gates one can modify the boundaries so that a rough boundary becomes smooth again and vice versa. In this procedure one removes qubits from the code at the west and south boundary and one add qubits at the north and east boundary so that the lattice is effectively shifted upwards. Instead of doing CNOT gates one can add ancilla qubits and immediately measure the new plaquette and star operators. What is important is that this rotation over 90 degrees can only be done gradually, in $O(L)$ steps as is sketched in Fig. \ref{fig:hadamard} so that the distance between two rough boundaries or two smooth boundaries remains $L$ (so as to protect the encoded qubit). The overall shift of the lattice can be repaired by either swapping qubits in the SW direction or just by keeping the shifted lattice and making sure other sheets connect to qubits on the shifted sheet. It means that some flexibility/non-locality in coupling structure is required at these boundaries. \\

It is simple to show that the Hadamard gate for the surface code requires a quantum circuit of depth scaling with $L$. For the Hadamard gate we have $\overline{H} \overline{X} \overline{H}=\overline{Z}$ and $\overline{H} \overline{Z} \overline{H}=\overline{X}$ and $\overline{X}$ and $\overline{Z}$ are the strings going from boundary to boundary.
Imagine $\overline{H}$ implemented by a constant-depth circuit; it implies that  $\overline{H} \overline{Z} \overline{H}$ is a fattened (by some constant factor) string going from the north to south boundary in Fig.~\ref{fig:sur}. The original $\overline{Z}$ was any string going from the north to south boundary and hence it is simple to see that the operator $\overline{H} \overline{Z} \overline{H}$ will commute with the original $\overline{Z}$ and therefore $\overline{H} \overline{Z} \overline{H}$ cannot be the logical $\overline{X}$ operator. This argument only fails when the depth of the circuit is of order $L$ so that the string might be completely spread over the lattice. 

\subsubsection{CNOT gate via Lattice Surgery}
\label{sec:surg}

\begin{figure}[htb]
\centering
\parbox{150pt}{
\Qcircuit @C=1em @R=1em {
& \qw & \multigate{1}{M_{ZZ}} & \qw  & \qw \\
\lstick{\ket{0}} & \multigate{1}{M_{XX}} & \ghost{M_{ZZ}} & \gate{H} & \meter \\
& \ghost{M_{XX}} & \qw & \qw & \qw
}
} \makebox[60pt][c]{=} \parbox{150pt}{
\Qcircuit @C=1em @R=3.5em {
& \ctrl{1} & \qw \\ 
& \targ & \qw
}
}
\caption{CNOT via 2-qubit quantum measurements. Here $M_{XX}$ measures the operator $X \otimes X$ etc. The ancilla qubit in the middle is discarded after the measurement disentangles it
from the other two input qubits. Each measurement has equal probability for outcome $\pm 1$ and Pauli corrections (not shown, see Eq.~(\ref{eq:cnot_cor})) depending on these measurement outcomes are done on the output target qubit.}
\label{fig:CNOTid}
\end{figure}
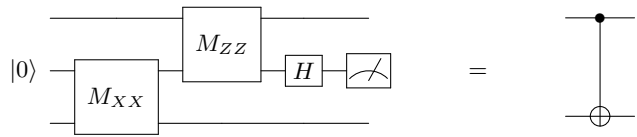

This construction for the logical CNOT gate is based on the circuit in Fig.~\ref{fig:CNOTid} which implements the CNOT gate through 2-qubit parity measurements, originally described in \cite{gottesman:higherd}.  In the dislocation encoding of \cite{HG:dislocation} this quantum circuit is similarly used to reduce a CNOT to the measurement of logical $XX$ and $ZZ$ operators. For the dislocation encoding of \cite{HG:dislocation} one also has the ability to measure the logical $ZX$ and $YZ$, that is, any product of two logical Pauli operators. One can then observe that any single (logical) qubit Hadamard $H$ gate or the $S$ gate ($SXS^{\dagger}=Y$) can be absorbed into either the following logical single-qubit Pauli measurement (if the logical qubit is to be measured) or a modified two-qubit logical Pauli measurement of a CNOT gate. This implies that in such a scheme executing such gates does not cost any additional time.\\

To verify the CNOT circuit one can consider the evolution of the input $\ket{c}_1 \ket{0}_2  \ket{t}_3$ for bits $c=0,1$ and $t=0,1$ explicitly (here 1 denotes the top qubit in the Figure). For $M_{XX}=+1$, we have a bit $b_{xx}=0$ and $M_{XX}=-1$ corresponds to $b_{xx}=1$ etc.
We have the overall evolution
\begin{eqnarray}
\ket{c}_1 \ket{0}_2 \ket{t}_3 \rightarrow \ket{c}_1 Z_2^{b_x}\ket{+}_2  Z_3^{b_{xx}} X_3^{b_{zz}}\ket{c \oplus t}_3.
\label{eq:cnot_cor}
\end{eqnarray}
We observe the logic of the CNOT gate on qubits $1$ and $3$ in addition to corrective Pauli's $Z_3^{b_{xx}} X_3^{b_{zz}}$ which depend on the outcomes $b_{xx}$ and $b_{zz}$ of the measurements $M_{XX}$ and $M_{ZZ}$ respectively. The measurement $M_X$ on the second qubit ensures that no information leaks to that qubit so that the CNOT gate properly works on any superposition of inputs.\\

\begin{figure}[htb]
    \centering
    \includegraphics[width=1\hsize]{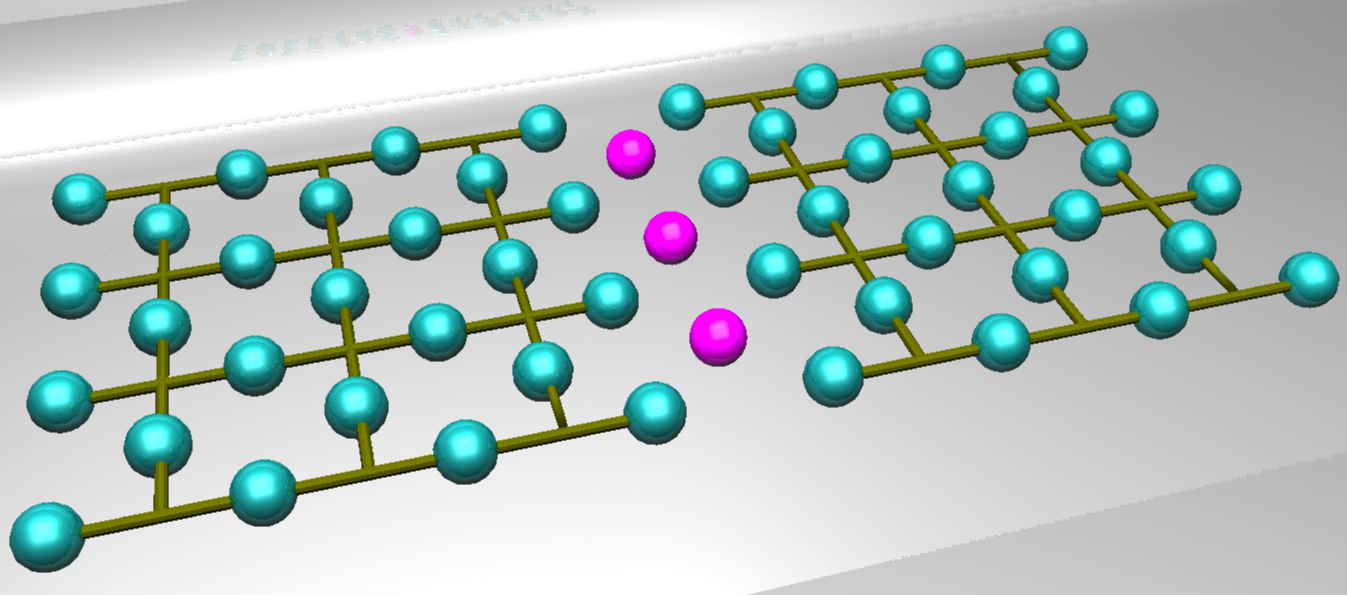} \caption{(Color Online) Picture from \cite{horsman+:suture}: two sheets (with (blue) qubits on the edges) are merged at their rough boundary by placing a row of (pink) ancilla qubits in the $\ket{0}$ state at their boundary and measuring the parity checks of the entire sheet. For a similar smooth merge, the ancillary qubits in between the two sheets are prepared in the $\ket{+}$ state, see the INT and C sheets in Fig.~\ref{fig:horsman}.}
\label{fig:horsman_merge}
\end{figure}

This circuit identity implies that we can realize a logical CNOT gate if we have the capability of projectively measuring the operators $\overline{X} \otimes \overline{X}$ and $\overline{Z} \otimes \overline{Z}$ of two qubits encoded in different sheets.
The capability to prepare a sheet in $\ket{\overline{0}}$ and the measurement $M_{\overline{X}}$ was discussed before. The realization of such joint measurement, say, $\overline{X} \otimes \overline{X}$ is possible by temporarily merging the two sheets, realizing the measurement and then splitting the sheets as follows. Consider two sheets laid out as in Fig.~\ref{fig:horsman_merge} where a row of 
ancillary qubits is prepared in $\ket{0}$ between the sheets. We realize a {\em rough} merge between the sheets by including the parity checks, plaquette and star operators, at this boundary. If the parity check measurements are perfect, the new weight-4 plaquette $Z$-checks have $+1$ eigenvalue as the ancilla qubits are prepared in $\ket{0}$. The 4 new star boundary checks have random $\pm 1$ eigenvalues subject to the constraint that the product of these boundary checks equals the product of $\overline{X}$s of the two sheets. Hence a perfect measurement would let us do a $\overline{X} \otimes \overline{X}$ measurement. As the parity check measurements are imperfect, one needs to repeat the procedure in the usual way to reliably infer the sign of $\overline{X} \otimes \overline{X}$. 

\begin{figure}[htb]
    \centering
    \includegraphics[width=1\hsize]{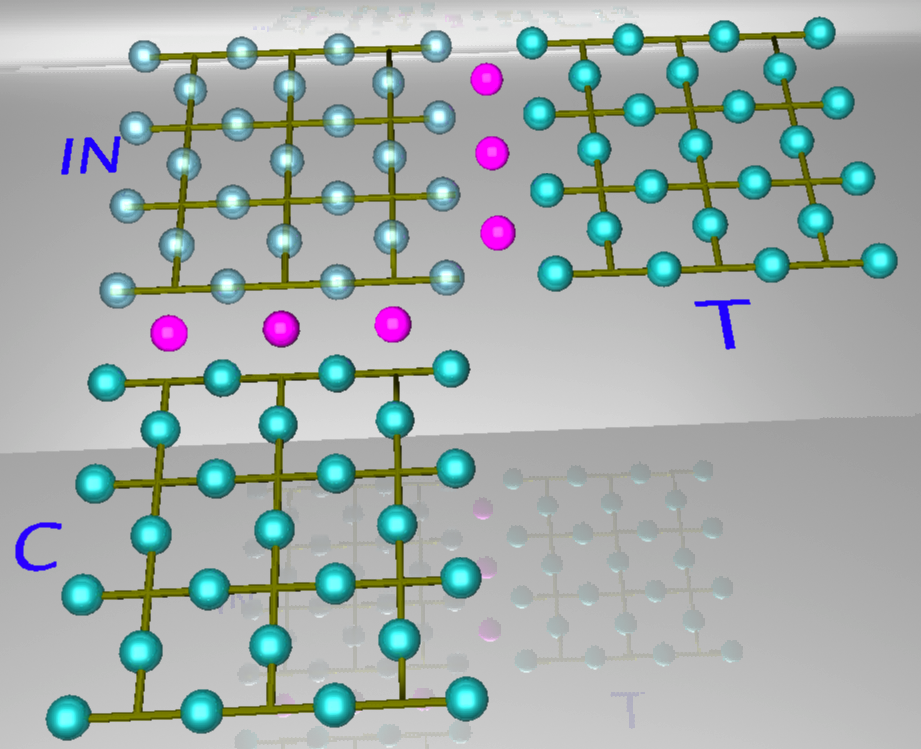} \caption{(Color Online) Picture from \cite{horsman+:suture}: using an ancilla (INT) qubit sheet we can do a CNOT between the control (C) and target sheet (T) by a sequence of mergings and splittings between the sheets.}
\label{fig:horsman}
\end{figure}

We are however not yet done as we wish to realize a projective $\overline{X} \otimes \overline{X}$ measurement on the qubits encoded in two {\em separate} sheets. This means that we should split the two sheets again: we can do this by reversing the merge operation and measure the ancillary qubits in the $Z$-basis and stop measuring the 4 boundary $X$-checks. Again, if the parity check measurements are perfect, the eigenvalues of the plaquette $Z$-checks at the boundary of both sheets will take random values, but both are correlated with the outcome of the $Z$-measurement on the ancillary qubits. Hence the individual $\overline{X}$ eigenvalues of the separate sheets may be randomized, but they are correlated so that $\overline{X} \otimes \overline{X}$ remains fixed. Similarly, a smooth merging and splitting (as between qubits C and INT in Fig.~\ref{fig:horsman}) with the ancillary qubits prepared and measured in the $X$-basis accomplishes a $\overline{Z} \otimes \overline{Z}$ measurement.\\

The procedure for a CNOT in Fig.~\ref{fig:horsman} then consists of: first a preparation of the INT qubit in $\ket{\overline{0}}$, then a rough merge and split of qubits T and INT followed by a smooth merge and split between qubits INT and C followed by a final $M_{\overline{X}}$ measurement of qubit INT.

\subsubsection{Topological Qubits and CNOT via Braiding}
\label{sec:braiding}

A different way of encoding multiple qubits and realizing a CNOT gate was first proposed in \cite{RHG:threshold, BM:deform}. In this method one considers a single sheet for the whole computation in which holes are made which encode logical qubits. By moving holes around, or `deforming the stabilizer', one can execute a CNOT gate. This method is also the one that is analyzed in \cite{fowler:practical} with the goal of giving a detailed overview of the procedures and practical space-time overhead. One possible disadvantage of this method is that it has an additional qubit overhead. A distance-3 smooth hole qubit (see the description below) costs many more than 13 physical qubits. A detailed comparative overhead analysis has not yet been performed between the separate sheet layout+lattice surgery scheme and this scheme. \\

\begin{figure}[htb]
    \centering
    \includegraphics[width=1\hsize]{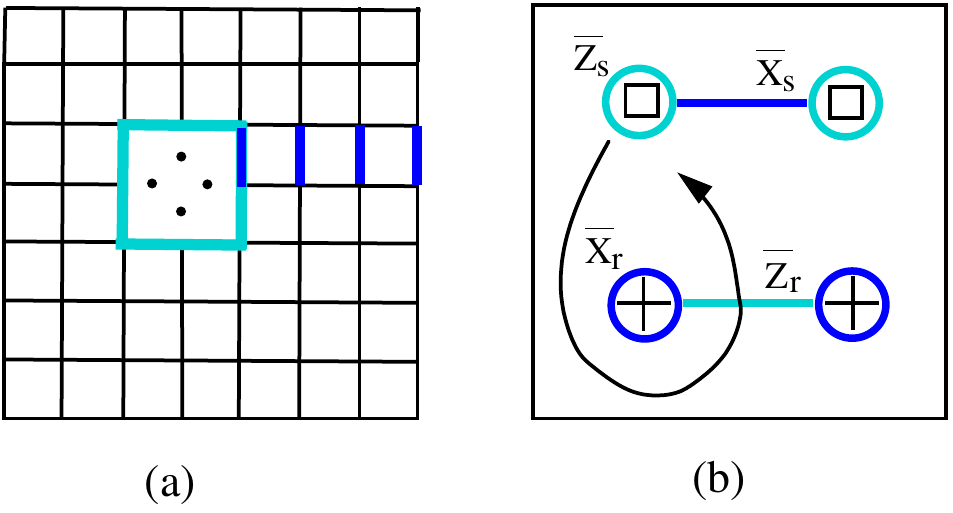} \caption{(Color Online) (a) A smooth hole is created by removing a block of plaquette operators and the star operators acting on qubits in the interior of the block. The $Z$-loop around the hole is $\overline{Z}$ while $\overline{X}$ is an $X$-string to the boundary. The qubits inside the hole (4 in the picture) are decoupled from the lattice. (b) Two smooth holes can make one smooth qubit and two rough holes can make one rough qubit so that moving a smooth hole around a rough hole realizes a CNOT gate.}
\label{fig:surhole}
\end{figure}

In order to see how to encode multiple qubits, we start with a simple square sheet with all smooth boundaries which encodes no qubits, Fig.~\ref{fig:surhole}(a) \footnote{On a $L \times L$ lattice there are $2L(L+1)$ qubits, $L^2+(L+1)^2$ stabilizer checks and one linear dependency between the star operators, hence zero encoded qubits.}. To encode qubits one makes a {\em hole} in the lattice, that is, one removes some checks from the stabilizer ${\cal S}$. This is a change in topology which affects the code space dimension. In stabilizer terms: when we remove one plaquette, say, $B_{p_*}$ for some $p_*$ from the stabilizer ${\calS}$, then $B_{p_*}$ is no longer an element in $\calS$ but still commutes with ${\cal S}$, therefore $B_{p_*}$ is a logical operator. The matching logical operator which anti-commutes with it starts at the hole and goes to the boundary. This encoded qubit has poor distance namely $d=4$ as $B_{p*}$ is of weight 4. We can modify this procedure in two ways such that logical qubits have a large distance and its logical operators do not relate to the boundary.  The particular choice of logical qubits will allow one to do a CNOT by moving holes. \\

To get a logical qubit with large distance we simply make a bigger hole. We remove all, say, $k^2$ plaquette operators in a block (and all $(k-1)^2$ star operators acting in the interior of this block) and modify the star operators at the boundary to be of weight 3, no longer acting on the qubits in the interior, see Fig.~\ref{fig:surhole}(a). The qubits in the interior of the block are now decoupled from the code qubits. The procedure creates one qubit with $\overline{Z}$ equal to any $Z$-loop around the hole. The $\overline{X}$ operator is a $X$-string which starts at the boundary and ends at the hole. Clearly, the distance is the minimum of the perimeter of the hole and the distance to the boundary. We call this a {\em smooth} hole as the hole boundary is smooth. Of course, we could do an identical procedure to the star operators, removing a cluster of stars and a smaller subset of plaquette operators and adapting the plaquette operators at the boundary. Such qubit will be called a {\em rough} hole and its $\overline{X}$ operator is an $X$-string around the hole (a string on the dual lattice) and $\overline{Z}$ is a $Z$-string to the boundary. 

In order to be independent of the boundary, we use two smooth holes to define one {\em smooth or primal qubit} and use two rough holes to define one {\em rough or dual qubit} as follows. Consider two smooth holes $1,2$ and define a new smooth qubit as $\ket{\overline{0}}_{s}=\ket{\overline{0}, \overline{0}}_{1,2}$ and $\ket{\overline{1}}_s=\ket{\overline{1}, \overline{1}}_{1,2}$. For this smooth qubit $s$ we have $\overline{Z}_{s}=\overline{Z}_i, i=1,2$ (we can deform $\overline{Z}_1$ into $\overline{Z}_2$ by plaquette operators) and $\overline{X}_{s}=\overline{X}_1 \overline{X}_2$ which we can deform to an $X$-string which connects the two holes, see Fig.~(\ref{fig:surhole})(b). The distance of this smooth qubit is the minimum of the distance between the holes and the perimeter of one of the holes (assuming the boundary is sufficiently far away). Similarly, we can create a {\em rough} qubit by taking two rough holes and defining 
\begin{eqnarray}
\ket{\overline{0}}_{r}=\frac{1}{\sqrt{2}}\left(\ket{\overline{0},\overline{0}}_{3,4}+\ket{\overline{1},\overline{1}}_{3,4}\right), \nonumber \\ 
\ket{\overline{1}}_{r}=\frac{1}{\sqrt{2}}\left(\ket{\overline{0},\overline{1}}_{3,4}+\ket{\overline{1},\overline{0}}_{3,4}\right).
\nonumber
\end{eqnarray}
With this choice $\overline{X}_{r}$ is the loop $\overline{X}_3$ (or equivalently $\overline{X}_4$) while $\overline{Z}_{r}=\overline{Z}_1 \overline{Z}_2$ is equivalent to the $Z$-string connecting the holes.\\

Imagine moving one smooth hole around a rough hole as in Fig.~\ref{fig:surhole}(b). After the move, the $X$-string connecting the smooth holes will additionally go around the rough hole enacting the transformation $\overline{X}_s \rightarrow \overline{X}_r \otimes \overline{X}_s$. This can be understood by noting that an $\overline{X}$-string with some endpoints $a$ and $b$ which loops around a rough hole is equivalent (modulo stabilizer operators) to an $\overline{X}$-loop around the rough hole disconnected from a direct $\overline{X}$-string between the endpoints $a$ and $b$.
Similarly, the $Z$-string $\overline{Z}_r$ connecting the rough holes will, after the move, wind around the smooth hole, leading to the transformation $\overline{Z}_r \rightarrow \overline{Z}_s \otimes \overline{Z}_r$. The loops $\overline{Z}_s$
and $\overline{X}_r$ are not changed by the move. This action precisely corresponds to the action of a CNOT with smooth qubit as control and rough qubit as target \footnote{The action of the CNOT in the Heisenberg representation is $X_c \otimes I_t \rightarrow X_c \otimes X_t$, $I_c \otimes X_t \rightarrow I_c \otimes X_t$, $Z_c \otimes I_t \rightarrow Z_c \otimes I_t$ and $I_c \otimes Z_t \rightarrow Z_c \otimes Z_t$ where $X_c$ ($X_t$) stands for Pauli $X$ on control qubit $c$ (target qubit $t$).}. \\

The ability to do a CNOT with a smooth qubit as control and a rough qubit as target qubit seems limited as all such gates commute. However one can use the one-bit teleportation circuits in Fig.~\ref{fig:onebit_tele} to convert a smooth qubit into a rough qubit and a rough qubit into a smooth qubit, using {\em only} CNOTs with smooth qubits as controls. We have already shown how to realize the other components in the one-bit teleportation circuit such as $M_{\overline{X}}$ and $M_{\overline{Z}}$. Thus by composing these circuits we can do a CNOT between smooth qubits alone (or rough qubits alone).\\

How is the braiding done using elementary gate operations? The advantage of realizing topological gates in stabilizer codes (as opposed to braiding of Majorana fermions or non-Abelian anyons in quantum Hall systems) is that braiding can be realized by changing where we measure the parity checks, or {\em deforming the code}. For example, one can enlarge the hole in Fig.~\ref{fig:surhole} to include, say, 2 more plaquettes and 3 more qubits in the interior. We stop measuring those two plaquette checks and the star checks in the interior, modify the star boundary measurements and measure the qubits in the interior in the $X$-basis. The modified weight-3 boundary checks will have random $\pm 1$ eigenvalues as their previous eigenstates were perfectly entangled with the qubits in the interior.
This corresponds to a high $Z$-error rate around the modified boundary. By repeating the measurement to increase the confidence in their outcome one can correct these $Z$-errors, but of course we may partially complete a $\overline{Z}$-loop this way. The protection against a full $\overline{Z}$-loop around the hole is thus provided by {\em the part of the hole boundary which remains fixed}.


This implies that the hole can safely be moved and braided in the following {\em caterpillar} manner. One first enlarges the hole (while keeping its `back-end' fixed providing the protection) so that it reaches its new position (``the caterpillar stretches out the front part of its body to a new position"). In terms of parity check measurements it means that from one time-step to the next one, one switches from measuring the small hole to the large hole parity checks. Due this extension errors will occur along the path over which the hole is moved and if error correction is noisy we should not act immediately to infer the new Pauli frame, but repeat the new check measurements to make this new frame more robust. Then as a last step, we shrink the hole to its new position and corroborate the new measurement record by repetition ("the caterpillar brings its rear-end next to its front-end again"). In \cite{fowler:practical} Figs.~19-23 depict the enlargement of the hole and its subsequent shrinkage and its effect on the logical operators. \\

Alternatively, one can move the hole by a sequence of small translations, so that the hole never becomes large. The speed at which the hole can then be safely moved is determined by the time it takes to establish a new Pauli frame (eliminate errors) after a small move. Details of hole moving schemes are discussed in e.g. \cite{fowler+:unisurf, fowler:practical}.

\subsection{Different 2D Code Constructions}
\label{sec:alt}

In this section we discuss a few 2D quantum codes which are variations of the surface code. These codes may have advantages over the surface code depending on physical hardware constraints. These codes are two competitive examples of 2D subsystem codes, as well as a surface code using harmonic oscillators instead of qubits.

\subsubsection{Bacon-Shor Code}
\label{sec:BS}

\begin{figure}[htb]
    \includegraphics[width=0.6\hsize]{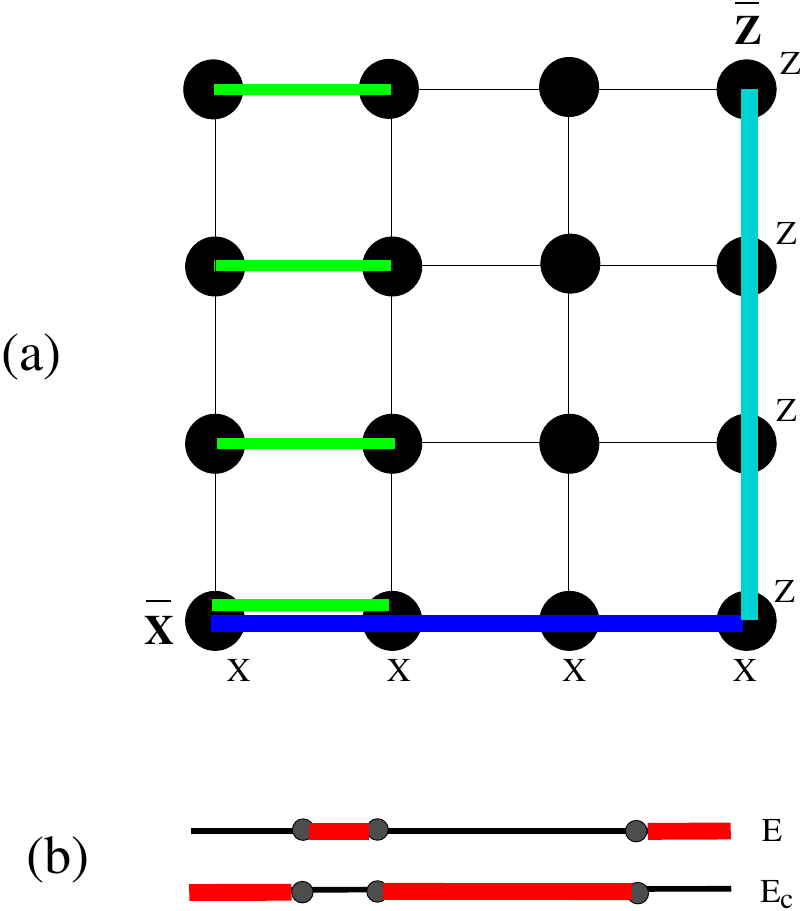} \caption{(Color Online) (a) $[[16,1,4]]$ Bacon-Shor code with $\overline{X}$, a row of $X$s, and $\overline{Z}$, a column of $Z$s. The stabilizer generators are double columns of $Z$s, ${\bf Z}_{||,i}$ (one is depicted) and double rows of $X$s, ${\bf X}_{=,j}$.(b) Decoding for $X$ errors (or $Z$ errors in the orthogonal direction). Black dots denote the places where the double column parity checks ${\bf Z}_{||,i}$ have eigenvalue $-1$ (defects). The $X$ error string $E$ has $X$ errors in the fattened (red) region and no errors elsewhere and $E_c$ is its complement. Clearly the string $E$ has lower weight than $E_c$ and is chosen as the likely error.}
\label{fig:BS}
\end{figure}

An interesting family of subsystem codes are the Bacon-Shor codes \cite{bacon:mem}. For the $[[m^2,1,m]]$ Bacon-Shor code the qubits are laid out in a 2D $m \times m$ square array, see Figs.~\ref{fig:BS1} and \ref{fig:BS}. The stabilizer parity checks are the double $Z$-column operators ${\bf Z}_{||,i}$ for columns $i=1,\ldots m-1$ and double $X$-row operators ${\bf X}_{=,j}$ for rows $j=1,\ldots m-1$. 

It is also possible to work with asymmetric Bacon-Shor codes with qubits in an $n \times m$ array. Asymmetric codes can have better performance when, say, $Z$ errors are more likely than $X$ errors (when $T_2 \ll T_1$), see \cite{BP:bs}. The gauge group ${\cal G}$ (see Section \ref{sec:ssc2}) is generated by weight-2 vertical $XX$ links and horizontal $ZZ$ links and contains the parity checks. The bare logical operators (which commute with $\calG$ but are not in $\calS$) are the single $Z$-column $\overline{Z}$ and a single $X$-row $\overline{X}$.

\begin{figure}[htb]
    \centering
    \includegraphics[width=1\hsize]{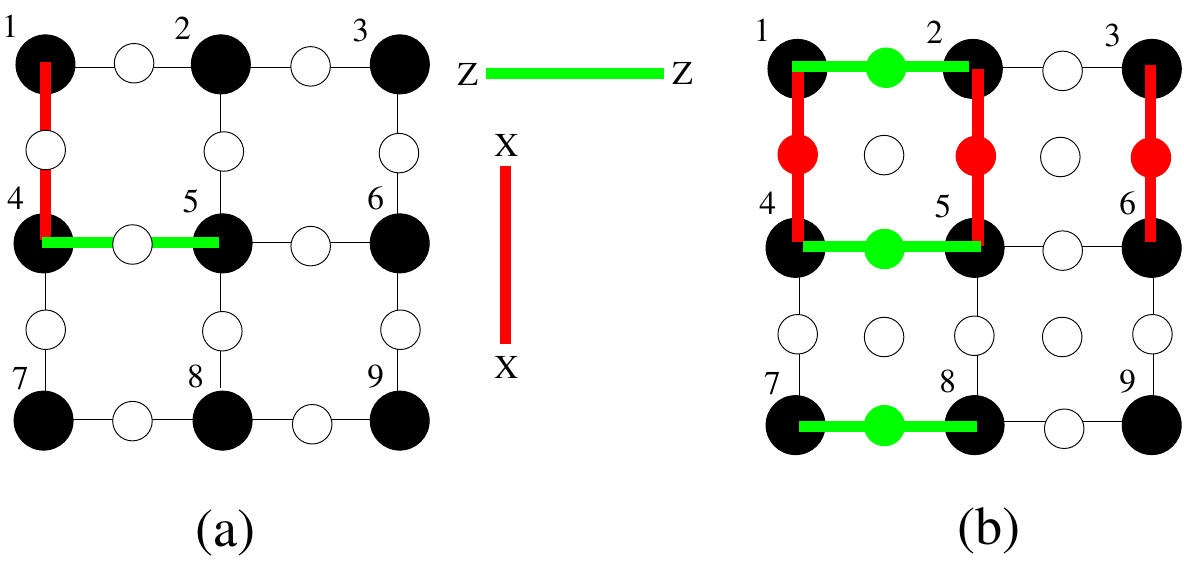} \caption{(Color Online) (a) In order to measure the $XX$ and $ZZ$ operators one can place ancilla qubits (open dots) in between the data qubits. Such ancilla qubit interacts with the two adjacent data qubits to collect the syndrome. (b) Alternatively, to measure ${\bf Z}_{||,i}$ one can prepare a 3-qubit entangled cat state $\frac{1}{\sqrt{2}}(\ket{000}+\ket{111})$ ((green) vertical line of dots) which interacts locally with the adjacent system qubits. ${\bf X}_{=,1}$ could be measured by preparing a cat state for the ancilla qubits placed at, say, the horizontal line of (red) dots. The ancilla qubits at the open dots can be used to prepare the cat states.}
\label{fig:BSancilla}
\end{figure}

Consider the correction of $X$ errors sprinkled on the lattice, assuming for the moment that the parity check measurement of ${\bf Z}_{||,i}$ is noise-free. For each column we note that an even number of $X$ errors is a product of the vertical $XX$ gauge operators and therefore does not affect the state of the logical qubit. This means that per column only the parity of the number of $X$ errors is relevant. The double column operator ${\bf Z}_{||,i}$ determines whether this parity flips from column $i$ to column $i+1$. The interpretation of the eigenvalues of ${\bf Z}_{||,i}$ is then the same as for a 1D repetition code (or 1D Ising model) with parity checks $Z_i Z_{i+1}$. Double columns where ${\bf Z}_{||,i} \equiv Z_i Z_{i+1}=-1$ are {\em defects} marking the end-points of $X$-strings (domain walls in the 1D Ising model). Minimum-weight decoding is very simple as it corresponds to choosing the minimum weight one between two possible $X$-error strings: $E$ or the complement string $E_c$ which both have the faulty double columns defects as end-points, see Fig. \ref{fig:BS}(b). The code can thus correct all errors of weight at most $\lfloor \frac{m}{2} \rfloor$ for odd $m$. Higher-weight errors can also be corrected as long they induce a low density of defects on the boundary. Note however that the number of syndrome bits scales as $m$ whereas the number of errors scales with $m^2$. This means that in the limit $m \rightarrow \infty$ the noise-free pseudo-threshold $p_c(m) \rightarrow 0$ as the fraction of uncorrectable errors will grow with $m$. So, how do we choose $m$ in order to minimize the logical error rate $\overline{p}(p,m)$? In \cite{NP:bs} the authors find that the optimally-sized Bacon-Shor code for equal $X$ and $Z$ error rate $p$ is given by $m=\frac{\ln 2}{4 p}$ and for that optimal choice they can bound the logical $X$ (or $Z$) error rate as $\overline{p}(p) \lesssim \exp(-0.06/p)$.\\

How does acquire the non-local parity check values? One can either measure the $XX$ and $ZZ$ gauge operators and use this information to get the eigenvalues of ${\bf X}_{=,i}$ and ${\bf Z}_{||,j}$, or one measures the parity checks directly. The first method has the advantage of being fully local: the ancilla qubits for measuring $XX$ and $ZZ$ can be placed in between the data qubits, see Fig.\ref{fig:BSancilla}(a). In the second method we can prepare an $m$-qubit cat state (Shor error correction), see e.g. \cite{BP:bs}. We could measure ${\bf Z}_{||,1}$ using the circuit in Fig.~\ref{fig:paritycheck}(a) with a single ancilla qubit in the $\ket{+}$ state and controlled-phase gates ($CZ$) gates. However, a single $X$ error on the ancilla qubit can feed back to the code qubits and cause multiple $Z$ errors making the procedure non fault-tolerant. In addition, the interaction between the ancilla qubit and the code qubits is non-local. Instead, we encode the ancilla qubit $\ket{+}$ using the repetition code, i.e. we prepare the $m$-qubit cat state $\frac{1}{\sqrt{2}}(\ket{00\ldots 0}+\ket{11 \ldots 1})$ such that a $CZ$-gate acts between one cat qubit and one code qubit. The $m$-qubit cat state, which itself is stabilized by $Z_i Z_{i+1}$ and $X_1 \ldots X_m$, can be made by preparing $\ket{+}^{\otimes m}$ and measuring $Z_i Z_{i+1}$ using local ancilla qubits. The $Z_i Z_{i+1}$ eigenvalues are recorded to provide the Pauli frame. In \cite{BP:bs} further details of this scheme are given, including estimates of the noise threshold for asymmetric Bacon-Shor codes which shows that the Bacon-Shor codes may be competitive, depending on further detailed numerical analysis, with the 2D surface code.

Consider now the first method of directly measuring $XX$ and $ZZ$: what happens when the local $XX$ and $ZZ$ checks are measured inaccurately? The good news is this only causes {\em local} errors on the system qubits. The bad news is that if the measurement outcome of, say, $XX$ has some probability of error $q$, then the error probability for a nonlocal stabilizer check ${\bf X}_{=,i}$ will approximately be $mq$. This is a disadvantage of the Bacon-Shor code. Researchers \cite{AC:bs, BP:bs} have sought to improve the fault-tolerance of the parity check measurements by replacing the preparation of simple single qubit ancillas by fault-tolerant ones (methods by Steane and Knill). In \cite{AC:bs} a best noise-threshold of $p_c \approx 0.02\%$ was numerically obtained for the (concatenated) $[[25,1,5]]$ code. \cite{NP:bs} have considered an alternative way of making the syndrome more robust, namely by simple repetition of the $XX$ and $ZZ$ measurements and a collective processing of the information (as is done for the surface code). We can view the effect of repetition as extending the 1D line of defects to a 2D lattice of defects, as in Fig.~\ref{fig:3ddec}, so that minimum weight decoding corresponds to finding a minimum weight matching of defect end-points. The error rate for vertical (black) links representing the parity check errors scales with $m$ while the error rate for horizontal links (when one column has an even and the other column has an odd number of errors) scales, for low $p$, also with $m$.

In \cite{NP:bs} the authors estimate that the optimal size for the Bacon-Shor code is then $m \approx 0.014/p$ and that for this choice, the logical error rate $\overline{p}(p) \lesssim \exp(-0.0068/p)$. Hence for an error rate of $p=5 \times 10^{-4}$, we can choose $m=28$ giving a logical $X$ (or $Z$) error rate of $\overline{p}\approx 1.25 \times 10^{-6}$. This does not compare favorably with the logical error rate for the surface code with $L=28$, which, using the empirical formula 
$\overline{p} \approx 0.03 \left(\frac{p}{p_c}\right)^{L/2}$ for even $L$ in \cite{fowler:practical}, is much lower than $10^{-6}$.

\subsubsection{Surface Code with Harmonic Oscillators}
\label{sec:surface_osc}

In this section we discuss whether it is possible to encode quantum information in a 2D lattice of coupled harmonic oscillators. We start by defining a continuous-variable version of the surface code which encodes an oscillator in a 2D array of oscillators.
Then we discuss how to modify this construction so that we concatenate the qubit-into-oscillator code described in Section \ref{sec:qiq} with the regular surface code and express the checks of the surface code in terms of operators on the local oscillators.
This scheme may be of interest if the qubit encoded in the oscillator has a sufficiently low error rate which we want to improve upon by further surface code encoding. For example, one can imagine a set of 2D or 3D microwave cavities each of which by itself encodes a qubit which we couple in a 2D array. \\

It is possible to define a qudit stabilizer surface code, see e.g. \cite{BB:qudit_surface}, where the elementary constituents on the edges of the lattice are qudits with internal dimension $d$ and the code encodes one or several qudits. Here we will just focus on the special case when we take $d \rightarrow\infty$ and each edge is represented by a harmonic oscillator with conjugate variables $\hat{p}, \hat{q}$. The goal of such continuous-variable surface code is to encode a non-local oscillator into an 2D array of oscillators such that the code states are protected against local shifts in $\hat{p}$ and $\hat{q}$. In addition, one can imagine using continuous-variable graph states to prepare such encoded states and observe anyonic statistics \cite{zhang+:anyon_kitaev}. \\

\begin{figure}[htb]
    \centering
    \includegraphics[width=1\hsize]{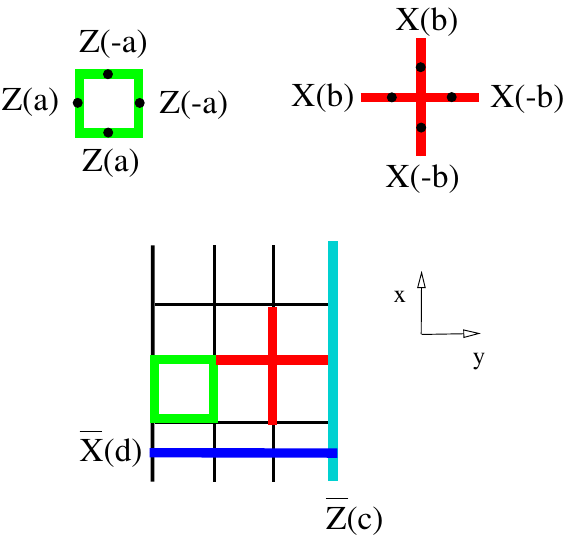} \caption{(Color Online) Small example of the oscillator surface code where oscillators on the edges are locally coupled with plaquette and star operators so as to define an encoded oscillator with logical, non-local, displacements $\overline{X}(d)$ and $\overline{Z}(c)$. The realization of the four-oscillator interaction will require strong 4-mode squeezing in either position (at plaquettes) or momenta (at stars).}
\label{fig:surf_osc}
\end{figure}

To get a surface code, we replace Pauli $X$ by $X(b)=\exp(2 \pi i b\, \hat{p})$ and Pauli $Z$ by $Z(a)=\exp(2 \pi i a\, \hat{q})$ with real parameters such that $Z^{\dagger}(a)=Z^{-1}(a)=Z(-a)$ etc. It follows that for any two oscillators 1 and 2, we have 
\begin{equation}
\forall a,b,\;[Z_{1}(a) Z_{2}(-a), X_{1}(b) X_{2}(b)]=0. 
\label{eq:comm}
\end{equation}
In the bulk of the surface code lattice, a plaquette operator centered at site $u$ can be chosen as $B_u(a)=Z_{u-\hat{x}}(a)Z_{u+\hat{x}}(-a)Z_{u-\hat{y}}(a)Z_{u+\hat{y}}(-a)$ while a star operator at site $s$ is equal to $A_s(b)=X_{s-\hat{x}}(-b) X_{s+\hat{x}}(b) X_{s-\hat{y}}(b) X_{s+\hat{y}}(-b)$, see Fig.~\ref{fig:surf_osc}.

Here $Z_{u-\hat{x}}(a)=\exp(2 \pi i a \,\hat{q}_{u-\hat{x}})$ where $\hat{q}_{u-\hat{x}}$ is the position variable of the oscillator at site $u-\hat{x}$ ($\hat{x}$ and $\hat{y}$ are orthogonal unit-vectors on the lattice). One can observe from the Figure and Eq.~(\ref{eq:comm}) that $B_u(a)$ and $B_u^{\dagger}(a)$ commute with $A_s(b)$ and $A_s^{\dagger}(b)$ for all $a,b$ in the bulk and at the boundary.

We can define Hermitian operators with real eigenvalues in the interval $[-1,1]$ as $H_u(a)=\frac{1}{2}(B_u(a)+B_u^{\dagger}(a))=\cos(2 \pi a (q_{u-\hat{x}}-q_{u+\hat{x}}+q_{u-\hat{y}}-q_{u+\hat{y}}))$ and $H_s(b)=\frac{1}{2}(A_s(b) +A_s^{\dagger}(b))=\cos(2 \pi b (-p_{s-\hat{x}}+p_{s+\hat{x}}+p_{s-\hat{y}}-p_{s+\hat{y}}))$. We now define the code space as the $+1$ eigenspace of all $H_u(a), H_s(b)$ for all $a$ and $b$.
It follows that a state in the code space is a delta-function in the positions of the oscillators around all plaquettes $u$, that is, $\delta(q_{u-\hat{x}}-q_{u+\hat{x}}+q_{u-\hat{y}}-q_{u+\hat{y}})$ for every $u$, while it is a delta-function in the momenta of the oscillators $\delta(-p_{s-\hat{x}}+p_{s+\hat{x}}+p_{s-\hat{y}}-p_{s+\hat{y}})$ located at all stars $s$. \\

One can compare such highly-entangled code state with its simpler cousin, the 2-mode EPR-state. In the 2-mode case we have two commuting operators namely $Z_1(a) Z_2(-a)$ and $X_1(b) X_1(b)$ on oscillator 1 and 2. The single state which is the $+1$ eigenstate of $\cos(2 \pi a (q_1-q_2))$ and $\cos(2 \pi b (p_1+p_2))$ for all $a,b$ is the two-mode infinitely-squeezed EPR state $\delta(p_1+p_2) \delta(q_1-q_2)$.

Unlike in the two-mode case, the oscillator surface code space is not one-dimensional, but infinite-dimensional as it encodes a non-local oscillator. The operators $\overline{Z}(c)=\exp(2 \pi i c \sum_{i \in \gamma_1} \hat{q}_i)$ where the path $\gamma_1$ runs straight from north to south commute with all $H_u(a),H_s(b)$, see Fig.~\ref{fig:surf_osc}. Similarly,  we have $\overline{X}(d)=\exp(2 \pi i d\sum_{j \in \gamma_2} \hat{p}_j)$ where $\gamma_2$ runs straight from east to west. As $\overline{Z}(c) \overline{X}(d)=e^{-i (2 \pi c)(2 \pi d)} \overline{X}(d) \overline{Z}(c)$, we can interpret $\overline{Z}(c)$ and $\overline{X}(d)$ as phase-space displacements of the encoded oscillator with logical position and momentum $\overline{p}=\sum_{i \in \gamma_2} p_i $ and $\overline{q}=\sum_{i \in \gamma_1} q_i$. We can deform these non-unique logical operators to follow deformed paths, e.g. multiply $\overline{Z}(c)$ by $B_p(c)$ plaquettes (note that if we multiply by $B_p(c')$ with $c' \neq c$ we get an operator with the union of supports).\\

How would one use such a code to encode quantum information and what protection would it offer? As its qubit incarnation, a sufficiently-low density of independent errors on the lattice can be corrected. For the array of oscillators or bosonic modes, one would instead expect that each oscillator $i$ will independently suffer from small dephasing, photon loss etc., that is, errors which can be expanded into small shifts $Z_i(e)X_i(e')$ with $|e|, |e'| \ll 1$, see Section \ref{sec:qiq}. This means that the likelihood for logical errors of the form $\overline{Z}(c)\overline{X}(d)$ for small $c,d$ will be high which relates of course to the fact that we are attempting to encode a continuous variable rather than a discrete amount of information. \\

However, one can imagine using only a 2-dimensional subspace, in particular the codewords of the GKP qubit-into-oscillator code,  see \ref{sec:qiq} for each oscillator in the array. One can also view this as a concatenation of the GKP code and the surface code in which we express the surface code plaquette and star operators in terms of the operators on the elementary oscillators in the array. One could prepare the encoded states $\ket{\overline{0}},\ket{\overline{1}},\ket{\overline{+}},\ket{\overline{-}}$ of the surface code by preparing each local oscillator in the qubit-into-oscillator logical states $\ket{0},\ket{1},\ket{+},\ket{-}$ and subsequently projecting onto the perfectly correlated momenta and position subspace. For example, the state $\ket{0}$ of a local oscillator $i$ is an eigenstate of $S_q(\alpha)=e^{2 \pi i \hat{q}_i/\alpha}$, $S_p=e^{-2 i \hat{p}_i\alpha}$ and the local $\overline{Z}_i=e^{i \pi\hat{q}_i/\alpha}$. This implies that after projecting onto the space with $H_u(a)=1,H_s(b)=1$ for all $a,b$,  it will be an eigenstate of $\overline{Z}(1/(2 \alpha))=e^{i\pi \sum_{i \in \gamma_1} \hat{q}_i/\alpha}$, i.e. the encoded $\ket{\overline{0}}$.

\subsubsection{Subsystem Surface Code}
\label{sec:SSC}

It is clear that codes for which one has to measure high-weight parity checks are disadvantageous: it requires that an ancilla qubit couples to many data qubits through noisy gates leading to a large error rate on the syndrome,  leading in turn to a lower noise threshold. Can we have a 2D stabilizer code which has weight-2 or weight-3 parity checks? The answer is no: one can prove that 2D qubit codes defined as eigenspaces of at most 3-local (involving at most 3 qubits) mutually commuting terms are trivial (with $O(1)$ distance) as quantum codes \cite{AE:top}. In addition any stabilizer code with only weight-2 checks can be shown to be trivial.\\

Such results do not hold for subsystem codes: the Bacon-Shor code shows that it is possible to have only 2-qubit non-commuting parity checks.  However, the Bacon-Shor code is not a topological subsystem code as the stabilizer checks are nonlocal on the 2D lattice and its asymptotic noise threshold is vanishing. Several topological subsystem codes have been proposed in which weight-2 parity checks are measured \cite{bombin:topsub}, but the asymptotic noise threshold for such codes is typically quite a bit lower than for the surface code, see e.g. \cite{SBT:top}. The question is whether it is possible to find a 2D subsystem code with checks of weight less than 4 which has a noise threshold which is similar to the surface code. Such a subsystem code may be of high interest if it is considerably easier to realize a weight-3 check in the physical hard-ware than a weight-4 check.\\

\begin{figure}[htb]
    \centering
    \includegraphics[width=1\hsize]{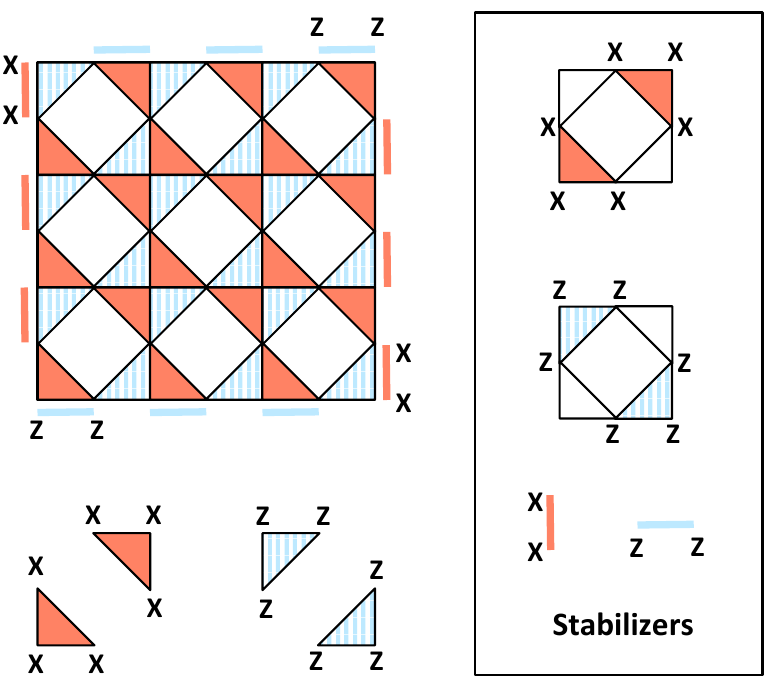} \caption{(Color Online) Picture from \cite{BDPS:3qubitcheck}: subsystem surface code on a lattice of size $L \times L$ with $L^2$ square plaquettes (depicted is $L=3$). The qubits live on the edges and vertices of the plaquettes and are acted upon by weight-3 $X$ and $Z$-triangle operators (which are modified to become weight-2 operators at the boundary). The stabilizer checks are weight 6 except at the boundary.}
\label{fig:surfsub}
\end{figure}

In \cite{BDPS:3qubitcheck} a topological subsystem code was proposed --a subsystem surface code-- in which the non-commuting parity checks are of weight-3 and the stabilizer generators are of weight 6, see Fig.~\ref{fig:surfsub}. More precisely, the gauge group $\calG$ is generated by the triangle operators $XXX$ and $ZZZ$, including cut-off weight-2 operators at the boundary. The stabilizer group $\calS=\calG \cap \calC(\calG)$ is generated by weight-6 plaquette operators (at the boundary $\rightarrow$ weight-2 operators). By measuring, say, the $Z$-triangles we can deduce the eigenvalues of the $Z$-plaquettes which are used to do error correction.\\

For a $L \times L$ lattice one has a total of $3L^2+4L+1$ qubits and $2L^2+4L$ independent stabilizer generators which gives $L^2+1$ qubits. One of these qubits is the logical qubit whose $\overline{Z}$ and $\overline{X}$ commute with all $Z$ and $X$-triangles. Similar as in the surface code, a vertical $Z$-line through $2L$ qubits can realize $\overline{Z}$ while a horizontal $X$-line realizes $\overline{X}$.  The logical operators for the $L^2$ gauge qubits, one for each plaquette, are pairs of triangle operators on a plaquette generating the group $\calG$. One can multiply, say, the vertical $Z$-line by $Z$-triangles to become a $\overline{Z}$ which acts only on $L$ qubits: in \cite{BDPS:3qubitcheck} it is indeed proved that the distance of the code is $L$. Note that such weight-$L$ $\overline{Z}$ acts on the logical qubit {\em and} the irrelevant gauge qubits.\\

For a code with distance $L=3$ one thus needs $41$ elementary qubits, substantially more than for the surface code. Multiple qubits can be encoded in this subsystem code by making holes as for the surface code. One can expect that braiding and lattice surgery methods for this code can be established in the same way as for the surface code. The interesting feature of this code are its relatively-high noise threshold obtained by reduced-weight parity checks (at the price of a bit more overhead). Decoding of stabilizer syndrome information is done by interpreting the syndrome as defects on a virtual lattice which can processed, similar as for the surface code, by minimum weight matching of defects or by RG decoding. For noise-free perfect error correction and independent $X,Z$ noise, the authors report a maximum threshold of $p_c \approx 7\%$ (compare with $11\%$ for the surface code). For noisy error correction the threshold depends on how single errors with probability $p$ in the parity check circuit affect the error rate on the virtual lattice. Modeling this effective noise-rate on the virtual lattice, the authors find a noise threshold of $p_c \approx 0.6\%$.\\

It is not surprising that decoding for the subsystem surface codes can be done using the decoding method for the surface code. It was proved in \cite{BDP:uni} that any 2D topological subsystem or stabilizer code can be locally mapped onto copies of the toric code. The upshot is that for any such code one can find, after removing some errors by local correction, a virtual lattice with toric code parity checks and an underlying effective error model. An example of another 2D topological subsystem code which may be analyzed this way is a concatenation of the $[[4,2,2]]$ code with the surface code. If we use the $[[4,2,2]]$ code as subsystem code then the concatenated code has weight-2 and weight-8 check operators. The scheme may be of interest if the weight-2 checks can be measured fast and with high accuracy.

\subsection{Decoding and (Direct) Parity Check Measurements}
\label{sec:disc_pract}

One can ask whether quantum error correction for the surface or other topological codes in $D=2$ or higher is possible by purely local means. The dissipative correction procedure of stabilizer pumping described in Section \ref{sec:ec_ft} is an example of a purely local error correction mechanism which does not use any communication. We can consider what such a mechanism does if we apply it to the toric code discussed in \ref{sec:surfcode}. Imagine a single $X$ error occurs. For the toric code, such an $X$ error is heralded by two odd-parity $Z$-checks, two defects. The error can be corrected by applying any small $X$-string which terminates at these defects. However, the mechanism described in \ref{sec:ec_ft} which applies an $X$ correction at a fixed qubit for every defect, will have the effect of moving the single $X$-error around or it will create more $X$ errors. It will utterly fail at removing errors. In \cite{DKP:encode} it was shown that it is possible to use such a very, but assumed to be perfect, local dissipative decoder to efficiently encode a quantum state in a toric code quantum memory. In this decoder the corrections are chosen such that the errors are pushed in a certain direction where they can mutually annihilate and a simple input state determines what state is encoded by the dissipative evolution. In such extremely local form of dissipative error correction there is absolutely no guarantee that the corrections which are applied are minimum weight corrections.\\

It is clear that an engineered dissipative dynamics for quantum error correction should at least correlate the parities of neighboring checks before applying any corrections. As an error string is only heralded by its two end-point defects, longer error strings require correlating the parity checks in a larger neighborhood, and hence more communication and delay in order to annihilate the error string. Said differently, one needs to dissipatively engineer the action of a non-local minimum-weight matching or RG decoder. \\

One can in fact view the classical non-local minimum-weight matching decoder as a source of computational power which jump-starts our quantum memory. Note that the RG decoder is non-local (even allowing for parallel processing of clusters on the lattice by local automata) as the maximum number of recursions $r$ scales as $O(\log L)$ leading to a maximum cluster size proportional to the linear size of the lattice. The lowest levels of the RG decoder are of course local and will provide some measure of protection by local means. 

In \cite{thesis:harrington} the author devises a scheme for doing purely local quantum error correction for the surface code by a 2D cellular automaton assuming independent local errors; the mere existence of such a scheme is nontrivial as it has to deal with noise and communication delays. Other examples of a local dissipative surface code decoder are the proposals in \cite{fuji+:decoder} and \cite{herold+:ca_decode}, but both these decoders assume noise-free parity-check measurements and noise-free classical processing in the cellular automaton. 

It was claimed in \cite{fowler:claim} that one can do minimum weight matching in O(1) parallel time {\em on average} taking into account finite speed of communication between processing cells. Establishing this in full rigor is an important fundamental question since it would be problematic to have a syndrome processing rate $r_{\rm proc}$ which fundamentally depends on $L$: if this were the case, there would always be a large enough $L$ such that one runs into the backlog problem discussed in Section \ref{sec:backlog}. 

The advantage of a 2D local on-chip decoder over extracting the syndrome data record to the classical world and using standard classical processing is that such on-chip decoder can (1) potentially lead to faster decoding (and this is important to avoid a syndrome record backlog), as it uses parallelism in full with dedicated hardware and (2) it avoids a large data stream having to come out of the quantum device. It is of interest to explore whether one can build such a local cellular automaton decoder out of reliable classical, and sufficiently fast (CMOS) logic at low, O(10) mK, temperature.\\

One should contrast the challenge of designing a fast decoder for the 2D surface code with the local decoder for the 4D toric code \cite{dennis+:top}, see the description of the 4D toric code in Section \ref{sec:homo}. The local redundancy of the error syndrome of the 4D toric code has the following consequence. A non-trivial error syndromes will form a closed loop which is the boundary of an error cluster: this is similar to a domain of flipped-spin surrounded by a domain wall in a 2D ferromagnetic Ising model. Then the idea of a local decoder, which, unlike in the 2D case, does not require communication over lengths depending on $L$, is then as follows. One removes an error cluster by locally shrinking the length of the non-trivial error syndrome loop. Thus there is a purely locally-defined ``defect energy" function whose minimization leads to the shrinking of the error cluster and thus to error correction. The local decoder could for example be entirely realized as a quantum circuit, requiring no quantum measurement. In this local decoding quantum circuit one should also expect errors to occur at a certain rate which means that clusters of errors on the data qubits can be sometimes grow instead of shrink. For sufficiently low error rates, one may expect that more errors are locally removed rather than added, either by decoherence or by incorrect decoding, such that a logical error is exponentially (exponential in some function of the block size $n$) rare and the quantum information is protected.\\

The absence of a local cost function which a local decoder can minimize, is a generic property of 2D stabilizer codes and is directly related to the string-like nature of the error excitations which have observable defects only at their 0-dimensional boundary. It has been proven that all 2D stabilizer codes \cite{BT:mem} have string-like logical operators and this directly ties in with the lack of self-correction for these models, see Section \ref{sec:self}.

\subsubsection{Parity Check Measurements and Their Implementation}
\label{sec:parityQED}

In a variety of physical systems (parity check) measurements are implemented as weak continuous measurements in time rather than a sequence of strong projective measurements. Examples of weak continuous qubit measurements are the measurement of a spin qubit in a semi-conducting quantum dot through a quantum point contact \cite{elzerman+:qpc} and the measurement of a superconducting transmon qubit through homodyne measurement of a microwave cavity field with which it interacts. 

For short time scales such continuous weak measurements suffer from inevitable shot-noise as the current or voltage is carried by low numbers of quanta (electrons or photons); this noise averages out on longer times scales revealing the signal. The shot noise thus bounds the rate at which parity information can be gathered. The effect of leakage of qubits appears in such measurement traces either as a different output signal (detection of leakage) or, in the worst case, leads to a similar, and thus non-trustworthy, output signal which makes the parity check record unreliable for as long as a qubit occupies a leaked state.\\

The idea of realizing the surface code in superconducting circuit-QED systems using ancilla qubits for measurement, laying out a possible way to couple transmon qubits and resonators, was considered in \cite{divincenzo_arch}. A scalable surface code architecture was proposed and a basic unit implemented in \cite{barends+:xmon} (see also \cite{barends+:jos_surf}). The optimal way of using superconducting transmon qubits to realize a surface code architecture is a subject of current ongoing research, see e.g.   \cite{GFG:surf_arch} for a direct comparison between three different architectures which differ in how transmon qubits are coupled to microwave resonators. In a transmon-qubit based architecture it is important to show how one can deal with leakage errors since transmon qubits are weakly anharmonic multi-level systems. An important feature of a physical parity check measurement scheme is whether it allows leakage on data or ancilla qubits to be detected or whether leakage goes undetected. Recent papers such as \cite{GF:leakage, SCG:leakage} have started to consider how to handle leakage in a surface code architecture.\\

In order to reduce qubit overhead and possibly make better use of the given physical interactions, e.g. cavity-atom (in cavity QED) or cavity-superconducting qubit (in circuit-QED) interactions, it is worthwhile to consider the idea of a {\em direct} parity check measurements instead of a parity measurement that is realized with several two-qubit gates and an ancilla as in Fig.~\ref{fig:paritycheck}. 

Any mechanism through which a probe pulse (modeled as a simple coherent state $\ket{\alpha(t)}$) picks up a $\pi$ phase shift depending on a qubit state being $\ket{0}$ or $\ket{1}$ could function as the basis for such direct parity measurement. As long as the imprint of multiple qubits is through the addition of such phase shifts, we have $\alpha(t) \rightarrow \alpha(t) e^{i \pi P}$ where $P$ is the parity of the qubits. Homodyne detection of such a probe pulse in time, that is, the continuous measurement of $\langle a(t)+a^{\dagger}(t) \rangle=(-1)^P 2 \langle \alpha(t)\rangle$ (assuming $\alpha=\alpha^*$), could then realize a weak continuous parity measurement. 

This kind of set-up is natural for strong light-matter interactions in cavity-QED and circuit-QED where the state of the qubit can alter the refractive index of the medium (cavity) on which a probe pulse impinges. The challenge is to obtain phase shifts as large as $\pi$ and ensure that these probe pulses do not contain more information about the qubits than their parity as this would lead to additional dephasing inside the odd/even parity subspace.\\

In the cavity-QED setting \cite{KBSM:parity} considered the realization of a continuous weak two-qubit parity measurement on two multi-level atoms each contained in a different cavity (for possible improvements on this scheme, see \cite{nielsen:2qubitparity}). \cite{LGB:2qubit_parity} considered a direct two-qubit parity measurement of two transmon qubits dispersively coupled to a single microwave cavity (circuit-QED setting). Similarly \cite{divsol:parity} and \cite{NG:stab} have considered direct 3 or more qubit parity check measurements for transmon qubits coupled to 2D or 3D microwave cavities.\\

\cite{kerckhoffetal:qmem, KPCM:bs} have developed the interesting idea of a fully autonomous quantum memory which runs with fixed, time independent, input driving fields. In this approach, it is imagined that qubits are encoded in multi-level atoms coupled to the standing electromagnetic modes of (optical) cavities. Both $Z$- as well as $X$-parity checks of the qubits are continuously obtained via probe pulses applied to these cavities. These probe pulses are to be subsequently processed via photonic switches to coherently perform continuous quantum error correction.

\subsection{Topological Order and Self-Correction}
\label{sec:self}

A different route towards protecting quantum information is based on passive Hamiltonian engineering. In this approach quantum information is encoded in an eigenspace, typically ground space, of a many-body, topologically-ordered, Hamiltonian. 
There is no completely rigorous definition of topological order in the literature. At an intuitive level it means that there does not exist a local order parameter or observable which distinguishes different degenerate ground states. Such property is immediately obtained when the groundspace is the code space of a quantum error correcting code with macroscopic (meaning scaling as some function of the system size) distance as follows. 

The quantum error correction conditions, Eq.~(\ref{eq:QEC}) in Sec. \ref{sec:phys_codes} can be slightly reformulated, see Theorem 3 in \cite{gottesman:reviewQEC}: A code $C$ can correct a set of errors $E \in {\cal E}$ if and only if for all states $\ket{\overline{\psi}}$ in the code space $C$ we have
\begin{equation}
\bra{\overline{\psi}} E^{\dagger} E \ket{\overline{\psi}}=c(E),
\end{equation}
where the constant $c(E)$ is independent of $\ket{\overline{\psi}}$. If the set of errors ${\cal E}$ is a set of errors which act locally on $O(1)$ qubits not scaling with system size, such that $E^{\dagger} E$ are local observables, then this condition precisely captures the intuitive idea of topological order. Thus if we devise a physical system with a Hamiltonian such that the ground space corresponds to the code space of a code which can correct any set of local errors, such system would be topologically-ordered. 

The simplest examples of such systems are $D$-dimensional stabilizer codes with macroscopic distances scaling with system-size. For such codes we can define a many-body qubit Hamiltonian $H_{\rm topo}=-\Delta \sum_i S_i$ where $S_i$ is a set of (overcomplete) stabilizer generators. A consequence of topological order is that the ground-space degeneracy of the Hamiltonian $H$ is insensitive to weak local perturbations. This feature has been rigorously proved for stabilizer codes in \cite{BHM:stab} under a slightly sharpened form of the quantum error correction conditions referred to as local topological order. It has not yet been established whether subsystem stabilizer code Hamiltonians of the form
$H=-\Delta \sum_i G_i$ with local generators $G_i$ of the gauge group $\calG$, also have an eigenspace degeneracy which is insensitive to weak perturbations. \\

If we store quantum information passively in a physical system described by some effective Hamiltonian $H_{\rm topo}$, we assume no physical mechanism which actively removes error excitations. Rather we invoke the argument that the presence of a sufficiently large energy gap above the ground-space in the Hamiltonian will exponentially (as $\exp(-\Delta/T)$) suppress error excitations at sufficiently low temperature $T$.  Whether this is practically sufficient, depends on the empirical value of $\Delta/T$ (and the uniformity of this value across a physical sample). \\

One may consider how to engineer a physical system such that it has the effective, say, 4-qubit interactions of the surface code between nearby qubits in a 2D array \cite{kitaev:anyon_pert}. The strength of this approach is that the protection is built into the hardware instead of being imposed dynamically, negating for example the need for control lines for time-dependent pulses. The challenge of this approach is that it requires one-, two- and three-qubit terms in the effective Hamiltonian to be small: the elementary qubits of the many-body system should therefore have approximately degenerate levels $\ket{0}$ and $\ket{1}$. However, in order to encode information in, say, the ground-space of such Hamiltonian, one will need to lift this degeneracy to be able to address these levels. Another challenging aspect of such Hamiltonian engineering is that the desired, say, 4-body interactions will typically be arrived at perturbatively. This means that their strength and therefore the gap of the topologically-ordered Hamiltonian compared with the temperature $T$ may be small, leading to inevitable error excitations. \cite{DI:qec} reviews several ideas for the topological protection of quantum information in superconducting systems while \cite{gladchenko+:nature} demonstrates their experimental feasibility. Another example is the proposal to realize the parity checks of the surface code through Majorana fermion tunneling between 2D arrays of superconducting islands, each supporting 4 Majorana bound states with fixed parity \cite{TFD:majsurf}. \\

The information stored in such passive, topologically-ordered many-body system is, at sufficiently low-temperature, protected by a non-zero energy gap. Research has been devoted to the question whether the $T=0$ topological phase can genuinely extend to non-zero temperature $T > 0$ 
\cite{dennis+:top}. The same question has also been approached from a dynamical perspective with the notion of a self-correcting quantum memory \cite{bacon:mem} (see also the notion of thermal fragility discussed in \cite{NO:thermal}).

A self-correcting quantum memory is a quantum memory in which the accumulation of error excitations over time, which can in turn lead to logical errors, is energetically disfavored due to the presence of macroscopic energy barriers. In this approach it is assumed that the quantum system is in contact with a thermal heat-bath which is both a source of error excitations as well as error correction depending on the energy of the error excitations and the temperature of the bath.
The difference with the active quantum error correction approach is thus that for active quantum error correction we strive to actively engineer part of the environment which should perform the error correction (via parity check measurements).
For a passive thermal memory one expects the rate of logical errors to scale empirically with an Arrhenius law as $A\exp(-E_{\rm barrier}/kT)$ where $E_{\rm barrier}$ is the height of the energy barrier and $A$ is an entropic prefactor. In order to achieve self-correction, we want a logical qubit encoded in such quantum memory to have a coherence time $\tau(T,n)$ which grows with the size $n$ (the elementary qubits of the memory) for some temperature $0 < T < T_c$. \\

One can study the question of self-correction for Hamiltonians $H_{\rm topo}=-\Delta \sum_i S_i$ related to $D$-dimensional stabilizer (or subsystem) codes. For stabilizer codes,  Pauli errors map the ground-space onto excited eigenstates with energy at least $2 \Delta$. The energy barrier associated with such a Hamiltonian is defined as the minimum energy that has to be expended in order to perform any logical operator by means of a sequence of local $O(1)$-weight Pauli errors \cite{BT:mem}. The application of each local Pauli error maps an energy eigenstate onto a new energy eigenstate: a sequence of such operators describes a path through the energy landscape. One can consider all sequences of local errors which result in overall executing a logical operator. The energy barrier of the logical operator is then given by the minimum over all paths of the maximum energy barrier on each path.\\

One important finding concerning self-correcting quantum memories is that a finite temperature `quantum memory phase'  based on macroscopic energy barriers is unlikely to exist for genuinely local 2D quantum systems. One can prove that any 2D stabilizer code has an energy barrier $E_{\rm barrier}=O(1)$: this result is obtained by showing that there always exist string-like logical operators for a 2D stabilizer code \cite{BT:mem}. The surface code with its string-like logical $\overline{X}$ and $\overline{Z}$ operators which run between boundaries provides a good example of this generic behavior.

The 3D toric code on a lattice of $n=O(L^3)$ qubits, as discussed in Section \ref{sec:homo}, has a surface-like logical $\overline{X}$ operator (element in $H^1(T_3,\mathbb{Z}_2)$) and thus an energy barrier $E_{\rm barrier}\sim L $ for the logical $\overline{X}$. But the logical $\overline{Z}$ (element in $H_1(T_3,\mathbb{Z}_2)$) is string-like and has a $O(1)$ energy barrier. One can view the 3D toric code as a model for storing a classical bit passively in a thermal environment \cite{CC:3Dtoric}.

The 4D toric code on a cubic lattice with linear dimension $L$ has been shown to be a good example of finite-temperature topological order or a self-correcting memory with a coherence time $\tau(T < T_c, L) \sim \exp(O(L))$,  see \cite{dennis+:top, AHHH:qmem}. The properties of the 4D toric code which make this possible are the fact that both logical operators are surface-like and error clusters are surrounded by closed non-trivial syndrome loops as discussed in Section \ref{sec:homo} and Section \ref{sec:disc_pract}.\\

For three-dimensional stabilizer codes which are translationally-invariant {\em and} for which the number of encoded qubits does not depend on the lattice size, it has been shown that there always exist string-like logical operators and thus the energy barrier is again $O(1)$ \cite{yoshida}.  For homological codes defined on three-dimensional manifolds this result can be understood by invoking (Poincar\'e) duality. For a $D$-dimensional manifold $M$ the $k$th cohomology group $H^k(M, \mathbb{Z}_2)$ is isomorphic to the homology group $H_{D-k}(M,\mathbb{Z}_2)$ (as $i$-simplices are mapped to $(n-i)$-simplices on the dual lattice). Thus in three dimensions, the presence of a surface-like logical operator in, say, $H_2(M,\mathbb{Z}_2)$) also implies the presence of a matching string-like logical operator in $H^2(M,\mathbb{Z}_2) \simeq H_1(M,\mathbb{Z}_2)$. \\

Given this duality perspective, it should be considered surprising that it is possible to construct 3-dimensional stabilizer codes which have an energy barrier which scales as a function of $L$.  Such codes have to avoid Yoshida's no-go result by either being non-translationally-invariant or encoding a number of qubits which does depend on the lattice size (or both). The first example of such 3D code was the Haah code with $E_{\rm barrier}\geq c \log L$ \cite{haah:nostring}, \cite{BH:barrier} for which all logical operators are fractal (instead of string or surface-like). For the Haah code the number of encoded qubits does non-trivially depend on the lattice size.

Another construction is Michnicki's welded-code \cite{michnicki} which breaks translational  invariance, and has an energy barrier $E_{\rm barrier}=O(L^{2/3})$ for a $n=O(L^3)$ system. For the Haah code it was shown in \cite{BH:3Dcube} that the existence of the energy barrier implies that $\tau(T,n) \sim L^{c/kT}$ as long as $L$ is below some critical temperature-dependent length scale and a similar result holds for the welded-code. 

It is an open question whether topological subsystem codes in 3D behave differently than stabilizer codes in terms of their self-correcting properties. 


We refer to e.g. \cite{wootton:review} and references therein for another overview of results in this area of research.

\section{Discussion}
\label{sec:prac}

The current qubit realizations seem perhaps awkwardly suited to constitute the elementary qubits of an error-correcting code. Most elementary qubits are realized as {\em non-degenerate} eigenlevels (in a higher-dimensional space), approximately described by some $H_0=-\frac{\omega}{2} Z$. The presence of $H_0$ immediately gives a handle on this qubit, i.e. processes which exchange energy with this qubit will drive it from $\ket{1}$ to $\ket{0}$ and vice versa (Rabi oscillations) and coupling of the qubit to other quantum degrees of freedom can be used for qubit read-out. Passive (time-independent) interactions with other quantum systems are intentionally weak and only lead to significant multiple-qubit interactions if we supply energy in the form of time-dependent AC or DC fields meeting resonance conditions. To drive, keep or project multiple qubits via local parity checks in a code space where they are highly entangled, active control at the elementary qubit level will thus be continuously needed, making the macroscopic coding overhead look daunting.\\

For such non-degenerate qubits typically all gates and preparation steps are realized in {\em the rotating frame}: the frame of reference in which the qubit state is no longer precessing around the $z$-axis on the Bloch-sphere due the presence of $H_0$. Any codeword $\ket{\overline{\psi}}$ is then only a fixed quantum state in this rotating frame while it is dynamically rotating under single-qubit $Z$ rotations in the lab frame. As measurements are only done in the lab frame, it is only $Z$-measurements which can be done directly while $X$-measurements typically require an active rotation (e.g. Hadamard) followed by a $Z$-measurement. 

Once elementary qubits are used for times much longer than their coherence time, i.e. when they are used together in a quantum memory, the question of stability of this lab reference frame or the stability of the qubit frequency $\omega$ becomes important. Due to $1/f$ noise and aging of materials from which qubits are constructed, the qubit frequency $\omega$ can systematically drift over longer times and average to new values which are different from short time-averages. This has two consequences: one is that one needs to determine the qubit frequency periodically, for example by taking qubits periodically off-line and measuring them. In this manner one can re-calibrate gates whose implementation depends on knowing the rotating frame. Secondly, shifts in qubit frequency also induce shifts in coherence times as these times depend on the noise power spectral density $S(\omega)$ at the qubit frequency. Such fluctuations of coherence times over longer time-scales have been observed. As an example we can take the results in \cite{metcalfe+:decoh} which report that the $T_1$ time of a superconducting `quantronium' qubit is changing every few seconds over a range of $1.4-1.8\mu$sec. It is clear that if elementary qubits are to be successfully used in a quantum memory, then fluctuations of the noise rate have to be such that one remains below the noise threshold of the code that is employed in the memory at all times. \\

Let us conclude by listing some issues on which we expect to see more progress from the perspective of coding theory. One question is the issue of minimizing qubit and computational overhead in a fault-tolerant computer. It is not clear that the surface code is the ideal platform for this because of its large overhead. It may be advantageous to consider architectures with non-local connects so that one can use a quantum LDPC code which does not make reference to spatial locality and which can escape the no-go results for low-dimensional stabilizer codes in allowing for, say, a transversal $T$ gate. In addition, quantum LDPC codes which are not restricted to $D$ dimensions allow for a constant encoding rate $k/n$. How much they can reduce overhead also depends on the numerical value of this rate which for various quantum LDPC codes has not yet been determined.\\

We can illustrate the issue of overhead due to the non-transversality of the $T$ gate by the following consideration. An efficient quantum algorithm on $N$ qubits takes ${\rm poly}(N)$ gates where ${\rm poly}(N)$ is typically not linear in $N$. For example, for Shor's factoring algorithm the bulk of the algorithm uses $O(N^3)$ Toffoli gates which are non-Clifford gates. If one uses ancillas to create such Toffoli gates (or $T$ gates for that matter which can be used to make Toffoli gates), it means that one needs at least $O(N^3)+N$ qubits. The size of the original quantum circuit in non-Clifford gates is thus converted to the number of logical qubits. As a concrete example, ~\cite{fowler:practical} estimates that, in order to factor a 2,000 bit number with the surface code architecture, using magic state distillation, only $6\%$ of the logical qubits are data qubits, all other's are logical ancillas for the $T$ gates. For this architecture, each logical qubit is already comprised of 14,500 physical qubits leading to a total of about 1 billion physical qubits. \\

One possible approach to reduce overhead is to choose the surface code as bottom code and a code with a transversal $T$ gate and a high rate as top code, as suggested in \cite{CDT:study}. In principle the choice of top code is not restricted by physical locality as one can implement a SWAP gate between the logical qubits of the bottom code using 3 CNOT gates, so any quantum LDPC code could be used. This SWAP gate will take a time which scales at least with $L$ (as one has to repeat syndrome measurements $O(L)$ times), hence more non-locality would lead to a slower computation. 

\section{Acknowledgements}
I would like to thank Ben Criger for quickly making some of the figures and Nikolas Breuckmann, David DiVincenzo and Daniel Gottesman for interesting discussions and feedback on this review. This research was supported in part by the Perimeter Institute for Theoretical Physics. Research at Perimeter Institute is supported by the Government of Canada through Industry Canada and by the Province of Ontario through the Ministry of Economic Development $\&$ Innovation. Funding is acknowledged through the EU via the programme ScaleQIT.

\bibliography{QEC_review}

\begin{thebibliography}{160}%
\makeatletter
\providecommand \@ifxundefined [1]{%
 \@ifx{#1\undefined}
}%
\providecommand \@ifnum [1]{%
 \ifnum #1\expandafter \@firstoftwo
 \else \expandafter \@secondoftwo
 \fi
}%
\providecommand \@ifx [1]{%
 \ifx #1\expandafter \@firstoftwo
 \else \expandafter \@secondoftwo
 \fi
}%
\providecommand \natexlab [1]{#1}%
\providecommand \enquote  [1]{``#1''}%
\providecommand \bibnamefont  [1]{#1}%
\providecommand \bibfnamefont [1]{#1}%
\providecommand \citenamefont [1]{#1}%
\providecommand \href@noop [0]{\@secondoftwo}%
\providecommand \href [0]{\begingroup \@sanitize@url \@href}%
\providecommand \@href[1]{\@@startlink{#1}\@@href}%
\providecommand \@@href[1]{\endgroup#1\@@endlink}%
\providecommand \@sanitize@url [0]{\catcode `\\12\catcode `\$12\catcode
  `\&12\catcode `\#12\catcode `\^12\catcode `\_12\catcode `\%12\relax}%
\providecommand \@@startlink[1]{}%
\providecommand \@@endlink[0]{}%
\providecommand \url  [0]{\begingroup\@sanitize@url \@url }%
\providecommand \@url [1]{\endgroup\@href {#1}{\urlprefix }}%
\providecommand \urlprefix  [0]{URL }%
\providecommand \Eprint [0]{\href }%
\providecommand \doibase [0]{http://dx.doi.org/}%
\providecommand \selectlanguage [0]{\@gobble}%
\providecommand \bibinfo  [0]{\@secondoftwo}%
\providecommand \bibfield  [0]{\@secondoftwo}%
\providecommand \translation [1]{[#1]}%
\providecommand \BibitemOpen [0]{}%
\providecommand \bibitemStop [0]{}%
\providecommand \bibitemNoStop [0]{.\EOS\space}%
\providecommand \EOS [0]{\spacefactor3000\relax}%
\providecommand \BibitemShut  [1]{\csname bibitem#1\endcsname}%
\let\auto@bib@innerbib\@empty
\bibitem [{\citenamefont {Aharonov}\ and\ \citenamefont
  {{Ben-Or}}(1997)}]{AB:faulttol}%
  \BibitemOpen
  \bibfield  {author} {\bibinfo {author} {\bibnamefont {Aharonov},
  \bibfnamefont {D}}, \ and\ \bibinfo {author} {\bibfnamefont {M.}~\bibnamefont
  {{Ben-Or}}}} (\bibinfo {year} {1997}),\ \bibfield  {title} {\enquote
  {\bibinfo {title} {Fault-tolerant quantum computation with constant error},}\
  }in\ \href@noop {} {\emph {\bibinfo {booktitle} {Proceedings of 29th
  STOC}}},\ pp.\ \bibinfo {pages} {176--188},\ \Eprint
  {http://arxiv.org/abs/\url{http://arxiv.org/abs/quant-ph/9611025}}
  {\url{http://arxiv.org/abs/quant-ph/9611025}} \BibitemShut {NoStop}%
\bibitem [{\citenamefont {Aharonov}\ and\ \citenamefont
  {Eldar}(2011)}]{AE:top}%
  \BibitemOpen
  \bibfield  {author} {\bibinfo {author} {\bibnamefont {Aharonov},
  \bibfnamefont {D}}, \ and\ \bibinfo {author} {\bibfnamefont {L.}~\bibnamefont
  {Eldar}}} (\bibinfo {year} {2011}),\ \bibfield  {title} {\enquote {\bibinfo
  {title} {On the complexity of commuting local {H}amiltonians, and tight
  conditions for topological order in such systems.}}\ }in\ \href@noop {}
  {\emph {\bibinfo {booktitle} {Proceedings of FOCS 2011}}}\ (\bibinfo
  {publisher} {IEEE})\ pp.\ \bibinfo {pages} {334--343},\ \bibinfo {note}
  {\url{http://arxiv.org/abs/1102.0770}}\BibitemShut {NoStop}%
\bibitem [{\citenamefont {{Aharonov}}\ \emph {et~al.}(2006)\citenamefont
  {{Aharonov}}, \citenamefont {{Kitaev}},\ and\ \citenamefont
  {{Preskill}}}]{AKP:shortrange}%
  \BibitemOpen
  \bibfield  {author} {\bibinfo {author} {\bibnamefont {{Aharonov}},
  \bibfnamefont {D}}, \bibinfo {author} {\bibfnamefont {A.}~\bibnamefont
  {{Kitaev}}}, \ and\ \bibinfo {author} {\bibfnamefont {J.}~\bibnamefont
  {{Preskill}}}} (\bibinfo {year} {2006}),\ \bibfield  {title} {\enquote
  {\bibinfo {title} {{Fault-Tolerant Quantum Computation with Long-Range
  Correlated Noise}},}\ }\href {\doibase 10.1103/PhysRevLett.96.050504}
  {\bibfield  {journal} {\bibinfo  {journal} {Phys. Rev. Lett.}\ }\textbf
  {\bibinfo {volume} {96}}~(\bibinfo {number} {5}),\ \bibinfo {eid}
  {050504}}\BibitemShut {NoStop}%
\bibitem [{\citenamefont {{Ahn}}\ \emph {et~al.}(2002)\citenamefont {{Ahn}},
  \citenamefont {{Doherty}},\ and\ \citenamefont {{Landahl}}}]{ADL:feedback}%
  \BibitemOpen
  \bibfield  {author} {\bibinfo {author} {\bibnamefont {{Ahn}}, \bibfnamefont
  {C}}, \bibinfo {author} {\bibfnamefont {A.~C.}\ \bibnamefont {{Doherty}}}, \
  and\ \bibinfo {author} {\bibfnamefont {A.~J.}\ \bibnamefont {{Landahl}}}}
  (\bibinfo {year} {2002}),\ \bibfield  {title} {\enquote {\bibinfo {title}
  {{Continuous quantum error correction via quantum feedback control}},}\
  }\href {\doibase 10.1103/PhysRevA.65.042301} {\bibfield  {journal} {\bibinfo
  {journal} {Phys.~Rev.~A}\ }\textbf {\bibinfo {volume} {65}}~(\bibinfo
  {number} {4}),\ \bibinfo {eid} {042301}}\BibitemShut {NoStop}%
\bibitem [{\citenamefont {{Alicea}}(2012)}]{alicea:review}%
  \BibitemOpen
  \bibfield  {author} {\bibinfo {author} {\bibnamefont {{Alicea}},
  \bibfnamefont {J}}} (\bibinfo {year} {2012}),\ \bibfield  {title} {\enquote
  {\bibinfo {title} {{New directions in the pursuit of Majorana fermions in
  solid state systems}},}\ }\href {\doibase 10.1088/0034-4885/75/7/076501}
  {\bibfield  {journal} {\bibinfo  {journal} {Reports on Progress in Physics}\
  }\textbf {\bibinfo {volume} {75}}~(\bibinfo {number} {7}),\ \bibinfo {eid}
  {076501}}\BibitemShut {NoStop}%
\bibitem [{\citenamefont {Alicki}\ \emph {et~al.}(2010)\citenamefont {Alicki},
  \citenamefont {Horodecki}, \citenamefont {Horodecki},\ and\ \citenamefont
  {Horodecki}}]{AHHH:qmem}%
  \BibitemOpen
  \bibfield  {author} {\bibinfo {author} {\bibnamefont {Alicki}, \bibfnamefont
  {R}}, \bibinfo {author} {\bibfnamefont {M.}~\bibnamefont {Horodecki}},
  \bibinfo {author} {\bibfnamefont {P.}~\bibnamefont {Horodecki}}, \ and\
  \bibinfo {author} {\bibfnamefont {R.}~\bibnamefont {Horodecki}}} (\bibinfo
  {year} {2010}),\ \bibfield  {title} {\enquote {\bibinfo {title} {On thermal
  stability of topological qubit in {K}itaev's {4D} model},}\ }\href {\doibase
  10.1142/S1230161210000023} {\bibfield  {journal} {\bibinfo  {journal} {Open
  Systems and Information Dynamics}\ }\textbf {\bibinfo {volume}
  {17}}~(\bibinfo {number} {01}),\ \bibinfo {pages} {1--20}}\BibitemShut
  {NoStop}%
\bibitem [{\citenamefont {Aliferis}(2007)}]{thesis:aliferis}%
  \BibitemOpen
  \bibfield  {author} {\bibinfo {author} {\bibnamefont {Aliferis},
  \bibfnamefont {P}}} (\bibinfo {year} {2007}),\ \emph {\bibinfo {title} {Level
  Reduction and the Quantum Threshold Theorem}},\ \href@noop {} {Ph.D. thesis}\
  (\bibinfo  {school} {CalTech}),\ \bibinfo {note}
  {\url{http://arxiv.org/abs/quant-ph/0703230}}\BibitemShut {NoStop}%
\bibitem [{\citenamefont {Aliferis}\ and\ \citenamefont {Cross}(2007)}]{AC:bs}%
  \BibitemOpen
  \bibfield  {author} {\bibinfo {author} {\bibnamefont {Aliferis},
  \bibfnamefont {P}}, \ and\ \bibinfo {author} {\bibfnamefont {A.}~\bibnamefont
  {Cross}}} (\bibinfo {year} {2007}),\ \bibfield  {title} {\enquote {\bibinfo
  {title} {Subsystem fault-tolerance with the {B}acon-{S}hor code},}\ }\href
  {\doibase 10.1103/PhysRevLett.98.220502} {\bibfield  {journal} {\bibinfo
  {journal} {Phys. Rev. Lett.}\ }\textbf {\bibinfo {volume} {98}},\ \bibinfo
  {pages} {220502}}\BibitemShut {NoStop}%
\bibitem [{\citenamefont {Aliferis}\ \emph {et~al.}(2006)\citenamefont
  {Aliferis}, \citenamefont {Gottesman},\ and\ \citenamefont
  {Preskill}}]{AGP:ft}%
  \BibitemOpen
  \bibfield  {author} {\bibinfo {author} {\bibnamefont {Aliferis},
  \bibfnamefont {P}}, \bibinfo {author} {\bibfnamefont {D.}~\bibnamefont
  {Gottesman}}, \ and\ \bibinfo {author} {\bibfnamefont {J.}~\bibnamefont
  {Preskill}}} (\bibinfo {year} {2006}),\ \bibfield  {title} {\enquote
  {\bibinfo {title} {Quantum accuracy threshold for concatenated distance-3
  codes},}\ }\href@noop {} {\bibfield  {journal} {\bibinfo  {journal} {Quantum
  Info. and Comput.}\ }\textbf {\bibinfo {volume} {6}},\ \bibinfo {pages}
  {97--165}}\BibitemShut {NoStop}%
\bibitem [{\citenamefont {{Aliferis}}\ \emph {et~al.}(2008)\citenamefont
  {{Aliferis}}, \citenamefont {{Gottesman}},\ and\ \citenamefont
  {{Preskill}}}]{AGK:post}%
  \BibitemOpen
  \bibfield  {author} {\bibinfo {author} {\bibnamefont {{Aliferis}},
  \bibfnamefont {P}}, \bibinfo {author} {\bibfnamefont {D.}~\bibnamefont
  {{Gottesman}}}, \ and\ \bibinfo {author} {\bibfnamefont {J.}~\bibnamefont
  {{Preskill}}}} (\bibinfo {year} {2008}),\ \bibfield  {title} {\enquote
  {\bibinfo {title} {{Accuracy threshold for postselected quantum
  computation}},}\ }\href@noop {} {\bibfield  {journal} {\bibinfo  {journal}
  {Quantum Inf. Comp.}\ }\textbf {\bibinfo {volume} {8}},\ \bibinfo {pages}
  {181--244}}\BibitemShut {NoStop}%
\bibitem [{\citenamefont {{Aliferis}}\ and\ \citenamefont
  {{Preskill}}(2009)}]{AP:fib}%
  \BibitemOpen
  \bibfield  {author} {\bibinfo {author} {\bibnamefont {{Aliferis}},
  \bibfnamefont {P}}, \ and\ \bibinfo {author} {\bibfnamefont {J.}~\bibnamefont
  {{Preskill}}}} (\bibinfo {year} {2009}),\ \bibfield  {title} {\enquote
  {\bibinfo {title} {{Fibonacci scheme for fault-tolerant quantum
  computation}},}\ }\href {\doibase 10.1103/PhysRevA.79.012332} {\bibfield
  {journal} {\bibinfo  {journal} {Phys.~Rev.~A}\ }\textbf {\bibinfo {volume}
  {79}}~(\bibinfo {number} {1}),\ \bibinfo {eid} {012332}}\BibitemShut
  {NoStop}%
\bibitem [{\citenamefont {Aliferis}\ and\ \citenamefont
  {Terhal}(2007)}]{AT:leakage}%
  \BibitemOpen
  \bibfield  {author} {\bibinfo {author} {\bibnamefont {Aliferis},
  \bibfnamefont {P}}, \ and\ \bibinfo {author} {\bibfnamefont {B.M.}\
  \bibnamefont {Terhal}}} (\bibinfo {year} {2007}),\ \bibfield  {title}
  {\enquote {\bibinfo {title} {Fault-tolerant quantum computation for local
  leakage faults},}\ }\href@noop {} {\bibfield  {journal} {\bibinfo  {journal}
  {Quantum Info. and Comput.}\ }\textbf {\bibinfo {volume} {7}},\ \bibinfo
  {pages} {139--156}}\BibitemShut {NoStop}%
\bibitem [{\citenamefont {{Aoki}}\ \emph {et~al.}(2009)\citenamefont {{Aoki}},
  \citenamefont {{Takahashi}}, \citenamefont {{Kajiya}}, \citenamefont
  {{Yoshikawa}}, \citenamefont {{Braunstein}}, \citenamefont {{van Loock}},\
  and\ \citenamefont {{Furusawa}}}]{aoki+:CVshor}%
  \BibitemOpen
  \bibfield  {author} {\bibinfo {author} {\bibnamefont {{Aoki}}, \bibfnamefont
  {T}}, \bibinfo {author} {\bibfnamefont {G.}~\bibnamefont {{Takahashi}}},
  \bibinfo {author} {\bibfnamefont {T.}~\bibnamefont {{Kajiya}}}, \bibinfo
  {author} {\bibfnamefont {J.-i.}\ \bibnamefont {{Yoshikawa}}}, \bibinfo
  {author} {\bibfnamefont {S.~L.}\ \bibnamefont {{Braunstein}}}, \bibinfo
  {author} {\bibfnamefont {P.}~\bibnamefont {{van Loock}}}, \ and\ \bibinfo
  {author} {\bibfnamefont {A.}~\bibnamefont {{Furusawa}}}} (\bibinfo {year}
  {2009}),\ \bibfield  {title} {\enquote {\bibinfo {title} {{Quantum error
  correction beyond qubits}},}\ }\href@noop {} {\bibfield  {journal} {\bibinfo
  {journal} {Nature Physics}\ }\textbf {\bibinfo {volume} {5}},\ \bibinfo
  {pages} {541}}\BibitemShut {NoStop}%
\bibitem [{\citenamefont {Bacon}(2006)}]{bacon:mem}%
  \BibitemOpen
  \bibfield  {author} {\bibinfo {author} {\bibnamefont {Bacon}, \bibfnamefont
  {D}}} (\bibinfo {year} {2006}),\ \bibfield  {title} {\enquote {\bibinfo
  {title} {Operator quantum error correcting subsystems for self-correcting
  quantum memories},}\ }\href@noop {} {\bibfield  {journal} {\bibinfo
  {journal} {Physical Review A}\ }\textbf {\bibinfo {volume} {73}},\ \bibinfo
  {pages} {012340}}\BibitemShut {NoStop}%
\bibitem [{\citenamefont {{Barends}}\ \emph {et~al.}(2013)\citenamefont
  {{Barends}}, \citenamefont {{Kelly}}, \citenamefont {{Megrant}},
  \citenamefont {{Sank}}, \citenamefont {{Jeffrey}}, \citenamefont {{Chen}},
  \citenamefont {{Yin}}, \citenamefont {{Chiaro}}, \citenamefont {{Mutus}},
  \citenamefont {{Neill}}, \citenamefont {{O'Malley}}, \citenamefont
  {{Roushan}}, \citenamefont {{Wenner}}, \citenamefont {{White}}, \citenamefont
  {{Cleland}},\ and\ \citenamefont {{Martinis}}}]{barends+:xmon}%
  \BibitemOpen
  \bibfield  {author} {\bibinfo {author} {\bibnamefont {{Barends}},
  \bibfnamefont {R}}, \bibinfo {author} {\bibfnamefont {J.}~\bibnamefont
  {{Kelly}}}, \bibinfo {author} {\bibfnamefont {A.}~\bibnamefont {{Megrant}}},
  \bibinfo {author} {\bibfnamefont {D.}~\bibnamefont {{Sank}}}, \bibinfo
  {author} {\bibfnamefont {E.}~\bibnamefont {{Jeffrey}}}, \bibinfo {author}
  {\bibfnamefont {Y.}~\bibnamefont {{Chen}}}, \bibinfo {author} {\bibfnamefont
  {Y.}~\bibnamefont {{Yin}}}, \bibinfo {author} {\bibfnamefont
  {B.}~\bibnamefont {{Chiaro}}}, \bibinfo {author} {\bibfnamefont
  {J.}~\bibnamefont {{Mutus}}}, \bibinfo {author} {\bibfnamefont
  {C.}~\bibnamefont {{Neill}}}, \bibinfo {author} {\bibfnamefont
  {P.}~\bibnamefont {{O'Malley}}}, \bibinfo {author} {\bibfnamefont
  {P.}~\bibnamefont {{Roushan}}}, \bibinfo {author} {\bibfnamefont
  {J.}~\bibnamefont {{Wenner}}}, \bibinfo {author} {\bibfnamefont {T.~C.}\
  \bibnamefont {{White}}}, \bibinfo {author} {\bibfnamefont {A.~N.}\
  \bibnamefont {{Cleland}}}, \ and\ \bibinfo {author} {\bibfnamefont {J.~M.}\
  \bibnamefont {{Martinis}}}} (\bibinfo {year} {2013}),\ \bibfield  {title}
  {\enquote {\bibinfo {title} {{Coherent Josephson Qubit Suitable for Scalable
  Quantum Integrated Circuits}},}\ }\href {\doibase
  10.1103/PhysRevLett.111.080502} {\bibfield  {journal} {\bibinfo  {journal}
  {Phys. Rev. Lett.}\ }\textbf {\bibinfo {volume} {111}}~(\bibinfo {number}
  {8}),\ \bibinfo {eid} {080502}}\BibitemShut {NoStop}%
\bibitem [{\citenamefont {{Barends}}\ \emph {et~al.}(2014)\citenamefont
  {{Barends}}, \citenamefont {{Kelly}}, \citenamefont {{Megrant}},
  \citenamefont {{Veitia}}, \citenamefont {{Sank}}, \citenamefont {{Jeffrey}},
  \citenamefont {{White}}, \citenamefont {{Mutus}}, \citenamefont {{Fowler}},
  \citenamefont {{Campbell}}, \citenamefont {{Chen}}, \citenamefont {{Chen}},
  \citenamefont {{Chiaro}}, \citenamefont {{Dunsworth}}, \citenamefont
  {{Neill}}, \citenamefont {{O'Malley}}, \citenamefont {{Roushan}},
  \citenamefont {{Vainsencher}}, \citenamefont {{Wenner}}, \citenamefont
  {{Korotkov}}, \citenamefont {{Cleland}},\ and\ \citenamefont
  {{Martinis}}}]{barends+:jos_surf}%
  \BibitemOpen
  \bibfield  {author} {\bibinfo {author} {\bibnamefont {{Barends}},
  \bibfnamefont {R}}, \bibinfo {author} {\bibfnamefont {J.}~\bibnamefont
  {{Kelly}}}, \bibinfo {author} {\bibfnamefont {A.}~\bibnamefont {{Megrant}}},
  \bibinfo {author} {\bibfnamefont {A.}~\bibnamefont {{Veitia}}}, \bibinfo
  {author} {\bibfnamefont {D.}~\bibnamefont {{Sank}}}, \bibinfo {author}
  {\bibfnamefont {E.}~\bibnamefont {{Jeffrey}}}, \bibinfo {author}
  {\bibfnamefont {T.~C.}\ \bibnamefont {{White}}}, \bibinfo {author}
  {\bibfnamefont {J.}~\bibnamefont {{Mutus}}}, \bibinfo {author} {\bibfnamefont
  {A.~G.}\ \bibnamefont {{Fowler}}}, \bibinfo {author} {\bibfnamefont
  {B.}~\bibnamefont {{Campbell}}}, \bibinfo {author} {\bibfnamefont
  {Y.}~\bibnamefont {{Chen}}}, \bibinfo {author} {\bibfnamefont
  {Z.}~\bibnamefont {{Chen}}}, \bibinfo {author} {\bibfnamefont
  {B.}~\bibnamefont {{Chiaro}}}, \bibinfo {author} {\bibfnamefont
  {A.}~\bibnamefont {{Dunsworth}}}, \bibinfo {author} {\bibfnamefont
  {C.}~\bibnamefont {{Neill}}}, \bibinfo {author} {\bibfnamefont
  {P.}~\bibnamefont {{O'Malley}}}, \bibinfo {author} {\bibfnamefont
  {P.}~\bibnamefont {{Roushan}}}, \bibinfo {author} {\bibfnamefont
  {A.}~\bibnamefont {{Vainsencher}}}, \bibinfo {author} {\bibfnamefont
  {J.}~\bibnamefont {{Wenner}}}, \bibinfo {author} {\bibfnamefont {A.~N.}\
  \bibnamefont {{Korotkov}}}, \bibinfo {author} {\bibfnamefont {A.~N.}\
  \bibnamefont {{Cleland}}}, \ and\ \bibinfo {author} {\bibfnamefont {J.~M.}\
  \bibnamefont {{Martinis}}}} (\bibinfo {year} {2014}),\ \bibfield  {title}
  {\enquote {\bibinfo {title} {{Superconducting quantum circuits at the surface
  code threshold for fault tolerance}},}\ }\href {\doibase 10.1038/nature13171}
  {\bibfield  {journal} {\bibinfo  {journal} {Nature}\ }\textbf {\bibinfo
  {volume} {508}},\ \bibinfo {pages} {500--503}}\BibitemShut {NoStop}%
\bibitem [{\citenamefont {{Barreiro}}\ \emph {et~al.}(2011)\citenamefont
  {{Barreiro}}, \citenamefont {{M{\"u}ller}}, \citenamefont {{Schindler}},
  \citenamefont {{Nigg}}, \citenamefont {{Monz}}, \citenamefont {{Chwalla}},
  \citenamefont {{Hennrich}}, \citenamefont {{Roos}}, \citenamefont
  {{Zoller}},\ and\ \citenamefont {{Blatt}}}]{barreiro+:stab}%
  \BibitemOpen
  \bibfield  {author} {\bibinfo {author} {\bibnamefont {{Barreiro}},
  \bibfnamefont {J~T}}, \bibinfo {author} {\bibfnamefont {M.}~\bibnamefont
  {{M{\"u}ller}}}, \bibinfo {author} {\bibfnamefont {P.}~\bibnamefont
  {{Schindler}}}, \bibinfo {author} {\bibfnamefont {D.}~\bibnamefont {{Nigg}}},
  \bibinfo {author} {\bibfnamefont {T.}~\bibnamefont {{Monz}}}, \bibinfo
  {author} {\bibfnamefont {M.}~\bibnamefont {{Chwalla}}}, \bibinfo {author}
  {\bibfnamefont {M.}~\bibnamefont {{Hennrich}}}, \bibinfo {author}
  {\bibfnamefont {C.~F.}\ \bibnamefont {{Roos}}}, \bibinfo {author}
  {\bibfnamefont {P.}~\bibnamefont {{Zoller}}}, \ and\ \bibinfo {author}
  {\bibfnamefont {R.}~\bibnamefont {{Blatt}}}} (\bibinfo {year} {2011}),\
  \bibfield  {title} {\enquote {\bibinfo {title} {{An open-system quantum
  simulator with trapped ions}},}\ }\href {\doibase 10.1038/nature09801}
  {\bibfield  {journal} {\bibinfo  {journal} {Nature}\ }\textbf {\bibinfo
  {volume} {470}},\ \bibinfo {pages} {486--491}}\BibitemShut {NoStop}%
\bibitem [{\citenamefont {{Bell}}\ \emph {et~al.}(2014)\citenamefont {{Bell}},
  \citenamefont {{Herrera-Mart{\'{\i}}}}, \citenamefont {{Tame}}, \citenamefont
  {{Markham}}, \citenamefont {{Wadsworth}},\ and\ \citenamefont
  {{Rarity}}}]{bell+:graph}%
  \BibitemOpen
  \bibfield  {author} {\bibinfo {author} {\bibnamefont {{Bell}}, \bibfnamefont
  {B~A}}, \bibinfo {author} {\bibfnamefont {D.~A.}\ \bibnamefont
  {{Herrera-Mart{\'{\i}}}}}, \bibinfo {author} {\bibfnamefont {M.~S.}\
  \bibnamefont {{Tame}}}, \bibinfo {author} {\bibfnamefont {D.}~\bibnamefont
  {{Markham}}}, \bibinfo {author} {\bibfnamefont {W.~J.}\ \bibnamefont
  {{Wadsworth}}}, \ and\ \bibinfo {author} {\bibfnamefont {J.~G.}\ \bibnamefont
  {{Rarity}}}} (\bibinfo {year} {2014}),\ \bibfield  {title} {\enquote
  {\bibinfo {title} {{Experimental demonstration of a graph state quantum
  error-correction code}},}\ }\href {\doibase 10.1038/ncomms4658} {\bibfield
  {journal} {\bibinfo  {journal} {Nature Communications}\ }\textbf {\bibinfo
  {volume} {5}},\ \bibinfo {eid} {3658}}\BibitemShut {NoStop}%
\bibitem [{\citenamefont {Bennett}\ \emph {et~al.}(1996)\citenamefont
  {Bennett}, \citenamefont {DiVincenzo}, \citenamefont {Smolin},\ and\
  \citenamefont {Wootters}}]{bdsw}%
  \BibitemOpen
  \bibfield  {author} {\bibinfo {author} {\bibnamefont {Bennett}, \bibfnamefont
  {CH}}, \bibinfo {author} {\bibfnamefont {D.P.}\ \bibnamefont {DiVincenzo}},
  \bibinfo {author} {\bibfnamefont {J.A.}\ \bibnamefont {Smolin}}, \ and\
  \bibinfo {author} {\bibfnamefont {W.K.}\ \bibnamefont {Wootters}}} (\bibinfo
  {year} {1996}),\ \bibfield  {title} {\enquote {\bibinfo {title} {Mixed state
  entanglement and quantum error correction},}\ }\href@noop {} {\bibfield
  {journal} {\bibinfo  {journal} {Phys. Rev. A}\ }\textbf {\bibinfo {volume}
  {54}},\ \bibinfo {pages} {3824--3851}}\BibitemShut {NoStop}%
\bibitem [{\citenamefont {{B{\'e}ny}}\ and\ \citenamefont
  {{Oreshkov}}(2010)}]{BO:approximate}%
  \BibitemOpen
  \bibfield  {author} {\bibinfo {author} {\bibnamefont {{B{\'e}ny}},
  \bibfnamefont {C}}, \ and\ \bibinfo {author} {\bibfnamefont {O.}~\bibnamefont
  {{Oreshkov}}}} (\bibinfo {year} {2010}),\ \bibfield  {title} {\enquote
  {\bibinfo {title} {{General Conditions for Approximate Quantum Error
  Correction and Near-Optimal Recovery Channels}},}\ }\href {\doibase
  10.1103/PhysRevLett.104.120501} {\bibfield  {journal} {\bibinfo  {journal}
  {Phys. Rev. Lett.}\ }\textbf {\bibinfo {volume} {104}}~(\bibinfo {number}
  {12}),\ \bibinfo {eid} {120501}}\BibitemShut {NoStop}%
\bibitem [{\citenamefont {{Bombin}}(2010{\natexlab{a}})}]{bombin:twist}%
  \BibitemOpen
  \bibfield  {author} {\bibinfo {author} {\bibnamefont {{Bombin}},
  \bibfnamefont {H}}} (\bibinfo {year} {2010}{\natexlab{a}}),\ \bibfield
  {title} {\enquote {\bibinfo {title} {{Topological Order with a Twist: Ising
  Anyons from an Abelian Model}},}\ }\href {\doibase
  10.1103/PhysRevLett.105.030403} {\bibfield  {journal} {\bibinfo  {journal}
  {Phys. Rev. Lett.}\ }\textbf {\bibinfo {volume} {105}}~(\bibinfo {number}
  {3}),\ \bibinfo {eid} {030403}}\BibitemShut {NoStop}%
\bibitem [{\citenamefont {{Bombin}}(2010{\natexlab{b}})}]{bombin:topsub}%
  \BibitemOpen
  \bibfield  {author} {\bibinfo {author} {\bibnamefont {{Bombin}},
  \bibfnamefont {H}}} (\bibinfo {year} {2010}{\natexlab{b}}),\ \bibfield
  {title} {\enquote {\bibinfo {title} {{Topological subsystem codes}},}\ }\href
  {\doibase 10.1103/PhysRevA.81.032301} {\bibfield  {journal} {\bibinfo
  {journal} {Phys. Rev. A}\ }\textbf {\bibinfo {volume} {81}}~(\bibinfo
  {number} {3}),\ \bibinfo {eid} {032301}}\BibitemShut {NoStop}%
\bibitem [{\citenamefont {{Bombin}}(2013)}]{bombin:gauge}%
  \BibitemOpen
  \bibfield  {author} {\bibinfo {author} {\bibnamefont {{Bombin}},
  \bibfnamefont {H}}} (\bibinfo {year} {2013}),\ \bibfield  {title} {\enquote
  {\bibinfo {title} {{Gauge Color Codes}},}\ }\href@noop {} {\bibfield
  {journal} {\bibinfo  {journal} {ArXiv e-prints}\ }}\Eprint
  {http://arxiv.org/abs/1311.0879} {arXiv:1311.0879 [quant-ph]} \BibitemShut
  {NoStop}%
\bibitem [{\citenamefont {{Bombin}}\ \emph
  {et~al.}(2012{\natexlab{a}})\citenamefont {{Bombin}}, \citenamefont
  {{Andrist}}, \citenamefont {{Ohzeki}}, \citenamefont {{Katzgraber}},\ and\
  \citenamefont {{Martin-Delgado}}}]{bombin+:depol_transition}%
  \BibitemOpen
  \bibfield  {author} {\bibinfo {author} {\bibnamefont {{Bombin}},
  \bibfnamefont {H}}, \bibinfo {author} {\bibfnamefont {R.~S.}\ \bibnamefont
  {{Andrist}}}, \bibinfo {author} {\bibfnamefont {M.}~\bibnamefont {{Ohzeki}}},
  \bibinfo {author} {\bibfnamefont {H.~G.}\ \bibnamefont {{Katzgraber}}}, \
  and\ \bibinfo {author} {\bibfnamefont {M.~A.}\ \bibnamefont
  {{Martin-Delgado}}}} (\bibinfo {year} {2012}{\natexlab{a}}),\ \bibfield
  {title} {\enquote {\bibinfo {title} {{Strong Resilience of Topological Codes
  to Depolarization}},}\ }\href {\doibase 10.1103/PhysRevX.2.021004} {\bibfield
   {journal} {\bibinfo  {journal} {Phys. Rev. X}\ }\textbf {\bibinfo {volume}
  {2}}~(\bibinfo {number} {2}),\ \bibinfo {eid} {021004}}\BibitemShut {NoStop}%
\bibitem [{\citenamefont {{Bombin}}\ \emph
  {et~al.}(2012{\natexlab{b}})\citenamefont {{Bombin}}, \citenamefont
  {{Duclos-Cianci}},\ and\ \citenamefont {{Poulin}}}]{BDP:uni}%
  \BibitemOpen
  \bibfield  {author} {\bibinfo {author} {\bibnamefont {{Bombin}},
  \bibfnamefont {H}}, \bibinfo {author} {\bibfnamefont {G.}~\bibnamefont
  {{Duclos-Cianci}}}, \ and\ \bibinfo {author} {\bibfnamefont {D.}~\bibnamefont
  {{Poulin}}}} (\bibinfo {year} {2012}{\natexlab{b}}),\ \bibfield  {title}
  {\enquote {\bibinfo {title} {{Universal topological phase of two-dimensional
  stabilizer codes}},}\ }\href {\doibase 10.1088/1367-2630/14/7/073048}
  {\bibfield  {journal} {\bibinfo  {journal} {New Journal of Physics}\ }\textbf
  {\bibinfo {volume} {14}}~(\bibinfo {number} {7}),\ \bibinfo {pages}
  {073048}}\BibitemShut {NoStop}%
\bibitem [{\citenamefont {{Bombin}}\ and\ \citenamefont
  {{Martin-Delgado}}(2006)}]{BM:colorcodes}%
  \BibitemOpen
  \bibfield  {author} {\bibinfo {author} {\bibnamefont {{Bombin}},
  \bibfnamefont {H}}, \ and\ \bibinfo {author} {\bibfnamefont {M.~A.}\
  \bibnamefont {{Martin-Delgado}}}} (\bibinfo {year} {2006}),\ \bibfield
  {title} {\enquote {\bibinfo {title} {{Topological Quantum Distillation}},}\
  }\href {\doibase 10.1103/PhysRevLett.97.180501} {\bibfield  {journal}
  {\bibinfo  {journal} {Phys. Rev. Lett.}\ }\textbf {\bibinfo {volume}
  {97}}~(\bibinfo {number} {18}),\ \bibinfo {eid} {180501}}\BibitemShut
  {NoStop}%
\bibitem [{\citenamefont {{Bombin}}\ and\ \citenamefont
  {{Martin-Delgado}}(2007)}]{BM:transT}%
  \BibitemOpen
  \bibfield  {author} {\bibinfo {author} {\bibnamefont {{Bombin}},
  \bibfnamefont {H}}, \ and\ \bibinfo {author} {\bibfnamefont {M.~A.}\
  \bibnamefont {{Martin-Delgado}}}} (\bibinfo {year} {2007}),\ \bibfield
  {title} {\enquote {\bibinfo {title} {{Topological Computation without
  Braiding}},}\ }\href {\doibase 10.1103/PhysRevLett.98.160502} {\bibfield
  {journal} {\bibinfo  {journal} {Phys. Rev. Lett.}\ }\textbf {\bibinfo
  {volume} {98}}~(\bibinfo {number} {16}),\ \bibinfo {eid}
  {160502}}\BibitemShut {NoStop}%
\bibitem [{\citenamefont {{Bombin}}\ and\ \citenamefont
  {{Martin-Delgado}}(2009)}]{BM:deform}%
  \BibitemOpen
  \bibfield  {author} {\bibinfo {author} {\bibnamefont {{Bombin}},
  \bibfnamefont {H}}, \ and\ \bibinfo {author} {\bibfnamefont {M.~A.}\
  \bibnamefont {{Martin-Delgado}}}} (\bibinfo {year} {2009}),\ \bibfield
  {title} {\enquote {\bibinfo {title} {{Quantum measurements and gates by code
  deformation}},}\ }\href {\doibase 10.1088/1751-8113/42/9/095302} {\bibfield
  {journal} {\bibinfo  {journal} {Jour. of Phys. A: Math. Gen.}\ }\textbf
  {\bibinfo {volume} {42}}~(\bibinfo {number} {9}),\ \bibinfo {pages}
  {095302}}\BibitemShut {NoStop}%
\bibitem [{\citenamefont {Braunstein}(1998)}]{braunstein:CV}%
  \BibitemOpen
  \bibfield  {author} {\bibinfo {author} {\bibnamefont {Braunstein},
  \bibfnamefont {S}}} (\bibinfo {year} {1998}),\ \bibfield  {title} {\enquote
  {\bibinfo {title} {Error correction for continuous quantum variables},}\
  }\href@noop {} {\bibfield  {journal} {\bibinfo  {journal} {Phys. Rev. Lett.}\
  }\textbf {\bibinfo {volume} {80}},\ \bibinfo {pages} {4084}}\BibitemShut
  {NoStop}%
\bibitem [{\citenamefont {{Bravyi}}\ \emph {et~al.}(2013)\citenamefont
  {{Bravyi}}, \citenamefont {{Duclos-Cianci}}, \citenamefont {{Poulin}},\ and\
  \citenamefont {{Suchara}}}]{BDPS:3qubitcheck}%
  \BibitemOpen
  \bibfield  {author} {\bibinfo {author} {\bibnamefont {{Bravyi}},
  \bibfnamefont {S}}, \bibinfo {author} {\bibfnamefont {G.}~\bibnamefont
  {{Duclos-Cianci}}}, \bibinfo {author} {\bibfnamefont {D.}~\bibnamefont
  {{Poulin}}}, \ and\ \bibinfo {author} {\bibfnamefont {M.}~\bibnamefont
  {{Suchara}}}} (\bibinfo {year} {2013}),\ \bibfield  {title} {\enquote
  {\bibinfo {title} {{Subsystem surface codes with three-qubit check
  operators}},}\ }\href@noop {} {\bibfield  {journal} {\bibinfo  {journal}
  {Quantum Info. and Comput.}\ }\textbf {\bibinfo {volume} {13}}~(\bibinfo
  {number} {11/12}),\ \bibinfo {pages} {0963--0985}}\BibitemShut {NoStop}%
\bibitem [{\citenamefont {{Bravyi}}\ and\ \citenamefont
  {{Haah}}(2011)}]{BH:barrier}%
  \BibitemOpen
  \bibfield  {author} {\bibinfo {author} {\bibnamefont {{Bravyi}},
  \bibfnamefont {S}}, \ and\ \bibinfo {author} {\bibfnamefont {J.}~\bibnamefont
  {{Haah}}}} (\bibinfo {year} {2011}),\ \bibfield  {title} {\enquote {\bibinfo
  {title} {{Energy Landscape of 3D Spin Hamiltonians with Topological
  Order}},}\ }\href {\doibase 10.1103/PhysRevLett.107.150504} {\bibfield
  {journal} {\bibinfo  {journal} {Phys. Rev. Lett.}\ }\textbf {\bibinfo
  {volume} {107}}~(\bibinfo {number} {15}),\ \bibinfo {eid}
  {150504}}\BibitemShut {NoStop}%
\bibitem [{\citenamefont {{Bravyi}}\ and\ \citenamefont
  {{Haah}}(2013)}]{BH:3Dcube}%
  \BibitemOpen
  \bibfield  {author} {\bibinfo {author} {\bibnamefont {{Bravyi}},
  \bibfnamefont {S}}, \ and\ \bibinfo {author} {\bibfnamefont {J.}~\bibnamefont
  {{Haah}}}} (\bibinfo {year} {2013}),\ \bibfield  {title} {\enquote {\bibinfo
  {title} {{Analytic and numerical demonstration of quantum self-correction in
  the 3D Cubic Code}},}\ }\href@noop {} {\bibfield  {journal} {\bibinfo
  {journal} {Phys. Rev. Lett.}\ }\textbf {\bibinfo {volume} {111}},\ \bibinfo
  {pages} {200501}}\BibitemShut {NoStop}%
\bibitem [{\citenamefont {{Bravyi}}\ \emph
  {et~al.}(2010{\natexlab{a}})\citenamefont {{Bravyi}}, \citenamefont
  {{Hastings}},\ and\ \citenamefont {{Michalakis}}}]{BHM:stab}%
  \BibitemOpen
  \bibfield  {author} {\bibinfo {author} {\bibnamefont {{Bravyi}},
  \bibfnamefont {S}}, \bibinfo {author} {\bibfnamefont {M.~B.}\ \bibnamefont
  {{Hastings}}}, \ and\ \bibinfo {author} {\bibfnamefont {S.}~\bibnamefont
  {{Michalakis}}}} (\bibinfo {year} {2010}{\natexlab{a}}),\ \bibfield  {title}
  {\enquote {\bibinfo {title} {{Topological quantum order: Stability under
  local perturbations}},}\ }\href {\doibase 10.1063/1.3490195} {\bibfield
  {journal} {\bibinfo  {journal} {Journal of Mathematical Physics}\ }\textbf
  {\bibinfo {volume} {51}}~(\bibinfo {number} {9}),\ \bibinfo {pages}
  {093512}}\BibitemShut {NoStop}%
\bibitem [{\citenamefont {Bravyi}\ and\ \citenamefont
  {Kitaev}(2005)}]{BK:magicdistill}%
  \BibitemOpen
  \bibfield  {author} {\bibinfo {author} {\bibnamefont {Bravyi}, \bibfnamefont
  {S}}, \ and\ \bibinfo {author} {\bibfnamefont {A.}~\bibnamefont {Kitaev}}}
  (\bibinfo {year} {2005}),\ \bibfield  {title} {\enquote {\bibinfo {title}
  {Universal quantum computation with ideal {C}lifford gates and noisy
  ancillas},}\ }\href@noop {} {\bibfield  {journal} {\bibinfo  {journal} {Phys.
  Rev. A}\ }\textbf {\bibinfo {volume} {71}},\ \bibinfo {pages}
  {022316}}\BibitemShut {NoStop}%
\bibitem [{\citenamefont {{Bravyi}}\ and\ \citenamefont
  {{Koenig}}(2013)}]{BK:uni}%
  \BibitemOpen
  \bibfield  {author} {\bibinfo {author} {\bibnamefont {{Bravyi}},
  \bibfnamefont {S}}, \ and\ \bibinfo {author} {\bibfnamefont {R.}~\bibnamefont
  {{Koenig}}}} (\bibinfo {year} {2013}),\ \bibfield  {title} {\enquote
  {\bibinfo {title} {{Classification of topologically protected gates for local
  stabilizer codes}},}\ }\href@noop {} {\bibfield  {journal} {\bibinfo
  {journal} {Phys. Rev. Lett.}\ }\textbf {\bibinfo {volume} {110}},\ \bibinfo
  {pages} {170503}}\BibitemShut {NoStop}%
\bibitem [{\citenamefont {{Bravyi}}\ \emph {et~al.}(2011)\citenamefont
  {{Bravyi}}, \citenamefont {{Leemhuis}},\ and\ \citenamefont
  {{Terhal}}}]{BLT:chamon}%
  \BibitemOpen
  \bibfield  {author} {\bibinfo {author} {\bibnamefont {{Bravyi}},
  \bibfnamefont {S}}, \bibinfo {author} {\bibfnamefont {B.}~\bibnamefont
  {{Leemhuis}}}, \ and\ \bibinfo {author} {\bibfnamefont {B.~M.}\ \bibnamefont
  {{Terhal}}}} (\bibinfo {year} {2011}),\ \bibfield  {title} {\enquote
  {\bibinfo {title} {{Topological order in an exactly solvable 3D spin
  model}},}\ }\href {\doibase 10.1016/j.aop.2010.11.002} {\bibfield  {journal}
  {\bibinfo  {journal} {Annals of Physics}\ }\textbf {\bibinfo {volume}
  {326}},\ \bibinfo {pages} {839--866}}\BibitemShut {NoStop}%
\bibitem [{\citenamefont {{Bravyi}}\ \emph
  {et~al.}(2010{\natexlab{b}})\citenamefont {{Bravyi}}, \citenamefont
  {{Poulin}},\ and\ \citenamefont {{Terhal}}}]{BPT:tradeoff}%
  \BibitemOpen
  \bibfield  {author} {\bibinfo {author} {\bibnamefont {{Bravyi}},
  \bibfnamefont {S}}, \bibinfo {author} {\bibfnamefont {D.}~\bibnamefont
  {{Poulin}}}, \ and\ \bibinfo {author} {\bibfnamefont {B.~M.}\ \bibnamefont
  {{Terhal}}}} (\bibinfo {year} {2010}{\natexlab{b}}),\ \bibfield  {title}
  {\enquote {\bibinfo {title} {{Tradeoffs for Reliable Quantum Information
  Storage in 2D Systems}},}\ }\href {\doibase 10.1103/PhysRevLett.104.050503}
  {\bibfield  {journal} {\bibinfo  {journal} {Phys. Rev. Lett.}\ }\textbf
  {\bibinfo {volume} {104}}~(\bibinfo {number} {5}),\ \bibinfo {eid}
  {050503}}\BibitemShut {NoStop}%
\bibitem [{\citenamefont {Bravyi}\ and\ \citenamefont {Terhal}(2009)}]{BT:mem}%
  \BibitemOpen
  \bibfield  {author} {\bibinfo {author} {\bibnamefont {Bravyi}, \bibfnamefont
  {S}}, \ and\ \bibinfo {author} {\bibfnamefont {B.~M.}\ \bibnamefont
  {Terhal}}} (\bibinfo {year} {2009}),\ \bibfield  {title} {\enquote {\bibinfo
  {title} {A no-go theorem for a two-dimensional self-correcting quantum memory
  based on stabilizer codes},}\ }\href@noop {} {\bibfield  {journal} {\bibinfo
  {journal} {New Journal of Physics}\ }\textbf {\bibinfo {volume} {11}},\
  \bibinfo {pages} {043029}}\BibitemShut {NoStop}%
\bibitem [{\citenamefont {Bravyi}\ and\ \citenamefont
  {Kitaev}(1998)}]{BK:surface}%
  \BibitemOpen
  \bibfield  {author} {\bibinfo {author} {\bibnamefont {Bravyi}, \bibfnamefont
  {S~B}}, \ and\ \bibinfo {author} {\bibfnamefont {A.~Yu}\ \bibnamefont
  {Kitaev}}} (\bibinfo {year} {1998}),\ \href@noop {} {\enquote {\bibinfo
  {title} {Quantum codes on a lattice with boundary},}\ }\Eprint
  {http://arxiv.org/abs/\url{http://arxiv.org/abs/quant-ph/9811052}}
  {\url{http://arxiv.org/abs/quant-ph/9811052}} \BibitemShut {NoStop}%
\bibitem [{\citenamefont {{Brell}}\ \emph {et~al.}(2014)\citenamefont
  {{Brell}}, \citenamefont {{Burton}}, \citenamefont {{Dauphinais}},
  \citenamefont {{Flammia}},\ and\ \citenamefont {{Poulin}}}]{brell+:nonab}%
  \BibitemOpen
  \bibfield  {author} {\bibinfo {author} {\bibnamefont {{Brell}}, \bibfnamefont
  {C~G}}, \bibinfo {author} {\bibfnamefont {S.}~\bibnamefont {{Burton}}},
  \bibinfo {author} {\bibfnamefont {G.}~\bibnamefont {{Dauphinais}}}, \bibinfo
  {author} {\bibfnamefont {S.~T.}\ \bibnamefont {{Flammia}}}, \ and\ \bibinfo
  {author} {\bibfnamefont {D.}~\bibnamefont {{Poulin}}}} (\bibinfo {year}
  {2014}),\ \bibfield  {title} {\enquote {\bibinfo {title} {{Thermalization,
  Error Correction, and Memory Lifetime for Ising Anyon Systems}},}\ }\href
  {\doibase 10.1103/PhysRevX.4.031058} {\bibfield  {journal} {\bibinfo
  {journal} {Phys. Rev. X}\ }\textbf {\bibinfo {volume} {4}}~(\bibinfo {number}
  {3}),\ \bibinfo {eid} {031058}}\BibitemShut {NoStop}%
\bibitem [{\citenamefont {{Brooks}}\ and\ \citenamefont
  {{Preskill}}(2013)}]{BP:bs}%
  \BibitemOpen
  \bibfield  {author} {\bibinfo {author} {\bibnamefont {{Brooks}},
  \bibfnamefont {P}}, \ and\ \bibinfo {author} {\bibfnamefont {J.}~\bibnamefont
  {{Preskill}}}} (\bibinfo {year} {2013}),\ \bibfield  {title} {\enquote
  {\bibinfo {title} {{Fault-tolerant quantum computation with asymmetric
  {B}acon-{S}hor codes}},}\ }\href@noop {} {\bibfield  {journal} {\bibinfo
  {journal} {Phys. Rev. A}\ }\textbf {\bibinfo {volume} {87}}~(\bibinfo
  {number} {3}),\ \bibinfo {eid} {032310}}\BibitemShut {NoStop}%
\bibitem [{\citenamefont {{Bullock}}\ and\ \citenamefont
  {{Brennen}}(2007)}]{BB:qudit_surface}%
  \BibitemOpen
  \bibfield  {author} {\bibinfo {author} {\bibnamefont {{Bullock}},
  \bibfnamefont {S~S}}, \ and\ \bibinfo {author} {\bibfnamefont {G.~K.}\
  \bibnamefont {{Brennen}}}} (\bibinfo {year} {2007}),\ \bibfield  {title}
  {\enquote {\bibinfo {title} {{Qudit surface codes and gauge theory with
  finite cyclic groups}},}\ }\href {\doibase 10.1088/1751-8113/40/13/013}
  {\bibfield  {journal} {\bibinfo  {journal} {Jour. of Phys. A: Math. Gen.}\
  }\textbf {\bibinfo {volume} {40}},\ \bibinfo {pages}
  {3481--3505}}\BibitemShut {NoStop}%
\bibitem [{\citenamefont {{Castelnovo}}\ and\ \citenamefont
  {{Chamon}}(2008)}]{CC:3Dtoric}%
  \BibitemOpen
  \bibfield  {author} {\bibinfo {author} {\bibnamefont {{Castelnovo}},
  \bibfnamefont {C}}, \ and\ \bibinfo {author} {\bibfnamefont {C.}~\bibnamefont
  {{Chamon}}}} (\bibinfo {year} {2008}),\ \bibfield  {title} {\enquote
  {\bibinfo {title} {{Topological order in a three-dimensional toric code at
  finite temperature}},}\ }\href {\doibase 10.1103/PhysRevB.78.155120}
  {\bibfield  {journal} {\bibinfo  {journal} {\prb}\ }\textbf {\bibinfo
  {volume} {78}}~(\bibinfo {number} {15}),\ \bibinfo {eid}
  {155120}}\BibitemShut {NoStop}%
\bibitem [{\citenamefont {{Chase}}\ \emph {et~al.}(2008)\citenamefont
  {{Chase}}, \citenamefont {{Landahl}},\ and\ \citenamefont
  {{Geremia}}}]{CLG:feedback_filter}%
  \BibitemOpen
  \bibfield  {author} {\bibinfo {author} {\bibnamefont {{Chase}}, \bibfnamefont
  {B~A}}, \bibinfo {author} {\bibfnamefont {A.~J.}\ \bibnamefont {{Landahl}}},
  \ and\ \bibinfo {author} {\bibfnamefont {J.}~\bibnamefont {{Geremia}}}}
  (\bibinfo {year} {2008}),\ \bibfield  {title} {\enquote {\bibinfo {title}
  {{Efficient feedback controllers for continuous-time quantum error
  correction}},}\ }\href {\doibase 10.1103/PhysRevA.77.032304} {\bibfield
  {journal} {\bibinfo  {journal} {\pra}\ }\textbf {\bibinfo {volume}
  {77}}~(\bibinfo {number} {3}),\ \bibinfo {eid} {032304}}\BibitemShut
  {NoStop}%
\bibitem [{\citenamefont {Cross}\ \emph
  {et~al.}(2009{\natexlab{a}})\citenamefont {Cross}, \citenamefont {Smith},
  \citenamefont {Smolin},\ and\ \citenamefont {Zeng}}]{cross+:cws}%
  \BibitemOpen
  \bibfield  {author} {\bibinfo {author} {\bibnamefont {Cross}, \bibfnamefont
  {A}}, \bibinfo {author} {\bibfnamefont {G.}~\bibnamefont {Smith}}, \bibinfo
  {author} {\bibfnamefont {J.A.}\ \bibnamefont {Smolin}}, \ and\ \bibinfo
  {author} {\bibfnamefont {Bei}\ \bibnamefont {Zeng}}} (\bibinfo {year}
  {2009}{\natexlab{a}}),\ \bibfield  {title} {\enquote {\bibinfo {title}
  {Codeword stabilized quantum codes},}\ }\href {\doibase
  10.1109/TIT.2008.2008136} {\bibfield  {journal} {\bibinfo  {journal}
  {Information Theory, IEEE Transactions on}\ }\textbf {\bibinfo {volume}
  {55}}~(\bibinfo {number} {1}),\ \bibinfo {pages} {433--438}}\BibitemShut
  {NoStop}%
\bibitem [{\citenamefont {Cross}\ \emph
  {et~al.}(2009{\natexlab{b}})\citenamefont {Cross}, \citenamefont
  {DiVincenzo},\ and\ \citenamefont {Terhal}}]{CDT:study}%
  \BibitemOpen
  \bibfield  {author} {\bibinfo {author} {\bibnamefont {Cross}, \bibfnamefont
  {A~W}}, \bibinfo {author} {\bibfnamefont {D.~P.}\ \bibnamefont {DiVincenzo}},
  \ and\ \bibinfo {author} {\bibfnamefont {B.~M.}\ \bibnamefont {Terhal}}}
  (\bibinfo {year} {2009}{\natexlab{b}}),\ \bibfield  {title} {\enquote
  {\bibinfo {title} {A comparative code study for quantum fault tolerance},}\
  }\href@noop {} {\bibfield  {journal} {\bibinfo  {journal} {Quantum Info. and
  Comput.}\ }\textbf {\bibinfo {volume} {9}}~(\bibinfo {number} {7}),\ \bibinfo
  {pages} {541--572}}\BibitemShut {NoStop}%
\bibitem [{\citenamefont {{Dengis}}\ \emph {et~al.}(2014)\citenamefont
  {{Dengis}}, \citenamefont {{K{\"o}nig}},\ and\ \citenamefont
  {{Pastawski}}}]{DKP:encode}%
  \BibitemOpen
  \bibfield  {author} {\bibinfo {author} {\bibnamefont {{Dengis}},
  \bibfnamefont {J}}, \bibinfo {author} {\bibfnamefont {R.}~\bibnamefont
  {{K{\"o}nig}}}, \ and\ \bibinfo {author} {\bibfnamefont {F.}~\bibnamefont
  {{Pastawski}}}} (\bibinfo {year} {2014}),\ \bibfield  {title} {\enquote
  {\bibinfo {title} {{An optimal dissipative encoder for the toric code}},}\
  }\href {\doibase 10.1088/1367-2630/16/1/013023} {\bibfield  {journal}
  {\bibinfo  {journal} {New Journal of Physics}\ }\textbf {\bibinfo {volume}
  {16}}~(\bibinfo {number} {1}),\ \bibinfo {eid} {013023}}\BibitemShut
  {NoStop}%
\bibitem [{\citenamefont {Dennis}\ \emph {et~al.}(2002)\citenamefont {Dennis},
  \citenamefont {Kitaev}, \citenamefont {Landahl},\ and\ \citenamefont
  {Preskill}}]{dennis+:top}%
  \BibitemOpen
  \bibfield  {author} {\bibinfo {author} {\bibnamefont {Dennis}, \bibfnamefont
  {E}}, \bibinfo {author} {\bibfnamefont {A.}~\bibnamefont {Kitaev}}, \bibinfo
  {author} {\bibfnamefont {A.}~\bibnamefont {Landahl}}, \ and\ \bibinfo
  {author} {\bibfnamefont {J.}~\bibnamefont {Preskill}}} (\bibinfo {year}
  {2002}),\ \bibfield  {title} {\enquote {\bibinfo {title} {Topological quantum
  memory},}\ }\href@noop {} {\bibfield  {journal} {\bibinfo  {journal} {J.
  Math. Phys.}\ }\textbf {\bibinfo {volume} {43}},\ \bibinfo {pages}
  {4452--4505}}\BibitemShut {NoStop}%
\bibitem [{\citenamefont {{DiVincenzo}}(2009)}]{divincenzo_arch}%
  \BibitemOpen
  \bibfield  {author} {\bibinfo {author} {\bibnamefont {{DiVincenzo}},
  \bibfnamefont {D~P}}} (\bibinfo {year} {2009}),\ \bibfield  {title} {\enquote
  {\bibinfo {title} {{Fault-tolerant architectures for superconducting
  qubits}},}\ }\href {\doibase 10.1088/0031-8949/2009/T137/014020} {\bibfield
  {journal} {\bibinfo  {journal} {Physica Scripta Volume T}\ }\textbf {\bibinfo
  {volume} {137}}~(\bibinfo {number} {1}),\ \bibinfo {pages}
  {014020}}\BibitemShut {NoStop}%
\bibitem [{\citenamefont {{DiVincenzo}}\ and\ \citenamefont
  {{Aliferis}}(2007)}]{DA:slow}%
  \BibitemOpen
  \bibfield  {author} {\bibinfo {author} {\bibnamefont {{DiVincenzo}},
  \bibfnamefont {D~P}}, \ and\ \bibinfo {author} {\bibfnamefont
  {P.}~\bibnamefont {{Aliferis}}}} (\bibinfo {year} {2007}),\ \bibfield
  {title} {\enquote {\bibinfo {title} {{Effective fault-tolerant quantum
  computation with slow measurements}},}\ }\href {\doibase
  10.1103/PhysRevLett.98.020501} {\bibfield  {journal} {\bibinfo  {journal}
  {Phys. Rev. Lett.}\ }\textbf {\bibinfo {volume} {98}}~(\bibinfo {number}
  {2}),\ \bibinfo {eid} {020501}}\BibitemShut {NoStop}%
\bibitem [{\citenamefont {DiVincenzo}\ and\ \citenamefont
  {Solgun}(2013)}]{divsol:parity}%
  \BibitemOpen
  \bibfield  {author} {\bibinfo {author} {\bibnamefont {DiVincenzo},
  \bibfnamefont {D~P}}, \ and\ \bibinfo {author} {\bibfnamefont
  {F.}~\bibnamefont {Solgun}}} (\bibinfo {year} {2013}),\ \bibfield  {title}
  {\enquote {\bibinfo {title} {Multi-qubit parity measurement in circuit
  quantum electrodynamics},}\ }\href@noop {} {\bibfield  {journal} {\bibinfo
  {journal} {New Journal of Physics}\ }\textbf {\bibinfo {volume}
  {15}}~(\bibinfo {number} {7}),\ \bibinfo {pages} {075001}}\BibitemShut
  {NoStop}%
\bibitem [{\citenamefont {Dou\c{c}ot}\ and\ \citenamefont
  {Ioffe}(2012)}]{DI:qec}%
  \BibitemOpen
  \bibfield  {author} {\bibinfo {author} {\bibnamefont {Dou\c{c}ot},
  \bibfnamefont {B}}, \ and\ \bibinfo {author} {\bibfnamefont {L.}~\bibnamefont
  {Ioffe}}} (\bibinfo {year} {2012}),\ \bibfield  {title} {\enquote {\bibinfo
  {title} {{Physical implementation of protected qubits }},}\ }\href@noop {}
  {\bibfield  {journal} {\bibinfo  {journal} {Reports on Progress in Physics}\
  }\textbf {\bibinfo {volume} {75}},\ \bibinfo {pages} {072001}}\BibitemShut
  {NoStop}%
\bibitem [{\citenamefont {Duclos-Cianci}\ and\ \citenamefont
  {Poulin}(2010)}]{DP:renorm}%
  \BibitemOpen
  \bibfield  {author} {\bibinfo {author} {\bibnamefont {Duclos-Cianci},
  \bibfnamefont {G}}, \ and\ \bibinfo {author} {\bibfnamefont {D.}~\bibnamefont
  {Poulin}}} (\bibinfo {year} {2010}),\ \bibfield  {title} {\enquote {\bibinfo
  {title} {A renormalization group decoding algorithm for topological quantum
  codes},}\ }in\ \href@noop {} {\emph {\bibinfo {booktitle} {Proceedings of
  Information Theory Workshop (ITW)}}}\ (\bibinfo  {publisher} {IEEE})\ pp.\
  \bibinfo {pages} {1--5},\ \bibinfo {note}
  {\url{http://arxiv.org/abs/1006.1362}}\BibitemShut {NoStop}%
\bibitem [{\citenamefont {{Duclos-Cianci}}\ and\ \citenamefont
  {{Poulin}}(2014)}]{DP:3Ddecoding}%
  \BibitemOpen
  \bibfield  {author} {\bibinfo {author} {\bibnamefont {{Duclos-Cianci}},
  \bibfnamefont {G}}, \ and\ \bibinfo {author} {\bibfnamefont {D.}~\bibnamefont
  {{Poulin}}}} (\bibinfo {year} {2014}),\ \bibfield  {title} {\enquote
  {\bibinfo {title} {{Fault-tolerant renormalization group decoder for Abelian
  topological codes}},}\ }\href@noop {} {\bibfield  {journal} {\bibinfo
  {journal} {Quantum Info. and Comput.}\ }\textbf {\bibinfo {volume}
  {14}}~(\bibinfo {number} {9})}\BibitemShut {NoStop}%
\bibitem [{\citenamefont {{Eastin}}\ and\ \citenamefont
  {{Knill}}(2009)}]{EK:nogo}%
  \BibitemOpen
  \bibfield  {author} {\bibinfo {author} {\bibnamefont {{Eastin}},
  \bibfnamefont {B}}, \ and\ \bibinfo {author} {\bibfnamefont {E.}~\bibnamefont
  {{Knill}}}} (\bibinfo {year} {2009}),\ \bibfield  {title} {\enquote {\bibinfo
  {title} {{Restrictions on transversal encoded quantum gate sets}},}\ }\href
  {\doibase 10.1103/PhysRevLett.102.110502} {\bibfield  {journal} {\bibinfo
  {journal} {Phys. Rev. Lett.}\ }\textbf {\bibinfo {volume} {102}}~(\bibinfo
  {number} {11}),\ \bibinfo {eid} {110502}}\BibitemShut {NoStop}%
\bibitem [{\citenamefont {Edmonds}(1965)}]{edmonds}%
  \BibitemOpen
  \bibfield  {author} {\bibinfo {author} {\bibnamefont {Edmonds}, \bibfnamefont
  {Jack}}} (\bibinfo {year} {1965}),\ \bibfield  {title} {\enquote {\bibinfo
  {title} {Paths, trees, and flowers},}\ }\href {\doibase
  http://dx.doi.org/10.4153/CJM-1965-045-4} {\bibfield  {journal} {\bibinfo
  {journal} {Canad. J. Math.}\ }\textbf {\bibinfo {volume} {17}},\ \bibinfo
  {pages} {449--467}}\BibitemShut {NoStop}%
\bibitem [{\citenamefont {Elzerman}\ \emph {et~al.}(2003)\citenamefont
  {Elzerman}, \citenamefont {Hanson}, \citenamefont {Greidanus}, \citenamefont
  {Willems~van Beveren}, \citenamefont {De~Franceschi}, \citenamefont
  {Vandersypen}, \citenamefont {Tarucha},\ and\ \citenamefont
  {Kouwenhoven}}]{elzerman+:qpc}%
  \BibitemOpen
  \bibfield  {author} {\bibinfo {author} {\bibnamefont {Elzerman},
  \bibfnamefont {J~M}}, \bibinfo {author} {\bibfnamefont {R.}~\bibnamefont
  {Hanson}}, \bibinfo {author} {\bibfnamefont {J.~S.}\ \bibnamefont
  {Greidanus}}, \bibinfo {author} {\bibfnamefont {L.~H.}\ \bibnamefont
  {Willems~van Beveren}}, \bibinfo {author} {\bibfnamefont {S.}~\bibnamefont
  {De~Franceschi}}, \bibinfo {author} {\bibfnamefont {L.~M.~K.}\ \bibnamefont
  {Vandersypen}}, \bibinfo {author} {\bibfnamefont {S.}~\bibnamefont
  {Tarucha}}, \ and\ \bibinfo {author} {\bibfnamefont {L.~P.}\ \bibnamefont
  {Kouwenhoven}}} (\bibinfo {year} {2003}),\ \bibfield  {title} {\enquote
  {\bibinfo {title} {Few-electron quantum dot circuit with integrated charge
  read out},}\ }\href {\doibase 10.1103/PhysRevB.67.161308} {\bibfield
  {journal} {\bibinfo  {journal} {Phys. Rev. B}\ }\textbf {\bibinfo {volume}
  {67}},\ \bibinfo {pages} {161308}}\BibitemShut {NoStop}%
\bibitem [{\citenamefont {Fowler}(2015)}]{fowler:claim}%
  \BibitemOpen
  \bibfield  {author} {\bibinfo {author} {\bibnamefont {Fowler}, \bibfnamefont
  {A}}} (\bibinfo {year} {2015}),\ \bibfield  {title} {\enquote {\bibinfo
  {title} {Minimum weight perfect matching of fault-tolerant topological
  quantum error correction in average {O(1)} parallel time},}\ }\href@noop {}
  {\bibfield  {journal} {\bibinfo  {journal} {Quantum Inf. and Comp.}\ }\textbf
  {\bibinfo {volume} {15}},\ \bibinfo {pages} {0145--0158}}\BibitemShut
  {NoStop}%
\bibitem [{\citenamefont {{Fowler}}(2013{\natexlab{a}})}]{fowler:surface_bad}%
  \BibitemOpen
  \bibfield  {author} {\bibinfo {author} {\bibnamefont {{Fowler}},
  \bibfnamefont {A~G}}} (\bibinfo {year} {2013}{\natexlab{a}}),\ \bibfield
  {title} {\enquote {\bibinfo {title} {{Accurate simulations of planar
  topological codes cannot use cyclic boundaries}},}\ }\href@noop {} {\bibfield
   {journal} {\bibinfo  {journal} {Phys. Rev. A}\ }\textbf {\bibinfo {volume}
  {87}},\ \bibinfo {pages} {062320}}\BibitemShut {NoStop}%
\bibitem [{\citenamefont {{Fowler}}(2013{\natexlab{b}})}]{fowler:correlated}%
  \BibitemOpen
  \bibfield  {author} {\bibinfo {author} {\bibnamefont {{Fowler}},
  \bibfnamefont {A~G}}} (\bibinfo {year} {2013}{\natexlab{b}}),\ \bibfield
  {title} {\enquote {\bibinfo {title} {{Optimal complexity correction of
  correlated errors in the surface code}},}\ }\href@noop {} {\bibfield
  {journal} {\bibinfo  {journal} {ArXiv e-prints}\ }}\Eprint
  {http://arxiv.org/abs/1310.0863} {arXiv:1310.0863 [quant-ph]} \BibitemShut
  {NoStop}%
\bibitem [{\citenamefont {{Fowler}}\ \emph
  {et~al.}(2012{\natexlab{a}})\citenamefont {{Fowler}}, \citenamefont
  {{Mariantoni}}, \citenamefont {{Martinis}},\ and\ \citenamefont
  {{Cleland}}}]{fowler:practical}%
  \BibitemOpen
  \bibfield  {author} {\bibinfo {author} {\bibnamefont {{Fowler}},
  \bibfnamefont {A~G}}, \bibinfo {author} {\bibfnamefont {M.}~\bibnamefont
  {{Mariantoni}}}, \bibinfo {author} {\bibfnamefont {J.~M.}\ \bibnamefont
  {{Martinis}}}, \ and\ \bibinfo {author} {\bibfnamefont {A.~N.}\ \bibnamefont
  {{Cleland}}}} (\bibinfo {year} {2012}{\natexlab{a}}),\ \bibfield  {title}
  {\enquote {\bibinfo {title} {{Surface codes: Towards practical large-scale
  quantum computation}},}\ }\href {\doibase 10.1103/PhysRevA.86.032324}
  {\bibfield  {journal} {\bibinfo  {journal} {\pra}\ }\textbf {\bibinfo
  {volume} {86}}~(\bibinfo {number} {3}),\ \bibinfo {eid} {032324}}\BibitemShut
  {NoStop}%
\bibitem [{\citenamefont {{Fowler}}\ \emph {et~al.}(2009)\citenamefont
  {{Fowler}}, \citenamefont {{Stephens}},\ and\ \citenamefont
  {{Groszkowski}}}]{fowler+:unisurf}%
  \BibitemOpen
  \bibfield  {author} {\bibinfo {author} {\bibnamefont {{Fowler}},
  \bibfnamefont {A~G}}, \bibinfo {author} {\bibfnamefont {A.~M.}\ \bibnamefont
  {{Stephens}}}, \ and\ \bibinfo {author} {\bibfnamefont {P.}~\bibnamefont
  {{Groszkowski}}}} (\bibinfo {year} {2009}),\ \bibfield  {title} {\enquote
  {\bibinfo {title} {{High-threshold universal quantum computation on the
  surface code}},}\ }\href {\doibase 10.1103/PhysRevA.80.052312} {\bibfield
  {journal} {\bibinfo  {journal} {Phys. Rev. A}\ }\textbf {\bibinfo {volume}
  {80}}~(\bibinfo {number} {5}),\ \bibinfo {pages} {052312}}\BibitemShut
  {NoStop}%
\bibitem [{\citenamefont {{Fowler}}\ \emph
  {et~al.}(2012{\natexlab{b}})\citenamefont {{Fowler}}, \citenamefont
  {{Whiteside}}, \citenamefont {{McInnes}},\ and\ \citenamefont
  {{Rabbani}}}]{fowler+:software}%
  \BibitemOpen
  \bibfield  {author} {\bibinfo {author} {\bibnamefont {{Fowler}},
  \bibfnamefont {A~G}}, \bibinfo {author} {\bibfnamefont {A.~C.}\ \bibnamefont
  {{Whiteside}}}, \bibinfo {author} {\bibfnamefont {A.~L.}\ \bibnamefont
  {{McInnes}}}, \ and\ \bibinfo {author} {\bibfnamefont {A.}~\bibnamefont
  {{Rabbani}}}} (\bibinfo {year} {2012}{\natexlab{b}}),\ \bibfield  {title}
  {\enquote {\bibinfo {title} {{Topological code Autotune}},}\ }\href {\doibase
  10.1103/PhysRevX.2.041003} {\bibfield  {journal} {\bibinfo  {journal}
  {Physical Review X}\ }\textbf {\bibinfo {volume} {2}}~(\bibinfo {number}
  {4}),\ \bibinfo {eid} {041003}}\BibitemShut {NoStop}%
\bibitem [{\citenamefont {Fowler}\ \emph {et~al.}(2012)\citenamefont {Fowler},
  \citenamefont {Whiteside},\ and\ \citenamefont {Hollenberg}}]{FWH:timing}%
  \BibitemOpen
  \bibfield  {author} {\bibinfo {author} {\bibnamefont {Fowler}, \bibfnamefont
  {Austin~G}}, \bibinfo {author} {\bibfnamefont {Adam~C.}\ \bibnamefont
  {Whiteside}}, \ and\ \bibinfo {author} {\bibfnamefont {Lloyd C.~L.}\
  \bibnamefont {Hollenberg}}} (\bibinfo {year} {2012}),\ \bibfield  {title}
  {\enquote {\bibinfo {title} {Towards practical classical processing for the
  surface code: Timing analysis},}\ }\href {\doibase
  10.1103/PhysRevA.86.042313} {\bibfield  {journal} {\bibinfo  {journal} {Phys.
  Rev. A}\ }\textbf {\bibinfo {volume} {86}},\ \bibinfo {pages}
  {042313}}\BibitemShut {NoStop}%
\bibitem [{\citenamefont {{Freedman}}\ and\ \citenamefont
  {{Hastings}}(2014)}]{FH:hypergraphs}%
  \BibitemOpen
  \bibfield  {author} {\bibinfo {author} {\bibnamefont {{Freedman}},
  \bibfnamefont {M~H}}, \ and\ \bibinfo {author} {\bibfnamefont {M.~B.}\
  \bibnamefont {{Hastings}}}} (\bibinfo {year} {2014}),\ \bibfield  {title}
  {\enquote {\bibinfo {title} {{Quantum Systems on non-k-hyperfinite complexes:
  a generalization of classical statistical mechanics on expander graphs}},}\
  }\href@noop {} {\bibfield  {journal} {\bibinfo  {journal} {Quantum Inf.
  Comput.}\ }\textbf {\bibinfo {volume} {14}},\ \bibinfo {pages}
  {144}}\BibitemShut {NoStop}%
\bibitem [{\citenamefont {Freedman}\ and\ \citenamefont
  {Meyer}(2001)}]{FM:surface}%
  \BibitemOpen
  \bibfield  {author} {\bibinfo {author} {\bibnamefont {Freedman},
  \bibfnamefont {M~H}}, \ and\ \bibinfo {author} {\bibfnamefont {D.~A.}\
  \bibnamefont {Meyer}}} (\bibinfo {year} {2001}),\ \bibfield  {title}
  {\enquote {\bibinfo {title} {Projective plane and planar quantum codes},}\
  }\href {\doibase 10.1007/s102080010013} {\bibfield  {journal} {\bibinfo
  {journal} {Found. Comput. Math.}\ }\textbf {\bibinfo {volume} {1}}~(\bibinfo
  {number} {3}),\ \bibinfo {pages} {325--332}},\ \Eprint
  {http://arxiv.org/abs/\url{http://arxiv.org/abs/quant-ph/9810055}}
  {\url{http://arxiv.org/abs/quant-ph/9810055}} \BibitemShut {NoStop}%
\bibitem [{\citenamefont {{Fujii}}\ \emph {et~al.}(2014)\citenamefont
  {{Fujii}}, \citenamefont {{Negoro}}, \citenamefont {{Imoto}},\ and\
  \citenamefont {{Kitagawa}}}]{fuji+:decoder}%
  \BibitemOpen
  \bibfield  {author} {\bibinfo {author} {\bibnamefont {{Fujii}}, \bibfnamefont
  {K}}, \bibinfo {author} {\bibfnamefont {M.}~\bibnamefont {{Negoro}}},
  \bibinfo {author} {\bibfnamefont {N.}~\bibnamefont {{Imoto}}}, \ and\
  \bibinfo {author} {\bibfnamefont {M.}~\bibnamefont {{Kitagawa}}}} (\bibinfo
  {year} {2014}),\ \bibfield  {title} {\enquote {\bibinfo {title}
  {{Measurement-Free Topological Protection Using Dissipative Feedback}},}\
  }\href {\doibase 10.1103/PhysRevX.4.041039} {\bibfield  {journal} {\bibinfo
  {journal} {Physical Review X}\ }\textbf {\bibinfo {volume} {4}}~(\bibinfo
  {number} {4}),\ \bibinfo {eid} {041039}},\ \Eprint
  {http://arxiv.org/abs/1401.6350} {arXiv:1401.6350 [quant-ph]} \BibitemShut
  {NoStop}%
\bibitem [{\citenamefont {{Ghosh}}\ and\ \citenamefont
  {{Fowler}}(2015)}]{GF:leakage}%
  \BibitemOpen
  \bibfield  {author} {\bibinfo {author} {\bibnamefont {{Ghosh}}, \bibfnamefont
  {J}}, \ and\ \bibinfo {author} {\bibfnamefont {A.~G.}\ \bibnamefont
  {{Fowler}}}} (\bibinfo {year} {2015}),\ \bibfield  {title} {\enquote
  {\bibinfo {title} {{Leakage-resilient approach to fault-tolerant quantum
  computing with superconducting elements}},}\ }\href {\doibase
  10.1103/PhysRevA.91.020302} {\bibfield  {journal} {\bibinfo  {journal}
  {\pra}\ }\textbf {\bibinfo {volume} {91}}~(\bibinfo {number} {2}),\ \bibinfo
  {eid} {020302}},\ \Eprint {http://arxiv.org/abs/1406.2404} {arXiv:1406.2404
  [quant-ph]} \BibitemShut {NoStop}%
\bibitem [{\citenamefont {{Ghosh}}\ \emph {et~al.}(2012)\citenamefont
  {{Ghosh}}, \citenamefont {{Fowler}},\ and\ \citenamefont
  {{Geller}}}]{GFG:surf_arch}%
  \BibitemOpen
  \bibfield  {author} {\bibinfo {author} {\bibnamefont {{Ghosh}}, \bibfnamefont
  {J}}, \bibinfo {author} {\bibfnamefont {A.~G.}\ \bibnamefont {{Fowler}}}, \
  and\ \bibinfo {author} {\bibfnamefont {M.~R.}\ \bibnamefont {{Geller}}}}
  (\bibinfo {year} {2012}),\ \bibfield  {title} {\enquote {\bibinfo {title}
  {{Surface code with decoherence: An analysis of three superconducting
  architectures}},}\ }\href {\doibase 10.1103/PhysRevA.86.062318} {\bibfield
  {journal} {\bibinfo  {journal} {\pra}\ }\textbf {\bibinfo {volume}
  {86}}~(\bibinfo {number} {6}),\ \bibinfo {eid} {062318}}\BibitemShut
  {NoStop}%
\bibitem [{\citenamefont {{Gladchenko}}\ \emph {et~al.}(2009)\citenamefont
  {{Gladchenko}}, \citenamefont {{Olaya}}, \citenamefont {{Dupont-Ferrier}},
  \citenamefont {{Dou{\c c}ot}}, \citenamefont {{Ioffe}},\ and\ \citenamefont
  {{Gershenson}}}]{gladchenko+:nature}%
  \BibitemOpen
  \bibfield  {author} {\bibinfo {author} {\bibnamefont {{Gladchenko}},
  \bibfnamefont {S}}, \bibinfo {author} {\bibfnamefont {D.}~\bibnamefont
  {{Olaya}}}, \bibinfo {author} {\bibfnamefont {E.}~\bibnamefont
  {{Dupont-Ferrier}}}, \bibinfo {author} {\bibfnamefont {B.}~\bibnamefont
  {{Dou{\c c}ot}}}, \bibinfo {author} {\bibfnamefont {L.~B.}\ \bibnamefont
  {{Ioffe}}}, \ and\ \bibinfo {author} {\bibfnamefont {M.~E.}\ \bibnamefont
  {{Gershenson}}}} (\bibinfo {year} {2009}),\ \bibfield  {title} {\enquote
  {\bibinfo {title} {{Superconducting nanocircuits for topologically protected
  qubits}},}\ }\href@noop {} {\bibfield  {journal} {\bibinfo  {journal} {Nature
  Physics}\ }\textbf {\bibinfo {volume} {5}},\ \bibinfo {pages}
  {48--53}}\BibitemShut {NoStop}%
\bibitem [{\citenamefont {{Glancy}}\ and\ \citenamefont
  {{Knill}}(2006)}]{GK:osc}%
  \BibitemOpen
  \bibfield  {author} {\bibinfo {author} {\bibnamefont {{Glancy}},
  \bibfnamefont {S}}, \ and\ \bibinfo {author} {\bibfnamefont {E.}~\bibnamefont
  {{Knill}}}} (\bibinfo {year} {2006}),\ \bibfield  {title} {\enquote {\bibinfo
  {title} {{Error analysis for encoding a qubit in an oscillator}},}\ }\href
  {\doibase 10.1103/PhysRevA.73.012325} {\bibfield  {journal} {\bibinfo
  {journal} {Phys.~Rev.~A}\ }\textbf {\bibinfo {volume} {73}}~(\bibinfo
  {number} {1}),\ \bibinfo {eid} {012325}}\BibitemShut {NoStop}%
\bibitem [{\citenamefont {Gottesman}(1997)}]{thesis:gottesman}%
  \BibitemOpen
  \bibfield  {author} {\bibinfo {author} {\bibnamefont {Gottesman},
  \bibfnamefont {D}}} (\bibinfo {year} {1997}),\ \emph {\bibinfo {title}
  {Stabilizer Codes and Quantum Error Correction}},\ \href@noop {} {Ph.D.
  thesis}\ (\bibinfo  {school} {CalTech}),\ \bibinfo {note}
  {\url{http://arxiv.org/abs/quant-ph/9705052}}\BibitemShut {NoStop}%
\bibitem [{\citenamefont {Gottesman}(1999a)}]{gottesman:higherd}%
  \BibitemOpen
  \bibfield  {author} {\bibinfo {author} {\bibnamefont {Gottesman},
  \bibfnamefont {D}}} (\bibinfo {year} {1999a}),\ \bibfield  {title} {\enquote
  {\bibinfo {title} {{Fault-tolerant quantum computation with
  higher-dimensional systems}},}\ }\href {\doibase
  10.1016/S0960-0779(98)00218-5} {\bibfield  {journal} {\bibinfo  {journal}
  {Chaos, Solitons and Fractals}\ }\textbf {\bibinfo {volume} {10}},\ \bibinfo
  {pages} {1749--1758}},\ \Eprint
  {http://arxiv.org/abs/\url{http://arxiv.org/abs/quant-ph/9802007}}
  {\url{http://arxiv.org/abs/quant-ph/9802007}} \BibitemShut {NoStop}%
\bibitem [{\citenamefont {{Gottesman}}(1999b)}]{gottesman:heisenberg}%
  \BibitemOpen
  \bibfield  {author} {\bibinfo {author} {\bibnamefont {{Gottesman}},
  \bibfnamefont {D}}} (\bibinfo {year} {1999b}),\ \bibfield  {title} {\enquote
  {\bibinfo {title} {{The Heisenberg representation of quantum computers}},}\
  }in\ \href@noop {} {\emph {\bibinfo {booktitle} {Group22: Proceedings of the
  XXII International Colloquium on Group Theoretical Methods in Physics}}},\
  pp.\ \bibinfo {pages} {32--43},\ \bibinfo {note}
  {\url{http://arxiv.org/abs/quant-ph/9807006}}\BibitemShut {NoStop}%
\bibitem [{\citenamefont {{Gottesman}}(2000)}]{gottesman:local}%
  \BibitemOpen
  \bibfield  {author} {\bibinfo {author} {\bibnamefont {{Gottesman}},
  \bibfnamefont {D}}} (\bibinfo {year} {2000}),\ \bibfield  {title} {\enquote
  {\bibinfo {title} {{Fault-tolerant quantum computation with local gates}},}\
  }\href {\doibase 10.1080/09500340008244046} {\bibfield  {journal} {\bibinfo
  {journal} {Journal of Modern Optics}\ }\textbf {\bibinfo {volume} {47}},\
  \bibinfo {pages} {333--345}}\BibitemShut {NoStop}%
\bibitem [{\citenamefont {{Gottesman}}(2009)}]{gottesman:reviewQEC}%
  \BibitemOpen
  \bibfield  {author} {\bibinfo {author} {\bibnamefont {{Gottesman}},
  \bibfnamefont {D}}} (\bibinfo {year} {2009}),\ \bibfield  {title} {\enquote
  {\bibinfo {title} {{An Introduction to Quantum Error Correction and
  Fault-Tolerant Quantum Computation}},}\ }\href@noop {} {\bibfield  {journal}
  {\bibinfo  {journal} {ArXiv e-prints}\ }}\Eprint
  {http://arxiv.org/abs/0904.2557} {arXiv:0904.2557 [quant-ph]} \BibitemShut
  {NoStop}%
\bibitem [{\citenamefont {{Gottesman}}(2013)}]{gottesman:overhead}%
  \BibitemOpen
  \bibfield  {author} {\bibinfo {author} {\bibnamefont {{Gottesman}},
  \bibfnamefont {D}}} (\bibinfo {year} {2013}),\ \bibfield  {title} {\enquote
  {\bibinfo {title} {{Fault-Tolerant Quantum Computation with Constant
  Overhead}},}\ }\href@noop {} {\bibfield  {journal} {\bibinfo  {journal}
  {ArXiv e-prints}\ }}\Eprint {http://arxiv.org/abs/1310.2984} {arXiv:1310.2984
  [quant-ph]} \BibitemShut {NoStop}%
\bibitem [{\citenamefont {{Gottesman}}\ and\ \citenamefont
  {{Chuang}}(1999)}]{GC:gatetele}%
  \BibitemOpen
  \bibfield  {author} {\bibinfo {author} {\bibnamefont {{Gottesman}},
  \bibfnamefont {D}}, \ and\ \bibinfo {author} {\bibfnamefont {I.~L.}\
  \bibnamefont {{Chuang}}}} (\bibinfo {year} {1999}),\ \bibfield  {title}
  {\enquote {\bibinfo {title} {{Demonstrating the viability of universal
  quantum computation using teleportation and single-qubit operations}},}\
  }\href {\doibase 10.1038/46503} {\bibfield  {journal} {\bibinfo  {journal}
  {Nature}\ }\textbf {\bibinfo {volume} {402}},\ \bibinfo {pages}
  {390--393}}\BibitemShut {NoStop}%
\bibitem [{\citenamefont {Gottesman}\ \emph {et~al.}(2001)\citenamefont
  {Gottesman}, \citenamefont {Kitaev},\ and\ \citenamefont {Preskill}}]{GKP}%
  \BibitemOpen
  \bibfield  {author} {\bibinfo {author} {\bibnamefont {Gottesman},
  \bibfnamefont {D}}, \bibinfo {author} {\bibfnamefont {A.Yu.}\ \bibnamefont
  {Kitaev}}, \ and\ \bibinfo {author} {\bibfnamefont {J.}~\bibnamefont
  {Preskill}}} (\bibinfo {year} {2001}),\ \bibfield  {title} {\enquote
  {\bibinfo {title} {Encoding a qubit in an oscillator},}\ }\href@noop {}
  {\bibfield  {journal} {\bibinfo  {journal} {Phys. Rev. A}\ }\textbf {\bibinfo
  {volume} {64}},\ \bibinfo {pages} {012310}}\BibitemShut {NoStop}%
\bibitem [{\citenamefont {{Guth}}\ and\ \citenamefont
  {{Lubotzky}}(2014)}]{GL:hyperbole}%
  \BibitemOpen
  \bibfield  {author} {\bibinfo {author} {\bibnamefont {{Guth}}, \bibfnamefont
  {L}}, \ and\ \bibinfo {author} {\bibfnamefont {A.}~\bibnamefont
  {{Lubotzky}}}} (\bibinfo {year} {2014}),\ \bibfield  {title} {\enquote
  {\bibinfo {title} {{Quantum error-correcting codes and 4-dimensional
  arithmetic hyperbolic manifolds}},}\ }\href@noop {} {\bibfield  {journal}
  {\bibinfo  {journal} {Journal of Math. Phys.}\ }\textbf {\bibinfo {volume}
  {55}},\ \bibinfo {pages} {082202}}\BibitemShut {NoStop}%
\bibitem [{\citenamefont {{Haah}}(2011)}]{haah:nostring}%
  \BibitemOpen
  \bibfield  {author} {\bibinfo {author} {\bibnamefont {{Haah}}, \bibfnamefont
  {J}}} (\bibinfo {year} {2011}),\ \bibfield  {title} {\enquote {\bibinfo
  {title} {{Local stabilizer codes in three dimensions without string logical
  operators}},}\ }\href {\doibase 10.1103/PhysRevA.83.042330} {\bibfield
  {journal} {\bibinfo  {journal} {\pra}\ }\textbf {\bibinfo {volume}
  {83}}~(\bibinfo {number} {4}),\ \bibinfo {eid} {042330}}\BibitemShut
  {NoStop}%
\bibitem [{\citenamefont {Haroche}\ \emph {et~al.}(2007)\citenamefont
  {Haroche}, \citenamefont {Brune},\ and\ \citenamefont
  {Raimond}}]{HBM:parity}%
  \BibitemOpen
  \bibfield  {author} {\bibinfo {author} {\bibnamefont {Haroche}, \bibfnamefont
  {S}}, \bibinfo {author} {\bibfnamefont {M.}~\bibnamefont {Brune}}, \ and\
  \bibinfo {author} {\bibfnamefont {J.-M.}\ \bibnamefont {Raimond}}} (\bibinfo
  {year} {2007}),\ \bibfield  {title} {\enquote {\bibinfo {title} {Measuring
  the photon number parity in a cavity: from light quantum jumps to the
  tomography of non-classical field states},}\ }\href {\doibase
  10.1080/09500340701391118} {\bibfield  {journal} {\bibinfo  {journal}
  {Journal of Modern Optics}\ }\textbf {\bibinfo {volume} {54}}~(\bibinfo
  {number} {13-15}),\ \bibinfo {pages} {2101--2114}}\BibitemShut {NoStop}%
\bibitem [{\citenamefont {Haroche}\ and\ \citenamefont
  {Raimond}(2006)}]{book:haroche}%
  \BibitemOpen
  \bibfield  {author} {\bibinfo {author} {\bibnamefont {Haroche}, \bibfnamefont
  {S}}, \ and\ \bibinfo {author} {\bibfnamefont {J.-M.}\ \bibnamefont
  {Raimond}}} (\bibinfo {year} {2006}),\ \href@noop {} {\emph {\bibinfo {title}
  {Exploring the Quantum: Atoms, Cavities, and Photons}}}\ (\bibinfo
  {publisher} {Oxford Univ. Press},\ \bibinfo {address} {Oxford})\BibitemShut
  {NoStop}%
\bibitem [{\citenamefont {Harrington}(2004)}]{thesis:harrington}%
  \BibitemOpen
  \bibfield  {author} {\bibinfo {author} {\bibnamefont {Harrington},
  \bibfnamefont {J}}} (\bibinfo {year} {2004}),\ \emph {\bibinfo {title}
  {Analysis of quantum error-correcting codes: symplectic lattice codes and
  toric codes}},\ \href@noop {} {Ph.D. thesis}\ (\bibinfo  {school}
  {CalTech}),\ \bibinfo {note}
  {\url{http://thesis.library.caltech.edu/1747/}}\BibitemShut {NoStop}%
\bibitem [{\citenamefont {{Hastings}}(2013)}]{hastings:eff_decoder}%
  \BibitemOpen
  \bibfield  {author} {\bibinfo {author} {\bibnamefont {{Hastings}},
  \bibfnamefont {M~B}}} (\bibinfo {year} {2013}),\ \bibfield  {title} {\enquote
  {\bibinfo {title} {{Decoding in hyperbolic spaces: LDPC Codes with linear
  rate and efficient error correction}},}\ }\href@noop {} {\bibfield  {journal}
  {\bibinfo  {journal} {ArXiv e-prints}\ }}\Eprint
  {http://arxiv.org/abs/1312.2546} {arXiv:1312.2546 [quant-ph]} \BibitemShut
  {NoStop}%
\bibitem [{\citenamefont {{Hastings}}\ and\ \citenamefont
  {{Geller}}(2014)}]{HG:dislocation}%
  \BibitemOpen
  \bibfield  {author} {\bibinfo {author} {\bibnamefont {{Hastings}},
  \bibfnamefont {M~B}}, \ and\ \bibinfo {author} {\bibfnamefont
  {A.}~\bibnamefont {{Geller}}}} (\bibinfo {year} {2014}),\ \bibfield  {title}
  {\enquote {\bibinfo {title} {{Reduced space-time and time costs using
  dislocation codes and arbitrary ancillas}},}\ }\href@noop {} {\bibfield
  {journal} {\bibinfo  {journal} {ArXiv e-prints}\ }}\Eprint
  {http://arxiv.org/abs/1408.3379} {arXiv:1408.3379 [quant-ph]} \BibitemShut
  {NoStop}%
\bibitem [{\citenamefont {{Herold}}\ \emph {et~al.}(2014)\citenamefont
  {{Herold}}, \citenamefont {{Campbell}}, \citenamefont {{Eisert}},\ and\
  \citenamefont {{Kastoryano}}}]{herold+:ca_decode}%
  \BibitemOpen
  \bibfield  {author} {\bibinfo {author} {\bibnamefont {{Herold}},
  \bibfnamefont {M}}, \bibinfo {author} {\bibfnamefont {E.~T.}\ \bibnamefont
  {{Campbell}}}, \bibinfo {author} {\bibfnamefont {J.}~\bibnamefont
  {{Eisert}}}, \ and\ \bibinfo {author} {\bibfnamefont {M.~J.}\ \bibnamefont
  {{Kastoryano}}}} (\bibinfo {year} {2014}),\ \bibfield  {title} {\enquote
  {\bibinfo {title} {{Cellular-automaton decoders for topological quantum
  memories}},}\ }\href@noop {} {\bibfield  {journal} {\bibinfo  {journal}
  {ArXiv e-prints}\ }}\Eprint {http://arxiv.org/abs/1406.2338} {arXiv:1406.2338
  [quant-ph]} \BibitemShut {NoStop}%
\bibitem [{\citenamefont {{Horsman}}\ \emph {et~al.}(2012)\citenamefont
  {{Horsman}}, \citenamefont {{Fowler}}, \citenamefont {{Devitt}},\ and\
  \citenamefont {{Van Meter}}}]{horsman+:suture}%
  \BibitemOpen
  \bibfield  {author} {\bibinfo {author} {\bibnamefont {{Horsman}},
  \bibfnamefont {C}}, \bibinfo {author} {\bibfnamefont {A.~G.}\ \bibnamefont
  {{Fowler}}}, \bibinfo {author} {\bibfnamefont {S.}~\bibnamefont {{Devitt}}},
  \ and\ \bibinfo {author} {\bibfnamefont {R.}~\bibnamefont {{Van Meter}}}}
  (\bibinfo {year} {2012}),\ \bibfield  {title} {\enquote {\bibinfo {title}
  {{Surface code quantum computing by lattice surgery}},}\ }\href {\doibase
  10.1088/1367-2630/14/12/123011} {\bibfield  {journal} {\bibinfo  {journal}
  {New Journal of Physics}\ }\textbf {\bibinfo {volume} {14}}~(\bibinfo
  {number} {12}),\ \bibinfo {pages} {123011}}\BibitemShut {NoStop}%
\bibitem [{\citenamefont {{Jones}}(2013a)}]{jones:dist}%
  \BibitemOpen
  \bibfield  {author} {\bibinfo {author} {\bibnamefont {{Jones}}, \bibfnamefont
  {C}}} (\bibinfo {year} {2013a}),\ \bibfield  {title} {\enquote {\bibinfo
  {title} {{Multilevel distillation of magic states for quantum computing}},}\
  }\href@noop {} {\bibfield  {journal} {\bibinfo  {journal} {Phys. Rev. A}\
  }\textbf {\bibinfo {volume} {87}},\ \bibinfo {pages} {042305}}\BibitemShut
  {NoStop}%
\bibitem [{\citenamefont {Jones}(2013b)}]{thesis:jones}%
  \BibitemOpen
  \bibfield  {author} {\bibinfo {author} {\bibnamefont {Jones}, \bibfnamefont
  {C}}} (\bibinfo {year} {2013b}),\ \emph {\bibinfo {title} {Logic Synthesis
  for Fault-Tolerant Quantum Computers}},\ \href@noop {} {Ph.D. thesis}\
  (\bibinfo  {school} {Stanford}),\ \bibinfo {note}
  {\url{http://arxiv.org/abs/1310.7290}}\BibitemShut {NoStop}%
\bibitem [{\citenamefont {{Katzgraber}}\ and\ \citenamefont
  {{Andrist}}(2013)}]{KA:glass}%
  \BibitemOpen
  \bibfield  {author} {\bibinfo {author} {\bibnamefont {{Katzgraber}},
  \bibfnamefont {H~G}}, \ and\ \bibinfo {author} {\bibfnamefont {R.~S.}\
  \bibnamefont {{Andrist}}}} (\bibinfo {year} {2013}),\ \bibfield  {title}
  {\enquote {\bibinfo {title} {{Stability of topologically-protected quantum
  computing proposals as seen through spin glasses}},}\ }\href {\doibase
  10.1088/1742-6596/473/1/012019} {\bibfield  {journal} {\bibinfo  {journal}
  {Journal of Physics Conference Series}\ }\textbf {\bibinfo {volume}
  {473}}~(\bibinfo {number} {1}),\ \bibinfo {eid} {012019}}\BibitemShut
  {NoStop}%
\bibitem [{\citenamefont {{Kerckhoff}}\ \emph {et~al.}(2009)\citenamefont
  {{Kerckhoff}}, \citenamefont {{Bouten}}, \citenamefont {{Silberfarb}},\ and\
  \citenamefont {{Mabuchi}}}]{KBSM:parity}%
  \BibitemOpen
  \bibfield  {author} {\bibinfo {author} {\bibnamefont {{Kerckhoff}},
  \bibfnamefont {J}}, \bibinfo {author} {\bibfnamefont {L.}~\bibnamefont
  {{Bouten}}}, \bibinfo {author} {\bibfnamefont {A.}~\bibnamefont
  {{Silberfarb}}}, \ and\ \bibinfo {author} {\bibfnamefont {H.}~\bibnamefont
  {{Mabuchi}}}} (\bibinfo {year} {2009}),\ \bibfield  {title} {\enquote
  {\bibinfo {title} {{Physical model of continuous two-qubit parity measurement
  in a cavity-QED network}},}\ }\href {\doibase 10.1103/PhysRevA.79.024305}
  {\bibfield  {journal} {\bibinfo  {journal} {\pra}\ }\textbf {\bibinfo
  {volume} {79}}~(\bibinfo {number} {2}),\ \bibinfo {eid} {024305}}\BibitemShut
  {NoStop}%
\bibitem [{\citenamefont {{Kerckhoff}}\ \emph {et~al.}(2010)\citenamefont
  {{Kerckhoff}}, \citenamefont {{Nurdin}}, \citenamefont {{Pavlichin}},\ and\
  \citenamefont {{Mabuchi}}}]{kerckhoffetal:qmem}%
  \BibitemOpen
  \bibfield  {author} {\bibinfo {author} {\bibnamefont {{Kerckhoff}},
  \bibfnamefont {J}}, \bibinfo {author} {\bibfnamefont {H.~I.}\ \bibnamefont
  {{Nurdin}}}, \bibinfo {author} {\bibfnamefont {D.~S.}\ \bibnamefont
  {{Pavlichin}}}, \ and\ \bibinfo {author} {\bibfnamefont {H.}~\bibnamefont
  {{Mabuchi}}}} (\bibinfo {year} {2010}),\ \bibfield  {title} {\enquote
  {\bibinfo {title} {{Designing quantum memories with embedded control:
  photonic circuits for autonomous quantum error correction}},}\ }\href
  {\doibase 10.1103/PhysRevLett.105.040502} {\bibfield  {journal} {\bibinfo
  {journal} {Phys. Rev. Lett.}\ }\textbf {\bibinfo {volume} {105}}~(\bibinfo
  {number} {4}),\ \bibinfo {eid} {040502}}\BibitemShut {NoStop}%
\bibitem [{\citenamefont {{Kerckhoff}}\ \emph {et~al.}(2011)\citenamefont
  {{Kerckhoff}}, \citenamefont {{Pavlichin}}, \citenamefont {{Chalabi}},\ and\
  \citenamefont {{Mabuchi}}}]{KPCM:bs}%
  \BibitemOpen
  \bibfield  {author} {\bibinfo {author} {\bibnamefont {{Kerckhoff}},
  \bibfnamefont {J}}, \bibinfo {author} {\bibfnamefont {D.~S.}\ \bibnamefont
  {{Pavlichin}}}, \bibinfo {author} {\bibfnamefont {H.}~\bibnamefont
  {{Chalabi}}}, \ and\ \bibinfo {author} {\bibfnamefont {H.}~\bibnamefont
  {{Mabuchi}}}} (\bibinfo {year} {2011}),\ \bibfield  {title} {\enquote
  {\bibinfo {title} {{Design of nanophotonic circuits for autonomous subsystem
  quantum error correction}},}\ }\href {\doibase 10.1088/1367-2630/13/5/055022}
  {\bibfield  {journal} {\bibinfo  {journal} {New Journal of Physics}\ }\textbf
  {\bibinfo {volume} {13}}~(\bibinfo {number} {5}),\ \bibinfo {eid}
  {055022}}\BibitemShut {NoStop}%
\bibitem [{\citenamefont {Kitaev}(2003)}]{kitaev:top}%
  \BibitemOpen
  \bibfield  {author} {\bibinfo {author} {\bibnamefont {Kitaev}, \bibfnamefont
  {A}}} (\bibinfo {year} {2003}),\ \bibfield  {title} {\enquote {\bibinfo
  {title} {Fault-tolerant quantum computation by anyons},}\ }\href@noop {}
  {\bibfield  {journal} {\bibinfo  {journal} {Ann. Phys.}\ }\textbf {\bibinfo
  {volume} {303}},\ \bibinfo {pages} {2--30}}\BibitemShut {NoStop}%
\bibitem [{\citenamefont {{Kitaev}}(2006)}]{kitaev:mirror}%
  \BibitemOpen
  \bibfield  {author} {\bibinfo {author} {\bibnamefont {{Kitaev}},
  \bibfnamefont {A}}} (\bibinfo {year} {2006}),\ \bibfield  {title} {\enquote
  {\bibinfo {title} {{Protected qubit based on a superconducting current
  mirror}},}\ }\href@noop {} {\ }\bibinfo {note}
  {\url{http://arxiv.org/abs/cond-mat/0609441}}\BibitemShut {NoStop}%
\bibitem [{\citenamefont {Kitaev}(1997)}]{kitaev:survey}%
  \BibitemOpen
  \bibfield  {author} {\bibinfo {author} {\bibnamefont {Kitaev}, \bibfnamefont
  {A~Yu}}} (\bibinfo {year} {1997}),\ \bibfield  {title} {\enquote {\bibinfo
  {title} {Quantum computations: algorithms and error correction},}\
  }\href@noop {} {\bibfield  {journal} {\bibinfo  {journal} {Russian Math.
  Surveys}\ }\textbf {\bibinfo {volume} {52}},\ \bibinfo {pages}
  {1191--1249}}\BibitemShut {NoStop}%
\bibitem [{\citenamefont {Kitaev}(2006)}]{kitaev:anyon_pert}%
  \BibitemOpen
  \bibfield  {author} {\bibinfo {author} {\bibnamefont {Kitaev}, \bibfnamefont
  {Alexei}}} (\bibinfo {year} {2006}),\ \bibfield  {title} {\enquote {\bibinfo
  {title} {Anyons in an exactly solved model and beyond},}\ }\href@noop {}
  {\bibfield  {journal} {\bibinfo  {journal} {Annals of Physics}\ }\textbf
  {\bibinfo {volume} {321}}~(\bibinfo {number} {1}),\ \bibinfo {pages}
  {2--111}}\BibitemShut {NoStop}%
\bibitem [{\citenamefont {Knill}(2005)}]{knill:nature}%
  \BibitemOpen
  \bibfield  {author} {\bibinfo {author} {\bibnamefont {Knill}, \bibfnamefont
  {E}}} (\bibinfo {year} {2005}),\ \bibfield  {title} {\enquote {\bibinfo
  {title} {Quantum computing with realistically noisy devices},}\ }\href@noop
  {} {\bibfield  {journal} {\bibinfo  {journal} {Nature}\ }\textbf {\bibinfo
  {volume} {434}},\ \bibinfo {pages} {39--44}}\BibitemShut {NoStop}%
\bibitem [{\citenamefont {Knill}\ and\ \citenamefont
  {Laflamme}(1997)}]{kl:qec}%
  \BibitemOpen
  \bibfield  {author} {\bibinfo {author} {\bibnamefont {Knill}, \bibfnamefont
  {E}}, \ and\ \bibinfo {author} {\bibfnamefont {R.}~\bibnamefont {Laflamme}}}
  (\bibinfo {year} {1997}),\ \bibfield  {title} {\enquote {\bibinfo {title} {A
  theory of quantum error-correcting codes},}\ }\href@noop {} {\bibfield
  {journal} {\bibinfo  {journal} {Phys. Rev. A}\ }\textbf {\bibinfo {volume}
  {55}},\ \bibinfo {pages} {900--911}}\BibitemShut {NoStop}%
\bibitem [{\citenamefont {Knill}\ \emph {et~al.}(1998)\citenamefont {Knill},
  \citenamefont {Laflamme},\ and\ \citenamefont {Zurek}}]{KLZ:res}%
  \BibitemOpen
  \bibfield  {author} {\bibinfo {author} {\bibnamefont {Knill}, \bibfnamefont
  {E}}, \bibinfo {author} {\bibfnamefont {R.}~\bibnamefont {Laflamme}}, \ and\
  \bibinfo {author} {\bibfnamefont {W.}~\bibnamefont {Zurek}}} (\bibinfo {year}
  {1998}),\ \bibfield  {title} {\enquote {\bibinfo {title} {Resilient quantum
  computation},}\ }\href@noop {} {\bibfield  {journal} {\bibinfo  {journal}
  {Science}\ }\textbf {\bibinfo {volume} {279}},\ \bibinfo {pages}
  {342--345}}\BibitemShut {NoStop}%
\bibitem [{\citenamefont {Koch}\ \emph {et~al.}(2007)\citenamefont {Koch},
  \citenamefont {Yu}, \citenamefont {Gambetta}, \citenamefont {Houck},
  \citenamefont {Schuster}, \citenamefont {Majer}, \citenamefont {Blais},
  \citenamefont {Devoret}, \citenamefont {Girvin},\ and\ \citenamefont
  {Schoelkopf}}]{koch+:transmon}%
  \BibitemOpen
  \bibfield  {author} {\bibinfo {author} {\bibnamefont {Koch}, \bibfnamefont
  {J}}, \bibinfo {author} {\bibfnamefont {Terri~M.}\ \bibnamefont {Yu}},
  \bibinfo {author} {\bibfnamefont {J.}~\bibnamefont {Gambetta}}, \bibinfo
  {author} {\bibfnamefont {A.~A.}\ \bibnamefont {Houck}}, \bibinfo {author}
  {\bibfnamefont {D.~I.}\ \bibnamefont {Schuster}}, \bibinfo {author}
  {\bibfnamefont {J.}~\bibnamefont {Majer}}, \bibinfo {author} {\bibfnamefont
  {A.}~\bibnamefont {Blais}}, \bibinfo {author} {\bibfnamefont {M.~H.}\
  \bibnamefont {Devoret}}, \bibinfo {author} {\bibfnamefont {S.~M.}\
  \bibnamefont {Girvin}}, \ and\ \bibinfo {author} {\bibfnamefont {R.~J.}\
  \bibnamefont {Schoelkopf}}} (\bibinfo {year} {2007}),\ \bibfield  {title}
  {\enquote {\bibinfo {title} {Charge-insensitive qubit design derived from the
  {C}ooper pair box},}\ }\href {\doibase 10.1103/PhysRevA.76.042319} {\bibfield
   {journal} {\bibinfo  {journal} {Phys. Rev. A}\ }\textbf {\bibinfo {volume}
  {76}},\ \bibinfo {pages} {042319}}\BibitemShut {NoStop}%
\bibitem [{\citenamefont {{Koenig}}\ \emph {et~al.}(2010)\citenamefont
  {{Koenig}}, \citenamefont {{Kuperberg}},\ and\ \citenamefont
  {{Reichardt}}}]{KKR}%
  \BibitemOpen
  \bibfield  {author} {\bibinfo {author} {\bibnamefont {{Koenig}},
  \bibfnamefont {R}}, \bibinfo {author} {\bibfnamefont {G.}~\bibnamefont
  {{Kuperberg}}}, \ and\ \bibinfo {author} {\bibfnamefont {B.~W.}\ \bibnamefont
  {{Reichardt}}}} (\bibinfo {year} {2010}),\ \bibfield  {title} {\enquote
  {\bibinfo {title} {{Quantum computation with Turaev-Viro codes}},}\ }\href
  {\doibase 10.1016/j.aop.2010.08.001} {\bibfield  {journal} {\bibinfo
  {journal} {Annals of Physics}\ }\textbf {\bibinfo {volume} {325}},\ \bibinfo
  {pages} {2707--2749}}\BibitemShut {NoStop}%
\bibitem [{\citenamefont {{Kovalev}}\ and\ \citenamefont
  {{Pryadko}}(2013)}]{KP:badcodes}%
  \BibitemOpen
  \bibfield  {author} {\bibinfo {author} {\bibnamefont {{Kovalev}},
  \bibfnamefont {A~A}}, \ and\ \bibinfo {author} {\bibfnamefont {L.~P.}\
  \bibnamefont {{Pryadko}}}} (\bibinfo {year} {2013}),\ \bibfield  {title}
  {\enquote {\bibinfo {title} {{Fault-tolerance of quantum low-density parity
  check codes with sublinear distance scaling}},}\ }\href {\doibase
  10.1103/PhysRevA.87.020304} {\bibfield  {journal} {\bibinfo  {journal} {Phys.
  Rev. A}\ }\textbf {\bibinfo {volume} {87}}~(\bibinfo {number} {2}),\ \bibinfo
  {eid} {020304}}\BibitemShut {NoStop}%
\bibitem [{\citenamefont {{Kribs}}\ \emph {et~al.}(2005)\citenamefont
  {{Kribs}}, \citenamefont {{Laflamme}},\ and\ \citenamefont
  {{Poulin}}}]{KLP:qec_uni}%
  \BibitemOpen
  \bibfield  {author} {\bibinfo {author} {\bibnamefont {{Kribs}}, \bibfnamefont
  {D}}, \bibinfo {author} {\bibfnamefont {R.}~\bibnamefont {{Laflamme}}}, \
  and\ \bibinfo {author} {\bibfnamefont {D.}~\bibnamefont {{Poulin}}}}
  (\bibinfo {year} {2005}),\ \bibfield  {title} {\enquote {\bibinfo {title}
  {{Unified and generalized approach to quantum error correction}},}\ }\href
  {\doibase 10.1103/PhysRevLett.94.180501} {\bibfield  {journal} {\bibinfo
  {journal} {Phys. Rev. Lett.}\ }\textbf {\bibinfo {volume} {94}}~(\bibinfo
  {number} {18}),\ \bibinfo {eid} {180501}}\BibitemShut {NoStop}%
\bibitem [{\citenamefont {{Lalumi{\`e}re}}\ \emph {et~al.}(2010)\citenamefont
  {{Lalumi{\`e}re}}, \citenamefont {{Gambetta}},\ and\ \citenamefont
  {{Blais}}}]{LGB:2qubit_parity}%
  \BibitemOpen
  \bibfield  {author} {\bibinfo {author} {\bibnamefont {{Lalumi{\`e}re}},
  \bibfnamefont {K}}, \bibinfo {author} {\bibfnamefont {J.~M.}\ \bibnamefont
  {{Gambetta}}}, \ and\ \bibinfo {author} {\bibfnamefont {A.}~\bibnamefont
  {{Blais}}}} (\bibinfo {year} {2010}),\ \bibfield  {title} {\enquote {\bibinfo
  {title} {{Tunable joint measurements in the dispersive regime of cavity
  QED}},}\ }\href@noop {} {\bibfield  {journal} {\bibinfo  {journal} {Phys.
  Rev. A}\ }\textbf {\bibinfo {volume} {81}}~(\bibinfo {number} {4}),\ \bibinfo
  {eid} {040301}}\BibitemShut {NoStop}%
\bibitem [{\citenamefont {{Landahl}}\ \emph {et~al.}(2011)\citenamefont
  {{Landahl}}, \citenamefont {{Anderson}},\ and\ \citenamefont
  {{Rice}}}]{LAR:colorcodes}%
  \BibitemOpen
  \bibfield  {author} {\bibinfo {author} {\bibnamefont {{Landahl}},
  \bibfnamefont {A~J}}, \bibinfo {author} {\bibfnamefont {J.~T.}\ \bibnamefont
  {{Anderson}}}, \ and\ \bibinfo {author} {\bibfnamefont {P.~R.}\ \bibnamefont
  {{Rice}}}} (\bibinfo {year} {2011}),\ \bibfield  {title} {\enquote {\bibinfo
  {title} {{Fault-tolerant quantum computing with color codes}},}\ }\href@noop
  {} {\ }\bibinfo {note} {\url{http://arxiv.org/abs/1108.5738}}\BibitemShut
  {NoStop}%
\bibitem [{\citenamefont {Leghtas}\ \emph {et~al.}(2013)\citenamefont
  {Leghtas}, \citenamefont {Kirchmair}, \citenamefont {Vlastakis},
  \citenamefont {Schoelkopf}, \citenamefont {Devoret},\ and\ \citenamefont
  {Mirrahimi}}]{leghtas+:QEC}%
  \BibitemOpen
  \bibfield  {author} {\bibinfo {author} {\bibnamefont {Leghtas}, \bibfnamefont
  {Z}}, \bibinfo {author} {\bibfnamefont {G.}~\bibnamefont {Kirchmair}},
  \bibinfo {author} {\bibfnamefont {B.}~\bibnamefont {Vlastakis}}, \bibinfo
  {author} {\bibfnamefont {R.J.}\ \bibnamefont {Schoelkopf}}, \bibinfo {author}
  {\bibfnamefont {M.~H.}\ \bibnamefont {Devoret}}, \ and\ \bibinfo {author}
  {\bibfnamefont {M.}~\bibnamefont {Mirrahimi}}} (\bibinfo {year} {2013}),\
  \bibfield  {title} {\enquote {\bibinfo {title} {Hardware-efficient autonomous
  quantum memory protection},}\ }\href@noop {} {\bibfield  {journal} {\bibinfo
  {journal} {Phys. Rev. Lett.}\ }\textbf {\bibinfo {volume} {111}},\ \bibinfo
  {pages} {120501}}\BibitemShut {NoStop}%
\bibitem [{\citenamefont {{Leung}}\ \emph {et~al.}(1997)\citenamefont
  {{Leung}}, \citenamefont {{Nielsen}}, \citenamefont {{Chuang}},\ and\
  \citenamefont {{Yamamoto}}}]{leung+:approx}%
  \BibitemOpen
  \bibfield  {author} {\bibinfo {author} {\bibnamefont {{Leung}}, \bibfnamefont
  {D~W}}, \bibinfo {author} {\bibfnamefont {M.~A.}\ \bibnamefont {{Nielsen}}},
  \bibinfo {author} {\bibfnamefont {I.~L.}\ \bibnamefont {{Chuang}}}, \ and\
  \bibinfo {author} {\bibfnamefont {Y.}~\bibnamefont {{Yamamoto}}}} (\bibinfo
  {year} {1997}),\ \bibfield  {title} {\enquote {\bibinfo {title} {{Approximate
  quantum error correction can lead to better codes}},}\ }\href {\doibase
  10.1103/PhysRevA.56.2567} {\bibfield  {journal} {\bibinfo  {journal}
  {Phys.~Rev.~A}\ }\textbf {\bibinfo {volume} {56}},\ \bibinfo {pages}
  {2567--2573}}\BibitemShut {NoStop}%
\bibitem [{\citenamefont {{Levin}}\ and\ \citenamefont
  {{Wen}}(2005)}]{LW:models}%
  \BibitemOpen
  \bibfield  {author} {\bibinfo {author} {\bibnamefont {{Levin}}, \bibfnamefont
  {M~A}}, \ and\ \bibinfo {author} {\bibfnamefont {X.-G.}\ \bibnamefont
  {{Wen}}}} (\bibinfo {year} {2005}),\ \bibfield  {title} {\enquote {\bibinfo
  {title} {{String-net condensation: A physical mechanism for topological
  phases}},}\ }\href {\doibase 10.1103/PhysRevB.71.045110} {\bibfield
  {journal} {\bibinfo  {journal} {Phys.~Rev.~B}\ }\textbf {\bibinfo {volume}
  {71}}~(\bibinfo {number} {4}),\ \bibinfo {eid} {045110}}\BibitemShut
  {NoStop}%
\bibitem [{\citenamefont {{Levy}}\ \emph {et~al.}(2009)\citenamefont {{Levy}},
  \citenamefont {{Ganti}}, \citenamefont {{Phillips}}, \citenamefont
  {{Hamlet}}, \citenamefont {{Landahl}}, \citenamefont {{Gurrieri}},
  \citenamefont {{Carr}},\ and\ \citenamefont {{Carroll}}}]{levy+:elec}%
  \BibitemOpen
  \bibfield  {author} {\bibinfo {author} {\bibnamefont {{Levy}}, \bibfnamefont
  {J~E}}, \bibinfo {author} {\bibfnamefont {A.}~\bibnamefont {{Ganti}}},
  \bibinfo {author} {\bibfnamefont {C.~A.}\ \bibnamefont {{Phillips}}},
  \bibinfo {author} {\bibfnamefont {B.~R.}\ \bibnamefont {{Hamlet}}}, \bibinfo
  {author} {\bibfnamefont {A.~J.}\ \bibnamefont {{Landahl}}}, \bibinfo {author}
  {\bibfnamefont {T.~M.}\ \bibnamefont {{Gurrieri}}}, \bibinfo {author}
  {\bibfnamefont {R.~D.}\ \bibnamefont {{Carr}}}, \ and\ \bibinfo {author}
  {\bibfnamefont {M.~S.}\ \bibnamefont {{Carroll}}}} (\bibinfo {year} {2009}),\
  \bibfield  {title} {\enquote {\bibinfo {title} {The impact of classical
  electronics constraints on a solid-state logical qubit memory},}\ }in\
  \href@noop {} {\emph {\bibinfo {booktitle} {Proceedings of the 21st annual
  symposium on Parallelism in algorithms and architectures}}}\ (\bibinfo
  {publisher} {ACM},\ \bibinfo {address} {New York, NY, USA})\ pp.\ \bibinfo
  {pages} {166--168},\ \bibinfo {note}
  {\url{http://arxiv.org/abs/0904.0003}}\BibitemShut {NoStop}%
\bibitem [{\citenamefont {{Lidar}}(2014)}]{lidar:review}%
  \BibitemOpen
  \bibfield  {author} {\bibinfo {author} {\bibnamefont {{Lidar}}, \bibfnamefont
  {D~A}}} (\bibinfo {year} {2014}),\ \bibfield  {title} {\enquote {\bibinfo
  {title} {{Review of decoherence free subspaces, noiseless subsystems, and
  dynamical decoupling}},}\ }\href@noop {} {\bibfield  {journal} {\bibinfo
  {journal} {Adv. Chem. Phys.}\ }\textbf {\bibinfo {volume} {154}},\ \bibinfo
  {pages} {295--354}}\BibitemShut {NoStop}%
\bibitem [{\citenamefont {Lidar}\ and\ \citenamefont
  {Brun}(2013)}]{book:lidar_brun}%
  \BibitemOpen
  \bibinfo {editor} {\bibnamefont {Lidar}, \bibfnamefont {DA}}, \ and\ \bibinfo
  {editor} {\bibfnamefont {T.A.}\ \bibnamefont {Brun}},\ Eds. (\bibinfo {year}
  {2013}),\ \href {http://books.google.de/books?id=qWHRNAEACAAJ} {\emph
  {\bibinfo {title} {Quantum Error Correction}}}\ (\bibinfo  {publisher}
  {Cambridge University Press})\BibitemShut {NoStop}%
\bibitem [{\citenamefont {Lloyd}\ and\ \citenamefont {Slotine}(1998)}]{LS:CV}%
  \BibitemOpen
  \bibfield  {author} {\bibinfo {author} {\bibnamefont {Lloyd}, \bibfnamefont
  {S}}, \ and\ \bibinfo {author} {\bibfnamefont {J.-J.}\ \bibnamefont
  {Slotine}}} (\bibinfo {year} {1998}),\ \bibfield  {title} {\enquote {\bibinfo
  {title} {Analog quantum error correction},}\ }\href@noop {} {\bibfield
  {journal} {\bibinfo  {journal} {Phys. Rev. Lett.}\ }\textbf {\bibinfo
  {volume} {80}},\ \bibinfo {pages} {4088}}\BibitemShut {NoStop}%
\bibitem [{\citenamefont {{Menicucci}}(2014)}]{menicucci:ft}%
  \BibitemOpen
  \bibfield  {author} {\bibinfo {author} {\bibnamefont {{Menicucci}},
  \bibfnamefont {N~C}}} (\bibinfo {year} {2014}),\ \bibfield  {title} {\enquote
  {\bibinfo {title} {{Fault-tolerant measurement-based quantum computing with
  continuous-variable cluster states}},}\ }\href {\doibase
  10.1103/PhysRevLett.112.120504} {\bibfield  {journal} {\bibinfo  {journal}
  {Phys. Rev. Lett.}\ }\textbf {\bibinfo {volume} {112}}~(\bibinfo {number}
  {12}),\ \bibinfo {eid} {120504}}\BibitemShut {NoStop}%
\bibitem [{\citenamefont {Metcalfe}\ \emph {et~al.}(2007)\citenamefont
  {Metcalfe}, \citenamefont {Boaknin}, \citenamefont {Manucharyan},
  \citenamefont {Vijay}, \citenamefont {Siddiqi}, \citenamefont {Rigetti},
  \citenamefont {Frunzio}, \citenamefont {Schoelkopf},\ and\ \citenamefont
  {Devoret}}]{metcalfe+:decoh}%
  \BibitemOpen
  \bibfield  {author} {\bibinfo {author} {\bibnamefont {Metcalfe},
  \bibfnamefont {M}}, \bibinfo {author} {\bibfnamefont {E.}~\bibnamefont
  {Boaknin}}, \bibinfo {author} {\bibfnamefont {V.}~\bibnamefont
  {Manucharyan}}, \bibinfo {author} {\bibfnamefont {R.}~\bibnamefont {Vijay}},
  \bibinfo {author} {\bibfnamefont {I.}~\bibnamefont {Siddiqi}}, \bibinfo
  {author} {\bibfnamefont {C.}~\bibnamefont {Rigetti}}, \bibinfo {author}
  {\bibfnamefont {L.}~\bibnamefont {Frunzio}}, \bibinfo {author} {\bibfnamefont
  {R.~J.}\ \bibnamefont {Schoelkopf}}, \ and\ \bibinfo {author} {\bibfnamefont
  {M.~H.}\ \bibnamefont {Devoret}}} (\bibinfo {year} {2007}),\ \bibfield
  {title} {\enquote {\bibinfo {title} {Measuring the decoherence of a
  quantronium qubit with the cavity bifurcation amplifier},}\ }\href {\doibase
  10.1103/PhysRevB.76.174516} {\bibfield  {journal} {\bibinfo  {journal} {Phys.
  Rev. B}\ }\textbf {\bibinfo {volume} {76}},\ \bibinfo {pages}
  {174516}}\BibitemShut {NoStop}%
\bibitem [{\citenamefont {{Michnicki}}(2014)}]{michnicki}%
  \BibitemOpen
  \bibfield  {author} {\bibinfo {author} {\bibnamefont {{Michnicki}},
  \bibfnamefont {K~P}}} (\bibinfo {year} {2014}),\ \bibfield  {title} {\enquote
  {\bibinfo {title} {{3D Topological quantum memory with a power-law energy
  barrier}},}\ }\href {\doibase 10.1103/PhysRevLett.113.130501} {\bibfield
  {journal} {\bibinfo  {journal} {Phys. Rev. Lett.}\ }\textbf {\bibinfo
  {volume} {113}}~(\bibinfo {number} {13}),\ \bibinfo {eid}
  {130501}}\BibitemShut {NoStop}%
\bibitem [{\citenamefont {{Mirrahimi}}\ \emph {et~al.}(2013)\citenamefont
  {{Mirrahimi}}, \citenamefont {{Leghtas}}, \citenamefont {{Albert}},
  \citenamefont {{Touzard}}, \citenamefont {{Schoelkopf}}, \citenamefont
  {{Jiang}},\ and\ \citenamefont {{Devoret}}}]{mirra+:cats}%
  \BibitemOpen
  \bibfield  {author} {\bibinfo {author} {\bibnamefont {{Mirrahimi}},
  \bibfnamefont {M}}, \bibinfo {author} {\bibfnamefont {Z.}~\bibnamefont
  {{Leghtas}}}, \bibinfo {author} {\bibfnamefont {V.~V.}\ \bibnamefont
  {{Albert}}}, \bibinfo {author} {\bibfnamefont {S.}~\bibnamefont {{Touzard}}},
  \bibinfo {author} {\bibfnamefont {R.~J.}\ \bibnamefont {{Schoelkopf}}},
  \bibinfo {author} {\bibfnamefont {L.}~\bibnamefont {{Jiang}}}, \ and\
  \bibinfo {author} {\bibfnamefont {M.~H.}\ \bibnamefont {{Devoret}}}}
  (\bibinfo {year} {2013}),\ \bibfield  {title} {\enquote {\bibinfo {title}
  {{Dynamically protected cat-qubits: a new paradigm for universal quantum
  computation}},}\ }\href@noop {} {\bibfield  {journal} {\bibinfo  {journal}
  {ArXiv e-prints}\ }}\Eprint {http://arxiv.org/abs/1312.2017} {arXiv:1312.2017
  [quant-ph]} \BibitemShut {NoStop}%
\bibitem [{\citenamefont {{M{\"u}ller}}\ \emph {et~al.}(2011)\citenamefont
  {{M{\"u}ller}}, \citenamefont {{Hammerer}}, \citenamefont {{Zhou}},
  \citenamefont {{Roos}},\ and\ \citenamefont {{Zoller}}}]{muller+:pumping}%
  \BibitemOpen
  \bibfield  {author} {\bibinfo {author} {\bibnamefont {{M{\"u}ller}},
  \bibfnamefont {M}}, \bibinfo {author} {\bibfnamefont {K.}~\bibnamefont
  {{Hammerer}}}, \bibinfo {author} {\bibfnamefont {Y.~L.}\ \bibnamefont
  {{Zhou}}}, \bibinfo {author} {\bibfnamefont {C.~F.}\ \bibnamefont {{Roos}}},
  \ and\ \bibinfo {author} {\bibfnamefont {P.}~\bibnamefont {{Zoller}}}}
  (\bibinfo {year} {2011}),\ \bibfield  {title} {\enquote {\bibinfo {title}
  {{Simulating open quantum systems: from many-body interactions to stabilizer
  pumping}},}\ }\href {\doibase 10.1088/1367-2630/13/8/085007} {\bibfield
  {journal} {\bibinfo  {journal} {New Journal of Physics}\ }\textbf {\bibinfo
  {volume} {13}}~(\bibinfo {number} {8}),\ \bibinfo {pages}
  {085007}}\BibitemShut {NoStop}%
\bibitem [{\citenamefont {{Napp}}\ and\ \citenamefont
  {{Preskill}}(2013)}]{NP:bs}%
  \BibitemOpen
  \bibfield  {author} {\bibinfo {author} {\bibnamefont {{Napp}}, \bibfnamefont
  {J}}, \ and\ \bibinfo {author} {\bibfnamefont {J.}~\bibnamefont
  {{Preskill}}}} (\bibinfo {year} {2013}),\ \bibfield  {title} {\enquote
  {\bibinfo {title} {Optimal {B}acon-{S}hor codes},}\ }\href@noop {} {\bibfield
   {journal} {\bibinfo  {journal} {Quantum Info. and Comput.}\ }\textbf
  {\bibinfo {volume} {13}},\ \bibinfo {pages} {490--510}}\BibitemShut {NoStop}%
\bibitem [{\citenamefont {{Ng}}\ and\ \citenamefont
  {{Mandayam}}(2010)}]{MN:approximate}%
  \BibitemOpen
  \bibfield  {author} {\bibinfo {author} {\bibnamefont {{Ng}}, \bibfnamefont
  {H~K}}, \ and\ \bibinfo {author} {\bibfnamefont {P.}~\bibnamefont
  {{Mandayam}}}} (\bibinfo {year} {2010}),\ \bibfield  {title} {\enquote
  {\bibinfo {title} {{Simple approach to approximate quantum error correction
  based on the transpose channel}},}\ }\href {\doibase
  10.1103/PhysRevA.81.062342} {\bibfield  {journal} {\bibinfo  {journal}
  {\pra}\ }\textbf {\bibinfo {volume} {81}}~(\bibinfo {number} {6}),\ \bibinfo
  {eid} {062342}}\BibitemShut {NoStop}%
\bibitem [{\citenamefont {{Nielsen}}(2010)}]{nielsen:2qubitparity}%
  \BibitemOpen
  \bibfield  {author} {\bibinfo {author} {\bibnamefont {{Nielsen}},
  \bibfnamefont {A~E~B}}} (\bibinfo {year} {2010}),\ \bibfield  {title}
  {\enquote {\bibinfo {title} {{Fighting decoherence in a continuous two-qubit
  odd- or even-parity measurement with a closed-loop setup}},}\ }\href
  {\doibase 10.1103/PhysRevA.81.012307} {\bibfield  {journal} {\bibinfo
  {journal} {\pra}\ }\textbf {\bibinfo {volume} {81}}~(\bibinfo {number} {1}),\
  \bibinfo {eid} {012307}}\BibitemShut {NoStop}%
\bibitem [{\citenamefont {Nielsen}\ and\ \citenamefont
  {Chuang}(2000)}]{book:nielsen&chuang}%
  \BibitemOpen
  \bibfield  {author} {\bibinfo {author} {\bibnamefont {Nielsen}, \bibfnamefont
  {M~A}}, \ and\ \bibinfo {author} {\bibfnamefont {I.~L.}\ \bibnamefont
  {Chuang}}} (\bibinfo {year} {2000}),\ \href@noop {} {\emph {\bibinfo {title}
  {Quantum computation and quantum information}}}\ (\bibinfo  {publisher}
  {Cambridge University Press},\ \bibinfo {address} {Cambridge,
  U.K.})\BibitemShut {NoStop}%
\bibitem [{\citenamefont {{Nielsen}}\ and\ \citenamefont
  {{Poulin}}(2007)}]{NP:operator_qec}%
  \BibitemOpen
  \bibfield  {author} {\bibinfo {author} {\bibnamefont {{Nielsen}},
  \bibfnamefont {M~A}}, \ and\ \bibinfo {author} {\bibfnamefont
  {D.}~\bibnamefont {{Poulin}}}} (\bibinfo {year} {2007}),\ \bibfield  {title}
  {\enquote {\bibinfo {title} {{Algebraic and information-theoretic conditions
  for operator quantum error-correction}},}\ }\href@noop {} {\bibfield
  {journal} {\bibinfo  {journal} {Phys.~Rev.A}\ }\textbf {\bibinfo {volume}
  {75}},\ \bibinfo {pages} {064304(R)}}\BibitemShut {NoStop}%
\bibitem [{\citenamefont {Nigg}\ \emph {et~al.}(2014)\citenamefont {Nigg},
  \citenamefont {Müller}, \citenamefont {Martinez}, \citenamefont {Schindler},
  \citenamefont {Hennrich}, \citenamefont {Monz}, \citenamefont
  {Martin-Delgado},\ and\ \citenamefont {Blatt}}]{nigg:steanecode}%
  \BibitemOpen
  \bibfield  {author} {\bibinfo {author} {\bibnamefont {Nigg}, \bibfnamefont
  {D}}, \bibinfo {author} {\bibfnamefont {M.}~\bibnamefont {Müller}}, \bibinfo
  {author} {\bibfnamefont {E.~A.}\ \bibnamefont {Martinez}}, \bibinfo {author}
  {\bibfnamefont {P.}~\bibnamefont {Schindler}}, \bibinfo {author}
  {\bibfnamefont {M.}~\bibnamefont {Hennrich}}, \bibinfo {author}
  {\bibfnamefont {T.}~\bibnamefont {Monz}}, \bibinfo {author} {\bibfnamefont
  {M.~A.}\ \bibnamefont {Martin-Delgado}}, \ and\ \bibinfo {author}
  {\bibfnamefont {R.}~\bibnamefont {Blatt}}} (\bibinfo {year} {2014}),\
  \bibfield  {title} {\enquote {\bibinfo {title} {Quantum computations on a
  topologically encoded qubit},}\ }\href {\doibase 10.1126/science.1253742}
  {\bibfield  {journal} {\bibinfo  {journal} {Science}\ }\textbf {\bibinfo
  {volume} {345}}~(\bibinfo {number} {6194}),\ \bibinfo {pages}
  {302--305}}\BibitemShut {NoStop}%
\bibitem [{\citenamefont {Nigg}\ and\ \citenamefont {Girvin}(2013)}]{NG:stab}%
  \BibitemOpen
  \bibfield  {author} {\bibinfo {author} {\bibnamefont {Nigg}, \bibfnamefont
  {S~E}}, \ and\ \bibinfo {author} {\bibfnamefont {S.~M.}\ \bibnamefont
  {Girvin}}} (\bibinfo {year} {2013}),\ \bibfield  {title} {\enquote {\bibinfo
  {title} {Stabilizer quantum error correction toolbox for superconducting
  qubits},}\ }\href@noop {} {\bibfield  {journal} {\bibinfo  {journal} {Phys.
  Rev. Lett.}\ }\textbf {\bibinfo {volume} {110}},\ \bibinfo {pages}
  {243604}}\BibitemShut {NoStop}%
\bibitem [{\citenamefont {{Nussinov}}\ and\ \citenamefont
  {{Ortiz}}(2009)}]{NO:thermal}%
  \BibitemOpen
  \bibfield  {author} {\bibinfo {author} {\bibnamefont {{Nussinov}},
  \bibfnamefont {Z}}, \ and\ \bibinfo {author} {\bibfnamefont {G.}~\bibnamefont
  {{Ortiz}}}} (\bibinfo {year} {2009}),\ \bibfield  {title} {\enquote {\bibinfo
  {title} {{Symmetry and Topological Order}},}\ }\href {\doibase
  10.1073/pnas.0803726105} {\bibfield  {journal} {\bibinfo  {journal}
  {Proceedings of the National Academy of Science}\ }\textbf {\bibinfo {volume}
  {106}},\ \bibinfo {pages} {16944--16949}}\BibitemShut {NoStop}%
\bibitem [{\citenamefont {{Paetznick}}\ and\ \citenamefont
  {{Reichardt}}(2013)}]{PR:gauge}%
  \BibitemOpen
  \bibfield  {author} {\bibinfo {author} {\bibnamefont {{Paetznick}},
  \bibfnamefont {A}}, \ and\ \bibinfo {author} {\bibfnamefont {B.~W.}\
  \bibnamefont {{Reichardt}}}} (\bibinfo {year} {2013}),\ \bibfield  {title}
  {\enquote {\bibinfo {title} {{Universal fault-tolerant quantum computation
  with only transversal gates and error correction}},}\ }\href {\doibase
  10.1103/PhysRevLett.111.090505} {\bibfield  {journal} {\bibinfo  {journal}
  {Phys. Rev. Lett.}\ }\textbf {\bibinfo {volume} {111}}~(\bibinfo {number}
  {9}),\ \bibinfo {eid} {090505}}\BibitemShut {NoStop}%
\bibitem [{\citenamefont {{Poulin}}(2005)}]{poulin:stabsub}%
  \BibitemOpen
  \bibfield  {author} {\bibinfo {author} {\bibnamefont {{Poulin}},
  \bibfnamefont {D}}} (\bibinfo {year} {2005}),\ \bibfield  {title} {\enquote
  {\bibinfo {title} {{Stabilizer formalism for operator quantum error
  correction}},}\ }\href {\doibase 10.1103/PhysRevLett.95.230504} {\bibfield
  {journal} {\bibinfo  {journal} {Phys. Rev. Lett.}\ }\textbf {\bibinfo
  {volume} {95}}~(\bibinfo {number} {23}),\ \bibinfo {eid}
  {230504}}\BibitemShut {NoStop}%
\bibitem [{\citenamefont {{Poulin}}(2006)}]{poulin:message}%
  \BibitemOpen
  \bibfield  {author} {\bibinfo {author} {\bibnamefont {{Poulin}},
  \bibfnamefont {D}}} (\bibinfo {year} {2006}),\ \bibfield  {title} {\enquote
  {\bibinfo {title} {{Optimal and efficient decoding of concatenated quantum
  block codes}},}\ }\href {\doibase 10.1103/PhysRevA.74.052333} {\bibfield
  {journal} {\bibinfo  {journal} {Phys. Rev. A}\ }\textbf {\bibinfo {volume}
  {74}}~(\bibinfo {number} {5}),\ \bibinfo {eid} {052333}}\BibitemShut
  {NoStop}%
\bibitem [{\citenamefont {Preskill}(1998)}]{preskill:faulttol}%
  \BibitemOpen
  \bibfield  {author} {\bibinfo {author} {\bibnamefont {Preskill},
  \bibfnamefont {J}}} (\bibinfo {year} {1998}),\ \bibfield  {title} {\enquote
  {\bibinfo {title} {Fault-tolerant quantum computation},}\ }in\ \href@noop {}
  {\emph {\bibinfo {booktitle} {Introduction to Quantum Computation}}}\
  (\bibinfo  {publisher} {World Scientific, Singapore})\ pp.\ \bibinfo {pages}
  {213--269}\BibitemShut {NoStop}%
\bibitem [{\citenamefont {{Raussendorf}}\ and\ \citenamefont
  {{Harrington}}(2007)}]{RH:cluster2D}%
  \BibitemOpen
  \bibfield  {author} {\bibinfo {author} {\bibnamefont {{Raussendorf}},
  \bibfnamefont {R}}, \ and\ \bibinfo {author} {\bibfnamefont {J.}~\bibnamefont
  {{Harrington}}}} (\bibinfo {year} {2007}),\ \bibfield  {title} {\enquote
  {\bibinfo {title} {{Fault-tolerant quantum computation with high threshold in
  two dimensions}},}\ }\href {\doibase 10.1103/PhysRevLett.98.190504}
  {\bibfield  {journal} {\bibinfo  {journal} {Phys. Rev. Lett.}\ }\textbf
  {\bibinfo {volume} {98}}~(\bibinfo {number} {19}),\ \bibinfo {pages}
  {190504}}\BibitemShut {NoStop}%
\bibitem [{\citenamefont {{Raussendorf}}\ \emph {et~al.}(2007)\citenamefont
  {{Raussendorf}}, \citenamefont {{Harrington}},\ and\ \citenamefont
  {{Goyal}}}]{RHG:threshold}%
  \BibitemOpen
  \bibfield  {author} {\bibinfo {author} {\bibnamefont {{Raussendorf}},
  \bibfnamefont {R}}, \bibinfo {author} {\bibfnamefont {J.}~\bibnamefont
  {{Harrington}}}, \ and\ \bibinfo {author} {\bibfnamefont {K.}~\bibnamefont
  {{Goyal}}}} (\bibinfo {year} {2007}),\ \bibfield  {title} {\enquote {\bibinfo
  {title} {{Topological fault-tolerance in cluster state quantum
  computation}},}\ }\href {\doibase 10.1088/1367-2630/9/6/199} {\bibfield
  {journal} {\bibinfo  {journal} {New J. Phys.}\ }\textbf {\bibinfo {volume}
  {9}},\ \bibinfo {pages} {199--219}}\BibitemShut {NoStop}%
\bibitem [{\citenamefont {Reichardt}(2005)}]{reichardt:distill}%
  \BibitemOpen
  \bibfield  {author} {\bibinfo {author} {\bibnamefont {Reichardt},
  \bibfnamefont {Ben~W}}} (\bibinfo {year} {2005}),\ \bibfield  {title}
  {\enquote {\bibinfo {title} {Quantum universality by distilling certain one-
  and two-qubit states with stabilizer operations},}\ }\href@noop {} {\bibfield
   {journal} {\bibinfo  {journal} {Quant. Info. Proc.}\ }\textbf {\bibinfo
  {volume} {4}},\ \bibinfo {pages} {251--264}}\BibitemShut {NoStop}%
\bibitem [{\citenamefont {Shor}(1996)}]{shor:faulttol}%
  \BibitemOpen
  \bibfield  {author} {\bibinfo {author} {\bibnamefont {Shor}, \bibfnamefont
  {P~W}}} (\bibinfo {year} {1996}),\ \bibfield  {title} {\enquote {\bibinfo
  {title} {Fault-tolerant quantum computation},}\ }in\ \href@noop {} {\emph
  {\bibinfo {booktitle} {Proceedings of 37th FOCS}}},\ pp.\ \bibinfo {pages}
  {56--65}\BibitemShut {NoStop}%
\bibitem [{\citenamefont {Steane}(2003)}]{steane:overhead}%
  \BibitemOpen
  \bibfield  {author} {\bibinfo {author} {\bibnamefont {Steane}, \bibfnamefont
  {A}}} (\bibinfo {year} {2003}),\ \bibfield  {title} {\enquote {\bibinfo
  {title} {Overhead and noise threshold of fault-tolerant quantum error
  correction},}\ }\href@noop {} {\bibfield  {journal} {\bibinfo  {journal}
  {Phys. Rev. A}\ }\textbf {\bibinfo {volume} {68}}~(\bibinfo {number} {4}),\
  \bibinfo {pages} {42322--1--19}}\BibitemShut {NoStop}%
\bibitem [{\citenamefont {{Steane}}(1997)}]{steane:active}%
  \BibitemOpen
  \bibfield  {author} {\bibinfo {author} {\bibnamefont {{Steane}},
  \bibfnamefont {A~M}}} (\bibinfo {year} {1997}),\ \bibfield  {title} {\enquote
  {\bibinfo {title} {{Active stabilization, quantum computation, and quantum
  state synthesis}},}\ }\href {\doibase 10.1103/PhysRevLett.78.2252} {\bibfield
   {journal} {\bibinfo  {journal} {Phys. Rev. Lett.}\ }\textbf {\bibinfo
  {volume} {78}},\ \bibinfo {pages} {2252--2255}}\BibitemShut {NoStop}%
\bibitem [{\citenamefont {Steane}(1999)}]{steane:RM}%
  \BibitemOpen
  \bibfield  {author} {\bibinfo {author} {\bibnamefont {Steane}, \bibfnamefont
  {AM}}} (\bibinfo {year} {1999}),\ \bibfield  {title} {\enquote {\bibinfo
  {title} {Quantum {R}eed-{M}uller codes},}\ }\href {\doibase
  10.1109/18.771249} {\bibfield  {journal} {\bibinfo  {journal} {IEEE Trans.
  Inf. Theory}\ }\textbf {\bibinfo {volume} {45}}~(\bibinfo {number} {5}),\
  \bibinfo {pages} {1701--1703}}\BibitemShut {NoStop}%
\bibitem [{\citenamefont {{Suchara}}\ \emph {et~al.}(2011)\citenamefont
  {{Suchara}}, \citenamefont {{Bravyi}},\ and\ \citenamefont
  {{Terhal}}}]{SBT:top}%
  \BibitemOpen
  \bibfield  {author} {\bibinfo {author} {\bibnamefont {{Suchara}},
  \bibfnamefont {M}}, \bibinfo {author} {\bibfnamefont {S.}~\bibnamefont
  {{Bravyi}}}, \ and\ \bibinfo {author} {\bibfnamefont {B.}~\bibnamefont
  {{Terhal}}}} (\bibinfo {year} {2011}),\ \bibfield  {title} {\enquote
  {\bibinfo {title} {{Constructions and noise threshold of topological
  subsystem codes}},}\ }\href {\doibase 10.1088/1751-8113/44/15/155301}
  {\bibfield  {journal} {\bibinfo  {journal} {Jour. of Phys. A: Math. and
  Gen.}\ }\textbf {\bibinfo {volume} {44}}~(\bibinfo {number} {15}),\ \bibinfo
  {pages} {155301}}\BibitemShut {NoStop}%
\bibitem [{\citenamefont {{Suchara}}\ \emph {et~al.}(2014)\citenamefont
  {{Suchara}}, \citenamefont {{Cross}},\ and\ \citenamefont
  {{Gambetta}}}]{SCG:leakage}%
  \BibitemOpen
  \bibfield  {author} {\bibinfo {author} {\bibnamefont {{Suchara}},
  \bibfnamefont {M}}, \bibinfo {author} {\bibfnamefont {A.~W.}\ \bibnamefont
  {{Cross}}}, \ and\ \bibinfo {author} {\bibfnamefont {J.~M.}\ \bibnamefont
  {{Gambetta}}}} (\bibinfo {year} {2014}),\ \bibfield  {title} {\enquote
  {\bibinfo {title} {{Leakage suppression in the toric code}},}\ }\href@noop {}
  {\bibfield  {journal} {\bibinfo  {journal} {ArXiv e-prints}\ }}\Eprint
  {http://arxiv.org/abs/1410.8562} {arXiv:1410.8562 [quant-ph]} \BibitemShut
  {NoStop}%
\bibitem [{\citenamefont {{Suchara}}\ \emph {et~al.}(2013)\citenamefont
  {{Suchara}}, \citenamefont {{Faruque}}, \citenamefont {{Lai}}, \citenamefont
  {{Paz}}, \citenamefont {{Chong}},\ and\ \citenamefont
  {{Kubiatowicz}}}]{suchara:study}%
  \BibitemOpen
  \bibfield  {author} {\bibinfo {author} {\bibnamefont {{Suchara}},
  \bibfnamefont {M}}, \bibinfo {author} {\bibfnamefont {A.}~\bibnamefont
  {{Faruque}}}, \bibinfo {author} {\bibfnamefont {C.-Y.}\ \bibnamefont
  {{Lai}}}, \bibinfo {author} {\bibfnamefont {G.}~\bibnamefont {{Paz}}},
  \bibinfo {author} {\bibfnamefont {F.~T.}\ \bibnamefont {{Chong}}}, \ and\
  \bibinfo {author} {\bibfnamefont {J.}~\bibnamefont {{Kubiatowicz}}}}
  (\bibinfo {year} {2013}),\ \bibfield  {title} {\enquote {\bibinfo {title}
  {{Comparing the overhead of topological and concatenated quantum error
  correction}},}\ }\href@noop {} {\bibfield  {journal} {\bibinfo  {journal}
  {ArXiv e-prints}\ }}\Eprint {http://arxiv.org/abs/1312.2316} {arXiv:1312.2316
  [quant-ph]} \BibitemShut {NoStop}%
\bibitem [{\citenamefont {{Sun}}\ \emph {et~al.}(2014)\citenamefont {{Sun}},
  \citenamefont {{Petrenko}}, \citenamefont {{Leghtas}}, \citenamefont
  {{Vlastakis}}, \citenamefont {{Kirchmair}}, \citenamefont {{Sliwa}},
  \citenamefont {{Narla}}, \citenamefont {{Hatridge}}, \citenamefont
  {{Shankar}}, \citenamefont {{Blumoff}}, \citenamefont {{Frunzio}},
  \citenamefont {{Mirrahimi}}, \citenamefont {{Devoret}},\ and\ \citenamefont
  {{Schoelkopf}}}]{sun+:cat}%
  \BibitemOpen
  \bibfield  {author} {\bibinfo {author} {\bibnamefont {{Sun}}, \bibfnamefont
  {L}}, \bibinfo {author} {\bibfnamefont {A.}~\bibnamefont {{Petrenko}}},
  \bibinfo {author} {\bibfnamefont {Z.}~\bibnamefont {{Leghtas}}}, \bibinfo
  {author} {\bibfnamefont {B.}~\bibnamefont {{Vlastakis}}}, \bibinfo {author}
  {\bibfnamefont {G.}~\bibnamefont {{Kirchmair}}}, \bibinfo {author}
  {\bibfnamefont {K.~M.}\ \bibnamefont {{Sliwa}}}, \bibinfo {author}
  {\bibfnamefont {A.}~\bibnamefont {{Narla}}}, \bibinfo {author} {\bibfnamefont
  {M.}~\bibnamefont {{Hatridge}}}, \bibinfo {author} {\bibfnamefont
  {S.}~\bibnamefont {{Shankar}}}, \bibinfo {author} {\bibfnamefont
  {J.}~\bibnamefont {{Blumoff}}}, \bibinfo {author} {\bibfnamefont
  {L.}~\bibnamefont {{Frunzio}}}, \bibinfo {author} {\bibfnamefont
  {M.}~\bibnamefont {{Mirrahimi}}}, \bibinfo {author} {\bibfnamefont {M.~H.}\
  \bibnamefont {{Devoret}}}, \ and\ \bibinfo {author} {\bibfnamefont {R.~J.}\
  \bibnamefont {{Schoelkopf}}}} (\bibinfo {year} {2014}),\ \bibfield  {title}
  {\enquote {\bibinfo {title} {{Tracking photon jumps with repeated quantum
  non-demolition parity measurements}},}\ }\href {\doibase 10.1038/nature13436}
  {\bibfield  {journal} {\bibinfo  {journal} {Nature}\ }\textbf {\bibinfo
  {volume} {511}},\ \bibinfo {pages} {444--448}}\BibitemShut {NoStop}%
\bibitem [{\citenamefont {Svore}\ \emph {et~al.}(2005)\citenamefont {Svore},
  \citenamefont {Cross}, \citenamefont {Chuang},\ and\ \citenamefont
  {Aho}}]{SCCA:flowmap}%
  \BibitemOpen
  \bibfield  {author} {\bibinfo {author} {\bibnamefont {Svore}, \bibfnamefont
  {KM}}, \bibinfo {author} {\bibfnamefont {A.W.}\ \bibnamefont {Cross}},
  \bibinfo {author} {\bibfnamefont {I.L.}\ \bibnamefont {Chuang}}, \ and\
  \bibinfo {author} {\bibfnamefont {A.V.}\ \bibnamefont {Aho}}} (\bibinfo
  {year} {2005}),\ \bibfield  {title} {\enquote {\bibinfo {title} {A flow-map
  model for analyzing pseudothresholds in fault-tolerant quantum computing},}\
  }\href@noop {} {\bibfield  {journal} {\bibinfo  {journal} {Quantum Info. and
  Comput.}\ }\textbf {\bibinfo {volume} {6}}~(\bibinfo {number} {3}),\ \bibinfo
  {pages} {193--212}}\BibitemShut {NoStop}%
\bibitem [{\citenamefont {Svore}\ \emph {et~al.}(2007)\citenamefont {Svore},
  \citenamefont {DiVincenzo},\ and\ \citenamefont {Terhal}}]{SDT:local}%
  \BibitemOpen
  \bibfield  {author} {\bibinfo {author} {\bibnamefont {Svore}, \bibfnamefont
  {KM}}, \bibinfo {author} {\bibfnamefont {D.P.}\ \bibnamefont {DiVincenzo}}, \
  and\ \bibinfo {author} {\bibfnamefont {B.M.}\ \bibnamefont {Terhal}}}
  (\bibinfo {year} {2007}),\ \bibfield  {title} {\enquote {\bibinfo {title}
  {Noise threshold for a fault-tolerant two-dimensional lattice
  architecture},}\ }\href@noop {} {\bibfield  {journal} {\bibinfo  {journal}
  {Quantum Info. and Comput.}\ }\textbf {\bibinfo {volume} {7}},\ \bibinfo
  {pages} {297--318}}\BibitemShut {NoStop}%
\bibitem [{\citenamefont {{Terhal}}\ \emph {et~al.}(2012)\citenamefont
  {{Terhal}}, \citenamefont {{Hassler}},\ and\ \citenamefont
  {{DiVincenzo}}}]{TFD:majsurf}%
  \BibitemOpen
  \bibfield  {author} {\bibinfo {author} {\bibnamefont {{Terhal}},
  \bibfnamefont {B~M}}, \bibinfo {author} {\bibfnamefont {F.}~\bibnamefont
  {{Hassler}}}, \ and\ \bibinfo {author} {\bibfnamefont {D.~P.}\ \bibnamefont
  {{DiVincenzo}}}} (\bibinfo {year} {2012}),\ \bibfield  {title} {\enquote
  {\bibinfo {title} {{From Majorana fermions to topological order}},}\ }\href
  {\doibase 10.1103/PhysRevLett.108.260504} {\bibfield  {journal} {\bibinfo
  {journal} {Phys. Rev. Lett.}\ }\textbf {\bibinfo {volume} {108}}~(\bibinfo
  {number} {26}),\ \bibinfo {eid} {260504}}\BibitemShut {NoStop}%
\bibitem [{\citenamefont {Tillich}\ and\ \citenamefont
  {Z{\'e}mor}(2009)}]{TZ:codes}%
  \BibitemOpen
  \bibfield  {author} {\bibinfo {author} {\bibnamefont {Tillich}, \bibfnamefont
  {Jean-Pierre}}, \ and\ \bibinfo {author} {\bibfnamefont {Gilles}\
  \bibnamefont {Z{\'e}mor}}} (\bibinfo {year} {2009}),\ \bibfield  {title}
  {\enquote {\bibinfo {title} {Quantum {LDPC} codes with positive rate and
  minimum distance proportional to n$^{\mbox{$\frac{1}{2}$}}$},}\ }in\
  \href@noop {} {\emph {\bibinfo {booktitle} {Proceedings of the IEEE Symposium
  on Information Theory}}},\ pp.\ \bibinfo {pages} {799--803}\BibitemShut
  {NoStop}%
\bibitem [{\citenamefont {{van Handel}}\ and\ \citenamefont
  {{Mabuchi}}(2005)}]{HM:filter}%
  \BibitemOpen
  \bibfield  {author} {\bibinfo {author} {\bibnamefont {{van Handel}},
  \bibfnamefont {R}}, \ and\ \bibinfo {author} {\bibfnamefont {H.}~\bibnamefont
  {{Mabuchi}}}} (\bibinfo {year} {2005}),\ \bibfield  {title} {\enquote
  {\bibinfo {title} {{Optimal error tracking via quantum coding and continuous
  syndrome measurement}},}\ }\href@noop {} {\bibinfo  {journal} {eprint
  arXiv:quant-ph/0511221}\ }\BibitemShut {NoStop}%
\bibitem [{\citenamefont {Vandersypen}\ and\ \citenamefont
  {Chuang}(2005)}]{VC:rmp}%
  \BibitemOpen
\bibfield  {journal} {  }\bibfield  {author} {\bibinfo {author} {\bibnamefont
  {Vandersypen}, \bibfnamefont {L~M~K}}, \ and\ \bibinfo {author}
  {\bibfnamefont {I.~L.}\ \bibnamefont {Chuang}}} (\bibinfo {year} {2005}),\
  \bibfield  {title} {\enquote {\bibinfo {title} {{NMR} techniques for quantum
  control and computation},}\ }\href {\doibase 10.1103/RevModPhys.76.1037}
  {\bibfield  {journal} {\bibinfo  {journal} {Rev. Mod. Phys.}\ }\textbf
  {\bibinfo {volume} {76}},\ \bibinfo {pages} {1037--1069}}\BibitemShut
  {NoStop}%
\bibitem [{\citenamefont {{Vasconcelos}}\ \emph {et~al.}(2010)\citenamefont
  {{Vasconcelos}}, \citenamefont {{Sanz}},\ and\ \citenamefont
  {{Glancy}}}]{VSG:all_opt}%
  \BibitemOpen
  \bibfield  {author} {\bibinfo {author} {\bibnamefont {{Vasconcelos}},
  \bibfnamefont {H}}, \bibinfo {author} {\bibfnamefont {L.}~\bibnamefont
  {{Sanz}}}, \ and\ \bibinfo {author} {\bibfnamefont {S.}~\bibnamefont
  {{Glancy}}}} (\bibinfo {year} {2010}),\ \bibfield  {title} {\enquote
  {\bibinfo {title} {{All-optical generation of states for ''Encoding a qubit
  in an oscillator''}},}\ }\href@noop {} {\bibfield  {journal} {\bibinfo
  {journal} {Opt. Lett.}\ }\textbf {\bibinfo {volume} {35}}~(\bibinfo {number}
  {3261--3263})}\BibitemShut {NoStop}%
\bibitem [{\citenamefont {{Wang}}\ \emph {et~al.}(2003)\citenamefont {{Wang}},
  \citenamefont {{Harrington}},\ and\ \citenamefont
  {{Preskill}}}]{WHP:threshold}%
  \BibitemOpen
  \bibfield  {author} {\bibinfo {author} {\bibnamefont {{Wang}}, \bibfnamefont
  {C}}, \bibinfo {author} {\bibfnamefont {J.}~\bibnamefont {{Harrington}}}, \
  and\ \bibinfo {author} {\bibfnamefont {J.}~\bibnamefont {{Preskill}}}}
  (\bibinfo {year} {2003}),\ \bibfield  {title} {\enquote {\bibinfo {title}
  {{Confinement-Higgs transition in a disordered gauge theory and the accuracy
  threshold for quantum memory}},}\ }\href {\doibase
  10.1016/S0003-4916(02)00019-2} {\bibfield  {journal} {\bibinfo  {journal}
  {Annals of Physics}\ }\textbf {\bibinfo {volume} {303}},\ \bibinfo {pages}
  {31--58}}\BibitemShut {NoStop}%
\bibitem [{\citenamefont {{Wang}}\ \emph {et~al.}(2010)\citenamefont {{Wang}},
  \citenamefont {{Fowler}},\ and\ \citenamefont
  {{Hollenberg}}}]{wangetal:threshold}%
  \BibitemOpen
  \bibfield  {author} {\bibinfo {author} {\bibnamefont {{Wang}}, \bibfnamefont
  {D~S}}, \bibinfo {author} {\bibfnamefont {A.~G.}\ \bibnamefont {{Fowler}}}, \
  and\ \bibinfo {author} {\bibfnamefont {L.~C.~L.}\ \bibnamefont
  {{Hollenberg}}}} (\bibinfo {year} {2010}),\ \bibfield  {title} {\enquote
  {\bibinfo {title} {{Quantum computing with nearest neighbor interactions and
  error rates over 1\%}},}\ }\href@noop {} {\bibfield  {journal} {\bibinfo
  {journal} {ArXiv e-prints}\ }}\Eprint {http://arxiv.org/abs/1009.3686}
  {arXiv:1009.3686 [quant-ph]} \BibitemShut {NoStop}%
\bibitem [{\citenamefont {Weissman}(1988)}]{weissman:rmp}%
  \BibitemOpen
  \bibfield  {author} {\bibinfo {author} {\bibnamefont {Weissman},
  \bibfnamefont {M~B}}} (\bibinfo {year} {1988}),\ \bibfield  {title} {\enquote
  {\bibinfo {title} {$1/f$ noise and other slow, nonexponential kinetics in
  condensed matter},}\ }\href {\doibase 10.1103/RevModPhys.60.537} {\bibfield
  {journal} {\bibinfo  {journal} {Rev. Mod. Phys.}\ }\textbf {\bibinfo {volume}
  {60}},\ \bibinfo {pages} {537--571}}\BibitemShut {NoStop}%
\bibitem [{\citenamefont {Wiseman}\ and\ \citenamefont
  {Milburn}(2010)}]{book:WM}%
  \BibitemOpen
  \bibfield  {author} {\bibinfo {author} {\bibnamefont {Wiseman}, \bibfnamefont
  {H}}, \ and\ \bibinfo {author} {\bibfnamefont {G.J.}\ \bibnamefont
  {Milburn}}} (\bibinfo {year} {2010}),\ \href@noop {} {\emph {\bibinfo {title}
  {Quantum Measurement and Control}}}\ (\bibinfo  {publisher} {Cambridge
  University Press},\ \bibinfo {address} {Cambridge})\BibitemShut {NoStop}%
\bibitem [{\citenamefont {{Wootton}}(2012)}]{wootton:review}%
  \BibitemOpen
  \bibfield  {author} {\bibinfo {author} {\bibnamefont {{Wootton}},
  \bibfnamefont {J~R}}} (\bibinfo {year} {2012}),\ \bibfield  {title} {\enquote
  {\bibinfo {title} {{Quantum memories and error correction}},}\ }\href
  {\doibase 10.1080/09500340.2012.737937} {\bibfield  {journal} {\bibinfo
  {journal} {Journal of Modern Optics}\ }\textbf {\bibinfo {volume} {59}},\
  \bibinfo {pages} {1717--1738}}\BibitemShut {NoStop}%
\bibitem [{\citenamefont {{Wootton}}\ \emph {et~al.}(2014)\citenamefont
  {{Wootton}}, \citenamefont {{Burri}}, \citenamefont {{Iblisdir}},\ and\
  \citenamefont {{Loss}}}]{wootton+:nonab}%
  \BibitemOpen
  \bibfield  {author} {\bibinfo {author} {\bibnamefont {{Wootton}},
  \bibfnamefont {J~R}}, \bibinfo {author} {\bibfnamefont {J.}~\bibnamefont
  {{Burri}}}, \bibinfo {author} {\bibfnamefont {S.}~\bibnamefont {{Iblisdir}}},
  \ and\ \bibinfo {author} {\bibfnamefont {D.}~\bibnamefont {{Loss}}}}
  (\bibinfo {year} {2014}),\ \bibfield  {title} {\enquote {\bibinfo {title}
  {{Decoding non-Abelian topological quantum memories}},}\ }\href@noop {}
  {\bibfield  {journal} {\bibinfo  {journal} {Phys. Rev. X}\ }\textbf {\bibinfo
  {volume} {4}},\ \bibinfo {pages} {011051}}\BibitemShut {NoStop}%
\bibitem [{\citenamefont {{Yao}}\ \emph {et~al.}(2012)\citenamefont {{Yao}},
  \citenamefont {{Wang}}, \citenamefont {{Chen}}, \citenamefont {{Gao}},
  \citenamefont {{Fowler}}, \citenamefont {{Raussendorf}}, \citenamefont
  {{Chen}}, \citenamefont {{Liu}}, \citenamefont {{Lu}}, \citenamefont
  {{Deng}}, \citenamefont {{Chen}},\ and\ \citenamefont {{Pan}}}]{yao+:topo}%
  \BibitemOpen
  \bibfield  {author} {\bibinfo {author} {\bibnamefont {{Yao}}, \bibfnamefont
  {X-C}}, \bibinfo {author} {\bibfnamefont {T.-X.}\ \bibnamefont {{Wang}}},
  \bibinfo {author} {\bibfnamefont {H.-Z.}\ \bibnamefont {{Chen}}}, \bibinfo
  {author} {\bibfnamefont {W.-B.}\ \bibnamefont {{Gao}}}, \bibinfo {author}
  {\bibfnamefont {A.~G.}\ \bibnamefont {{Fowler}}}, \bibinfo {author}
  {\bibfnamefont {R.}~\bibnamefont {{Raussendorf}}}, \bibinfo {author}
  {\bibfnamefont {Z.-B.}\ \bibnamefont {{Chen}}}, \bibinfo {author}
  {\bibfnamefont {N.-L.}\ \bibnamefont {{Liu}}}, \bibinfo {author}
  {\bibfnamefont {C.-Y.}\ \bibnamefont {{Lu}}}, \bibinfo {author}
  {\bibfnamefont {Y.-J.}\ \bibnamefont {{Deng}}}, \bibinfo {author}
  {\bibfnamefont {Y.-A.}\ \bibnamefont {{Chen}}}, \ and\ \bibinfo {author}
  {\bibfnamefont {J.-W.}\ \bibnamefont {{Pan}}}} (\bibinfo {year} {2012}),\
  \bibfield  {title} {\enquote {\bibinfo {title} {{Experimental demonstration
  of topological error correction}},}\ }\href {\doibase 10.1038/nature10770}
  {\bibfield  {journal} {\bibinfo  {journal} {Nature}\ }\textbf {\bibinfo
  {volume} {482}},\ \bibinfo {pages} {489--494}}\BibitemShut {NoStop}%
\bibitem [{\citenamefont {Yoshida}(2011)}]{yoshida}%
  \BibitemOpen
  \bibfield  {author} {\bibinfo {author} {\bibnamefont {Yoshida}, \bibfnamefont
  {Beni}}} (\bibinfo {year} {2011}),\ \bibfield  {title} {\enquote {\bibinfo
  {title} {Feasibility of self-correcting quantum memory and thermal stability
  of topological order},}\ }\href@noop {} {\bibfield  {journal} {\bibinfo
  {journal} {Annals of Physics}\ }\textbf {\bibinfo {volume} {326}}~(\bibinfo
  {number} {10}),\ \bibinfo {pages} {2566 -- 2633}}\BibitemShut {NoStop}%
\bibitem [{\citenamefont {Zeng}\ \emph {et~al.}(2011)\citenamefont {Zeng},
  \citenamefont {Cross},\ and\ \citenamefont {Chuang}}]{ZCC:trans}%
  \BibitemOpen
  \bibfield  {author} {\bibinfo {author} {\bibnamefont {Zeng}, \bibfnamefont
  {Bei}}, \bibinfo {author} {\bibfnamefont {Andrew~W.}\ \bibnamefont {Cross}},
  \ and\ \bibinfo {author} {\bibfnamefont {Isaac~L.}\ \bibnamefont {Chuang}}}
  (\bibinfo {year} {2011}),\ \bibfield  {title} {\enquote {\bibinfo {title}
  {Transversality versus universality for additive quantum codes},}\ }\href
  {\doibase 10.1109/TIT.2011.2161917} {\bibfield  {journal} {\bibinfo
  {journal} {{IEEE} Transactions on Information Theory}\ }\textbf {\bibinfo
  {volume} {57}}~(\bibinfo {number} {9}),\ \bibinfo {pages}
  {6272--6284}}\BibitemShut {NoStop}%
\bibitem [{\citenamefont {{Zhang}}\ \emph {et~al.}(2008)\citenamefont
  {{Zhang}}, \citenamefont {{Xie}}, \citenamefont {{Peng}},\ and\ \citenamefont
  {{van Loock}}}]{zhang+:anyon_kitaev}%
  \BibitemOpen
  \bibfield  {author} {\bibinfo {author} {\bibnamefont {{Zhang}}, \bibfnamefont
  {J}}, \bibinfo {author} {\bibfnamefont {C.}~\bibnamefont {{Xie}}}, \bibinfo
  {author} {\bibfnamefont {K.}~\bibnamefont {{Peng}}}, \ and\ \bibinfo {author}
  {\bibfnamefont {P.}~\bibnamefont {{van Loock}}}} (\bibinfo {year} {2008}),\
  \bibfield  {title} {\enquote {\bibinfo {title} {{Anyon statistics with
  continuous variables}},}\ }\href {\doibase 10.1103/PhysRevA.78.052121}
  {\bibfield  {journal} {\bibinfo  {journal} {Phys. Rev. A}\ }\textbf {\bibinfo
  {volume} {78}}~(\bibinfo {number} {5}),\ \bibinfo {eid} {052121}}\BibitemShut
  {NoStop}%
\bibitem [{\citenamefont {{Zhou}}\ \emph {et~al.}(2000)\citenamefont {{Zhou}},
  \citenamefont {{Leung}},\ and\ \citenamefont {{Chuang}}}]{zhou+:gatetele}%
  \BibitemOpen
  \bibfield  {author} {\bibinfo {author} {\bibnamefont {{Zhou}}, \bibfnamefont
  {X}}, \bibinfo {author} {\bibfnamefont {D.~W.}\ \bibnamefont {{Leung}}}, \
  and\ \bibinfo {author} {\bibfnamefont {I.~L.}\ \bibnamefont {{Chuang}}}}
  (\bibinfo {year} {2000}),\ \bibfield  {title} {\enquote {\bibinfo {title}
  {{Methodology for quantum logic gate construction}},}\ }\href {\doibase
  10.1103/PhysRevA.62.052316} {\bibfield  {journal} {\bibinfo  {journal} {Phys.
  Rev. A}\ }\textbf {\bibinfo {volume} {62}}~(\bibinfo {number} {5}),\ \bibinfo
  {eid} {052316}}\BibitemShut {NoStop}%
\end{thebibliography}%

\end{document}